\documentclass[sinlgecolumn,preprintnumbers,amsmath,amssymb,superscriptaddress,longbibliography,nofootinbib, notitlepage]{revtex4-1}

\usepackage{graphicx}
\usepackage{bm}% bold math
\usepackage{dsfont}
\usepackage[usenames,dvipsnames]{xcolor}
\usepackage{pstricks}
\usepackage[tight]{subfigure}
\usepackage{verbatim}
\usepackage{units}
\usepackage{multirow}
\usepackage{enumitem}
\usepackage{mathrsfs}
\usepackage{leftidx}
\usepackage{xspace}
\usepackage{braket}
\usepackage{bbm}
\usepackage{cancel}
\usepackage{amsfonts,amssymb,amsmath}
\usepackage[normalem]{ulem}

\usepackage{xr}

\usepackage[a4paper]{hyperref}
\hypersetup{colorlinks=true,linktoc=all,linkcolor=blue,breaklinks=true,citecolor=blue,urlcolor=blue}
\usepackage[english]{babel}
\usepackage{cleveref}

\renewcommand{\P}{\mathcal{P}}

\begin{document}

\title{Quantum transport phenomena induced by time-dependent fields}
\author{Matteo Acciai}
\affiliation{Scuola Internazionale Superiore di Studi Avanzati (SISSA), Via Bonomea 365, 34136 Trieste, Italy}

\author{Liliana Arrachea}
\email[To whom correspondence should be addressed: ]{liliana.arrachea@ib.edu.ar, larrachea@unsam.edu.ar}
\affiliation{Departamento de Teoría de Sólidos Cuánticos y Sistemas Desordenados and Instituto de Nanociencia y Nanotecnología, Centro Atómico Bariloche and Instituto Balseiro, Av Bustillo, San Carlos de Bariloche, 8400, Argentina}

\author{Janine Splettstoesser}
\affiliation{Department of Microtechnology and Nanoscience (MC2),Chalmers University of Technology, S-412 96 G\"oteborg, Sweden}

\begin{abstract}
We present an overview of time-dependent transport phenomena in quantum systems, with a particular emphasis on steady-state regimes. We 
present the ideas after 
the main theoretical frameworks to study open-quantum systems out of equilibrium, that are useful to study 
quantum transport under time-dependent driving. We discuss the fundamentals of the key mechanisms such as dissipation, quantum pumping, noise, and energy conversion that are associated to the problem of quantum transport.

Our primary focus is on electronic systems, where decades of research have established a rich theoretical foundation and a wealth of experimental realizations. Topics of interest include quantum optics with electrons, high-precision electron spectroscopy, quantum electrical metrology, and the critical role of quantum fluctuations in transport and thermodynamics. We also extend the discussion to atomic, molecular, and optical systems, as well as nanomechanical platforms, which offer complementary perspectives and are currently experiencing rapid experimental development.
Finally, we examine the intersection of time-dependent transport and topological matter, a domain of active investigation. 

This review aims to gather the diverse approaches and emerging trends that define the current landscape of quantum transport research under time-dependent conditions, bridging theoretical insights with experimental advances across multiple physical platforms.
\end{abstract}
\maketitle

%%%%%%%%%%%%%%%%%%%%%%%%%%%%%%%%%%%%%%%%%%%%%%%%%%%%%
\section{Introduction}\label{ref_intro}
%%%%%%%%%%%%%%%%%%%%%%%%%%%%%%%%%%%%%%%%%%%%%%%%%%%%%

\subsection{Historical perspective}
Quantum transport, the study of charge, spin, and energy dynamics at the nanoscale, is a fundamental topic in condensed matter physics, in mesoscopic physics, and in the study of quantum technologies. Introductions to this broad field, technical details, and historical and topical overviews can for example be found in the books of Refs~\cite{Datta2005Jun,Imry2008Oct,Ihn2009Nov,Nazarov2009May,Ferry2009Aug}. The understanding of how quantum systems respond to time-dependent external fields is essential for applications ranging from ultrafast electronics and quantum computing to molecular electronics and topological materials.

The field of quantum transport emerged as a response to the need for a microscopic, quantum-mechanical understanding of charge and energy transport in nanoscale systems, where classical approaches like the Drude model or semiclassical Boltzmann transport theory fail. The behavior of electrons in confined geometries, involving phase coherence, interference, and quantization, gave rise to a new transport regime — mesoscopic physics.
The initial theoretical backbone of quantum transport theory was developed after the combined work of Landauer, Büttiker, and Imry. 

Rolf Landauer formulated the first theoretical proposal to calculate electronic currents in the coherent regime. This put forward the idea of identifying the electrical conductance 
with a transmission probability, as in the case of wave propagation \cite{Landauer1957Jul,Landauer1961Jul,Landauer1970Apr}. 
Building upon this foundation,  Markus B\"uttiker and Yoseph Imry 
extended the treatment to several terminals 
and formulated the scattering matrix theory for quantum transport~\cite{Buttiker1985Aug,Buttiker1986Mar,Buttiker1988Nov,Buttiker1993Dec,Imry1986Mar,Imry1999Mar}. 

These theoretical advances created a powerful synergy with the surge in fabrication precision of small conductors that emerged in the 1980s, driven by the development of techniques such as molecular beam epitaxy and nanolithography in semiconductors.
One of the most prominent examples is the quantum Hall effect \cite{Klitzing1980Aug}, where the edge states are the perfect example of quantum coherence in electron systems and the first example of topological modes. 
Other remarkable systems are quantum dots \cite{Kouwenhoven1998Jun,Kouwenhoven2001Jun,van2002electron} and mesoscopic rings \cite{Gefen1984,Gefen1984Jul,Webb1985Jun,Murat1986Jul}. 
A description of these devices has been thoroughly reviewed in Refs. \cite{Imry2008Oct,Beenakker1991Jan,Glazman2003}.
Later achievements took place after the integration of superconducting parts in these mesoscopic devices \cite{Beenakker1991Dec,Martin-Rodero2011Dec,janvier2015coherent}, 
nanotubes and graphene \cite{Bachtold1999Feb,Charlier2007May,Laird2015Jul} and molecular structures \cite{Scheer1998Jul,
Nitzan2003May,xn--Agrait-0yd2003Apr,
Zimbovskaya2011Dec}.
The first key questions that were addressed in the theory of quantum transport were the origin of resistance \cite{Buttiker1988Nov}, conductance fluctuations \cite{Lee1985Oct}, and the role of disorder \cite{altshuler_book}. This was followed by the investigation of
many-body interactions leading to Coulomb blockade \cite{Aleiner2002Mar} and Kondo effect \cite{Goldhaber-Gordon1998Jan}.
In the last years, the discovery of new topological materials hosting
the quantum spin Hall effect offer new playgrounds and additional possibilities \cite{Bernevig2006Dec,Konig2007Nov,Qi2011Oct,Sato2017May}.

%%%%%%%%%%%%%%%%%%%%%%%%%%%%%%%%%%%%%%%%%%%%%%%%%%%%%%%%%%
\subsection{Time-dependent quantum transport in electronic systems}
%%%%%%%%%%%%%%%%%%%%%%%%%%%%%%%%%%%%%%%%%%%%%%%%%%%%%%%%%%
One of the earliest time-dependent quantum transport problems to be investigated was the effective photon-assisted tunneling in superconductors generated by an oscillating barrier in superconducting structures \cite{tien1963multiphoton}. The photonic-like structure of the quantum mechanical description of periodically driven systems is at the heart of Floquet representation of the wave functions~\cite{Shirley1965May}. The impact of Floquet theory in the theoretical description of periodic ac driving in
tunneling processes has been thoroughly reviewed in Refs.~\cite{Grifoni1998Oct,Platero2004May}.  
Another fundamental problem that was initially addressed was the effect of a time-dependent magnetic flux threading a mesoscopic ring structure \cite{Buttiker1983Jul,Buttiker1985Aug,Landauer1986May,Gefen1987Aug}. A particularly interesting case arises when the magnetic flux varies linearly with time, which corresponds to a constant electromotive force around the ring. This effectively generates a uniform electric field along the circumference, leading to the acceleration of charge carriers. Due to the periodic boundary conditions inherent to the ring geometry, this results in Bloch-like oscillations of the current.
Unlike traditional conductors connected to macroscopic leads, where an applied voltage yields a steady-state (dc) current, the isolated ring configuration hosts intrinsically time-dependent (ac) currents. 

A major conceptual advancement was the theoretical and experimental study of the quantum capacitor, a mesoscopic system composed of a small quantum dot or cavity weakly coupled to a single electron reservoir through a quantum point contact. Only when subjected to a time-periodic gate voltage, transport can occur in such a system. Under specific conditions and with a proper driving protocol, this device can  emit and absorb single electrons in a controlled and coherent manner, operating as a source of quantized charge pulses. The mesoscopic capacitor was first proposed and analyzed by B\"uttiker and collaborators~\cite{Buttiker1993Jun,Buttiker1993Sep}, and later experimentally realized in the gigahertz regime~\cite{Gabelli2006Jul}, where it served as a prototype for on-demand single-electron emitters~\cite{Feve2007May}. The quantum capacitor became a fundamental building block in time-resolved quantum transport and enabled the exploration of quantum noise, dynamical Coulomb blockade, and ac admittance in the quantum regime.

The next significant milestone was the introduction of the concept of quantum pumping. In a quantum pump, net current is generated in the absence of any bias voltage, solely due to the cyclic, time-periodic modulation of system parameters such as gate voltages or coupling barriers. First experimental implementations were shown in Refs.~ \cite{Kouwenhoven1991Sep,Pothier1992Jan,Switkes1999Mar}, providing a significant boost for the theoretical investigation of this phenomenon.

More recently, attention has shifted toward electronic quantum optics, an emerging field where single-electron wave packets are manipulated and interfere in solid-state devices with a level of control analogous to that of single photons in optical setups. A key milestone was the generation of ``Levitons'' — minimal excitation states of the Fermi sea — first proposed by Levitov~\cite{Levitov1996Oct,Ivanov1997Sep,Keeling2006Sep} and later realized experimentally~\cite{Gabelli2013Feb,Dubois2013Oct}. These excitations are generated by applying Lorentzian-shaped voltage pulses to a contact, creating single, coherent, and noiseless electrons that propagate ballistically through quantum conductors.

These developments have enabled the design of electronic analogs of optical interferometers, such as the electronic Mach-Zehnder and Hong-Ou-Mandel interferometers, where quantum interference of electrons from independent sources can be observed. The coherence, entanglement, and statistics of single electrons can now be studied with unprecedented precision, paving the way for quantum information processing and quantum metrology in mesoscopic systems~\cite{Bauerle2018Apr}.

%%%%%%%%%%%%%%%%%%%%%%%%%%%%%%%%%%%%%%%%%%%%%%%%%%%%%%%%%%
\subsection{Time-dependent quantum transport meets thermodynamics and quantum information processing}
%%%%%%%%%%%%%%%%%%%%%%%%%%%%%%%%%%%%%%%%%%%%%%%%%%%%%%%%%%
In parallel with his seminal contributions to the foundation of the theory of quantum transport, 
Landauer also played a key role in identifying the thermodynamic limitation of information processing \cite{Landauer1961Jul,Landauer1991May,Landauer1996Jul}.
He predicted that the process of erasing a bit of information is associated with a change of entropy and a minimal heat exchange of $k_{\mathrm B} T \log 2$. 

In recent years, the field of thermodynamics has irrupted in the scenario of quantum systems. Starting from the fundamental problem of thermalization
in cold atoms,  the study of heat, work, energy conversion, and dissipation is an active avenue of research and a point of convergence for the community of condensed matter, statistical physics,
quantum information, and atomic, molecular, and optical (AMO) physics. The activity devoted to address fundamental questions related to the validity of laws that have been originally formulated 
for macroscopic systems, the possibility of generalizing classical machines to generate power or refrigerate in the quantum realm, and understanding the fluctuations in this context has become an active field of research; see recent reviews~\cite{Pekola2015Feb,Goold2016Feb,Vinjanampathy2016Oct,Pekola2015Feb,Benenti2017Jun,Binder:2018rix,Alicki2019Apr,Pekola2021Oct,Manzano2022Jun,arrachea2023energy,Campbell2025Apr}. 
Thermalization and equilibration of driven closed systems is another fundamental process, studied for example in the context of cold atoms in optical lattices and trapped ions \cite{Gogolin2016Apr}.
The interest in these problems is further fueled by the emergence of quantum technologies, which brings about an active discussion on the energetic aspects of these developments \cite{Auffeves2022Jun}, typically also requiring time-dependent operation.

%%%%%%%%%%%%%%%%%%%%%%%%%%%%%%%%%%%%%%%%%%%%%%%%%%%%%%%%%%
\subsection{Time-dependent quantum transport meets geometry and topology}
%%%%%%%%%%%%%%%%%%%%%%%%%%%%%%%%%%%%%%%%%%%%%%%%%%%%%%%%%%
A fundamental step in the topological characterization of quantum systems 
was done by Michael Berry, who identified a geometrical phase accumulated in the slow dynamics of a (quantum) system  when the evolution occurs along a closed loop in parameter space~\cite{Victor1984Mar}. 

Topological concepts are profoundly connected to quantum transport. In static systems, topological invariants such as the Chern number classify insulating phases, and are directly related to the Berry curvature and the Hall resistance. 
In time-dependent settings, the paradigmatic example is topological pumping. In particular, the model introduced by Thouless \cite{Thouless1983May}, which consists of a pump defined by a cyclic adiabatic modulation of parameters in a one-dimensional system, leads to the quantized transfer of an integer number of charges per cycle. 

More recently, time-dependent protocols have enabled access to Floquet topological phases, where systems driven by periodic time-dependent fields acquire effective Hamiltonians with nontrivial topological properties that are not present in the static counterpart. Moreover, these ideas were recently introduced to even more exotic regimes, such as higher-order Floquet topological phases and time-dependent topological insulators and superconductors. In these systems, time plays the role of a synthetic dimension, which enables the stabilization of topological modes
\cite{Rudner2020May,delaTorre2021Oct}.

In open time-dependent quantum systems, geometric concepts like the Berry curvature are also useful to describe non-topological pumping mechanisms, where the transported quantities are not quantized. 
This opens promising directions to further understand the mechanisms and protocols for realizing quantized charge or heat pumps in non-equilibrium quantum devices that leverage the robustness of topological protection. This is a very active research direction, not only in condensed matter systems but also in cold atoms \cite{Amico2022Nov,Citro2023Feb}.

%%%%%%%%%%%%%%%%%%%%%%%%%%%%%%%%%%%%%%%%%%%%%%%%%%%%%%%%%%
\subsection{Organization of the present review}
%%%%%%%%%%%%%%%%%%%%%%%%%%%%%%%%%%%%%%%%%%%%%%%%%%%%%%%%%%
This review is organized as follows. Basic concepts to describe transport induced by time-dependent driving of particles, charge and energy in the quantum realm are presented in {\bf section 2}. There, we also introduce notation and symbols used along the rest of the review. It is important to mention that we mostly focus on steady-state transport observables rather than on the transient dynamics, except for the case of the mesoscopic capacitor. 
{\bf Section 3} is devoted to an overview
of the main formal methods used to solve problems of time-dependent quantum transport. The basic mechanisms and regimes taking place in the problem of quantum transport due to time-dependent driving are described in {\bf section 4.}  This includes the paradigmatic problem of the quantum capacitor, where only pure time-dependent transport of charge and energy with a net dissipation of energy takes place, or of a driven qubit coupled to a single thermal bath, and the dissipation of energy in this system. Other mechanisms discussed here are pumping and energy conversion when the driving operates in combination with electrical and thermal biases. {\bf Section 5} is
devoted to review concrete problems recently studied in the field of time-dependent quantum transport and advances in the understanding of related phenomena.
In {\bf section 6} we present concluding remarks.

%%%%%%%%%%%%%%%%%%%%%%%%%%%%%%%%%%%%%%%%%%%%%%%%%%%%%%%%%%
\renewcommand{\arraystretch}{1.4}
\begin{table}[]
    \centering
    \begin{tabular}{|l|l|}
    \hline
    Mathematical symbol & Description\\\hline\hline
    $\hat{\bullet}$ & Operators indicated by hats\\\hline
    $\tilde{\bullet}$ & Operators in the interaction picture indicated by tilde\\\hline
    $\bar{\bullet}$ & Time-averages indicated by bar\\\hline
    $q$  & Quasiparticle charge (electrons or other)  \\\hline
    $t,t'$ & Time variables\\\hline
    $\tau$ & Time difference\\\hline
    $\tau_\mathrm{index}$ &  Characteristic traversal times\\\hline
    $\Omega$ & Driving frequency\\\hline
    $\mathcal{T}=2\pi/\Omega$ & Driving period\\\hline
    $V$, $V(t)=V_\mathrm{dc}+V_\mathrm{ac}(t)$ & Bias voltage (shape typically indicated as subscript, e.g. $V_{\rm Lor}$)\\\hline
    $\mathfrak{q}$ & Dimensionless charge per period injected by bias $\mathfrak{q}=qV_{\mathrm{dc}}/(\hbar\Omega)$ \\\hline
    $\sigma$ & Typical time width of an excitation generated by pulses \\\hline
    $V_\mathrm{g}$ & Gate voltage\\\hline
    $\mathfrak{i}(E)$ & Spectral current \\\hline
    $I^N(t), I^Q(t), I^E(t), I^c(t)$ & Particle-, heat-, energy-, charge currents \\\hline
    $d,r$ & Transmission/reflection amplitudes\\\hline
    $D,R$ & Transmission/reflection probabilities\\\hline
    $\underline{S}$, $S_{\alpha\gamma}$ & Scattering matrix and its components\\\hline
    $\underline{S}_F$, $S_{F,\alpha\gamma}$ & Floquet scattering matrix and its components\\\hline
    $\mathcal{S}_{\alpha\gamma}$ & Noise (indices sometimes omitted, see text)\\\hline
    $\alpha=1,...,M$ & Labeling of contacts\\\hline
    $\beta=1/(k_\mathrm{B} T)$ & Inverse temperature\\\hline
    $f_\alpha(E)$ & Fermi (or Bose) function of contact $\alpha$\\\hline
    $\Delta N_\alpha$ & Pumped charges\\\hline
    $\hat{H},\hat{H}_\mathrm{sys},\hat{H}_\mathrm{coup},\hat{H}_\alpha$ & Hamiltonian, system Hamiltonian, coupling Hamiltonian,\\
    & Hamiltonian of contact $\alpha$ \\\hline
    $\hat{H}_t$ & Hamiltonian $\hat{H}(t)$ with time frozen at $t$\\\hline    
   $w$ & Tunneling amplitude\\\hline 
   $N=\langle\hat{N}\rangle, N_\alpha = \langle\hat{N}_\alpha\rangle$ & particle number (operator) of central system and contacts \\\hline
   $\hat{a}$, $\hat{a}^\dagger$ &  Annihilation/creation operators of the environment (e.g., contacts) \\\hline
   $\hat{d}$, $\hat{d}^\dagger$ &  Annihilation/creation operators of the central system (e.g., a dot) \\\hline
   \end{tabular}
    \caption{Overview over the mathematical symbols that are most commonly used in this review.}
    \label{tab:my_label}
\end{table}
%%%%%%%%%%%%%%%%%%%%%%%%%%%%%%%%%%%%%%%%%%%%%%%%%%%%%%%%%%

%%%%%%%%%%%%%%%%%%%%%%%%%%%%%%%%%%%%%%%%%%%%%%%%%%%%%
\section{Basic Concepts}\label{ref_concepts}
%%%%%%%%%%%%%%%%%%%%%%%%%%%%%%%%%%%%%%%%%%%%%%%%%%%%%

\begin{figure}
    \centering
    \includegraphics[width=0.7\linewidth]{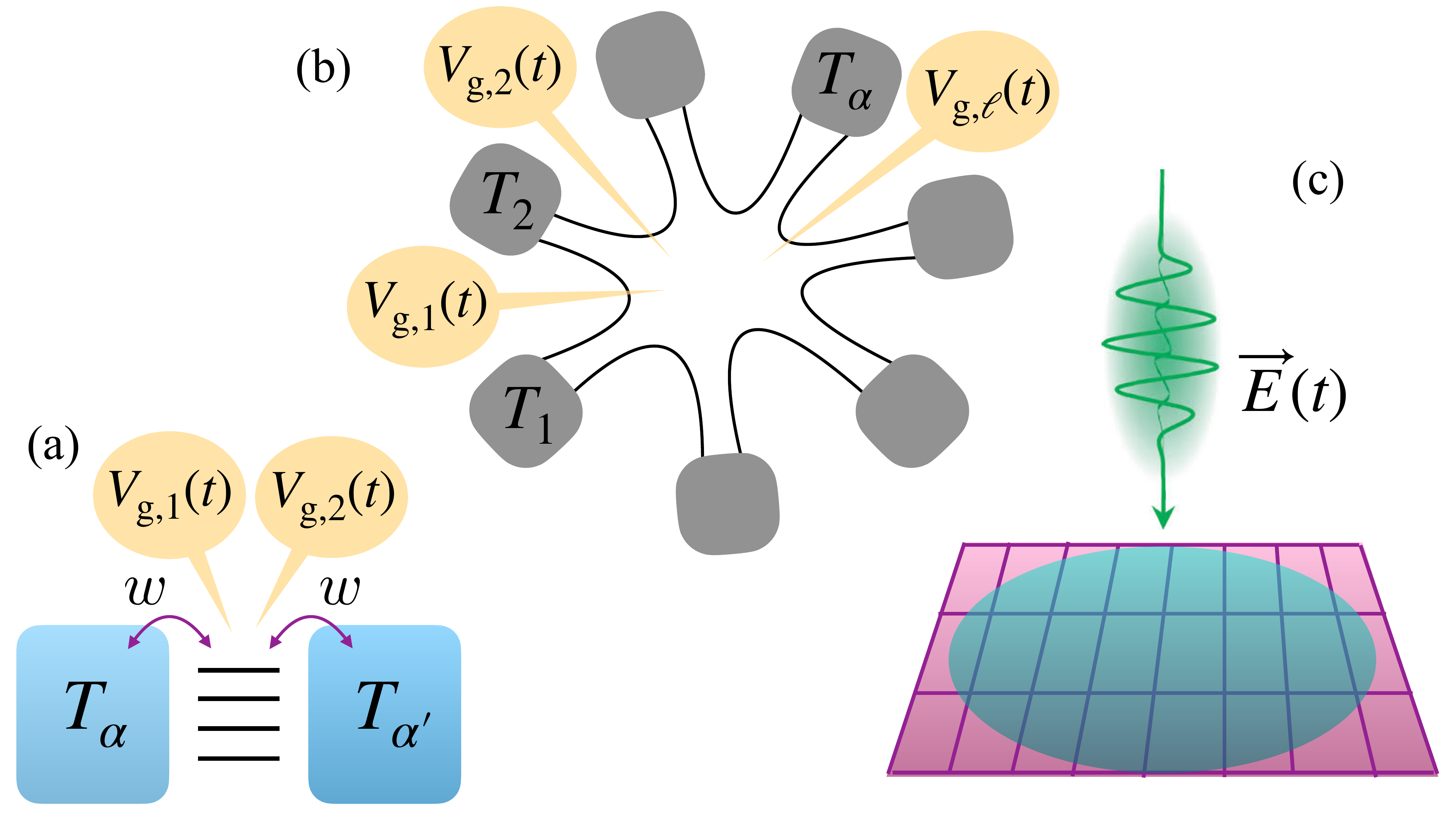}
    \caption{Examples of driven systems. (a) A two-terminal setup where a few-level quantum system is connected to macroscopic reservoirs with well-defined temperatures and/or chemical potentials, while locally driven by time-dependent gate voltages. (b) A generic multi-terminal setup described in terms of a continuum model and with locally applied time-dependent potentials. (c) Homogeneous system described by a lattice with an applied time-dependent electric field.}
    \label{fig:generic_setup}
\end{figure}

In this review, we discuss transport phenomena due to externally applied time-dependent driving. The systems we deal with are generic multi-terminal systems, as those indicated in Fig.~\ref{fig:generic_setup}, with a central conductor coupled to a series of $\alpha=1,2,3,...$ contacts.
%%%%%%%%%%%%%%%%%%%%%%%%%%%%%%%%%%%%%%%%%%%%%%%%%%%%%%
%%%%%%%%%%%%%%%%%%%%%%%%%%%%%%%%%%%%%%%%%%%%%%%%%%%%%%
Time-dependent driving can be applied either via the contacts, for example by applying time-dependent bias voltages, or via the central region, for example by applying time-dependent gate voltages. As a result, currents are flowing into the contacts due to particle and energy exchange. In this section, some of the main concepts that are relevant for these setups will be introduced.

%%%%%%%%%%%%%%%%%%%%%%%%%%%%%%%%%%%%%%%%%%%%%%%%%%%%%
\subsection{Model Hamiltonian and driving parameters}\label{ref_driving_parameters}
%%%%%%%%%%%%%%%%%%%%%%%%%%%%%%%%%%%%%%%%%%%%%%%%%%%%%

Two perspectives will be chosen to describe the system sketched in Fig.~\ref{fig:generic_setup}. On one hand, there is the perspective of the conductor behaving as an open quantum system, in contact with an environment. In this picture, the Hamiltonian of the central region provides the starting point of the analysis. The full Hamiltonian is then generically split into
\begin{equation}\label{hamil0}
    \hat{H}(t)=\sum_{\alpha=1}^{M} \left[ \hat{H}_{\alpha} +\hat{H}_{\rm coup, \alpha} \right] + \hat{H}_{\rm sys} (t),
\end{equation}
where the first two terms represent, respectively, the Hamiltonian of the reservoirs and the couplings between the reservoirs and the central system. It is
natural to represent the
reservoirs as a non-interacting gas of fermionic or bosonic excitations of the form
\begin{equation}
    \hat{H}_{\alpha}=\sum_{k,\alpha} \varepsilon_{\alpha k} \hat{a}^\dagger_{\alpha k}  \hat{a}_{\alpha k},
\end{equation}
where driving of the contact degrees of freedom can be added.
Furthermore, for central systems where the (typically time-dependent) Hamiltonian $\hat{H}_{\rm sys} (t)$ is expressed in terms of creation and annihilation operators
of particles acting on single-particle states labeled by $\ell$, it is usual to model
the coupling by tunneling Hamiltonians of the form
\begin{equation}
    \hat{H}_{\rm coup, \alpha}=\sum_{\alpha,k,\ell} \left(w_{\alpha k\ell} \hat{a}^\dagger_{\alpha k}  \hat{d}_{\ell} + \mathrm{H.c.}\right).
\end{equation}
Here, $\hat{a}_{\alpha k}^\dagger, \hat{a}_{\alpha k}$ are creation and annihilation operators acting on the degrees of freedom of the reservoirs and $\hat{d}_\ell^\dagger, \hat{d}_\ell$
in the central system\footnote{We use this notation throughout the review, see also Table~\ref{tab:my_label} When the division between system and reservoirs/environment is not applicable, creation and annihilation operators are generically indicated by $\hat c^\dagger$ and $\hat c$.}.  These operators satisfy fermionic or bosonic commutation relations, depending on the nature of the particles. The parameter $w_{\alpha k \ell}$ describes the tunneling amplitude and can possibly also be modulated in a time-dependent way.

Another perspective that can be taken to describe the system models quasiparticles incoming from the contacts by field operators
\begin{equation}
    \hat{\Psi}_\alpha (t,\boldsymbol{r})\ .
\end{equation}
These injected states are occupied following the boundary conditions imposed by the contacts $\alpha$. Transport of quasiparticles through the central region is then characterized by, e.g., scattering matrices or Green's functions depending on two time variables, see Secs.~\ref{sec:scattering_theory} and~\ref{sec:Greens_functions}.

There is also a third scenario, which we will briefly address in this review. This corresponds to 
quantum macroscopic systems under the effect of time-dependent driving, without separately considering external reservoirs. In such a case, the focus is mainly on local density of currents.

In this review, we focus on time-dependent driving due to classical driving fields. This is of relevance when quantum fluctuations in the driving field can be neglected.  For transport due to coupling to quantized fields, such as driven cavities or coupling to quantized phonon baths, see for example introductions and overviews provided in \cite{Haroche1995Sep,Gu2017Nov,Burkard2020Mar}.

We typically deal with two different situations: either the time-dependent driving is applied to the conductor, or it is applied via the contacts. In the first case, which can for example be realized by applying time-dependent gates or even (time-dependent) magnetic fields, the system Hamiltonian depends on a set of time-dependent parameters
\begin{equation}\label{param}
    \boldsymbol{X}(t)=
\left(X_1 (t), \ldots, X_M(t)\right)\ .
\end{equation}
In the second case, time-dependent driving is applied via bias voltages or even via modulated temperatures or spin polarizations. Rotating spin polarizations are of interest in the context of spin batteries, but will not be treated in this review; see instead for example Ref.~\cite{Manchon2019Sep} for a review.  
Via gauge transformations, the time-dependent driving of the bias voltage can in certain situations conveniently be treated on the same footing as the modulation of the local conductor. In order to implement time-dependent temperatures, which hence go along with a modulation of the macroscopic bath parameters and not of parameters naturally entering the contact Hamiltonians, different strategies can be pursued, see Sec.~\ref{sec:temperature bias}.

%%%%%%%%%%%%%%%%%%%%%%%%%%%%%%%%%%%%%%%%%%%%%%%%%%%%%
\subsection{Time-dependent transport observables}\label{ref_observable_definitions}
%%%%%%%%%%%%%%%%%%%%%%%%%%%%%%%%%%%%%%%%%%%%%%%%%%%%%
We now introduce observables of interest, related to currents detected in the contacts. 
Enabled by the coupling between contacts and conductor, particle currents can flow into each contact $\alpha$,
\begin{equation}\label{eq:particlecurrent}
    I^N_\alpha(t)=\frac{\partial}{\partial t}\langle \hat{N}_\alpha\rangle\ , \quad \ 
    \sum_\alpha I^N_\alpha(t)=-\frac{dN}{dt}.\
\end{equation}
As indicated by the second equation, this current fulfills charge conservation, namely the sum over the \textit{time-dependent} particle currents into all contacts equals the time-dependent decrease and increase of the number of particles $N$ on the central conductor.
In addition to or together with particle currents, also energy is exchanged between contacts and conductor leading to energy currents
\begin{equation}\label{eq:energycurrent}
 I^E_\alpha(t)=\frac{\partial}{\partial t}\langle \hat{H}_\alpha\rangle    ,\quad \ 
    \sum_\alpha I^E_\alpha(t)=\frac{dE}{dt}+\frac{dE_\text{coup}}{dt}\ .
\end{equation}
Here, the energy conservation law contains the fact that energy can also be stored in and released from the coupling. Furthermore, the change in energy due to an external driving is included in $E(t)$, in particular the power provided or received by time-dependent driving 
\begin{equation}\label{power}
P(t)=\bigg\langle \frac{\partial \hat{H}}{\partial t}\bigg\rangle\ .
\end{equation}
These currents, Eqs.~\eqref{eq:particlecurrent} and \eqref{eq:energycurrent}, are the constituents to obtain the charge current 
\begin{equation}\label{eq:chargecurrent}
    I^c_\alpha(t)=qI^N_\alpha(t),
\end{equation} 
with the charge $q$ of the transported quasiparticles, and the heat current
\begin{equation}\label{eq:heatcurrent}
I^Q_\alpha(t)= I_\alpha^E(t)-\mu_\alpha(t)I^N_\alpha(t)\ .
\end{equation}
The heat current corresponds to the excess energy current with respect to particles transported at the electrochemical potential, which in a macroscopic bath needs to be dissipated as heat. Note that the heat current does not fulfill its own conservation law since heat can be generated.
We note that also other types of currents could in general be of interest, such as spin currents, or even entropy currents. Those will however not be in the focus of this review.
In Eqs.~\eqref{eq:particlecurrent} to \eqref{eq:heatcurrent}, we have shown \textit{time-resolved} currents, which could even be transient currents or pure ac currents. In this review, we will often, but not always, focus on time-dependently generated \textit{directed} currents, where also the time-averaged currents are of interest. This is particularly relevant in the case of periodic driving, with frequency $\Omega$ and driving period $\mathcal{T}=2\pi/\Omega$, where the time average reads
\begin{equation}
\bar{I}_\alpha = \int_0^\mathcal{T}\frac{dt}{\mathcal{T}}I_\alpha(t).
\label{eq:time-averaging-def}
\end{equation}
The currents flowing into the contacts typically fluctuate and an additional observable of interest is hence the noise. It is obtained from the correlators between the fluctuations of these currents, $\Delta\hat{I}_\alpha(t)=\hat{I}_\alpha(t)-\langle\hat{I}_\alpha(t)\rangle$, which in the case of time-dependent driving depend on two times, 
\begin{equation}
    \mathcal{S}^{I I'}_{\alpha\beta}(t,t')=\langle \hat{I}_{\alpha}(t)\hat{I}_{\beta}'(t')\rangle\ -\braket{\hat{I}_\alpha(t)}\braket{\hat{I}_{\beta}'(t')}.
    \label{eq:noise-correlator-general}
\end{equation}
Here, both cross-correlators---between different currents or between currents into different contacts---and auto-correlators---between same types of currents into the same contact at different times---can be of interest. 
For stationary problems, the current-current correlation functions depend only on the time difference $\tau=t-t'$, and one can Fourier-transform with respect to this variable obtaining the (unsymmetrized) finite-frequency noise
\begin{equation}
    \mathcal{S}_{\alpha\beta}^{II'}(\omega)=\int_{-\infty}^{+\infty}d\tau\,e^{i\omega\tau}\mathcal{S}_{\alpha\beta}^{II'}(\tau)\,.
\end{equation} 
In the presence of time-dependent drivings of interest in this review, time-translation invariance is absent, and the correlation function~\eqref{eq:noise-correlator-general} thus depends both on $\tau$ and on the average time $\bar{t}=(t+t')/2$. In order to deal with this complication, it is standard practice to introduce a time-average over the variable $\bar{t}$, namely~\cite{Ferraro2013Nov}
\begin{equation}
    \mathcal{S}_{\alpha\beta}^{II'}(\omega)=\int_{-\infty}^{+\infty}d\tau\,e^{i\omega\tau}\overline{\mathcal{S}_{\alpha\beta}^{II'}(\bar{t}+\tau/2,\bar{t}-\tau/2)}^{\bar{t}}\,.\label{eq:single-frequency noise}
\end{equation}
This quantity is experimentally accessible in electronic transport measurements.
In the case of periodic drives, the time-averaging is done over one period, as shown for the current in Eq.~\eqref{eq:time-averaging-def}. In all other cases, the same definition can be adopted, but $\mathcal{T}$ can then be taken as a long measurement time.
In this review, we will mainly focus on the time-averaged, zero-frequency noise of a single current, which is often relevant in experiments, and can be written in the symmetric form
\begin{equation}\label{zero-frequency noise}
    \mathcal{S}_{\alpha\beta}^I=\frac{1}{2}\int_{0}^{\mathcal{T}}\frac{d\bar{t}}{\mathcal{T}}\int_{-\infty}^{+\infty}d\tau\,[\mathcal{S}_{\alpha\beta}^{II}(\bar{t}+\tau/2,\bar{t}-\tau/2)+\mathcal{S}_{\alpha\beta}^{II}(\bar{t}-\tau/2,\bar{t}+\tau/2)]\,.
\end{equation}
While we will not describe in detail how to explicitly calculate the noise exploiting the introduced methods, noise as a spectroscopy tool in time-dependent transport will be highlighted in Secs.~\ref{sec:noise and energy} and \ref{sec:interferometry}.

%%%%%%%%%%%%%%%%%%%%%%%%%%%%%%%%%%%%%%%%%%%%%%%%%%%%%
\subsection{Screening effects}\label{sec:screening}
%%%%%%%%%%%%%%%%%%%%%%%%%%%%%%%%%%%%%%%%%%%%%%%%%%%%%
The time-dependent modulation of external fields leads to charge accumulation and depletion in various sections of the driven conductor. This becomes obvious already in Eq.~\eqref{eq:chargecurrent}, where charge conservation involves time-dependent charge accumulation on the conductor region. This charge accumulation in turn results in screening effects going along with charge redistribution in the overall system.

From a perspective of microscopic modeling, taking care of such charge accumulation and screening effects, is a requirement in order to capture the effect of driving fields on the actual system parameters. This can be done from first principles, using density functional methods, or via capacitive models; see for instance Refs.~\cite{Pedersen1998Nov,Buttiker1993Dec,Battista2014Aug,Misiorny2018Feb}.

Accounting for charge accumulation, interaction effects, and charge redistribution due to induced currents is crucial in order to satisfy charge conservation and gauge invariance. However, to which extent these mechanisms are incorporated from the start, depends strongly on the choice of theoretical method. In particular, when interaction effects are neglected in the overall description or in parts of the model, it might be required to subsequently include screening effects, e.g., via self-consistent approaches~\cite{Buttiker1993Jun,Buttiker1993Dec,Christen1996Jul,Christen1996Sep,Pedersen1998Nov,Sanchez2013Jan,Meair2013Jan,Dashti2021Dec}, or to take into account displacement currents in the spirit of the Ramo-Shockley theorem~\cite{Blanter2000Sep}.  

%%%%%%%%%%%%%%%%%%%%%%%%%%%%%%%%%%%%%%%%%%%%%%%%%%%%%
\subsection{Definition of adiabatic-response regime}
%%%%%%%%%%%%%%%%%%%%%%%%%%%%%%%%%%%%%%%%%%%%%%%%%%%%%
In several parts of this review, we focus on the \textit{adiabatic-response regime}. It is  therefore convenient to introduce this concept already at this stage.

In open quantum systems, adiabatic response is associated with a slow dynamics, where the typical time scale ${\cal T}$ for the changes of a time-dependent Hamiltonian $\hat{H}(t)$ is much larger than the
typical time scale of the dynamics of the open quantum system. 

Transport observables like the currents introduced above can be obtained following two different ``philosophies". On the one hand, it is possible to start from the adiabatic evolution of a closed system, which means that the system stays in its eigenstate under the slow driving of parameters, while the eigenenergies can change in time. This approach has been extended to open quantum systems~\cite{Sarandy2004Dec,Sarandy2005Jan}, where the time evolution of the density operator is considered and where --importantly-- the adiabatic time evolution of the non-stationary modes yields the (geometric) contribution to the system's response and hence to quantum transport, see e.g. Refs.~\cite{Pluecker2017Apr,Pluecker2017Nov}.

On the other hand, one can start from the dynamics of an open quantum system, \textit{frozen} at time $t$. We use $\hat H_t$ to represent the time-frozen Hamiltonian. This defines a sequence of quasi-static \textit{equilibrium} states $\hat{\rho}_t$.  On top of this \textit{instantaneous} dynamics, one takes into account a first-order correction in the adiabatic expansion, accounting for the response of the system to the slow modulation. This adiabatic-response regime is characterized by response functions described by
$\hat{\rho}_t$ and its derivatives, which leads to a dynamics $\propto {\cal T}^{-1}$. These two approaches lead to equivalent results. In the present review, we mostly follow the idea of obtaining adiabatic-response transport as a correction to the instantaneous dynamics; references to methods and their applications are hence provided in the following sections.

Also in classical Carnot engines, the term ``adiabatic" is used. It defines the state evolution happening in the absence of heat exchange, which can often relate to a fast change of state where the system changes pressure and volume without exchanging heat with the environment. The connection between the adiabaticity of classical heat engines and the adiabatic response in quantum transport lies exactly in this absence of heat exchange with the environment. In the limit of slow driving of a quantum system, heat is possibly exchanged between contacts, but no heat is generated.

%%%%%%%%%%%%%%%%%%%%%%%%%%%%%%%%%%%%%%%%%%%%%%%%%%%%%
\section{Methods}\label{ref_methods}
%%%%%%%%%%%%%%%%%%%%%%%%%%%%%%%%%%%%%%%%%%%%%%%%%%%%%

%%%%%%%%%%%%%%%%%%%%%%%%%%%%%%%%%%%%%%%%%%%%%%%%%%%%%%
\subsection{Summary of scope and relations between the different methods}
%%%%%%%%%%%%%%%%%%%%%%%%%%%%%%%%%%%%%%%%%%%%%%%%%%%%%%

The study of time-dependent quantum transport relies on a variety of theoretical frameworks that provide different levels of approximation and applicability. The aim of this subsection is to provide a brief overview over the different methods that will be introduced in this review, with a focus on those that are not purely numerical.
In what follows, we hence provide a summary of the main assumptions beyond them, as well as the typical physical scenarios where they are used, their limitations and challenges. Details about each of these methods is then given in the subsequent sections.

%%%%%%%%%%%%%%%%%%%%%%%%%%%%%%%%%%%%%%%%%%%%%%%%%%%%%%
\subsubsection{Kubo linear response and adiabatic linear response}
\label{sec:kubo_general_summary}
%%%%%%%%%%%%%%%%%%%%%%%%%%%%%%%%%%%%%%%%%%%%%%%%%%%%%%
Linear-response theory, based on the Kubo formalism and adiabatic expansions, are generally powerful tools in the case of weak perturbations around equilibrium. The usual linear-response theory applies to situations where the amplitude of the
driving defines a small energy scale compared to the dynamics of the non-driven system. The adiabatic version of this theory implies a long time scale for the
driving, in comparison to the characteristic time for the  dynamics of the ``frozen" problem. In both cases, this description introduces
response functions, or susceptibilities, defined with respect to the equilibrium Hamiltonian. Hence, the concrete evaluation of the time-dependent observables like the charge and/or energy fluxes must be complemented with a many-body method to calculate these susceptibilities. Since such evaluation is done in equilibrium, there are many
well-established methods, starting from those based on Matsubara summations, equations of motion, as well as numerical ones like---for example---numerical renormalization group, density-matrix renormalization group, or quantum Monte Carlo. 
In cases where the transport takes place in the presence of a temperature bias, this formalism has to be complemented by Luttinger's Hamiltonian representation of the temperature bias.

%%%%%%%%%%%%%%%%%%%%%%%%%%%%%%%%%%%%%%%%%%%%%%%%%%%%%%
\subsubsection{Schr\"odinger equation in the Floquet 
representation}
%%%%%%%%%%%%%%%%%%%%%%%%%%%%%%%%%%%%%%%%%%%%%%%%%%%%%%
Beyond the linear-response regime, fully time-dependent methods are required. The Floquet representation of the Schr\"odinger equation provides a natural framework for systems under periodic driving, characterized by a frequency $\Omega$.
Like Bloch's theory for electrons in periodic lattices, Floquet theory introduces
a natural basis to expand the quantum states, containing explicitly the time-periodicity. This construction is very useful to define effective Hamiltonians, where energy quanta $\ell \hbar \Omega$ (with $\ell$ integer) are exchanged in the driving process.

 This approach is particularly adequate for systems described by lattice models isolated from reservoirs.
The full description relies on the solution of the time-dependent Schr\"odinger equation. 
The expansion in terms of Floquet states, and the dynamics determined by the exchange of Floquet quanta,  also
emerges naturally in the scattering matrix and Green's function description of open quantum systems under periodic driving.

%%%%%%%%%%%%%%%%%%%%%%%%%%%%%%%%%%%%%%%%%%%%%%%%%%%%%%
\subsubsection{Scattering matrix theory}
%%%%%%%%%%%%%%%%%%%%%%%%%%%%%%%%%%%%%%%%%%%%%%%%%%%%%%
Scattering matrix theory, widely used in mesoscopic physics, offers an intuitive approach to transport in regimes characterized by electronic phase coherence.
 It is particularly useful in problems described by models in the continuum and it is valid for  weakly interacting particles (mean-field interaction, linear Hamiltonians).
 
Rather than focusing on local properties like current density or electric fields inside the conductor, the scattering approach relates incoming and outgoing quantum states at the boundaries of a device, emphasizing the role of reservoirs and interfaces. In this picture, transport properties such as conductance, noise, and full counting statistics are determined by the probabilities for electrons to scatter among different channels, encoded in the elements of the scattering matrix. 
Originally formulated for  energy-conserving processes, the theory has been extended to  time-dependent periodic drives by introducing scattering 
 processes involving the exchange of Floquet quanta.

%%%%%%%%%%%%%%%%%%%%%%%%%%%%%%%%%%%%%%%%%%%%%%%%%%%%%%
\subsubsection{Non-equilibrium Green's functions}
%%%%%%%%%%%%%%%%%%%%%%%%%%%%%%%%%%%%%%%%%%%%%%%%%%%%%%
The non-equilibrium Green's function formalism offers a powerful and versatile framework for analyzing quantum transport, especially in situations where many-body interactions play essential roles. In contrast to scattering matrix approaches that focus on asymptotic states, this formalism provides a real-time description by defining the evolution on the Schwinger-Keldysh contour (forward evolution from the initial state in the past and backwards). This enables a systematic treatment of the many-body interactions in combination with the non-equilibrium properties, for example by means of
perturbative expansions or the solution of equations of motion. 

It is particularly adequate to analyze few-level quantum systems in contact with reservoirs represented by a continuum of non-interacting particles or quasiparticles.  In problems with periodic drivings, the exchange of Floquet quanta naturally emerges in the dynamics. In the limit of weak interactions it is possible to define a clear relation between this formalism and scattering matrix theory.

%%%%%%%%%%%%%%%%%%%%%%%%%%%%%%%%%%%%%%%%%%%%%%%%%%%%%%
\subsubsection{Master and rate equations}
%%%%%%%%%%%%%%%%%%%%%%%%%%%%%%%%%%%%%%%%%%%%%%%%%%%%%%
Quantum master equations provide a useful theoretical framework to describe the dynamics of open quantum systems coupled to external reservoirs. 
The focus is on the reduced density matrix of the system, which results from tracing out the degrees of freedom of the baths/reservoirs. Indeed, the master equation
is actually an equation of motion for this reduced density matrix. 
The procedure 
followed to derive it 
can be justified for weak coupling between the quantum system and the bath and for Markovian (short-memory) dynamics. Additional
assumptions like the secular approximation are usually introduced in order to obtain a Lindbladian structure, which guarantees a trace-preserving
and positive-defined evolution. When the focus is on the diagonal elements of the density matrix, these equations are reduced to classical rate equations.
Extensions of this formalism to generalized master equations can be derived from the dynamics on the Keldysh contour.

%%%%%%%%%%%%%%%%%%%%%%%%%%%%%%%%%%%%%%%%%%%%%%%%%%%%%%
\subsubsection{Other methods not covered in this review}
%%%%%%%%%%%%%%%%%%%%%%%%%%%%%%%%%%%%%%%%%%%%%%%%%%%%%%
Alongside analytical techniques, a wide range of numerical techniques have been developed to study time-dependent quantum transport in complex or interacting systems where exact solutions are not feasible.
In this direction, the time evolution in low-dimensional problems is efficiently described by means of time-dependent density matrix renormalization group~\cite{White2004Aug,Schollwock2011Jan}. 
Time-dependent density functional theory is another widely employed technique to model transport in atomic and molecular systems~\cite{marques2006time}. In this technique, many-body interactions are approximated by a mean-field density. The development of the exchange potential underlying the dynamics is a complex challenge in the presence of time-dependent driving.

The development of numerical methods to describe time-evolution is an active field of research, and includes tight-binding-based models~\cite{Kloss2021Feb},  
recent proposals on time-dependent variational Monte Carlo methods \cite{Sinibaldi2023Oct} and more efficient bases in the density-matrix renormalization group \cite{Paeckel2019Dec}.

%%%%%%%%%%%%%%%%%%%%%%%%%%%%%%%%%%%%%%%%%%%%%%%%%%%%%
\subsection{Linear response theory}
%%%%%%%%%%%%%%%%%%%%%%%%%%%%%%%%%%%%%%%%%%%%%%%%%%%%%
We start by reviewing the regime of time-dependent driving applied to the system, where this driving is weak, namely the amplitude is small or the driving is slow, in the sense illustrated in Sec.~\ref{sec:kubo_general_summary}.

%%%%%%%%%%%%%%%%%%%%%%%%%%%%%%%%%%%%%%%%%%%%%%%%%%%%%
\subsubsection{Kubo linear response}\label{sec:kubo}
%%%%%%%%%%%%%%%%%%%%%%%%%%%%%%%%%%%%%%%%%%%%%%%%%%%%%

A general way to tackle observables of interest in situations with time-dependent perturbations of \textit{small amplitude} is the Kubo linear-response formalism. Here, we follow the book by Bruus and Flensberg~\cite{bruus2004many} to summarize the main steps required to evaluate observables of interest.
We consider a time-dependent Hamiltonian like the one presented in Eq.~(\ref{hamil0}) and here decomposed as follows
\begin{equation}\label{h-kubo}
\hat{H}(t)=\hat{H}_0+\hat{H}'(t),
\end{equation}
where $\hat{H}_0$ does not depend on time. The time-dependent component is assumed to have the form
\begin{equation}\label{hf}
    \hat{H}'(t)=-\hat{\boldsymbol F}\cdot {\boldsymbol X}(t), 
\end{equation}
where the vector ${\hat{\boldsymbol F}}=(\hat{F}^1,\ldots, \hat{F}^M)$ contains a set of operators and ${\boldsymbol X}(t)$ is a vector collecting a set of time-dependent parameters like the ones defined in Eq.~(\ref{param}). This formalism focuses on  
{\em small} amplitudes of the time-dependent parameters, so that $\hat{H}'(t)$ can be regarded as a perturbation.

Considering a perturbation that is switched on at time $t_0$,
for a generic operator $\hat{O}$, the expectation values can be written as follows,
\begin{eqnarray}
{\langle \hat{O} \rangle}_0 &=& \mbox{Tr}\left[\hat{\rho}_0 \hat{O}\right], \;\;\;\;\;\; t< t_0 \nonumber\\
 \hat{\rho}_0&=&\frac{e^{-\beta \hat{H}_0}}{Z_0}=\frac{1}{Z_0}\sum_n |n\rangle \langle n| e^{-\beta E_n}
\end{eqnarray}
and
\begin{eqnarray}
\langle \hat{O}\rangle(t) &=& \mbox{Tr}\left[\hat{\rho}(t) \hat{O}\right], \;\;\;\;\;\; t>t_0 \nonumber\\
 \hat{\rho}(t)&=&\frac{1}{Z_0}\sum_n |n(t)\rangle \langle n(t)| e^{-\beta E_n},
 \label{eq:lin_resp_average_later_times}
\end{eqnarray}
where it is assumed that the eigenstates $|n(t)\rangle$ evolve preserving the Boltzmann distribution, such that the partition function $Z_0$ remains the same. 
In the Schr\"odinger picture the evolution of the states is
\begin{equation}
i \hbar\partial_t |n(t)\rangle=\hat{H}(t)|n(t)\rangle.
\end{equation}
It can be related to the time evolution in the interaction picture as
\begin{equation}
 |n(t)\rangle=e^{-\frac{i}{\hbar}\hat{H}_0t}|\tilde{n}(t)\rangle =e^{-\frac{i}{\hbar}\hat{H}_0t} \tilde{U}(t,t_0) |\tilde{n}(t_0)\rangle,
\end{equation}
where
\begin{equation}
    \tilde{U}(t,t_0)=\hat{\mathscr{T}}\exp\left\{-\frac{i}{\hbar}\int_{t_0}^t dt' \hat{H}^{\prime}_{H_0}(t')\right\}
    \label{eq:evolution_op_interaction_picture}
\end{equation}
and $\hat{\mathscr{T}}$ is the time-ordering operator, the tilde on states denotes the interaction picture representation. In this picture, operators evolve according to the unperturbed Hamiltonian, namely $\hat{\bullet}_{H_0}(t)\equiv e^{i\hat{H}_0t/\hbar}\,\hat{\bullet}\,e^{-i\hat{H}_0t/\hbar}$, while the evolution of states is dictated by the evolution operator in~\eqref{eq:evolution_op_interaction_picture}.
The crucial step in this formalism is to approximate the evolution operator at linear order in the perturbation part of the Hamiltonian, namely
\begin{eqnarray}
\label{evol-op-int}
 \tilde{U}(t,t_0) \simeq 1-\frac{i}{\hbar} \int_{t_0}^t dt' \hat{H}^{\prime}_{H_0}(t').
\end{eqnarray}
With this approximation, the evaluation of mean values in Eq.~\eqref{eq:lin_resp_average_later_times} leads to
\begin{eqnarray}
\label{olr}
\langle \hat{O}\rangle (t)&\simeq& 
{\langle \hat{O} \rangle}_0-\frac{i}{\hbar} \int_{t_0}^t dt' \Braket{  \left[\hat{O}_{H_0}(t),\hat{H}^{\prime}_{H_0}(t')\right]  }_0,
\end{eqnarray}
where $\langle \bullet \rangle_0\equiv \mbox{Tr}\left[\hat{\rho}(t_0) \, \bullet\right]$ is the expectation value with respect to the unperturbed state. This method can be used, for instance, to calculate currents and other expectation values in systems under weak time-dependent driving. 

It is usual to focus on the response long after the switching-on  process, in which case we can consider $t_0\rightarrow -\infty$.
For a Hamiltonian of the form of Eqs.~\eqref{h-kubo} and \eqref{hf}, Eq.~(\ref{olr}) reads
\begin{eqnarray}\label{olr-1}
\langle \hat{O}\rangle (t)&\simeq&  {\langle \hat{O} \rangle}_0-\sum_j \int_{-\infty}^{+\infty} dt' 
\chi_{O,F^j}(t-t') X_j(t'),
\end{eqnarray}
where
\begin{equation}
    \chi_{O,F^j}(t-t')=-\frac{i}{\hbar} \theta(t-t')\Braket{\left[\hat{O}_{H_0}(t),\hat{F}^{j}_{H_0}(t')\right]}_0 
    \label{eq:kubo_response}
\end{equation}
is the Kubo susceptibility or response function, and $\theta(\bullet)$ is Heaviside's step function. These response functions are obtained by evaluating average values with respect to an {\em equilibrium}, unperturbed Hamiltonian $\hat{H} _0$, which also governs the time evolution in the interaction picture. As a consequence, they are functions of $\tau=t-t'$. Hence, they can be Fourier-transformed with respect to this variable and have a cutoff in time governed by energy scales of the internal dynamics. In addition, they obey micro-reversibility and Onsager relations. 

An alternative derivation makes contact with work-fluctuation theorems~\cite{Monnai2005Aug,Talkner2007May,Talkner2007Jun,Campisi2009May,Campisi2011Jan}  and was proposed by Andrieux and Gaspard~\cite{Andrieux2008Jun}. 
The starting point is the following equilibrium identity
\begin{equation}\label{ida}
   {\langle \hat{O}_H(t) e^{-\beta \hat{H}_H (t)} e^{\beta \hat{H}_{0}}\rangle}_{0}= {\langle \hat{O} \rangle}_0,
\end{equation}
where the subscript $H$ denotes Heisenberg-picture operators, evolving according to $\hat{\bullet}_H(t)=\hat{U}^\dagger(t,t_0)\,\hat{\bullet}\,\hat{U}(t,t_0)$, with $\hat{U}(t,t_0)$ the time evolution operator generated by the full time-dependent Hamiltonian.
Equation~\eqref{ida} can be proven as follows:
\begin{eqnarray}
  {\langle \hat{O} \rangle}_{0}&=&  \frac{1}{Z_{0}}\mbox{Tr}\left[ e^{-\beta \hat{H}_0 } \hat{O}\right] 
  %\nonumber \\
  =\frac{1}{Z_{0}}\mbox{Tr}\left[ e^{-\beta \hat{H}_0 }\hat{U}(t,t_0) \hat{U}^{\dagger}(t,t_0) \hat{O} \hat{U}(t,t_0)\hat{U}^{\dagger}(t,t_0)\right]\nonumber\\
&=& \frac{1}{Z_{0}}\mbox{Tr}\left[  \hat{O}_H(t)
  e^{-\beta \hat{H}_H(t) } e^{\beta \hat{H}_{0} } e^{-\beta \hat{H}_{0} }\right]
  %\nonumber \\&=&
  ={\langle \hat{O}_H(t)
  e^{-\beta \hat{H}_H(t) } e^{\beta \hat{H}_{0} }\rangle}_{0}.
\end{eqnarray}
Next, one introduces the following quantity
\begin{equation}\label{defw}
    \hat{W}=\hat{H}_H(t)-\hat{H}_{H_0}=-\int_{t_0}^{t} dt' \hat{\boldsymbol F}_H(t')\cdot \dot{\boldsymbol X}(t')
    =  \int_{t_0}^{t} dt' \dot{\hat{\boldsymbol F}}_{H}(t')\cdot {\boldsymbol X}(\it{t}').
    %\dot{\hat{\bf F}}_H(t') \cdot {\bf X}(t').
\end{equation}
Here, the first equality can be proved by using the fact that in the Heisenberg picture the partial derivative of the Hamiltonian is equal to the total
derivative. Hence,
\begin{eqnarray}
\hat{W}=\int_{t_0}^{t} dt' \frac{d\hat{H}}{dt'}=\int_{t_0}^{t} dt' \frac{\partial \hat{H}}{\partial t'}= -\int_{t_0}^{t} dt' \hat{\boldsymbol F}_H(t')\cdot \dot{\boldsymbol X}(t').
\end{eqnarray}
Substituting Eq.~(\ref{defw}) in Eq.~(\ref{ida}), one thus finds
\begin{equation}\label{an1}
    {\langle \hat{O}\rangle}_{0}={\left\langle \hat{O}_H(t) e^{-\beta (\hat{H}_0+ \hat{W})} e^{\beta \hat{H}_0}\right\rangle}_{0}.
\end{equation}
The right-hand side of this equation can be simplified by relying on the following identity
\begin{equation}
\label{eq:id_exp_work}
    e^{-\beta (\hat{H}_0+\hat{W})} e^{\beta \hat{H}_0} = 1-\int_0^{\beta} du\, e^{-u (\hat{H}_0+\hat{W})} \hat{W} e^{u \hat{H}_0}\simeq 1-\int_0^{\beta} du\, e^{-u \hat{H}_0} \hat{W} e^{u \hat{H}_0},
\end{equation}
where the last expression is obtained as a first-order approximation in $\hat{W}$, which contains the driving parameters $\boldsymbol{X}$. Using this result, we have
\begin{eqnarray}
    e^{-\beta (\hat{H}_0+\hat{W})} e^{\beta \hat{H}_0} &\simeq&
    1-\int_{t_0}^t dt' {\boldsymbol X}(t')\cdot \int_0^\beta du e^{-u \hat{H}_0} \dot{{\hat{\boldsymbol F}}}({\it t}')e^{{\it u \hat{H}}_0} \nonumber \\
    & = & 1-\int_{t_0}^t dt' {\boldsymbol X}(t')\cdot \int_0^\beta du\, \dot{{\hat{\boldsymbol F}}}_{H_0}{\it(t'+i u\hbar)},
\end{eqnarray}
where we have used that at first order in the driving parameters the time evolution is governed by the free Hamiltonian, thus equivalent to the time evolution in the interaction picture ${\hat{\boldsymbol F}}(t)= e^{i \hat{H}_0 t/\hbar} {\hat{\boldsymbol F}} e^{-i \hat{H}_0 t/\hbar}=\hat{\boldsymbol F}_{H_0}(t)$. 
Substituting in Eq. (\ref{an1}) leads to
\begin{equation}
\label{kuboa}
\langle \hat{O}\rangle(t)= {\langle \hat{O} \rangle}_{0}+ \int_0^\beta du\int_{t_0}^t dt'
{\boldsymbol X}(t')\cdot  {\left\langle \hat{O}_{H_0}(-iu\hbar) \dot{{\hat{\boldsymbol F}}}_{{H}_0}{\it (t'-t)} \right\rangle}_0.
\end{equation}
While Eqs.~\eqref{kuboa} and~\eqref{olr-1} are not manifestly equivalent, it can be proved that the two expressions coincide, as detailed in Appendix~\ref{app:kubo_equivalence}.

%%%%%%%%%%%%%%%%%%%%%%%%%%%%%%%%%%%%%%%%%%
\subsubsection{From Kubo linear response to adiabatic linear response}\label{sec:adiabatic_kubo}
%%%%%%%%%%%%%%%%%%%%%%%%%%%%%%%%%%%%%%%%%%
Kubo's linear response formalism is particularly useful for describing the situation of slowly varying external driving forces. Following Ref.~\cite{Ludovico2016Feb},
we consider Eqs.~(\ref{h-kubo}) and~(\ref{hf}),
assuming a slow evolution. This means that the characteristic time for the changes in ${\boldsymbol X}(t)$ is much larger than any internal time scale for the quantum system coupled to the reservoirs. This type of evolution is identified as {\em adiabatic response}. Thus, we expand the time-dependent part of the Hamiltonian at linear order with respect to a reference time $t$, namely 
\begin{equation}\label{hf-ad}
    \hat{H}(t')\simeq \hat{H}_t-{\hat{\boldsymbol F}}\cdot \dot{{\boldsymbol X}}(t)(t'-t),
\end{equation}
where $\hat{H}_t\equiv \hat{H}_0- {\hat{\boldsymbol F}}\cdot \boldsymbol{X}(t)$ is the Hamiltonian with the parameters \emph{frozen} at time $t$. \textit{Adiabatic response} is hence the first-order correction to this \textit{frozen} evolution.
Then, we employ the usual Kubo formalism with $\hat{H}_0=\hat{H}_t$ and $\hat{H}'(t')=-{\hat{\boldsymbol F}}\cdot \dot{{\boldsymbol X}}(t)(t'-t)$. 
Substituting in Eq.~(\ref{olr}), the result is
\begin{eqnarray}\label{adia-ludo}
\langle \hat{O} \rangle (t)&\simeq& 
 {\langle \hat{O} \rangle}_t+\frac{i}{\hbar}\sum_j \int_{t_0}^t dt' (t'-t)\Braket{\left[\hat{O}_{H_t}(t),\hat{F}^j_{H_t}(t')\right]}_t\dot{X}_j(t),
\end{eqnarray}
where $\braket{\bullet}_t$ is the average with respect to the frozen Hamiltonian $\hat{H}_t$. The assumption underlying the above result is a short internal characteristic damping time compared to the characteristic time for change of the parameters ${\boldsymbol X}(t)$.

An alternative formulation, closer to the linear-response theory of Ref.~\cite{Andrieux2008Jun}, was proposed in Ref.~\cite{Campisi2012Sep}.
The steps are similar to those followed in the derivation of Eq.~\eqref{kuboa}. First, one starts with an identity similar to Eq.~\eqref{ida}, connecting the expectation value at the frozen time $t$ to the equilibrium one as follows
\begin{equation}\label{id}
   {\langle \hat{O}_H(t) e^{-\beta \hat{H}_H (t)} e^{\beta \hat{H}_{t_0}}\rangle}_{t_0}= e^{-\beta \Delta \mathcal{F}} {\langle \hat{O} \rangle}_t.
\end{equation}
where
$\Delta\mathcal{F}=-\ln(Z_{t}/Z_{t_0})/\beta$ is the free-energy difference. The proof of this identity is analogous to that of Eq.~\eqref{ida}:
\begin{eqnarray}
  {\langle \hat{O} \rangle}_t&=&  \frac{1}{Z_{t}}\mbox{Tr}\left[ e^{-\beta \hat{H}_t } \hat{O}\right] 
  =\frac{1}{Z_{t}}\mbox{Tr}\left[ \hat{U}^{\dagger}(t,t_0) \hat{O} \hat{U}(t,t_0)
 \hat{U}^{\dagger}(t,t_0) e^{-\beta \hat{H}_t }\hat{U}(t,t_0) e^{\beta \hat{H}_{t_0} } e^{-\beta \hat{H}_{t_0} }\right]\nonumber\\
&=& \frac{1}{Z_{t}}\mbox{Tr}\left[  \hat{O}_H(t)
  e^{-\beta \hat{H}_H(t) } e^{\beta \hat{H}_{t_0} } e^{-\beta \hat{H}_{t_0} }\right]
  =\frac{Z_{t_0}}{Z_{t}}{\left\langle \hat{O}_H(t)
  e^{-\beta \hat{H}_H(t) } e^{\beta \hat{H}_{t_0} }\right\rangle}_{t_0}.
\end{eqnarray}
It is then useful to introduce the operator
\begin{eqnarray}
    \hat{W}_{\rm dis} &=& \hat{W}-\Delta \mathcal{F} =- \int_{t_0}^{t} dt' \left[{\hat{\boldsymbol F}}_H(t')- \langle {\hat{\boldsymbol F}} \rangle_{t_0} \right]\cdot \dot{{\boldsymbol X}}(t').
\end{eqnarray}
As emphasized in Refs.~\cite{Talkner2007Jun,Campisi2012Sep}, it does not correspond to any quantum observable\footnote{Note that there is no observable for work, which is rather the result of a process and, as such, depends on the initial and final states. Therefore,  there is no operator that would yield the work as outcome of a single projective measurement.}, but it approaches the dissipated work in the classical limit~\cite{Campisi2011Jan}.
In terms of this quantity, Eq.~\eqref{id} becomes
\begin{equation}\label{cam1}
    {\langle \hat{O}\rangle_t=\left\langle \hat{O}_H(t) e^{-\beta \hat{H}_H(t)} e^{\beta\left[\hat{H}_H(t)-\hat{W}_{\rm dis}\right]}\right\rangle}_{t_0},
\end{equation}
and can be simplified by using a slight modification of the identity~\eqref{eq:id_exp_work}.
The result is 
\begin{equation}
    {\langle \hat{O}\rangle}_t={\left\langle \hat{O}_H(t) \left[ 1- \int_0^{\beta} du\, e^{-u \hat{H}_H(t)} \hat{W}_{\rm dis} e^{u \hat{H}_H(t)}\right] \right\rangle}_{t_0}.
\end{equation}
Hence
\begin{align}
{\langle \hat{O}_H(t) \rangle}_{t_0}- {\langle \hat{O}\rangle}_t &=   \int_0^{\beta} du \left\langle \hat{O}_H(t)e^{-u \hat{H}_H(t)} \hat{W}_{\rm dis} e^{u \hat{H}_H(t)}\right\rangle_{t_0}\nonumber\\
&= - \int_{t_0}^{t} dt'\int_0^{\beta} du {\left\langle \hat{O}_H(t)e^{-u \hat{H}_H(t)}
\Delta{\hat{\boldsymbol F}}(t') e^{u \hat{H}_H(t)}\right\rangle}_{t_0}\cdot \dot{{\boldsymbol X}}(t'),
\end{align}
where $\Delta{\hat{\boldsymbol F}}(t')={\hat{\boldsymbol F}}_H(t')- \langle {\hat{\boldsymbol F}} \rangle_{t_0}$. 
In the above expression, the integrand is rewritten as follows by using the cyclicity of the trace and the unitarity of the evolution operator:
\begin{equation}
    \mbox{Tr}\left[\hat{U}^{\dagger}(t, t_0) \rho(t) \hat{O} \hat{U}(t, t_0) e^{-u \hat{H}_t}\hat{U}(t, t') \Delta {\hat{\boldsymbol F}}\, \hat{U}^\dagger(t, t')  e^{u \hat{H}_t}\right] \cdot \dot{{\boldsymbol X}}(t').\label{eq:integrand_delta_O}
\end{equation}
Within the adiabatic-response regime of interest, the following approximations are introduced to simplify the final expression. (i) The exact density matrix is approximated by the frozen equilibrium density matrix: $ \rho(t)\simeq \rho_{t}$. This is because the full expression of Eq.~\eqref{eq:integrand_delta_O} is already first order in $ \dot{\boldsymbol{X}}(t')$. (ii)
A quick decay of the correlation function within the characteristic time for the variation of ${\boldsymbol X}(t)$ is assumed, so that the following
evolution is considered $\hat{U}(t,t')\simeq e^{-i \hat{H}_t (t'-t)/\hbar}$. (iii) The following approximation $\dot{{\boldsymbol X}}(t') \simeq \dot{{\boldsymbol X}}(t)$ is also justified, under the same hypothesis. Using these replacements in the previous expression, one finds
\begin{equation} \label{adia-cam}
   \langle \hat{O}\rangle (t) \simeq  \langle \hat{O}\rangle_t - \int_0^{\beta} du \int_{t_0}^{t} dt' {\left\langle \hat{O}(-iu \hbar) \Delta {\hat{\boldsymbol F}}(t'-t) \right\rangle}_{t} \cdot \dot{{\boldsymbol X}}(t).
\end{equation}
Similarly to Eq.~\eqref{kuboa}, in this expression the time evolution of $\hat{O}$ and $\Delta{\hat{\boldsymbol F}}$ should be calculated with respect to the frozen Hamiltonian $\hat{H}_t$. We show in Appendix~\ref{app:kubo_equivalence} that Eqs.~\eqref{adia-cam} and~\eqref{adia-ludo} are fully equivalent, even though it is not immediately evident.

%%%%%%%%%%%%%%%%%%%%%%%%%%%%%%%%%%%%%%%%%%%%%%%%%%%%%%%%%%%%
\subsubsection{Luttinger's formalism for the Hamiltonian representation of a temperature bias}\label{sec:temperature bias}
%%%%%%%%%%%%%%%%%%%%%%%%%%%%%%%%%%%%%%%%%%%%%%%%%%%%%%%%%%%%
Both the usual Kubo approach and the adiabatic version of linear-response theory rely on the Hamiltonian
representation of the non-equilibrium perturbation. In the case of a system coupled to reservoirs at different temperatures or
under the effect of a thermal gradient, it is necessary to introduce a Hamiltonian representation for the temperature bias, even
if it is stationary. This problem was originally addressed by Luttinger \cite{Luttinger1964Sep}. Here, we provide a summary of the main ideas; we also refer to a recent review in the context of stationary thermal and thermoelectric transport~\cite{Chernodub2022Sep}. 

Luttinger adapted early ideas proposed by Tolman and Ehrenfest in the context of general relativity \cite{Tolman1930Dec}.
There, temperature gradients were considered to compensate the energy flux generated by spatial changes of the gravitational field, hence restoring the equilibrium.
Luttinger considered a perturbation of the form,
\begin{equation}\label{lut1}
    \hat{H}'_{\phi_{\rm th}}(t)=\int d{\boldsymbol r} \; \hat{h}({\boldsymbol r},t) \phi_{\rm th}({\boldsymbol r},t),
\end{equation}
where $\hat{h}({\boldsymbol r},t)$ is the Hamiltonian density and $\nabla \phi_{\rm th}({\boldsymbol r},t)=-\nabla T/T$ is an inhomogeneous fictitious field representing the temperature bias.

More recently, these ideas were revisited in Refs.~\cite{Tatara2015May}, where a closer analogy was formulated between Luttinger's proposal and electromagnetism. Basically,  the energy current density is defined from the conservation law:
$\nabla \cdot \hat{{\boldsymbol j}}_E({\boldsymbol r},t) + \partial_t  \hat{h}({\boldsymbol r},t) =0$ and a {\em thermal vector potential} ${\boldsymbol A}_{\rm th}({\boldsymbol r},t)$ is introduced. In this way, in addition to Eq. (\ref{lut1})
the temperature bias can be introduced by the following perturbation,
\begin{equation}\label{lut2}
    \hat{H}'_{A_{\rm th}}(t)=\int d{\boldsymbol r}\, \hat{\boldsymbol j}_E({\bf r},t) \cdot {\boldsymbol A}_{\rm th}({\boldsymbol r},t), 
\end{equation}
with 
\begin{equation}
\partial_t {\boldsymbol A}_{\rm th}({\boldsymbol r},t) + \nabla \phi_{\rm th}({\boldsymbol r},t)=-\frac{\nabla T}{T}\ .
\end{equation}
Notice that the perturbation in a system with charge $q$ in an electrical field ${\boldsymbol E}({\boldsymbol r},t)$, with the associated scalar $\phi({\bf r},t)$ and vector potential ${\boldsymbol A}({\boldsymbol r},t)$, is expressed in the form 
\begin{equation}\label{em}
\hat{H}'(t)=q\int d{\boldsymbol r} \phi({\boldsymbol r},t) \hat{n}({\boldsymbol r},t) + q\int d{\boldsymbol r} {\boldsymbol A}({\boldsymbol r},t) \cdot \hat{{\boldsymbol j}}_N({\boldsymbol r},t),
\end{equation}
with $\nabla \cdot \hat{{\boldsymbol j}}_N({\boldsymbol r},t) + \partial_t  n({\boldsymbol r},t) =0$ and
\begin{equation}
\partial_t {\boldsymbol A}({\boldsymbol r},t) + \nabla \phi({\boldsymbol r},t)=-{\boldsymbol E}({\boldsymbol r},t).
\end{equation}
This is precisely the same structure of the thermal bias under the gauge-invariant representation of Luttinger's formulation.

It is also important to notice that this Hamiltonian approach to describe temperature biases can be extended to address the case of \emph{time-dependent} temperature biases. For instance, Ref.~\cite{Lopez2024Apr} considered heat and charge transport through a multi-level quantum dot coupled to reservoirs whose temperatures are modulated in time. This approach was also followed in the study of spin torques generated by heat currents \cite{Kohno2016Sep} and in the analysis of transients \cite{Eich2014Sep,Eich2016Apr}. The representation of the temperature bias as a time-dependent
vector potential was also used in the geometric description of adiabatic thermal machines \cite{Bhandari2020Oct}.
Another interesting recent development was reported in Ref.~\cite{Portugal2024Jun}, where a periodic temperature modulation was addressed by combining an approach similar to Luttinger's representation of the temperature bias and Floquet theory (see Sec.~\ref{sec:floquet}).

%%%%%%%%%%%%%%%%%%%%%%%%%%%%%%%%%%%%%%%%%%%%
\subsubsection{Particle and energy fluxes}\label{sub:par-en-flux}
%%%%%%%%%%%%%%%%%%%%%%%%%%%%%%%%%%%%%%%%%%%%
Linear response and adiabatic linear response can be used to calculate the time-dependent expectation values of different operators. In Secs.~\ref{sec:kubo} and \ref{sec:adiabatic_kubo}, this has been shown for generic operators $\hat{O}$. Here, we are in particular interested in the particle and energy currents entering the reservoir
$\alpha$. For the Hamiltonian expressed as in Eq. (\ref{hf}) they read in linear response
\begin{equation}
\begin{split}
    I^N_\alpha(t)&= \sum_{j}\int_{-\infty}^{+\infty} dt'\chi_{I^N_\alpha,F^j}(t-t') X_j(t'), \\
    I^E_\alpha(t)&= \sum_j \int_{-\infty}^{+\infty} dt' \chi_{I^E_\alpha,F^j}(t-t') X_j(t'),
\end{split}
\end{equation}
with the response function $\chi$ defined in Eq.~\eqref{eq:kubo_response}.
In the adiabatic linear-response approach one finds
\begin{equation}\label{eq:adia-n-e}
\begin{split}
     I^N_\alpha(t)&= I^N_{\alpha} \left[{\boldsymbol X}(t)\right]+\sum_{j}\Lambda_{I^N_\alpha,F^j} 
     \left[{\boldsymbol X}(t)\right] \dot{X}_j(t), \\
    I^E_\alpha(t)&= I^E_{\alpha} \left[{\boldsymbol X}(t)\right]+\sum_{j}\Lambda_{I^E_\alpha,F^j} 
     \left[{\bf X}(t)\right] \dot{X}_j(t),
\end{split}
\end{equation}
with 
\begin{equation}
  \Lambda_{I^{N/E}_\alpha,F^j} 
     \left[{\boldsymbol X}(t)\right]  =\int_{-\infty}^{+\infty} dt' \; (t-t') \; \chi_{I^{N/E}_\alpha,F^j}(t-t').
\end{equation}
This notation stresses that the mean values are calculated with respect to the equilibrium Hamiltonian corresponding to the
parameters ${\boldsymbol X}(t)$ frozen at the time $t$. 

Note that the electromagnetic perturbations and the effect of the temperature differences represented in terms of Luttinger's description,
as given in Eqs. (\ref{lut1}), (\ref{lut2}), (\ref{em}), have the structure of the Hamiltonian (\ref{hf}). Hence, the electrical and thermal potentials
can be simply identified as time-dependent  parameters $X_j(t)$.

%%%%%%%%%%%%%%%%%%%%%%%%%%%%%%%%%%%%%%%%%%%%
\subsubsection{Power generated by the driving}\label{sec:power}
%%%%%%%%%%%%%%%%%%%%%%%%%%%%%%%%%%%%%%%%%%%%
Another quantity of interest is the power developed by the driving forces. This is defined in 
Eq.~(\ref{power}). For a Hamiltonian of the form of Eq. (\ref{hf}) it can be expressed
as $P(t)=\sum_j P_j(t)$, where we have considered separately the power associated to each driving
parameter $P_j(t)=\langle \hat{F}^j\rangle(t) \dot{X}_j(t)$. Hence, the main goal is to calculate the time-dependent mean values $\langle \hat{F}^j\rangle(t)$, which
is usually named  ``reaction force''. This concept was originally introduced by Berry \cite{Victor1993Sep} in the context of slow driving 
and was adopted in several other places in the literature \cite{Bode2011Jul,Ludovico2016Jul,Fernandez-Alcazar2015Aug,Calvo2017Oct,Hopjan2018Jul,Bhandari2020Oct,Deghi2024Sep}.

Within linear response the result leads to the following expression 
\begin{equation}
    P_j(t)= \sum_{j'}\int_{-\infty}^{+\infty} dt'\dot{X}_{j}(t) \chi_{F^j,F^{j'}}(t-t') X_{j'}(t'),
\end{equation}
while for the adiabatic linear response the result is
\begin{equation}
    P_j(t)=  P^{\rm cons}_j(t)+\sum_{j'} \dot{X}_{j}(t) \Lambda_{F^j,F^{j'}}\left[{\boldsymbol X}(t)\right] \dot{X}_{j'}(t).
\end{equation}
Here, $P^{\rm cons}_j(t)=\langle \partial_{X_j} \hat{H} \rangle_t \dot{X}_j$ is identified as the {\em conservative} or {\em quasi-static} component of the power, since it is evaluated with respect to 
a sequence of equilibrium states defined by the frozen Hamiltonian. Over a cycle, this conservative component has zero average.

%%%%%%%%%%%%%%%%%%%%%%%%%%%%%%%%%%%%%%%%%%%%
\subsection{Floquet theory}
\label{sec:floquet}
%%%%%%%%%%%%%%%%%%%%%%%%%%%%%%%%%%%%%%%%%%%%

A second regime of main interest for the applied driving, complementing the one of linear response, is \textit{periodic} driving. In the following section, we assume that all time-dependent parameters are driven at some frequency $\Omega$ with driving period $\mathcal{T}=2\pi/\Omega$. This periodic-driving regime can conveniently be approached using Floquet theory.

%%%%%%%%%%%%%%%%%%%%%%%%%%%%%%%%%%%%%%%%%%%%
\subsubsection{General formalism}
%%%%%%%%%%%%%%%%%%%%%%%%%%%%%%%%%%%%%%%%%%%%
The aim of Floquet theory is to solve the time-dependent Schr\"odinger equation of a time-dependent Hamiltonian with period ${\cal T}$, $\hat{H}(t+{\cal T})=\hat{H}(t)$,
\begin{equation}
    i \hbar \frac{d}{d {\cal T}} |\psi(t)\rangle = \hat{H}(t)|\psi(t)\rangle.
\end{equation}
In what follows, we present a summary of the main ideas following Refs. \cite{santoro2019,Casati1989May,Grifoni1998Oct,Holthaus2015Nov,Bukov2015Mar}, without giving proofs.

Mathematically, there are similarities with the Bloch theory of spatially periodic systems. 
In such a case, and focusing on one dimension,
it is natural to rely on a quasimomentum $\hbar k$, with $k=2\pi /a$, with $a$ the lattice constant. In Floquet theory the counterpart is the quasienergy $\hbar \Omega= 2\pi\hbar /{\cal T}$.
An important aspect is the following   property of the evolution operator in the Schr\"odinger picture 
\begin{equation}\label{evol-op-sh}
 \hat{U}(t+n{\cal T},t_0) =\hat{U}(t,t_0) \left[\hat{U}(t_0+n{\cal T},t_0)\right]^n,
\end{equation}
which implies that the knowledge of $\hat{U}(t,t_0)$ for $t \in \left[t_0,t_0+{\cal T}\right]$ is enough to write the evolution operator at an arbitrary time $t+n{\cal T}$. 

A concrete procedure to make this property explicit is to represent the one-period evolution operator as $\hat{U}(t_0+{\cal T},t_0)=e^{-(i/\hbar)\hat{H}_{F_0} {\cal T}}$, with $\hat{H}_{F_0}$ a Hermitian operator.
In this way, the evolution operator between two arbitrary times $t_1$ and $t_2$ is expressed as
\begin{eqnarray}\label{ut21}
    \hat{U}(t_2,t_1) &=& \hat{U}(t_2,t_0+n{\cal T}) e^{-\frac{i}{\hbar}\hat{H}_{F_0} n{\cal T}}\hat{U}(t_0,t_1)\nonumber \\
    &\equiv& e^{-iK_F[t_0] (t_2)} e^{-\frac{i}{\hbar}\hat{H}_{F_0} (t_2-t_1)}e^{iK_F[t_0] (t_1)},
\end{eqnarray}
where we have introduced the definition of the stroboscopic  kick operator
$K_F[t_0]$. It is also useful to introduce a change of representation by defining
kick operators $K$  and an effective Hamiltonian $\tilde{H}$ as follows: 
\begin{equation}
e^{-iK_F[t_0] (t)}= e^{-i K(t)} e^{i K(t_0)},\;\;\;\;\;\;\;\;\;\tilde{H}=e^{i K t_0} \hat{H}_{F_0} e^{-i K t_0}.
\end{equation}
In this way, Eq. (\ref{ut21}) can be written as
\begin{equation}
    \hat{U}(t_2,t_1)= 
    e^{-i K t_2} e^{-\frac{i}{\hbar}\tilde{H}(t_2-t_1)}e^{i K t_1}.
\end{equation}
These definitions lead to the following expressions of the time-dependent mean values of generic observables
\begin{eqnarray}\label{oflo}
    \langle \hat{O}(t_2)\rangle = \mbox{Tr}\left[\tilde{\rho}(t_1) \; \hat{U}_F(t_2,t_1) \tilde{O}(t_2) \right],
\end{eqnarray}
where $\tilde{\rho}(t_1)= e^{i K t_1} \hat{\rho}e^{-i K t_1}$, $\tilde{O}(t_2)= e^{i K t_2} \hat{O}e^{-i K t_2}$ 
and $\hat{U}_F(t_2,t_1)=e^{-(i/\hbar)\tilde{H}(t_2-t_1)}$.

In addition, it is interesting to mention that a complete basis for the Hilbert space can be defined as follows 
\begin{equation}
   \hat{H}_{F_0} |u_{j}(t_0)\rangle= \varepsilon_j |u_{j}(t_0)\rangle,
\end{equation}
where the eigenenergies $\varepsilon_j $ are determined modulo $\hbar \Omega$ since $e^{-(i/\hbar)\varepsilon_j {\cal T}}= e^{-(i/\hbar)(\varepsilon_j +n \hbar \Omega){\cal T}}$.
The complete set of solutions of the time-dependent Schr\"odinger equation has the form
\begin{equation}
    |\psi_j(t)\rangle =e^{-\frac{i}{\hbar} \varepsilon_j (t-t_0)}|u_{j}(t)\rangle,
\end{equation}
with $|u_{j}(t+n {\cal T})\rangle=|u_{j}(t)\rangle$. 

An alternative approach which relies on the Fourier expansion of the periodic 
 Hamiltonian is the so-called  Shirley-Floquet approach. The starting point is the
 eigenvalue problem defined by the Schr\"odinger equation expressed in the basis $|u_{j}(t)\rangle$. It reads
 \begin{equation}
     \left[\hat{H}(t) -i\hbar \frac{d}{dt}\right]|u_{j}(t)\rangle
     = \varepsilon_j |u_{j}(t)\rangle.
 \end{equation}
 The Hamiltonian and the eigenstates are expanded in Fourier components as follows,
 \begin{eqnarray}
     \hat{H}(t) &=&\sum_n \hat{H}^{(n)} e^{-i n \Omega t},\nonumber \\
     |u_{j}(t)\rangle &=& \sum_n e^{-i n \Omega t} |u_{j}^{(n)} \rangle,
 \end{eqnarray}
 which leads to the following linear eigenvalue problem
 \begin{equation}
     \sum_{n'} \hat{H}^{(n')} |u_{j}^{(n-n')} \rangle-n \hbar \Omega|u_{j}^{(n)} \rangle=\varepsilon_j|u_{j}^{(n)} \rangle. 
 \end{equation}
 This representation has a structure akin to the Schr\"odinger equation of  a tight-binding Hamiltonian in real space. It defines 
 a practical way to solve the original time-dependent problem with numerical methods. To this end, a cutoff 
 in the number of Fourier components $n-n'$ coupled to the mode $n$  must be introduced.

%%%%%%%%%%%%%%%%%%%%%%%%%%%%%%%%%%%%%%%%%%%%
 \subsubsection{Particle flux}
%%%%%%%%%%%%%%%%%%%%%%%%%%%%%%%%%%%%%%%%%%%%
 The Floquet formalism is frequently used in lattice models, which typically have the structure \cite{D'Alessio2015Oct,Dehghani2015Apr}
 \begin{equation}\label{eq:lattice_H}
    \hat{H}(t)= \sum_{\ell,\ell'} h_{\ell,\ell'} (t) \; \hat{c}^{\dagger}_\ell \hat{c}_{\ell'} + \hat{H}_{\rm int},
 \end{equation}
 where the creation and annihilation operators $\hat{c}^{\dagger}_\ell, \hat{c}_{\ell'}$ obey fermionic or bosonic commutation relations, and
 $\hat{H}_{\rm int}$ is a many-body interaction.
 The particle flux is defined by analyzing the change in time of the local occupation at the site $\ell$,
 \begin{equation}
     \Braket{\frac{d \hat{N}_\ell}{dt}} =-\frac{i}{\hbar}\Braket{\left[\hat{N}_\ell,\hat{H} \right]}.
 \end{equation}
 Usually, $\left[\hat{N}_\ell,\hat{H}_{\rm int}\right]=0$
 commutes with the local particle density and the current between the sites $\ell$ and $\ell'$ are defined as
 \begin{equation}
     J_{\ell'\rightarrow \ell}(t) = 2 \mbox{Im}\left[h_{\ell,\ell'} (t) \; \langle \hat{c}^{\dagger}_\ell \hat{c}_{\ell'} \rangle \right].
 \end{equation}
 Floquet states can be used to calculate this mean value. Reference~\cite{Bukov2015Mar} presents a detailed discussion on using Eq.~(\ref{oflo}) to calculate the mean values and averages over time.

%%%%%%%%%%%%%%%%%%%%%%%%%%%%%%%%%%%%%%%%%%%%%%%%%%%%%%%%
\subsection{Scattering matrix theory}\label{sec:scattering_theory}
%%%%%%%%%%%%%%%%%%%%%%%%%%%%%%%%%%%%%%%%%%%%%%%%%%%%%%%%
The scattering matrix theory of coherent quantum transport was developed by Landauer, B\"uttiker, and Imry~\cite{Landauer1957Jul,Landauer1961Jul,Landauer1970Apr,Buttiker1985Aug,Buttiker1986Mar,Buttiker1988Nov,Buttiker1993Dec,Imry1986Mar,Imry1999Mar}. We summarize  the main ideas following \cite{Moskalets2002Jul,Moskalets2002Nov,Moskalets2011Sep}, focusing in particular on scattering theory for time-dependently driven conductors. 
This theory applies to conductors connected to $N_r$ reservoirs with well-defined temperatures and chemical potentials. It 
relies on the description of the wave functions of the particles injected from one reservoir and scattered into the same or a different reservoir as a consequence of the applied biases and the scattering properties of the conductor. 

%%%%%%%%%%%%%%%%%%%%%%%%%%%%%%%%%%%%%%%%%%%%%%%%%%%%%%%%
\subsubsection{Stationary case}
%%%%%%%%%%%%%%%%%%%%%%%%%%%%%%%%%%%%%%%%%%%%%%%%%%%%%%%%
We start by introducing the concepts of scattering theory in the time-independent, stationary case. The key object is the scattering matrix $\underline{S}$, characterizing the conductor and relating the outgoing and the incoming fluxes to each other. Orthonormal bases of single-particle wave functions $\left\{\psi_{\alpha}^{\rm (in/out)}\right\}$ 
in the reservoir $\alpha$ for the
incoming and outgoing particles, respectively, are considered. Assuming that they are plane waves with velocities $v_{\alpha}(E) =\hbar k_\alpha(E)/m$ along the longitudinal direction $x$,
the field operators for these particles are expressed as:
\begin{equation}
\begin{split}
\hat{\Psi}_\alpha(t,{\boldsymbol r})&=\frac{1}{\sqrt{2\pi}}\int_0^\infty \frac{dE}{\sqrt{\hbar v_{\alpha}(E)}} e^{-i\frac{E}{\hbar} t}\sum_{\alpha} \left\{\hat{a}_\alpha \psi_{\alpha}^{\rm (in)}(E,{\boldsymbol r}) +
\hat{b}_\alpha \psi_{\alpha}^{\rm (out)}(E,{\boldsymbol r})
\right\}, \\
\hat{\Psi}_\alpha^\dagger(t,{\boldsymbol r})&=\frac{1}{\sqrt{2\pi}}\int_0^\infty \frac{dE}{\sqrt{\hbar v_{\alpha}(E)}} e^{-i\frac{E}{\hbar} t}\sum_{\alpha} \left\{\hat{a}^\dagger_\alpha \psi_{\alpha}^{\rm (in)*}(E,{\boldsymbol r}) +
\hat{b}^\dagger_\alpha \psi_{\alpha}^{\rm (out)*}(E,{\boldsymbol r})
\right\},
\end{split}
\end{equation}
with ${\boldsymbol r}=(x,r_\perp)$, while $\hat{a}^\dagger_\alpha,\hat{b}^\dagger_\alpha$ create and  $\hat{a}_\alpha,\hat{b}_\alpha$ annihilate particles in the incoming/outgoing states, respectively. 
They obey fermionic/bosonic commutation relations, depending on the nature of the particles. 
The scattering matrix $\underline{S}$ is defined by the elements $S_{\gamma\alpha}$
relating the annihilation operators for the outgoing and incoming particles,
\begin{equation}
\label{scatt}
    \hat{b}_\gamma =\sum_\alpha S_{\gamma\alpha} \hat{a}_\alpha.
\end{equation}
Here, we have suppressed possible additional indices for the channels in each lead that would occur in a multi-channel setup. An extension to this case is straightforward, where one would have to replace $\hat{a}_\alpha\rightarrow\hat{a}_{\alpha n}$ and $S_{\gamma\alpha}\rightarrow S_{\gamma m\,\alpha n}$.  The scattering matrix is unitary,
\begin{equation}\label{unit-s}
    \underline{S}^{\dagger}\underline{S}=\underline{S}\ \underline{S}^{\dagger}=\mathbbm{1},
\end{equation}
with $\mathbbm{1}$ being the unit matrix with the same dimension as $\underline{S}$. In addition, it obeys micro-reversibility. This implies that, in the presence of a magnetic field $B$, the scattering matrix
satisfies $\underline{S}(B)=\underline{S}^\dagger (-B)$. 

Starting from the field operators, the particle current operator flowing in reservoir $\alpha$ reads
\begin{equation}\label{curop}
    \hat{I}^N_\alpha(t,x) =\frac{i\hbar}{2m} \int dr_{\perp}\left\{ \frac{\partial \hat{\Psi}^\dagger_\alpha (t,{\boldsymbol r})}{\partial x}  \hat{\Psi}_\alpha (t,{\boldsymbol r})
    -\hat{\Psi}^\dagger_\alpha (t,{\boldsymbol r}) \frac{\partial \hat{\Psi}_\alpha (t,{\boldsymbol r})}{\partial x} 
    \right\}.
\end{equation}
Using the definition of Eq. (\ref{scatt}), the mean value of the particle current defined in Eq. (\ref{curop}) can be expressed as follows
\begin{equation}
    I^N_\alpha=\frac{1}{h}\int_0^{\infty} dE \sum_{\gamma=1}^{N_r} |S_{\alpha \gamma}(E)|^2 \left[f_{\gamma}(E)-f_\alpha(E) \right],
\end{equation}
where $f_\alpha(E)$ are Fermi-Dirac or Bose-Einstein distribution functions, depending on the nature of the particles. 

%%%%%%%%%%%%%%%%%%%%%%%%%%%%%%%%%%%%%%%%%%%%%%%%%%%%%%%%
\subsubsection{Periodic time-dependent scatterer}\label{t-per-scatt}
%%%%%%%%%%%%%%%%%%%%%%%%%%%%%%%%%%%%%%%%%%%%%%%%%%%%%%%%
When the conductor is under the  effect of a time-periodic potential with a frequency $\Omega$, the scattering process 
at the conductor takes place with a gain or loss of energy quanta $\hbar \Omega$. Hence, the outgoing state at energy $E$ can be expressed as a superposition of states incoming at energies $E_n=E+n\hbar\Omega$, with $n$ an integer number. Consequently, the scattering matrix introduced in Eq. (\ref{scatt}) is generalized
and expressed in the Floquet representation by introducing a Floquet scattering matrix $\underline{S}_F$, whose components $S_{F,\gamma\alpha}(E,E_n)$ relate the incoming and outgoing states as follows
\begin{equation}
\label{s-floq}
    \hat{b}_{\gamma}(E)=\sum_\alpha \sum_{E_n>0} S_{F,\gamma\alpha}(E,E_n) \hat{a}_\alpha(E_n).
\end{equation}
This can be interpreted as the quantum-mechanical amplitude for an electron coming from reservoir $\alpha$ with energy $E_n$ to be scattered
into reservoir $\gamma$ with an exchange of $-n$ quanta of the oscillating conductor. In the
Floquet space, this matrix (now with an increased dimension due to the exchange of energy quanta) obeys an analogous unitary property as expressed for the static counterpart
in Eq. (\ref{unit-s})
\begin{equation}
    \sum_{E_n>0}\sum_\zeta S^*_{F,\zeta\alpha}(E_n,E_m)S_{F,\zeta\gamma}(E_n,E)=\delta_{m0}\delta_{\alpha\gamma}\,.
\end{equation}
With this property, the mean value of Eq. (\ref{curop}) can be calculated, and the result is
\begin{equation}
    I_\alpha^N(t)=\frac{1}{h}\int_0^{\infty}dE\sum_{\gamma}\sum_{E_n,E_\ell>0}e^{-i\ell\Omega t}S^*_{F,\alpha\gamma}(E,E_n)S_{F,\alpha\gamma}(E_\ell,E_n)[f_{\gamma}(E_n)-f_\alpha(E)]\,.
\end{equation}
A simpler expression is found for the directed particle current $\bar{I}_\alpha^N=\int_{0}^{\cal T}dt I_\alpha^N/\cal{T}$, which reads
\begin{eqnarray}\label{in-sf}
    \bar I_\alpha^N &=&\frac{1}{h} \int_0^\infty dE \sum_{\gamma} \sum_{E_n>0}|S_{F,\alpha \gamma}(E,E_n)|^2 \left[
    f_{\gamma}(E_n)- f_\alpha(E)\right]\,.
    \end{eqnarray}
Following a similar procedure, the time-resolved and directed heat currents into reservoir $\alpha$ can be calculated. Assuming the same chemical potential $\mu$ for all reservoirs, the result is
\begin{align}
    I_\alpha^Q(t) &= \frac{1}{h}\int_0^{\infty}dE\sum_{\gamma}\sum_{E_n,E_\ell>0}\left(E-\mu+\frac{\ell\hbar\Omega}{2}\right)e^{-i\ell\Omega t}S^*_{F,\alpha\gamma}(E,E_n)S_{F,\alpha\gamma}(E_\ell,E_n)\nonumber\\
    &\qquad\qquad\qquad\qquad\qquad\quad\times[f_{\gamma}(E_n)-f_\alpha(E)]\,,\\
    \bar I_\alpha^Q &=\frac{1}{h} \int_0^\infty dE \sum_{\gamma} \sum_{E_n>0}\left(E-\mu\right)|S_{F,\alpha \gamma}(E,E_n)|^2 \left[
    f_{\gamma}(E_n)- f_\alpha(E)\right].
\end{align}
In addition to the energy representation $\underline{S}_F(E,E_n)$ of the Floquet scattering matrix, it is often convenient to adopt a mixed time-energy representation, which is obtained from the following relation
\begin{equation}
    \underline{S}(t,E)=\sum_{n=-\infty}^{\infty}\underline{S}_F(E_n,E)e^{-in\Omega t}\iff\underline{S}_F(E_n,E)=\int_{0}^{\mathcal{T}}\frac{dt}{\mathcal{T}}\underline{S}(t,E)e^{in\Omega t}\,.
\end{equation}
In terms of this alternative representation, the time-dependent particle current is given by
\begin{equation}
    I_\alpha^N(t)=\frac{1}{h}\int_0^{\infty} dE\sum_{\gamma}\sum_{E_n>0}[f_{\gamma}(E)-f_\alpha(E_n)]\int_0^{\mathcal{T}}\frac{dt'}{\mathcal{T}}e^{in\Omega(t-t')}S^*_{\alpha\gamma}(t',E)S_{\alpha\gamma}(t,E)\,.
\label{eq:current_floquet_time-energy}
\end{equation}

%%%%%%%%%%%%%%%%%%%%%%%%%%%%%%%%%%
\subsubsection{Reservoirs with ac voltages}
\label{sec:Floquet_ac_voltages}
%%%%%%%%%%%%%%%%%%%%%%%%%%%%%%%%%%
In addition (or alternatively) to the time-dependent driving of the central conductor region, a paradigmatic situation corresponds to electron systems where time-dependent voltages are applied to
one or more reservoirs. Importantly, a way to model such a time-dependent bias voltage is by including the ac-part of the potential bias into the scattering matrix. Again, the approach relies on the Floquet representation of the scattering matrix and
Eq.~(\ref{s-floq}), see e.g. Ref.~\cite{Pedersen1998Nov,Rychkov2005Oct}, with
\begin{equation}
    \hat{a}_\alpha(E)=\sum_{\ell=-\infty}^\infty c_{\alpha, \ell} %e^{-i m \varphi_\alpha}
    \hat{a}'_\alpha(E-\ell\hbar \Omega),
\end{equation}
where $c_{\alpha,\ell}$ are Floquet coefficients defined as
\begin{eqnarray}
    c_{\alpha, \ell} & = & \int_0^\mathcal{T}\frac{dt}{\mathcal{T}}e^{-i\phi_\alpha(t)}e^{i \ell\Omega t}\,, \label{eq:photoassisted_amplitudes}\\
    \phi_\alpha(t) & = & \frac{q}{\hbar}\int_0^t dt'V_\alpha^\mathrm{ac}(t')\,.\label{eq:Floquet_Faraday_flux}
\end{eqnarray}
If the time-dependent bias voltage is of the form
$V_\alpha(t)=V_{\alpha}^0 \cos(\Omega t +\varphi_\alpha)$, the Floquet coefficients equal  $c_{\alpha, \ell}=J_\ell \left(\frac{qV_{\alpha}^0}{\hbar \Omega}\right)e^{-i\ell\varphi_\alpha}$, where $J_\ell(x)$ are Bessel functions of the first kind.
Substituting in Eq.~(\ref{curop}) and taking the average over one period of the current expectation value leads to the following expression of the
charge current, assuming spinless electrons and a single transport channel,
\begin{eqnarray}\label{iac}
    \overline{I}_\alpha^N &=&\frac{1}{h}\int_0^\infty dE \sum_{\gamma} \sum_{n=-\infty}^\infty f_{0,\gamma}(E-n\hbar \Omega) 
    \Bigg\{\sum_{m,\ell} S^*_{F,\alpha \gamma}(E,E_\ell)S_{F,\alpha \gamma}(E,E_m)
    \nonumber \\
    & & \;\;\;\;\;\;\; \times c^*_{\gamma,n+\ell}c_{\gamma,n+m}-\delta_{\alpha \gamma } |c_{\alpha, n}|^2
    \Bigg\},
\end{eqnarray}
with $f_{0,\gamma}(E) $ the Fermi function, depending on the temperature and the chemical potential of the reservoir $\alpha'$ without the effect of the ac voltage (indicated by the additional subscript $0$). Note that since the driven bias is here modeled in terms of a scattering coefficient, it is possible to rewrite the expression~\eqref{iac} in terms of a combined effective scattering matrix
\begin{eqnarray}
\label{in-sf_eff}
    \overline{I}_\alpha^N &=&\frac{1}{h} \int_0^\infty dE \sum_{\gamma} \sum_{E_n>0}|\tilde{S}_{F,\alpha \gamma}(E,E_n)|^2 \left[
    f_{\gamma}(E_n)- f_\alpha(E)\right]\,,
    \end{eqnarray}
with
\begin{eqnarray}
    \tilde{S}_{F,\alpha \gamma}(E,E_n)=\sum_{\ell}S_{F,\alpha \gamma}(E,E_\ell)c_{\alpha, \ell-n}.
\end{eqnarray}
Equation~\eqref{in-sf_eff} can be used to define a spectral current, which is nothing but the energy-resolved contribution to the particle current
\begin{equation}
    \mathfrak{i}_\alpha(E)=\sum_{\gamma} \sum_{E_n>0}|\tilde{S}_{F,\alpha \gamma}(E,E_n)|^2 \left[
    f_{\gamma}(E_n)- f_\alpha(E)\right]\,.
\label{eq:spectral_current_general}
\end{equation}
This quantity is especially useful in two-terminal transport geometries, for example when the scattering matrix defines the action of a single-electron source, as discussed in Sec.~\ref{sec:injection_target_drives}.

%%%%%%%%%%%%%%%%%%%%%%%%%%%%%%%%%%
\subsubsection{Adiabatic approximation}\label{adia-scatt}
%%%%%%%%%%%%%%%%%%%%%%%%%%%%%%%%%%
As previously discussed, for time-periodic transport, the adiabatic-response approximation applies to the situation where the period,
which for moderate driving amplitudes defines the typical time scale for the changes of the driving, is much larger than the typical time scale for the internal dynamics of the conductors. In terms of energy, this means that $\hbar \Omega $ is much smaller than the typical energy windows defined 
by the changes in $\underline{S}$ and the changes in the distribution functions of the reservoirs. As before, we focus on problems where
the time dependence enters through parameters ${\boldsymbol X}(t)$.
In the framework of the scattering matrix theory, the adiabatic approximation 
is implemented as an expansion of the Floquet scattering matrix up to first order in the driving frequency. The first crucial ingredient is the notion of \emph{frozen scattering matrix} $\underline{S}_0(t,E)$. It is defined as the stationary scattering matrix with the parameters $\boldsymbol{X}$ frozen at time $t$, namely $\underline{S}_0(t,E)=\underline{S}_0(E,\boldsymbol{X}(t))$. With this, the first-order expansion of the Floquet scattering matrix $\underline{S}_F$ reads~\cite{Moskalets2004May}
\begin{equation}
\underline{S}_{F}(E,E_n)\simeq \underline{S}_{0,n}(E)+\hbar \Omega \left[\frac{n}{2} \frac{\partial \underline{S}_{0,n}(E)}{\partial E} + \underline{A}_n(E)\right]\,,
\end{equation}
where
\begin{equation}
\underline{S}_{0,n}(E)=\int_0^{\cal T}\frac{dt}{\cal T}\,e^{in\Omega t} \underline{S}_0(E,{\boldsymbol X}(t))\quad\iff\quad \underline{S}_0(E,{\boldsymbol X}(t))=\sum_{n=-\infty}^\infty
\underline{S}_{0,n}(E) e^{-i n \Omega t}
\end{equation}
is the $n$-th Fourier component of the frozen scattering matrix. The quantities $\underline{A}_n$ are the Fourier components of a matrix $\underline{A}(t,E)$ satisfying
\begin{equation}
    \underline{S}_0^\dagger(t,E)\underline{A}(t,E)+\underline{A}^\dagger(t,E)\underline{S}_0(t,E) =  \frac{i}{2\Omega}\left[\frac{\partial \underline{S}_0^\dagger(t,E)}{\partial t}\frac{\partial \underline{S}_0(t,E)}{\partial E}-\frac{\partial \underline{S}_0^\dagger(t,E)}{\partial E}\frac{\partial \underline{S}_0(t,E)}{\partial t}\right].
\end{equation}
Introducing these expansions into Eq. (\ref{iac}), assuming reservoirs at the same temperature and preserving terms up to linear order in  $\hbar \Omega$ and $V_{\alpha,0}$, the following expression for the directed electron current is found
\begin{equation}
I_\alpha^N=\int_0^\infty dE \left(-\frac{\partial f(E)}{\partial E}\right)\left[I_\alpha^{\rm (pump)}(E)+
I_\alpha^{\rm (rect)}(E)\right].
\end{equation}
The first term is identified as {\em pumping} and describes the transport induced by the time-dependent variation of the parameters acting in the conductor
connected to the reservoirs and is $\propto \hbar \Omega$. The second component is identified as {\em rectification} and describes the transport induced by
the voltage bias applied at the reservoirs. 
These components read
\begin{eqnarray}
    I_\alpha^{\rm (pump)}(E) &=&  \frac{i}{2\pi} \overline{\left(\frac{\partial \underline{S}_0(t,E)}{\partial t} \underline{S}^\dagger_0(t,E)\right)_{\alpha \alpha}}\nonumber \\
    I_\alpha^{\rm (rect)}(E) &=& \frac{q}{h} \sum_\gamma \overline{\left( V_\gamma(t)-V_\alpha(t)\right) | S_{0,\alpha \gamma}(t,E)|^2}.
\end{eqnarray}

%%%%%%%%%%%%%%%%%%%%%%%%%%%%%%%%%%
\subsection{Green's function formalism}\label{sec:Greens_functions}
%%%%%%%%%%%%%%%%%%%%%%%%%%%%%%%%%%

%%%%%%%%%%%%%%%%%%%%%%%%%%%%%%%%%%
\subsubsection{General considerations}
%%%%%%%%%%%%%%%%%%%%%%%%%%%%%%%%%%
The non-equilibrium Schwinger-Keldysh Green's function formalism is a powerful method to treat many-body problems under the effect of time-dependent 
driving. Formally, it enables the combined treatment of many-body interactions and non-equilibrium effects. In the theory of quantum transport, it was
first introduced in problems without time-dependent drives, where transport is induced by means of dc voltage biases \cite{Caroli1971Jun,Caroli1972Jan,Pastawski1991Sep,Meir1992Apr,Wingreen1994Apr} and
then extended to time-dependent problems in the stationary regime \cite{Pastawski1992Aug,Wingreen1993Sep,Jauho1994Aug,Cuevas1996Sep} as well as in transients \cite{Tuovinen2014Feb,SeoaneSouto2015Sep,Covito2018May}. Here, we very briefly 
present the main ideas and focus on the implementation in time-periodic problems following \cite{Arrachea2005Sep}, as well as its relation to the scattering matrix theory \cite{Arrachea2006Dec}. Details of this theory can be studied in 
\cite{di2008electrical,haug2008quantum,rammer2011quantum,Stefanucci2025Jan,kamenev2023field}.

The starting point is a Hamiltonian of the form
\begin{equation}
    \hat{\cal H}(t)=\hat{H}+\hat{H}'(t),\;\;\;\;\; \hat{H}=\hat{H}_0+\hat{H}_{\rm int},
\end{equation}
where the static component $\hat{H}$ contains a single-particle term  $\hat{H}_0$ and a many-body interaction term $\hat{H}_{\rm int}$. The time-dependent part $\hat{H}'(t)$ is
considered to be switched on at time $t_0$.

The key concept in this theory is the time evolution along the Keldysh contour, which is defined by a forward path from $t_0$ to $\infty$ and closed by 
a backward piece from $\infty$ to $t_0$. 
The time-ordered single-particle Green's function over this contour reads
\begin{equation}
    G({\boldsymbol r}, t; {\boldsymbol r}', t)=-i \left\langle \mathscr{T}_C \left[ \hat{\psi}_{\cal H} ({\boldsymbol r}, t) \hat{\psi}^\dagger_{\cal H} ({\boldsymbol r}', t')\right]\right\rangle,
\end{equation}
where $\mathscr{T}_C$ denotes time ordering along the Keldysh contour of the  fermionic or bosonic field operators expressed in the Heisenberg representation with respect to
the full Hamiltonian ${\cal H}(t)$. The mean value $\langle \bullet \rangle \equiv {\rm Tr}\left[\hat{\rho}_H \bullet\right]$ is calculated with respect to the
density matrix of the equilibrium Hamiltonian $\hat{H}$. The combination with the many-body perturbation theory to treat $\hat{H}_{\rm int}$ is implemented by 
assuming that the interactions are adiabatically switched on at $t_0'=-\infty$. For problems where the transient introduced by switching on $\hat{H}'(t)$ are neglected, it is convenient to also extend the Keldysh contour to $t_0 \rightarrow -\infty$. This defines the so-called Schwinger-Keldysh contour $C$ which
consists of a path where time evolves forward from $-\infty$ to $+\infty$ followed by a path with a backward evolution from $+\infty$ to $-\infty$. 
Introducing the interaction picture for $\hat{H}_{\rm int}$, the Green's function can be written as follows 
\begin{equation}
    G({\boldsymbol r}, t; {\boldsymbol r}', t)=-i {\left\langle \mathscr{T}_C \left[ e^{-i\int_C d \tau \left( \hat{H}^{\rm int}_{H_0}(\tau) +\hat{H}^{\prime}_{H_0}(\tau)\right)}
    \hat{\psi}_{H_0} ({\boldsymbol r}, t) \hat{\psi}^\dagger_{H_0} ({\boldsymbol r}', t')\right]\right\rangle}_0,
\end{equation}
where the operators are expressed in the Heisenberg representation with respect to the non-interacting Hamiltonian $\hat{H}_0$ and the mean value
$\langle \bullet \rangle_0$ is calculated with respect to the equilibrium density of this Hamiltonian. As in usual perturbation theory, the expansion of the exponential in combination with Wick's theorem leads to terms at different orders in the interactions $\hat{H}_{\rm int}$ and
$\hat{H}^{\prime}(t)$, which can be represented by Feynman diagrams. Defining the non-interacting Green's function
\begin{equation}
G_0({\boldsymbol r}, t; {\boldsymbol r}', t')=-i {\left\langle \mathscr{T}_C \left[ 
    \hat{\psi}_{H_0} ({\boldsymbol r}, t) \hat{\psi}^\dagger_{H_0} ({\boldsymbol r}', t')\right]\right\rangle}_0
\end{equation}
and suitably collecting the higher-order terms of this expansion, the Dyson equation is obtained as in usual perturbation theory. Introducing the shorthand notation $j\equiv ({\boldsymbol r}_j, t_j)$, it has the following structure
    \begin{equation}\label{dyson4}
        G(1,1')=G_0(1,1')+\int_{C} dt_2 d t_3\int d{\boldsymbol r}_3 d{\boldsymbol r}_2 G_0(1,3) \Sigma(3,2) G(2,1'),
    \end{equation}
    where $\Sigma(3,2)$ is the self-energy in the Schwinger-Keldysh contour. In general the self-energy is a complicated function of the interactions and $G$.

    As a consequence of the fact that $C$ contains a forward $+$ and a backward $-$ path, each of the time arguments of these functions has an implicit index $+,-$.
    Therefore, $G, G_0, \Sigma$ have the structure of a  $2 \times 2$ matrix  in these indices. The following notation is introduced 
    \begin{eqnarray}
    & & G(1^+,2^-)\equiv G^>(1,2) = -i \langle  \hat{\psi}_{\cal H} ({\boldsymbol r}_1, t_1) \hat{\psi}^\dagger_{\cal H} ({\boldsymbol r}_2, t_2)\rangle \nonumber \\
    & & G(1^-,2^+)\equiv G^<(1,2)= \mp i \langle \hat{\psi}^\dagger_{\cal H} ({\boldsymbol r}_2, t_2) \hat{\psi}_{\cal H} ({\boldsymbol r}_1, t_1) \rangle,
    \label{eq:greater_lesser_GF}
    \end{eqnarray}
    where the prefactor in the last line $\mp$ applies to bosons and fermions. The functions indicated with the symbols $>,<$ are, respectively named greater and lesser Green's functions. It is also useful to define retarded and advanced Green's functions
    \begin{equation}
        G^r(1,2)=-i \theta(t_1-t_2)\left[G^>(1,2) -G^<(1,2)  \right]= \left[G^a(2,1)\right]^\dagger.
    \end{equation}
    Using properties of the contour-ordered Green's functions known as Langreth theorem, it can be shown that the convolution of two contour-ordered functions of the form $G(1,1')=\int_C G_1(1,2)G_2(2,1') $ can be decomposed into the following identities for the real-time Green's functions:
    \begin{eqnarray}
    G^{r/a}(1,1') &=&\int dt_2 G^{r/a}_1(1,2)G^{r/a}_2(2,1') ,\nonumber \\
    G^{</>}(1,1') &=&\int dt_2 \left[G^{</>}_1(1,2)G^{a}_2(2,1')+ G^{r}_1(1,2)G^{</>}_2(2,1')\right].
    \end{eqnarray}
    Using these relations, the Dyson equation~(\ref{dyson4}) is reduced to the following set of equations
    \begin{eqnarray}\label{dyson-realt}
        G^{r/a}&=&G_0^{r/a}+ G_0^{r/a}\Sigma^{r/a} G^{r/a} = G_0^{r/a}+ G^{r/a}\Sigma^{r/a} G_0^{r/a}, \nonumber \\
         G^{</>}&=&\left( 1+ G^{r}\Sigma^{r} \right)G_0^{</>}\left( 1+ \Sigma^{a} G^{a}\right)
         + G^{r} \Sigma^{</>} G^{a},    
    \end{eqnarray}
    where, for simplicity, we have omitted the indices of the Green's functions and the integrals in the products.  In many problems, the first term
    of the second equation vanishes in the long-time limit, where the transient is damped. 

    This approach is particularly useful to evaluate the particle and energy fluxes in systems modeled by Hamiltonians with spatial discretization. In particular, notice that the mean values
    of the operators defined 
    as 
    \begin{eqnarray}\label{curop-tun}
        \hat{I}^N_\alpha(t) & = &  - \frac{i}{\hbar} \left[ \hat{N}_{\alpha},\hat{H}(t)\right]= - \frac{1}{\hbar}\sum_{\ell,k}\left[ i w_{\alpha k\ell}\hat{a}^\dagger_{\alpha k}  \hat{d}_{\ell} + \mathrm{H.c.}\right],\\
    \label{ecurrop-tun}
        \hat{I}^E_\alpha(t) & = &  - \frac{i}{\hbar} \left[ \hat{H}_{\alpha},\hat{H}(t)\right]= - \frac{1}{\hbar}\sum_{\ell,k}\varepsilon_{\alpha k}\left[ i w_{\alpha k\ell}\hat{a}^\dagger_{\alpha k}  \hat{d}_{\ell} + \mathrm{H.c.}\right],
    \end{eqnarray}
can be directly expressed in terms of lesser Green's functions as follows
    \begin{eqnarray}
       I^N_\alpha(t)&=&-\frac{2}{\hbar}\sum_{\ell,k} \mbox{Re}\left[ G^<_{\ell,\alpha k}(t,t) w_{\alpha k\ell}\right] ,\nonumber\\
       I^E_\alpha(t)&=&-\frac{2}{\hbar}\sum_{\ell,k} \mbox{Re}\left[ G^<_{\ell,\alpha k}(t,t) \varepsilon_{\alpha k} w_{\alpha k\ell}\right].
    \end{eqnarray}
The explicit calculation of the Green's functions depends on the details of the Hamiltonian and, crucially, on the presence or absence of many-body interactions.

%%%%%%%%%%%%%%%%%%%%%%%%%%%%%%%%%%%%%%%%%%%%%%%%%%%%%%%%%%%%%
\subsubsection{Systems without many-body interactions}
%%%%%%%%%%%%%%%%%%%%%%%%%%%%%%%%%%%%%%%%%%%%%%%%%%%%%%%%%%%%%
In systems where  the Hamiltonian of the driven part can be expressed as a bilinear form of creation and annihilation operators, 
\begin{equation}
    \hat{H}_{\rm sys}(t)=\sum_{\ell,\ell'} V_{\ell,\ell'}\left[{\boldsymbol X}(t)\right] \hat{d}^\dagger_\ell \hat{d}_{\ell '},
\end{equation}
with 
\begin{equation}
V_{\ell,\ell'}\left[{\boldsymbol X}(t)\right]=\sum_{m=-\infty}^{+\infty}V_{\ell,\ell'}^{(m)} e^{-i m \Omega t},
\end{equation}
the Green's functions can be exactly calculated. Following \cite{Arrachea2005Sep,Arrachea2006Dec}, it is convenient to introduce the Fourier-Floquet representation for the retarded Green's function as follows
\begin{eqnarray}
    \underline{G}^r(t,t')&=&\int \frac{d\varepsilon}{2\pi} \underline{G}(t,\varepsilon)e^{-i\frac{\varepsilon}{\hbar}(t-t')},\nonumber \\
   \underline{G}(t,\varepsilon) &=&\sum_{m=-\infty}^{+\infty}e^{im \Omega t} \underline{\mathscr G}(m,\varepsilon),
\end{eqnarray}
where $\underline{G}$ denotes a matrix structure in the indices $\ell$ of the driven system. Consequently, a matrix  $\underline{V}^{(m)}$ is defined and 
the Dyson equation can be expressed as
\begin{equation}\label{dy-flo}
    \underline{G}^r(t,\varepsilon)=\underline{G}^0(\varepsilon)+\sum_{m\neq 0}e^{-im\Omega t } \underline{G}^r(t,\varepsilon+m \hbar \Omega)\underline{V}^{(m)}\underline{G}^0(\varepsilon),
\end{equation}
where $\underline{G}^0(\varepsilon)$ is the Green's function of the static system  described by  $\underline{V}^{(0)}$ and in contact to the reservoirs.

Describing the coupling to the reservoirs in terms of the spectral matrix with elements 
\begin{equation}\label{gamma}
    \Gamma_{\alpha,\ell,\ell'}(\varepsilon)=2 \pi \sum_{k_\alpha}w_{\alpha k \ell } \delta(\varepsilon-\varepsilon_{k_\alpha}) w^*_{\alpha k\ell' },
\end{equation}
and using properties of the Green's functions,
the directed particle and energy currents (averaged over one period) can be expressed as 
    \begin{eqnarray}\label{in-ie}
       I^N_\alpha&=&\sum_{\alpha'} \sum_m \int \frac{d\varepsilon}{2\pi} T_{\alpha,\alpha'}(m,\varepsilon)       \left[ f_{\alpha'}(\varepsilon)-f_\alpha (\varepsilon+m \hbar \Omega)\right],\\
       I^E_\alpha&=&\sum_{\alpha'} \sum_m \int \frac{d\varepsilon}{2\pi} \left(\varepsilon +m \hbar \Omega\right)  T_{\alpha,\alpha'}(m,\varepsilon)
       \left[ f_{\alpha'}(\varepsilon)-f_\alpha (\varepsilon+m\hbar\Omega)\right], 
       \nonumber
    \end{eqnarray}
    with
    \begin{equation}
        T_{\alpha,\alpha'}(m,\varepsilon) = \mbox{Tr}\left[\underline{\Gamma}_{\alpha}(\varepsilon+m \hbar \Omega)\underline{\mathscr G}(m,\varepsilon) \underline{\Gamma}_{\alpha'}(\varepsilon)
       \underline{\mathscr G}^{\dagger}(m,\varepsilon)\right]. 
    \end{equation}
For reservoirs with constant density of states and smoothly connected to the driven device, the functions of Eq. (\ref{gamma}) are constant and one can define a translation between  the
Green's functions and the scattering matrix~\cite{Arrachea2006Dec}
\begin{equation}\label{s-g-trans}
S_{F,\alpha,\alpha'}(E_m,E_n)=\delta_{\alpha,\alpha'}\delta_{m,n}-i \sqrt{\Gamma_\alpha } \underline{\Pi}_\alpha  \underline{\mathscr G}(m-n,\varepsilon + n\Omega)\underline{\Pi}_{\alpha'} \sqrt{\Gamma_{\alpha'}},
\end{equation}
where the operators $\underline{\Pi}_{\alpha}$ and $\underline{\Pi}_{\alpha'}$ project the indices of the central device on those entering the tunneling matrix elements $w_{\alpha k \ell}$ and $w_{\alpha' k\ell}$, respectively.
In this way, the expression for $I^N_\alpha$ in Eq. (\ref{in-ie}) can be shown to be equivalent to the expression in terms of the Floquet scattering matrix given by Eq. (\ref{in-sf}). A similar procedure can be followed to derive an expression like Eq. (\ref{iac}) in terms of Green's functions for an electron system under the effect of an ac bias voltage. 

Similar and equivalent expressions have been derived for the currents by calculating the equation of motion and introducing the representation 
of the retarded Green's function given in Eq. (\ref{dy-flo}) \cite{Platero2004May,Kohler2005Feb}.
The expression for the energy current given in Eq. (\ref{in-ie}) is also valid for phononic/photonic 
systems under periodic driving described by bilinear Hamiltonians \cite{Arrachea2012Sep}.

%%%%%%%%%%%%%%%%%%%%%%%%%%%%%%%%%%%%%%%%%%%%%%%%%%%%%%%%%
\subsubsection{Adiabatic approximation}\label{sec:greens_adiabatic}
%%%%%%%%%%%%%%%%%%%%%%%%%%%%%%%%%%%%%%%%%%%%%%%%%%%%%%%%%
As in the case of the scattering matrix, the Floquet retarded Green's function can be expanded for small $\Omega$. This is accomplished starting from 
 Eq. (\ref{dy-flo}). 
 
Keeping terms up to linear order in $\Omega$ (equivalent to ${\cal O}( \dot{\boldsymbol{X}})$), one finds
\begin{equation}\label{dyad}
    \underline{G}^r(t,\varepsilon)\simeq\underline{G}^0(\varepsilon)+ \underline{G}^r(t,\varepsilon)\underline{V}(t)\underline{G}^0(\varepsilon) + i\hbar  \partial_{\varepsilon} \underline{G}^r(t,\varepsilon)
    \frac{d\underline{V}(t)}{dt}\underline{G}^0(\varepsilon).
\end{equation}
Introducing the definition of the {\em frozen} Green's function\footnote{Note that we use the superscript 0 for the frozen time-evolution in other parts of this review instead of using $f$. But we refrain from doing this here in order not to confuse with the unperturbed evolution, here indicated by a zero.}, corresponding to the static problem defined by the parameters fixed at time $t$,
\begin{equation}
     \underline{G}^f(t,\varepsilon)=\left[\underline{G}^0(\varepsilon)^{-1}-\underline{V}(t)\right]^{-1},
\end{equation}
the exact solution of Eq. (\ref{dyad}) up to linear order in $\Omega$ or $\dot{\boldsymbol{X}}$ reads
\begin{eqnarray}\label{gad}
    \underline{G}^r(t,\varepsilon)&\simeq&\underline{G}^f(t,\varepsilon)+  \frac{i\hbar}{2}\left[  \frac{\partial^2\underline{G}^f(t,\varepsilon)}{\partial t\partial \varepsilon} +\underline{A}(t,\varepsilon)\right],
    \nonumber \\
    \underline{A}(t,\varepsilon)&=&  \partial_{\varepsilon} \underline{G}^f(t,\varepsilon)\underline{V}(t)\underline{G}^f(t,\varepsilon)-\underline{G}^f(t,\varepsilon)\underline{V}(t)\partial_{\varepsilon} \underline{G}^f(t,\varepsilon),
\end{eqnarray}
which can be related to the adiabatic approximation of the scattering matrix discussed in Sec. \ref{adia-scatt}.

%%%%%%%%%%%%%%%%%%%%%%%%%%%%%%%%%%%%%%%%%%%%%%%%%%%%%%%%%%%%%%%%%%%%%%%%%
\subsubsection{Problems with many-body interactions}
%%%%%%%%%%%%%%%%%%%%%%%%%%%%%%%%%%%%%%%%%%%%%%%%%%%%%%%%%%%%%%%%%%%%%%%%%
Up to here, we have focused on Green's function methods for a quadratic Hamiltonian, meaning that many-body interactions are at most taken into account at the mean-field level. However, many-body interactions not only affect the shape of the instantaneous Green's function through the self-energy, but also have a strong impact on how approximations for specific driving regimes are done. In this case, Eq. (\ref{dy-flo}) must hence be generalized to include the effect of many-body terms. 

In certain cases, even strongly interacting systems can be mapped to noninteracting ones, such as in the Kondo-regime~\cite{Aono2004Sep,Arrachea2008Apr}, allowing for a description of time-dependent transport exploiting tools from the noninteracting theory. Furthermore, corrections represented by a self-energy $\underline{\Sigma}(t,t')$
can be solved at some level of approximation, for specific regimes, such as Hartree-Fock or Hubbard approximations starting from an equation of motion approach for the two-time Green's function~\cite{Vovchenko2013Nov}. Alternatively, renormalization group approaches have been used, such as the functional renormalization group approach, to study periodic driving in interacting systems with moderately strong coupling~\cite{Kennes2012Feb,Eissing2016Dec,Eissing2016Jan}.

Formulas for \textit{adiabatic} pumping through noninteracting systems, generally expressed in terms of quasi-stationary Green's functions and their derivatives, have been set up in Refs.~\cite{Splettstoesser2005Dec,Sela2006Apr,Splettstoesser2007Jun,Fioretto2008Jun}. There, approximations have been carried out on how the time dependence of the self-energy is treated. These approaches involve an ``average-time" approximation~\cite{Splettstoesser2005Dec,Splettstoesser2007Jun}, which basically neglects vertex corrections~\cite{Sela2006Apr}. It nonetheless remains applicable not only for noninteracting, but also for large classes of interacting systems, in particular when temperature is zero and mapping to a Fermi liquid is possible or when the coupling to the environment is treated perturbatively~\cite{Fioretto2008Jun}. Extensions to nonadiabatic transport have also been achieved based on these approaches~\cite{Hasegawa2022Jun}. 
This formalism can be  also used to solve driven qubits in strong coupling to reservoirs by introducing the representation of spins to Majorana fermions \cite{Portugal2021Nov}.

%%%%%%%%%%%%%%%%%%%%%%%%%%%%%%%%%%%%%%%%%%%%%%%%%%
\subsubsection{Theory of electronic coherence}\label{sec_electron-coherence}
%%%%%%%%%%%%%%%%%%%%%%%%%%%%%%%%%%%%%%%%%%%%%%%%%
The Green's function formalism is very useful to describe the properties of nonequilibrium states that are generated by single- or few-electron sources. This is of special interest in the context of electronic quantum optics, which we will discuss in detail in Sec.~\ref{sec_QOE}. 
Here, we present the basic concepts of the formalism. Typically, it is applied to one-dimensional systems, such as the edge modes of the quantum Hall effect, that play the role of waveguides for electron propagation.

One of the key concepts in electronic quantum optics is the notion of electron coherence. It is defined in close analogy to Glauber's optical coherence~\cite{Glauber1963Jun}. An immediate difference compared to optical coherence is that we can have here a coherence function for both electrons and holes. Explicitly, the single-electron and single-hole coherences (or first-order electron and hole coherences) associated with a state characterized by a many-body density matrix $\hat \rho$ are defined as
\begin{equation}
\begin{split}
    \mathcal{G}^<(x,t;x't')&=\mathrm{Tr}[\rho\,\hat{\psi}^\dagger(x',t')\hat{\psi}(x,t)]\equiv\Braket{\hat{\psi}^\dagger(x',t')\hat{\psi}(x,t)}_\rho,\\
    \mathcal{G}^>(x,t;x't')&=\mathrm{Tr}[\rho\,\hat{\psi}(x,t)\hat{\psi}^\dagger(x',t')]\equiv\Braket{\hat{\psi}(x,t)\hat{\psi}^\dagger(x',t')}_\rho,
\end{split}
\label{eq:electron_coherence_def}
\end{equation}
where $\hat{\psi}(x,t)$ is the electronic field operator in the Heisenberg representation. Although a standard notation in the literature is $\mathcal{G}^{e/h}$ for electron and hole coherences, respectively, our choice emphasizes that these functions are basically the non-equilibrium lesser and greater Keldysh Green's functions, cf. Eq.~\eqref{eq:greater_lesser_GF}. From the definitions above, the following symmetry property immediately follows
\begin{align}
    &\mathcal{G}^\gtrless(x,t;x',t')={[\mathcal{G}^\gtrless(x',t';x,t)]}^*\,,
\label{eq:electron_coherence_symmetry}
\end{align}
while the anti-commutation of fermionic operators implies
\begin{align}
    &\mathcal{G}^<(x,t;x',t)+\mathcal{G}^>(x,t;x',t)=\delta(x-x')\,.
\end{align}
In many relevant situations, especially when dealing with local observables, the single-electron coherence is evaluated at a given position, so it is sufficient to use a local version of the more general definition above, where $x=x'$ and the spatial variable is dropped for notational convenience. Such simplification is even not necessary in the case of chiral conductors with linear dispersion, where the space and time variables only appear in the combination $x-v_F t$, where $v_F$ is the Fermi velocity or the characteristic propagation velocity.
In the following, we will assume such a situation unless otherwise specified.

The single-electron coherence can be represented in three different ways: we have a time representation, an energy representation, and a mixed time-energy representation.
The time representation directly stems from the definition~\eqref{eq:electron_coherence_def}. The energy (or, more properly, frequency) representation is obtained by performing a double Fourier transform
\begin{equation}
    \tilde{\mathcal{G}}^\gtrless(\omega,\omega')=\int_{-\infty}^{+\infty}dt\int_{-\infty}^{+\infty}dt'\,\mathcal{G}^\gtrless(t,t')e^{i(\omega t-\omega' t')}.
\end{equation}
Using the operator decomposition
\begin{equation}
    \hat{\psi}(t)=\frac{1}{\sqrt{2\pi v_F}}\int_{-\infty}^{+\infty}d\omega\, \hat{c}(\omega)e^{-i\omega t},
\label{eq:1D_fermionic_field_decomposition}
\end{equation}
where $v_F$ is the Fermi velocity, one finds
\begin{equation}
    \tilde{\mathcal{G}}^<(\omega,\omega')=\frac{2\pi}{v_F}\Braket{\hat{c}^\dagger(\omega)\hat{c}(\omega')}_\rho\,,
\end{equation}
showing that the energy representation is best suited to obtain the energy distribution function, which can be accessed by taking the diagonal limit $\omega=\omega'$.

The mixed representation relies on the notion of Wigner function, which (adapting it to our context), can be defined as~\cite{Ferraro2013Nov}
\begin{subequations}
\begin{align}
    W^\gtrless(t,\omega)&=v_F\int_{-\infty}^{\infty} d\tau\, \mathcal{G}^\gtrless\left(t+\frac{\tau}{2},t-\frac{\tau}{2}\right)e^{i\omega\tau}\label{eq:Wigner_definition_time}\\
    &=v_F\int_{-\infty}^{+\infty}\frac{d\xi}{2\pi}\,\tilde{\mathcal{G}}^\gtrless\left(\omega+\frac{\xi}{2},\omega-\frac{\xi}{2}\right)e^{-it\xi}.
    \label{eq:Wigner_definition_energy}
\end{align}
\label{eq:Wigner_definition}
\end{subequations}
By using Eq.~\eqref{eq:electron_coherence_symmetry}, one can show that this quantity is real. Moreover, in a chiral conductor, there is the additional property
\begin{equation}
    W^<(t,\omega)+W^>(t,-\omega)=1\,.
\end{equation}
The Wigner function representation is particularly useful because it allows one to see in a transparent way both the temporal profile of a given few-electron state, as well as its energy content. Moreover, the marginal distributions of the Wigner function yield the time-dependent charge current and the electronic energy distribution:
\begin{align}
    I^c(t)&=q\int_{-\infty}^{+\infty}\frac{d\omega}{2\pi}W^<(t,\omega),\\
    f_e(\omega)&=\overline{W^<(t,\omega)}^t\,.
\end{align}
The time average in the second equation can be a simple integration $\int_{\mathbb{R}}dt$ for states with a finite number of extra particles (or holes) on top of the Fermi sea, or rather an average over a single period $\int_{-\mathcal{T}/2}^{\mathcal{T}/2}\frac{dt}{\mathcal{T}}$ when dealing with periodic sources (see the next paragraph). Moreover, strictly speaking, the true energy distribution is obtained by replacing $\omega\to\hbar\omega$. In Sec.~\ref{sec_QOE}, we find it simpler to keep the frequency variable instead of the energy $\varepsilon=\hbar\omega$.

\paragraph{Electron coherence for periodic states}
In the presence of periodic sources, which is the most common situation in experiments, the first-order coherence functions inherit the periodicity property
\begin{equation}
    \mathcal{G}^\gtrless(t+\mathcal{T},t'+\mathcal{T})=\mathcal{G}^\gtrless(t,t')\,.
\label{eq:electron_coherence_periodicity}
\end{equation}
As a result, it is possible to perform a decomposition into a Fourier series with respect to the average time $\bar{t}=(t+t')/2$ and a Fourier transform in the time difference $\tau=t-t'$:
\begin{equation}
    \mathcal{G}^\gtrless(t,t')=\sum_{\ell\in\mathbb{Z}}e^{-i\ell\Omega\bar{t}}\int_{-\infty}^{+\infty}\frac{d\omega}{2\pi}g^\gtrless_\ell(\omega)e^{-i\omega\tau}\,.
\end{equation}
Moreover, the $\mathcal{T}$-periodicity in $\bar{t}$ means that the Wigner function is periodic and can be represented as
\begin{equation}
    W^\gtrless(t,\omega)=v_F\sum_{\ell\in\mathbb{Z}}g^\gtrless_\ell(\omega)e^{-i\ell\Omega t}\,,
\end{equation}
showing that $g^\gtrless_\ell(\omega)$ are nothing but the harmonics of the Wigner function. Finally, the energy representation reads
\begin{equation}
    \tilde{\mathcal{G}}^\gtrless(\omega,\omega')=\sum_{\ell\in\mathbb{Z}}\delta(\omega-\omega'-\ell\Omega)g^\gtrless_\ell\left(\frac{\omega+\omega'}{2}\right)\,.
\end{equation}
The above decompositions make clear that, for periodic sources, the problem is reduced to the calculation of the harmonics $g^\gtrless_\ell(\omega)$. For noninteracting electrons, this task can be tackled by relying on the Floquet approach presented in Section~\ref{t-per-scatt}. Explicitly, the field operators $\hat{\psi}(t)$ that enter the Wigner function calculation, are given in this framework by the decomposition in Eq.~\eqref{eq:1D_fermionic_field_decomposition}, where the operators $\hat{c}(\omega)$ play the role of the $\hat{b}(E)$ operators in Eq.~\eqref{s-floq} (suppressing the indices $\alpha$, $\gamma$ that are not needed in this case). This leads to the result~\cite{Ferraro2013Nov}
\begin{equation}
    g_\ell^<(\omega)=\sum_{m\in\mathbb{Z}}S_F(\omega_\ell,\omega_m)S_F^*(\omega_{\ell+m},\omega_{-\ell})f\left[\omega+\Omega\left(m+\frac{\ell}{2}\right)\right]\,.
\end{equation}
The details of the final expression thus depend on the Floquet scattering matrix characterizing the source.

In the simple case of a periodic classical drive, i.e., a voltage $V(t)$ applied to an ohmic contact, the previous equation can be expressed in a simpler form by using the coefficients in Eq.~\eqref{eq:photoassisted_amplitudes}. One gets
\begin{equation}
    g^<_\ell(\omega)=\sum_{m\in\mathbb{Z}}c_\ell^* c_{\ell+m}f\left[\omega-\frac{qV_{\rm dc}}{\hbar}-\Omega\left(m+\frac{\ell}{2}\right)\right]\,,
\end{equation}
where $V_{\rm dc}=\int_0^{\mathcal{T}} dt\,V(t)/\mathcal{T}$ is the dc component of the voltage.

%%%%%%%%%%%%%%%%%%%%%%%%%%%%%%%%%%%%%%%%%%%%%%%%%%%%%%%%%%%%%%%%%%%%%%%%%
\subsection{Master and rate equations}
%%%%%%%%%%%%%%%%%%%%%%%%%%%%%%%%%%%%%%%%%%%%%%%%%%%%%%%%%%%%%%%%%%%%%%%%%
The methods of the previous sections do not rely on the type of coupling between the system and the baths. We now turn to present 
master and rate equation approaches. These are useful in problems where the system is typically weakly coupled to baths and are valid even when its Hamiltonian 
cannot be expressed in terms of bilinear products of creation and annihilation operators.

There are several routes to derive quantum master equations. These are used in  atomic, optical, as well as in condensed matter physics in problems identified as ``open quantum systems''. This concept applies to few-level or discrete-level quantum systems coupled to one or more baths containing many degrees of freedom. 
The goal is to describe the dynamics of the reduced density matrix for the central system as follows
\begin{equation}\label{liou}
    \frac{d \hat{\rho}(t)}{dt}={\cal L}(t) \hat{\rho}(t),
\end{equation}
being ${\cal L}(t)$ the Liouvillian operator (strictly speaking it is a superoperator acting on operators in Liouville space), which depends on the Hamiltonian of the system as well as its coupling to the reservoirs. Concrete examples for this general operator are given below in Eqs.~\eqref{lind} and \eqref{seve}.

%%%%%%%%%%%%%%%%%%%%%%%%%%%%%%%%%%%%%%%%%%%%%%%%%%%%%%%%%%%%%%%%%%%%%%%%%
\subsubsection{Stationary case}
%%%%%%%%%%%%%%%%%%%%%%%%%%%%%%%%%%%%%%%%%%%%%%%%%%%%%%%%%%%%%%%%%%%%%%%%%
Here we start with presenting the derivation for the standard case of stationary systems, namely in the absence of time-dependent driving, following textbooks 
\cite{breuer2002theory,Lidar2019Feb}. It is based on a structure of the coupling between the system and bath of the form 
\begin{equation}
\hat{H}_{\rm coup}=g \sum_{\nu} \hat{A}_{\nu} \hat{B}_{\nu},
\end{equation}
where $\hat{A}_{\nu},\;\hat{B}_{\nu} $ are different operators, labeled by $\nu$ and associated to the system and bath, respectively, and $g$ is a characteristic coupling between them.
For simplicity, we summarize here the procedure by considering a single bath and we omit the contact index $\alpha$ (which could otherwise be included in the list labeled by $\nu$). The extension to several baths will be recovered in the end. 

The evolution of the density matrix for the system coupled to the bath starts from the initial condition (at $t=0$) where these systems are assumed to be decoupled, that is
$\hat{\rho}_{SB}(0)=\hat{\rho}_B(0)\otimes \hat{\rho}(0)$. It is convenient to express the time evolution in 
the interaction picture,
\begin{equation}
    \tilde{A}_\nu(t)=e^{\frac{i}{\hbar} \hat{H}_S t } \hat{A}_\nu e^{-\frac{i}{\hbar} \hat{H}_S t },\;\;\;\;\tilde{B}_\nu(t)=e^{\frac{i}{\hbar} \hat{H}_B t } \hat{B}_\nu e^{-\frac{i}{\hbar} \hat{H}_B t }.
\end{equation}
Integrating the Liouville equation of motion for the density matrix in the interaction picture $i\hbar d \tilde{\rho}(t)/dt=[H_{\rm coup},\tilde{\rho}(t)]$, plugging it back into \eqref{liou}, and tracing over the bath degrees of freedom, $\tilde{\rho}(t)=\mbox{Tr}_B[\tilde{\rho}_{SB}(t)]$, one finds for the density operator in the interaction picture
\begin{equation}
    \frac{d}{dt}\tilde{\rho}(t)=-\mbox{Tr}_B\left\{\left[\tilde{H}_{\rm coup}(t),\int_0^t d\tau 
    \left[\tilde{H}_{\rm coup}(t-\tau),\tilde{\rho}_{SB}(t-\tau)\right]\right]\right\}.
\end{equation}
The next step is to take the Born approximation, $\tilde{\rho}_{SB}(t)\simeq \tilde{\rho}(t) \otimes \rho_B(0)$, valid for weak system-bath coupling. This leads to
\begin{equation}\label{born}
    \frac{d}{dt}\tilde{\rho}(t)=-g^2 \sum_{\nu,\nu'}\int_0^t d\tau \left\{{\cal B}_{\nu,\nu'}(\tau)
    \left[\tilde{A}_\nu(t),\tilde{A}_{\nu'}(t-\tau)\tilde{\rho}(t-\tau)\right] + {\rm h.c.}\right\},
\end{equation} 
where we define the correlation function of the bath
\begin{equation}
  {\cal B}_{\nu,\nu'}(\tau)=\mbox{Tr}\left[ \tilde{B}_\nu(\tau) B_{\nu'} \rho_B\right].
\end{equation}
As a next step, a Markov approximation is performed, assuming that the correlation time characterizing the bath
is very short. Introducing this ``short-memory'' approximation into Eq. (\ref{born}), leads to the
{\rm Redfield master equation}, which is \textit{local in time} and has the form of Eq.~(\ref{liou}) with
\begin{equation}\label{redfield}
  {\cal L}(t)\tilde{\rho}(t)=   -g^2 \sum_{\nu,\nu'}\int_0^t d\tau \left\{{\cal B}_{\nu,\nu'}(\tau)
    \left[\tilde{A}_\nu(t),\tilde{A}_{\nu'}(t-\tau)\right] \tilde{\rho}(t)+ {\rm h.c.}\right\}.
\end{equation}
This first-order differential equation can be solved given an initial condition, and the stationary value can be found in the long-time limit. 
However, Eq.~\eqref{redfield} has the short-coming that it is not possible to guarantee the
positivity of the solution. Instead, the positivity property can be proved for master equations that have {\em Lindblad} form. This form can be
obtained from Eq.~(\ref{redfield}) after the rotating wave approximations (also named secular approximation). 
This approximation consists in representing 
\begin{equation} \label{arep}
     \tilde{A}_\nu(t)=\sum_{m,n}
    e^{-i(\varepsilon_n-\varepsilon_m)\frac{t}{\hbar}}|m\rangle \langle  m|
    \hat{A}_\nu|n\rangle \langle n|\equiv \sum_{\hbar\omega=\varepsilon_n-\varepsilon_m} \hat{A}_\nu(\omega) e^{-i \omega t}
    = \sum_{\hbar\omega=\varepsilon_n-\varepsilon_m} \hat{A}^\dagger_\nu(\omega) e^{i \omega t},
\end{equation}
where $\hat{H}_S |n\rangle =\varepsilon_n |n\rangle$, and in substituting these expressions in Eq. (\ref{redfield}). This 
generates oscillatory terms $\propto e^{i(\omega-\omega')t} \left[\hat{A}^\dagger_{\nu}(\omega), \hat{A}_{\nu}(\omega')\right]$.
The rotating wave or secular approximation consists in neglecting the fast oscillatory terms while preserving only those where 
$\omega=\omega'$.
The result,
after transforming back to the Schr\"odinger picture, is the {\em Lindblad-Davies} master equation, which has the same structure as
 Eq. (\ref{liou}) with  the following action of the Liouvillian 
 \begin{eqnarray}\label{lind}
    {\cal L}(t)\hat{\rho}(t)&=&-i\left[\hat{H}_S+\hat{H}_\mathrm{Lamb},\hat{\rho}\right]\nonumber \\
    & & +\sum_{\omega} \sum_{\nu \nu'} \left( \hat{L}_{\nu'}(\omega) 
    \hat{\rho} \hat{L}^\dagger_\nu(\omega)  -\frac{1}{2}\left\{\hat{L}^\dagger_\nu(\omega) \hat{L}_{\nu'}(\omega),\hat{\rho}\right\}\right).
\end{eqnarray}
Here, we have introduced the definitions for the so-called Lindblad jump operators
\begin{eqnarray}\label{def:jump}
        \hat{L}_{\nu'}(\omega) &=& g \sqrt{ \gamma(\omega) }\hat{A}_{\nu'}(\omega)
\end{eqnarray}
which introduce transitions between the states of the central system due to the coupling to the environment. They hence represent the non-unitary and dissipative effects introduced by the coupling. The definition of the jump operators contains the spectral function for the bath
\begin{eqnarray}\label{defi:spectral_bath}
    \gamma_{\nu \nu'}(\omega) &=&\int_{-\infty}^\infty d\tau\, {\cal B}_{\nu,\nu'}({\tau})\,e^{i\omega\tau}.
\end{eqnarray}
In the jump operators~\eqref{def:jump}, we have assumed $\gamma_{\nu,\nu'}(\omega) \simeq \gamma(\omega)$.  Furthermore, we have introduced  the Lamb-shift Hamiltonian
\begin{eqnarray}
    \hat{H}_\mathrm{Lamb}&=&g^2 \sum_{\nu,\nu'}S_{\nu,\nu'}(\omega) \hat{A}^\dagger_{\nu}(\omega)\hat{A}_{\nu'}(\omega)\\
    S_{\nu,\nu'}(\omega) &= & \int_{-\infty}^\infty \frac{d\omega}{2\pi}\gamma_{\nu,\nu'}(\omega') {\cal P}\left(\frac{1}{\omega-\omega'}\right)\nonumber
\end{eqnarray}
where ${\cal P}(\bullet)$ denotes the principal value.  The Lamb-shift introduces a correction of the original Hamiltonian for the system due to the coupling to the environment, in the unitary part of the dynamics of Eq.~\eqref{lind}.

In the case of {\em several reservoirs}, each has an associated dissipator ${\cal D}_\alpha$ defined from jump operators $\hat{L}_{\nu,\alpha}$ which describe
transitions between the levels of $\hat{H}_S$ due to the coupling to the reservoir $\alpha$. The full evolution is described by
\begin{equation}\label{seve}
\frac{d\hat{\rho}(t)}{dt}=-i\left[\hat{H}_S+\hat{H}_\mathrm{Lamb},\hat{\rho}(t)\right]+
\sum_{\alpha} {\cal D}_\alpha \;\hat{\rho}(t).
\end{equation}
Importantly, the structure
of Eq.~\eqref{seve}, and hence of Eq.~(\ref{lind}), can be proved to guarantee a time evolution which preserves the trace and the positivity of $\hat{\rho}$.

The Lindblad equation~\eqref{lind} gets the form of a standard rate equation, when the dynamics of the off-diagonal elements of the density matrix (coherences) decouples from the dynamics of the diagonal elements (populations). The populations are hence given by
\begin{equation}
p_a=\hat{\rho}_{aa}=\mbox{Tr}[\hat{\rho} \Pi_a],
\end{equation}
where $\Pi_a=|a\rangle \langle a|$, and $\hat{H}_S|a\rangle =\varepsilon_a|a\rangle$.
The corresponding equation of motion is a {\em rate equation}. Explicitly, after performing the trace in Eqs. (\ref{liou}) and (\ref{lind}), we get after some algebra
\begin{equation}\label{rate}
    \dot{p}_a=\sum_{a'}\left[ W(a|a')p_{a'}-W(a'|a)p_{a} \right],
\end{equation}
being $W(a|a')=\sum_{\nu \nu',\omega} \gamma_{\nu \nu'}(\varepsilon_{a'}-\varepsilon_a) \langle a'|\hat{A}_\nu (\omega)|a\rangle \langle a|\hat{A}_{\nu'}(\omega) |a'\rangle$.
In the case of a thermal bath, the correlation function satisfies the Kubo-Martin-Schwinger condition 
\begin{equation}
    \langle \hat B_{\nu}(\tau) \hat B_{\nu'}(0)\rangle =\langle  \hat B_{\nu'}(0) \hat B_{\nu}(\tau+i\beta\hbar)\rangle = \langle \hat B_{\nu'}(-\tau-i\beta\hbar) \hat B_{\nu}(0)\rangle,
\end{equation}
which implies
\begin{equation}\label{kms}
    \gamma_{\nu \nu'}(-\varepsilon)=e^{-\beta \varepsilon} \gamma_{\nu' \nu}(\varepsilon).
\end{equation}
For a thermal reservoir, the property of Eq. (\ref{kms}), implies for the transition rates the detailed balance relation $W(a|a')=e^{-\beta(\varepsilon_a-\varepsilon_{a'})}W(a'|a)$. This guarantees a Gibbs state as the stationary
solution of Eq. (\ref{rate}): $p_a=e^{-\beta \varepsilon_a}/Z$, $Z=\sum_a e^{-\beta \varepsilon_a}$. A similar reasoning can be followed to show that the full Lindblad master equation also has a Gibbs state as a stationary solution. 

Recently, in  Refs.~\cite{Majenz2013Jul,Mozgunov2020Feb}, improvements in the derivation of Eq. (\ref{lind}) have been presented with the goal of avoiding the secular approximation, which is less well-founded that the Born approximation justified by a weak system-bath coupling and the Markov approximation justified by fast correlation time for the bath. The crucial step~\cite{Majenz2013Jul,Mozgunov2020Feb} is to substitute the representation of Eq. (\ref{def:jump}) by a {\em coarse-grained} version,
\begin{eqnarray}\label{leps}
\hat{L}_\nu(\varepsilon)&=&\sqrt{\frac{\gamma(\varepsilon)}{2\pi t_a}}\int_{-t_a/2}^{t_a/2} dt \; e^{i\varepsilon \frac{t}{\hbar} }\hat{A}_\nu(t) =\sum_\omega h(\varepsilon,\omega)\hat{A}_\nu(\omega), \\
h(\varepsilon,\omega) &=& \sqrt{\frac{\gamma(\varepsilon) t_a}{2\pi}}\mbox{sinc}\left[\frac{t_a(\varepsilon -\hbar\omega)}{2}\right],\nonumber
\end{eqnarray}
with $\mbox{sinc}(x)=\sin(x)/x$ and $t_a$ a phenomenological parameter that can de adjusted according to the characteristics of the bath. 
An alternative proposal was formulated in Ref.~\cite{Nathan2020Sep}, which is based on a particular decomposition of the bath correlation function.

Although the previous arguments assume that the jump operators act on the eigenstates of $\hat H_S$, namely we here treat \textit{global} master equations, there are also proposals for \textit{local} master equations, where these operators act on states of the basis of a subsystem of $\hat H_S$. In recent years, intensive discussions have been held about the validity of these approaches~\cite{DeChiara2018Nov,Cattaneo2019Nov,Dann2021Apr} and in particular its consistency with thermodynamics.

%%%%%%%%%%%%%%%%%%%%%%%%%%%%%%%%%%%%%%%%%%%%%%%%%%%%%%%%%%%
\subsubsection{Time-dependent driving}
%%%%%%%%%%%%%%%%%%%%%%%%%%%%%%%%%%%%%%%%%%%%%%%%%%%%%%%%%%%
In the case of time-dependent driving, where we assume that it is the local quantum system that is driven in time, the Hamiltonian becomes time-dependent: $\hat{H}_S \rightarrow \hat{H}_{\rm sys}(t)$. The Lindblad master equation discussed for the stationary case in the previous section is then generalized (see Ref.~\cite{Mozgunov2020Feb}) by substituting the definition of Eq. (\ref{leps}) for the jump operators by 
\begin{eqnarray}
    \hat{L}_\nu(t,\varepsilon)&=&\sqrt{\frac{\gamma(\varepsilon)}{2\pi t_a}}\int_{-t_a/2}^{t_a/2} dt_1 \; e^{i\varepsilon \frac{t_1}{\hbar} }\hat{A}_\nu(t+t_1,t),
\end{eqnarray}
with $\hat{A}_\nu(t',t)=\hat U^\dagger(t',t) \hat{A}_\nu \hat U(t',t)$ and $U(t',t)={\mathscr{T}}\exp[-\frac{i}{\hbar}\int_t^{t'} \hat{H}_{\rm sys}(s) ds]$.  The resulting master equation reads
    \begin{eqnarray}\label{lindt}
    \frac{d\hat{\rho}(t)}{dt}&=&-i\left[\hat{H}_{\rm sys}(t)+\hat{H}_{\rm Lamb}(t),\hat{\rho}(t)\right]\nonumber \\
    & & +\int d\varepsilon \sum_{\nu \nu'} \left( \hat{L}_{\nu'}(t,\varepsilon) 
    \hat{\rho} \hat{L}^\dagger_{\nu}(t,\varepsilon)  -\frac{1}{2}\left\{\hat{L}^\dagger_{\nu}(t,\varepsilon) \hat{L}_{\nu'}(t,\varepsilon),\hat{\rho}(t)\right\}\right).
\end{eqnarray}
This master equation is the starting point for evaluating the dynamics of the driven system. For periodic driving, a Floquet approach has been used~\cite{Mori2023Mar,Grifoni1998Oct}.
Recently, other non-adiabatic quantum master equations~\cite{Dann2018Nov} have been proposed
under the assumption that there is a time scale separation between bath times and driving.

In the following, we provide details on the slow-driving regime, where an adiabatic approximation can be carried out~\cite{Sarandy2004Dec,Albash2012Dec,Kamleitner2013Apr,Cavina2017Aug,Scandi2019Oct,Bhandari2021Jul}.

%%%%%%%%%%%%%%%%%%%%%%%%%%%%%%%%%%%%%%%%%%%%%%%%%%%
\subsubsection{Adiabatic approximation}\label{sec:ME_adiabatic}
%%%%%%%%%%%%%%%%%%%%%%%%%%%%%%%%%%%%%%%%%%%%%%%%%%%
As in previous sections, we focus on a set of parameters ${\boldsymbol X}(t)$ slowly changing in time controlling the dynamics of the driven system $\hat{H}_{\rm sys}[{\boldsymbol X}(t)]$, which is here chosen to be the local open quantum system. 
In a quasistatic description, the parameters are frozen at their values at a  time $t$, $\boldsymbol{X}_t$ and it is possible to find the solution of
\begin{equation}\label{frozen}
    \frac{d\hat{\rho}_t}{dt}={\cal L}_t \hat{\rho}_t,
\end{equation}
with ${\cal L}_t={\cal L} [{\boldsymbol X}_t]$. The stationary state  corresponds to
${\cal L}_t \hat{\rho}_t=0$ and the basis used to express $\hat{\rho}_t$ is the set of eigenstates  $\hat{H}_{\rm sys}[{\boldsymbol X}_t]|a({\boldsymbol X}_t)\rangle =\varepsilon_a({\boldsymbol X}_t) |a({\boldsymbol X}_t)\rangle$.
The adiabatic approximation is the first-order correction at ${\cal O}(\dot{{\boldsymbol X}})$ of this solution,
\begin{equation}
    \hat{\rho}(t)=\hat{\rho}_t+ \delta_{{\boldsymbol X}} \hat{\rho}_t \cdot \dot{{\boldsymbol X}}(t).
\end{equation}
Importantly, $\delta_{{\boldsymbol X}} \hat{\rho}_t $ depends on two effects.
\begin{description}
\item[(i)] The first effect is the change in time of the basis states. 
The modified state can be expressed in the instantaneous basis as follows:
$|\partial_{{\boldsymbol X}} a({\boldsymbol X}_t)\rangle=\sum_{a'\neq a}{\cal A}_{a,a'} |a'({\boldsymbol X}_t)\rangle$, with 
${\cal A}_{a,a'}=\langle a'({\boldsymbol X}_t)|\partial_{{\boldsymbol X}} a({\boldsymbol X}_t)\rangle$. Using the properties
\begin{eqnarray}
& &\langle a'({\boldsymbol X}_t)  |\partial_{{\boldsymbol X}} a({\boldsymbol X}_t)\rangle  =-\langle \partial_{{\boldsymbol X}} a'({\boldsymbol X}_t)  | a({\boldsymbol X}_t)\rangle,\nonumber \\
& & \varepsilon_a\langle \partial_{{\boldsymbol X}} a'({\boldsymbol X}_t) |a({\boldsymbol X}_t)\rangle+
 \varepsilon_{a'}\langle  a'({\boldsymbol X}_t) |\partial_{{\boldsymbol X}}a({\boldsymbol X}_t)\rangle+ \langle  a'({\boldsymbol X}_t) |\partial_{{\boldsymbol X}}\hat{H}_{\rm sys}[{\boldsymbol X}(t)]|a({\boldsymbol X}_t)\rangle=0\nonumber
\end{eqnarray}
we find
\begin{equation}
 {\cal A}_{a,a'}=\left\langle a'({\boldsymbol X}_t)\left|\frac{\partial_{{\boldsymbol X}}\hat{H}_{\rm sys}[{\boldsymbol X}(t)]}{\varepsilon_{a}-\varepsilon_{a'}}\right|a({\boldsymbol X}_t)\right\rangle.
\end{equation}

\item[(ii)] The second effect is due to the change in time of the Liouvillian, $\partial_{{\boldsymbol X}} {\cal L} [{\boldsymbol X}_t]$. 
\end{description}
The combination of the two effects can be expressed as
\begin{equation}
  \delta_{{\boldsymbol X}} \hat{\rho}_t=   \hat{\cal A}\hat{\rho}_t +\hat{\rho}_t\hat{\cal A}^\dagger- {\cal L}_t^{-1} [{\boldsymbol X}_t] \partial_{{\boldsymbol X}} {\cal L}_t [{\boldsymbol X}_t].
\end{equation}
It is interesting to notice that the first terms are related to the unitary dynamics. These are the ones taken into account in the 
{\em adiabatic perturbation theory} for closed systems \cite{Weinberg2017May}. Instead, the second term is related to the non-unitary
dynamics.

\subsubsection{Calculation of particle and energy fluxes}
In the case of the energy flux, we can notice that the rate of change of the internal energy 
stored in the driven system
contains two terms
\begin{equation}\label{master-h}
\frac{d \mbox{Tr}[\hat{\rho}(t) \hat{H}_{\rm sys}(t)]}{dt}= \mbox{Tr}\left[\frac{ d \hat{\rho}(t)}{dt} \hat{H}_{\rm sys}(t)\right]+\mbox{Tr}\left[ \hat{\rho}(t) \frac{\hat{H}_{\rm sys}(t)}{dt}\right].
\end{equation}
The first term can be related to the energy or heat flux
\begin{equation}
    \mbox{Tr}\left[\frac{ d\hat{\rho}(t)}{dt} \hat{H}_{\rm sys}(t)\right]=\sum_{\alpha =1}^{N_r} \mbox{Tr}\left[{\cal D}_\alpha(t) \hat{\rho}(t)\hat{H}_{\rm sys}(t)\right].
\end{equation}
Assuming that the coupling between the system and baths is weak, we can neglect the energy temporarily stored in the coupling, and the right-hand side of
this equation is identified as the sum of the energy fluxes entering the reservoirs
\begin{equation}
    I^E_\alpha(t)= \mbox{Tr}[{\cal D}_\alpha(t) \hat{\rho}(t) \hat{H}_{\rm sys}(t)].
\end{equation}
The second term of Eq. (\ref{master-h}) describes the power developed by the driving,
\begin{equation}
P(t)=\mbox{Tr}\left[ \hat{\rho}(t) \frac{\hat{H}_{\rm sys}(t)}{dt}\right]\,.
\end{equation}
We can follow a similar reasoning to define the particle current in setups where there is exchange of particles between the driven system
and the reservoirs. 
Details for the case of a driven quantum dot using Lindblad master equation can be found in Ref.~\cite{Schulenborg2024Mar}.
From the master equation we can analyze the change in the number of particles stored in the central system,
\begin{equation}
\mbox{Tr}\left[\frac{ d \hat{\rho}(t)}{dt} \hat{N}\right]= \sum_{\alpha} \mbox{Tr}[{\cal D}_\alpha(t) \hat{\rho}(t)\hat{N}].
\end{equation}
From this equation, we can identify the particle flux from each reservoir as
\begin{equation}
    I^N_\alpha(t)= \mbox{Tr}[{\cal D}_\alpha(t) \hat{\rho}(t) \hat{N}].
\end{equation}

It is important to notice that these fluxes are second order in the system-bath coupling. ''First-principles'' calculation of the current up to second order in perturbation theory
with respect to this parameter verify these definitions \cite{Konig1996Mar,Konig1996Dec,Splettstoesser2006Aug,Splettstoesser2008May,Winkler2013Apr,Kashuba2012Jun,Bhandari2020Oct,Bhandari2021Jul,Reina-Galvez2021Dec}.
In these calculations it is important to properly account for the fermionic sign in the different terms of ${\cal D}_\alpha(t)$.

%%%%%%%%%%%%%%%%%%%%%%%%%%%%%%%%%%%%%%%%%%%%%%%%%%%%%%%%%%%%%%%%%%%%%%%%%
\subsubsection{Generalized master equation approach from perturbation theory}\label{sec:ME_perturbative}
%%%%%%%%%%%%%%%%%%%%%%%%%%%%%%%%%%%%%%%%%%%%%%%%%%%%%%%%%%%%%%%%%%%%%%%%%

An alternative approach for the analysis of the dynamics starting from the density matrix of the central system is a real-time diagrammatic approach, perturbative in the tunnel coupling, but without making any further  approximations concerning the many-body interactions in the central system~\cite{Konig1996Mar,Konig1996Dec}. 
The goal of procedure is  the calculation of the mean values of the matrix elements of the reduced density matrix with respect to the full 
many-body state as functions of time. 
This leads to a master equation for the full density matrix elements in lowest order in the tunnel coupling, but allows to systematically include higher orders in the tunnel coupling which are not treated by a standard master equation approach
\begin{equation}\label{eq:generalized_master}
    \dot{\boldsymbol{P}} = -iL\boldsymbol{P} +\int_{-\infty}^t dt' W(t,t') \boldsymbol{P}(t').
\end{equation}
Here, the vector $\boldsymbol{P}$ includes also off-diagonal elements of the density matrix. The Kernel $W(t,t')$ is the transition matrix due to tunneling processes which can in principle contain tunneling events in arbitrarily high order, while the Liouvillian $L$ is local in time and contains the matrix elements of the commutator of the density matrix with the system Hamiltonian.

In the stationary limit the kernel $W$ depends on a time-difference and the integral yields its zero-frequency Laplace transform. It can be calculated using a diagrammatic approach. For weak coupling, meaning that the time-scale on which the density matrix changes due to tunneling is much smaller than the bath time scale, which typically means that the tunnel coupling is much smaller than temperature, $\Gamma/(k_\mathrm{B} T)\ll1$, a perturbation expansion can be performed~\cite{Konig1996Mar,Konig1996Dec,Koenig-thesis}. In lowest order in the tunneling coupling transition matrix elements connecting diagonal elements of the density matrix simply coincide with results from Fermi's golden rule. 

In the presence of slow time-dependent driving, namely in adiabatic response, Eq.~\eqref{eq:generalized_master}, can be further expanded order by order in a small driving parameter, accounting for higher orders in the tunneling coupling and for coherences~\cite{Splettstoesser2006Aug,Splettstoesser2008May,Winkler2013Apr,Kashuba2012Jun}. For this, the time scale imposed by the driving needs to be small with respect to the time-scale on which the density matrix changes and small compared to the support of the kernel~\cite{Riwar2014}. In lowest order in the tunnel coupling, this concretely means $\Omega\delta X/(k_\mathrm{B}T\Gamma$), where $\delta X$ stands for the amplitude of a time-dependent parameter~\cite{Reckermann2010Jun}. With this, the expansion of Eq.~\eqref{eq:generalized_master} involves an expansion justified when the support of the kernel is short compared to the driving time
\begin{equation}\label{eq:generalized_master_exp1}
     \dot{\boldsymbol{P}} = -iL\boldsymbol{P} +\int_{-\infty}^t dt' W(t,t') \left(\boldsymbol{P}(t)+(t'-t)\boldsymbol{\dot{P}}(t)\right)
\end{equation}
together with an expansion of the density matrix itself $\boldsymbol{P}(t)=\boldsymbol{P}^{(0)}(t)+\boldsymbol{P}^{(1)}(t)+...$ and with an expansion of the kernel elements $W(t,t')=W^{(0)}(t,t')+W^{(1)}(t,t')+...$. The latter means that the time dependence of the parameters $X(t')=X(t)+(t'-t)dX/dt$ is considered in the time evolution of the irreducible kernel elements in contributions starting from order $(1)$~\cite{Splettstoesser2006Aug}. It turns out that these corrections to the irreducible kernels only start to contribute to the generalized master equation~\eqref{eq:generalized_master}, when going beyond the sequential-tunneling regime, as analyzed in detail in Ref.~\cite{Splettstoesser2006Aug}.

Instead, at lowest order in the tunnel coupling, only the instantaneous contribution to the kernel enters the master equation: both the expansion of the kernel itself and the expansion around the support time of the kernel given in the last expression in Eq.~\eqref{eq:generalized_master} contribute only in higher orders in the tunnel coupling~\cite{Splettstoesser2006Aug}. Therefore, a re-summation scheme~\cite{Cavaliere2009Sep,Riwar2016Jun} allows to set up a master equation
\begin{equation}\label{eq:ME_resummation}
    \dot{\boldsymbol{P}}(t)=-iL\boldsymbol{P}+\mathcal{W}_t^{0,1}\boldsymbol{P}
\end{equation}
as long as the driving remains slow with respect to the bath time scales~\cite{Riwar2014}, $\Omega\delta X/(k_\mathrm{B}T)^2\ll 1$. Here, $\mathcal{W}_t^{0,1}$ is the zero-frequency Laplace transform of the instantaneous (denoted by the first superscript) kernel with parameters taken at time $t$ (as indicated by the subscript) and evaluated in first order in the tunnel coupling $\Gamma$ (denoted by the second superscript). 
Importantly, the internal dynamics of the system itself can still be fast, such as Landau-Zener-St\"uckelberg transitions in a few-level system. Carefully accounting for the time scale separation of these dynamics~\cite{Riwar2016Jun}, this phenomenon can still be treated within the framework provided by Eq.~\eqref{eq:ME_resummation}. Implementing in this framework Born and Markov approximations, as well as the secular approximation 
leads to the same 
 Liouvillian of the Lindblad equation \cite{Koller2010Dec}.

The non-zero eigenvalues of the kernel $W$ provide the time-scales of the response of the system to external perturbations. Understanding them is therefore key to understanding the dynamics of a time-dependently driven system, see also Sec.~\ref{sec:beyondperturbative}.

%%%%%%%%%%%%%%%%%%%%%%%%%%%%%%%%%%%%%%%%%%%%%%%%%%%%%%%%%%%%%%%%%%%%%%%%%
\subsubsection{Density matrix approaches beyond perturbation theory}\label{sec:beyondperturbative}
%%%%%%%%%%%%%%%%%%%%%%%%%%%%%%%%%%%%%%%%%%%%%%%%%%%%%%%%%%%%%%%%%%%%%%%%%

The time evolution of a (possibly driven) system can more generally be calculated from the time evolution of the density matrix. For a closed system, the time-evolution of the density matrix is given by
\begin{equation}\label{eq:time-evolution_von_Neumann}
    \frac{d \hat{\rho}(t)}{dt}=-\frac{i}{\hbar}\left[\hat{H},\hat{\rho}\right],
\end{equation}
Splitting the Hamiltonian into system-, bath- and coupling terms, allows to write down the time evolution of the reduced system density matrix in the interaction picture. Considering the interaction order by order perturbatively on the Keldysh contour, allows for the treatment described in the previous Sec.~\ref{sec:ME_perturbative}. However, more involved resummation schemes have been developed, where even strong tunnel coupling can be treated in the dynamics of the system, see for example~\cite{Schoeller2009Feb,Saptsov2012Dec,Saptsov2014Jul}. These approaches can for example treat short-time dynamics after quenches for strongly coupled systems.

This more general treatment of the time-evolution operator of the reduced density matrix, reveals a dissipative symmetry, coined fermionic duality, which relates different decay modes and their decay rates to each other~\cite{Saptsov2014Jul,Schulenborg2016Feb,Schulenborg2018}. While valid for the time-evolution operator at arbitrary orders in the tunnel coupling, in lowest order perturbation theory, this symmetry takes the simple form, now for the superoperator $W$ acting on the density operator in Liouville space
\begin{eqnarray}
    W^\dagger = -\Gamma - \mathcal{P}W^\mathrm{dual}\mathcal{P}
\end{eqnarray}
where $\Gamma$ is the lumped sum over all coupling constants, $\mathcal{P}$ is the parity superoperator, while the superscript ``dual" means that we are considering an operator for a fictitious dual system in which all energies ---including the Coulomb interaction energy--- are inverted. This mapping to a dual system is insightful to understand effects in the dynamics of interacting systems that seemingly stem from attractive interaction~\cite{Schulenborg2016Feb}.

%%%%%%%%%%%%%%%%%%%%%%%%%%%%%%%%%%%%%%%%%%%%%%%%%%%%%
\section{Mechanisms}
%%%%%%%%%%%%%%%%%%%%%%%%%%%%%%%%%%%%%%%%%%%%%%%%%%%%%

%%%%%%%%%%%%%%%%%%%%%%%%%%%%%%%%%%%%%%%%%%%%%%%%%%%%%
\subsection{The quantum capacitor: Pure time-dependent charge transport and energy dissipation}\label{intro-ac}
%%%%%%%%%%%%%%%%%%%%%%%%%%%%%%%%%%%%%%%%%%%%%%%%%%%%%

A quantum or mesoscopic capacitor is the most basic system, in which ac currents can be observed while steady-state charge or particle currents do not exist. It consists of an electron cavity coupled to a single electron reservoir and driven by a single ac gate voltage. The basic idea has been introduced by B\"uttiker and coworkers
 in Refs. \cite{Buttiker1993Jun,Buttiker1993Sep}. This device has been later the subject of many other theoretical and experimental works, see details in this section; for applications of the mesoscopic capacitor in single-particle control and quantum optics with electrons, see Secs.~\ref{sec_single-particle-control} and~\ref{sec_QOE}. 
 The nice feature of this device is the fact that it can be regarded as a realization of an RC circuit in the quantum realm. Its dynamics combines the fundamental aspects of storage of charge and energy and of energy dissipation. 
 It is also a beautiful playground to implement and benchmark different approaches. In fact, this is one of the few problems that can be solved by recourse to many different methods. 

Here we introduce  the quantum capacitor  following Refs.~\cite{Splettstoesser2010Apr,Mora2010Sep,Filippone2020Jul} in the limit of a tunnel-coupled cavity. It is described by a Hamiltonian with three components that represent the reservoir, the driven cavity, and the coupling between both systems. 
It reads 
\begin{equation}\label{tot-ham}
\hat{H}(t)=\hat{H}_{\rm res}+\hat{H}_{\rm sys}(t) + \hat{H}_{\rm coup},
\end{equation}
with
\begin{subequations}
\label{eq:q-cap-all}
\begin{align}
\label{q-cap}
\hat{H}_{\rm res} &= \sum_{k} \varepsilon_k \hat{a}^{\dagger}_k \hat{a}_k,\\
\hat{H}_{\rm sys}(t) &= \sum_n \varepsilon_{n} \hat{d}^{\dagger}_n \hat{d}_n + U \left(\hat{N} - \frac{V_{\rm g}(t) C}{q}  \right)^2,\label{eq:q-cap-b}\\
\hat{H}_{\rm coup}&= w \sum_{k,n}  \left(\hat{a}^{\dagger}_k \hat{d}_n + \hat{d}^{\dagger}_n \hat{a}_k \right),
\end{align}
\end{subequations}
where $k$ labels the degrees of freedom of the non-interacting reservoir and and $n$ labels the  electron states of the quantum cavity, which could include the spin degree of freedom or multiple orbital levels.
While the
first set defines a continuum of states, the states of the cavity are discrete and finite with a mean energy level spacing $\Delta$. The operator $\hat{N}=\sum_n \hat{d}^{\dagger}_n \hat{d}_n$ describes the total number of electrons in the cavity, the effect of the gate voltage $V_{\rm g}(t)$ is accounted for by a single driving parameter $ X(t)\equiv 2 U V_{\rm g}(t) C/q$, being $C$ the geometrical capacitance of the cavity, $q$ the electron charge and $U= q^2/(2 C)$ the charging energy in the constant interaction model~\cite{Glazman2003}. 
The tunneling amplitude is $w$, which we here assume to be independent of $k$ and $n$. 

%%%%%%%%%%%%%%%%%%%%%%%%%%%%%%%%%%%%%%%%%%%%%%%%%%%%%
\subsubsection{Charge dynamics in linear response}
\label{sec:capacitor_linear_response}
%%%%%%%%%%%%%%%%%%%%%%%%%%%%%%%%%%%%%%%%%%%%%%%%%%%%%
The driving induces tunneling processes between the cavity and the reservoir. We start by analyzing the charge response of the mesoscopic capacitor in the linear-response regime. This provides insightful expressions for the charge dynamics, since it is straightforward to draw analogies with the case of classical circuits. We denote the
occupation of the cavity for $V_{\rm g}=0$ by $\langle \hat{N}\rangle_0$.
Assuming a small amplitude  of the ac gate voltage, 
the dynamics of the charge in the cavity, namely of the net charge $q N(t)=q \left(\langle \hat{N}\rangle(t)-\langle \hat{N}\rangle_0\right)$, is described by the following Kubo linear-response equation for the Fourier-transformed quantity
\begin{equation}\label{q-qcap-lr}
    q N(\omega) = q^2 \chi_c(\omega) V_{\rm g}(\omega).
\end{equation}
Here, $\chi_c(\omega)$ is the Fourier transform of the response function $\chi_c (t-t')=-i/\hbar \theta(t-t') \langle \left[ \hat{N}(t),\hat{N}(t')\right]\rangle$, which describes the charge fluctuations in the cavity, with $\hat{N}(t)=e^{\frac{i}{\hbar}\hat{H}_0 t} \hat{N}e^{-\frac{i}{\hbar}\hat{H}_0 t}$ and $\hat{H}_0=\hat{H}(V_{\rm g}=0)$. This leads to the definition of the admittance $G(\omega)= -i q\omega N(\omega)/V_{\rm g} (\omega)= -i \omega q^2 \chi_c(\omega)$. In a classical linear RC circuit, this quantity
is related to the resistance and the capacitance as $G(\omega)= -i \omega C/\left(1- i\omega RC \right)$. Focusing on low frequency the admittance can be expanded up to second order in $\omega$, such that one gets $G(\omega)\simeq -i \omega C_\mu\left(1+i \omega RC_\mu \right)$, where $C_\mu$ is in general different from the geometrical capacitance $C$ appearing in Eq.\eqref{eq:q-cap-b}. This leads to
the following identification 
\begin{equation}
\label{rc}
    C_\mu=q^2 \chi_c(0) = q \frac{\partial \langle \hat{N}\rangle}{\partial V_{\rm g} (\omega=0)}, \;\;\;\;\; \;\; R= \frac{1}{q^2 \chi_c(0)^2}\frac{\mbox{Im}\left[\chi_c(\omega)\right]}{\omega}\bigg|_{\omega \rightarrow 0}\,.
\end{equation}
As pointed out in Refs.~\cite{Mora2010Sep,Filippone2011Oct,Filippone2012Sep}, 
at zero temperature, an interesting relation can be shown between the real and the imaginary part of the susceptibility. This  is referred to as ``Korringa-Shiba'' relation and has been proved in the framework of perturbation theory \cite{matveev1991quantum,glazman1990lifting} and in a Fermi-liquid description for the case of a single impurity \cite{Shiba1975Oct},
\begin{equation}\label{k-s}
    \mbox{Im}\left[\chi_c(\omega)\right] = \hbar \pi \omega \left\{\mbox{Re}\left[\chi_c(0)\right]\right\}^2.
\end{equation}
Remarkably, substituting this relation in Eq. (\ref{rc}) we get the expression for the celebrated quantum resistance originally introduced by B\"uttiker and coworkers in the framework of scattering-matrix theory for non-interacting electrons,
\begin{equation}
\label{rq}
    R_\mathrm{qu}= \frac{h}{2q^2}.
\end{equation}
The above reasoning implies that this result is valid also for systems with many-body interactions under
linear response and low frequencies provided that the relation of Eq. (\ref{k-s}) is satisfied. The validity of this relation is easily verified in non-interacting systems. It has been generalized and verified in quantum dots modeled by the interacting Anderson impurity model with and without magnetic field, by means of analytical and numerical methods~\cite{Filippone2011Oct,Filippone2012Sep,Filippone2013Jul,Lee2011May,Kashuba2012Jun,Romero2017Jun,TerrenAlonso2019Mar}.
The quantized resistance $R_\mathrm{qu}$ of the mesoscopic RC circuit has been experimentally demonstrated in Ref. \cite{Gabelli2006Jul}. Progress in the study of this fundamental quantum circuit will be
discussed in Section~\ref{sec:injection_confined}.

%%%%%%%%%%%%%%%%%%%%%%%%%%%%%%%%%%%%%%%%%%%%%%%%%%%%%
 \subsubsection{Energy dynamics}\label{sec:mechanism_energy_dynamics}
%%%%%%%%%%%%%%%%%%%%%%%%%%%%%%%%%%%%%%%%%%%%%%%%%%%%%

 In this section, we analyze the complementary aspect of the energy dynamics.
 Already in the linear response of this simple time-dependently driven system---the mesocopic capacitor---one can observe how the energy dynamics are influenced by time-dependent driving. Concretely, the Onsager coefficents, namely the charge and energy response of the mesoscopic capacitor due to a driving of potential and temperature are modified and Onsager reciprocity is broken as expected from the fact that time-reversal symmetry is broken~\cite{Lim2013Nov,Chen2015May}.
 
 More generally, while the charge dynamics are governed by $RC$-times in linear response or by a charge relaxation time when subject to an arbitrary driving, the energy decay only has the same decay dynamics in special cases, namely where many-body interactions can be neglected and where energy-exchange takes place via a discrete energy level. This special case is also referred to as the ``tight-coupling regime", see e.g.~\cite{Benenti2017Jun}. However, in general the energy dynamics are independent of the charge dynamics and can for example be governed by multiple decay modes, among which the dynamics related to interaction energies~\cite{Splettstoesser2010Apr,Contreras-Pulido2012Feb,Schulenborg2016Feb}. This competition between different decay modes, which can generally be observed in the energy or in other thermodynamically relevant quantities like the nonequilibrium free energy or the relative entropy, have recently been in the focus in the context of the so-called Mpemba effect~\cite{Mpemba1969May,Lu2017May,Bechhoefer2021Aug}, also in driven quantum-dot systems~\cite{Chatterjee2023Aug,Graf2025Apr}.
 
Further distinctive aspects of the energy dynamics compared to the charge dynamics are (i) that for the case of a single reservoir, the only dc mechanism is the \textit{dissipation} of energy and (ii) that temporary energy storage is possible in the lead-system \textit{coupling}. To highlight this, we here focus on the slow-driving, adiabatic-response regime, where the period of the ac driving is much longer than the typical time scale for the electron relaxation. We follow closely Refs.~\cite{Ludovico2016Jul,Ludovico2016Nov}.

Of special interest in this context are conservation laws, where the energy dynamics differ fundamentally from the charge dynamics. In fact, 
 the total number of particles $N_{\rm res}+{N}$ is conserved, 
where $\hat{N}_{\rm res}$ and $\hat{N}$ are the number operators in the reservoir and the few-level system, while
$N_{\rm res}(t)=\langle \hat{N}_{\rm res}\rangle(t)$ and ${N}(t)=\langle \hat{N}\rangle(t)$ are the corresponding the mean values. Taking into account that $\dot{N}_{\rm res}(t)=-\frac{i}{\hbar}\langle[\hat{N}_{\rm res},\hat{H}]\rangle$ and $\dot{N}(t)=-\frac{i}{\hbar}\langle[\hat{N},\hat{H}]\rangle$, the charge conservation law reads
\begin{equation}\label{part-cons}
    -\frac{i}{\hbar}\bigg\langle \!\!\left[\hat{N}_{\rm res},\hat{H}\right]\!\!\bigg\rangle-\frac{i}{\hbar}\bigg\langle \!\!\left[\hat{N},\hat{H}\right]\!\!\bigg\rangle=0.
\end{equation}
It is natural to identify the particle current into each subsystem as $I_{\rm res}^N(t) =  \dot{N}_{\rm res}(t)=-\frac{i}{\hbar}\langle[\hat{N}_{\rm res},\hat{H}]\rangle 
$ and $I_{\rm sys}^N(t) =  \dot{N}(t)=-\frac{i}{\hbar}\langle[\hat{N},\hat{H}]\rangle 
$ and we see that 
Eq.~(\ref{part-cons}) is a time-dependent continuity equation, which is a consequence of 
the particle conservation.
The total energy stored in the system is, instead, not conserved. In fact, the change in time of the mean value $\langle \hat{H}(t) \rangle$ is equal to the power delivered by the 
external driving sources $P(t)$. Hence,
\begin{equation}\label{pext}
    P(t)= \frac{d \langle \hat{H} \rangle }{dt}=-\frac{i}{\hbar}\bigg\langle\!\! \left[\hat{H}_{\rm res},\hat{H}\right]\!\!\bigg\rangle -\frac{i}{\hbar}\bigg\langle \!\!\left[\hat{H}_{\rm coup},\hat{H}\right]\!\!\bigg\rangle 
    -\frac{i}{\hbar}\bigg\langle \!\!\left[\hat{H}_{\rm sys},\hat{H}\right]\!\!\bigg\rangle
    + \bigg\langle \frac{\partial\hat{H}}{\partial t} \bigg\rangle,
\end{equation}
where  we can easily verify
\begin{equation}\label{en-cont}
     -\frac{i}{\hbar}\bigg\langle \!\!\left[\hat{H}_{\rm res},\hat{H}\right]\!\!\bigg\rangle -\frac{i}{\hbar}\bigg\langle \!\!\left[\hat{H}_{\rm coup},\hat{H}\right]\!\!\bigg\rangle 
    -\frac{i}{\hbar}\bigg\langle \!\!\left[\hat{H}_{\rm sys},\hat{H}\right]\!\!\bigg\rangle=0.
\end{equation}
Only in special cases, like in the weak-coupling regime, the energy storage in the coupling is negligible, see e.g. Ref.~\cite{Schulenborg2018}.
It is interesting to compare Eq. (\ref{part-cons}) with Eq. (\ref{en-cont}), where we see that in the second equation there is an extra term which
takes into account the rate of change of the energy stored in the contact. The role of this  term in the heat production and dissipation 
has been pointed out in Ref. \cite{Ludovico2014Apr} and was a subject of further discussion and debate addressed in Refs. \cite{Esposito2015Feb,Esposito2015Dec,Rossello2015Mar,Bruch2016Mar,Ludovico2016Jul,Ludovico2016Nov,Ludovico2018Jan,Dou2018Oct,Haughian2018Feb,Webb2024Jul,Kumar2024Aug}. In the adiabatic-response regime it was argued that, in order to satisfy the second law of thermodynamics, a meaningful definition for the time-dependent heat current in the reservoir is 
\begin{equation}\label{iqt-slow}
    I^Q_{\rm res}(t)=I^E_{\rm res}(t)+\frac{1}{2}\frac{d \langle \hat{H}_{\rm coup}\rangle}{dt} - \mu I_{\rm res}^N(t),\;\;\;\;\;\;{\rm slow\;driving,}
\end{equation}
where we have introduced the definitions of the energy flux in the reservoir $I^E_{\rm res}(t) =-i/\hbar\langle[\hat{H}_{\rm res},\hat{H}]\rangle$ and
the rate of change of the energy temporarily stored in the contact, $d \langle \hat{H}_{\rm coup}\rangle/dt= -i/\hbar\langle [\hat{H}_{\rm coup},\hat{H}]\rangle $. This term has been identified as ``energy reactance'' \cite{Ludovico2018Jan}, where the  factor 1/2 in front of it has been a matter of debate. Even if it is important in the instantaneous  dynamics, it does however not provide a net contribution to the average over time. Hence, upon 
time-average we get the usual definition of the heat flux,
\begin{equation}
    \overline{I}^Q_{\rm res}=\overline{I}^E_{\rm res}- \mu \overline{I}_{\rm res}^N.
\end{equation}
To address the issue of the prefactor of the energy reactance, it is instructive to  solve the problem of a slowly driven non-interacting single-level quantum dot where the system Hamiltonian is linear, $\hat{H}_\mathrm{sys}=\sum_n \epsilon_n \hat{d}_n^\dagger \hat{d}_n$.
For this model,
Schwinger-Keldysh and scattering-matrix theory \cite{Ludovico2014Apr} lead to the same result  under slow driving by adopting the definition of Eq. (\ref{iqt-slow}) for the time-dependent heat current. Furthermore, in this limit it was shown that
\begin{equation}\label{joule}
I^Q_{\rm res}(t)=R_\mathrm{qu} \left[I^c_{\rm res}(t)\right]^2,
\end{equation}
is satisfied, where $I^c_{\rm res}(t)= q I^N_{\rm res}(t)$ is the charge current and $R_\mathrm{qu}$ is the resistance quantum introduced by B\"uttiker and coworkers \cite{Buttiker1993Jun,Buttiker1993Sep}, as
defined in Eq. (\ref{rq}). Interestingly, Eq. (\ref{joule}) has the form of an instantaneous Joule law for the energy that is dissipated as heat in the reservoir. This result, along with the perfect matching between Green's function and scattering-matrix approach provide support for the
definition given in Eq. (\ref{iqt-slow}). 

A complementary point of view is to analyze the energy dissipation. This can be done explicitly in the model of the interacting quantum capacitor presented in Eq.~\eqref{eq:q-cap-all} by solving it within the linear response formalism \cite{Filippone2011Oct,Filippone2012Sep}. 
The power developed by the driving is 
\begin{equation}
    P(t)=  \left\langle\frac{\partial \hat{H}}{\partial t}\right\rangle  =
    -e \dot{V}_g(t) \langle \hat{N}\rangle (t)
\end{equation}
Considering periodic driving, $V_g(t)= V_0 \cos (\omega t)$, and substituting the expressions for the charge fluctuation and the current given by Eqs. (\ref{q-qcap-lr}), the net power over a period can be calculated. 
For the Anderson impurity model, using the Korringa-Siba relation given in Eq. (\ref{k-s}), one finds
\begin{equation}\label{pow-jou}
    \overline{P}= \frac{V_0^2}{2}\omega \mbox{Im}\left[ \chi_c(\omega)\right]= R_\mathrm{qu} \overline{\left[{I}^c_{\rm res}\right]^2}.
\end{equation}
This indicates that the Joule law of Eq. (\ref{joule}) provides also the description of the net energy dissipation in systems with many-body interactions where 
the relation of Eq. (\ref{k-s}) is satisfied. The above result strictly applies to spinless fermions. It can be generalized to electrons with spin, to systems with magnetic field and to large amplitudes of the driving potential \cite{Kashuba2012Jun,Filippone2012Sep,Filippone2013Jul,Romero2017Jun,TerrenAlonso2019Mar} or to Luttinger liquids as shown by bosonization technique~\cite{Sukhorukov2016Mar,Slobodeniuk2013Oct,Spanslatt2024Aug}. The main conclusion of these generalizations is related to the fact that
each channel behaves as in Eq. (\ref{pow-jou}), while the spin and other degrees of freedom define multiple channels that respond in parallel.
The extension to systems with a superconducting lead has been also studied in Refs. \cite{Arrachea2018Jul,Baran2019Aug,Ortmanns2023May}.

%%%%%%%%%%%%%%%%%%%%%%%%%%%%%%%%%%%%%%%%%
\subsubsection{Driven qubit and dissipation} \label{sec_qubit_dissipation}
%%%%%%%%%%%%%%%%%%%%%%%%%%%%%%%%%%%%%%%%%
A similarly simple driven system coupled to baths can also be analyzed in the context of bosonic reservoirs. The simplest device is here a driven qubit represented as a spin-boson model. The structure of the
Hamiltonian is the same as in Eq. (\ref{tot-ham}). The driven system is a two-level system represented as
\begin{equation}
    \hat{H}_{\rm sys}(t)=-\boldsymbol{B}(t)\cdot \hat{\boldsymbol{\sigma}},
\end{equation}
where $\hat{\boldsymbol{\sigma}}=\left(\hat{\sigma}_x,\hat{\sigma}_y,\hat{\sigma}_z\right)$ is a vector of Pauli matrices $\hat{\sigma}_j$ and $\boldsymbol{B}(t)$ contains  time-dependent parameters, leading to a time-dependent level splitting. The reservoir has the same structure as $\hat{H}_{\rm res}$ in Eq. (\ref{q-cap}), but it consists in a set of bosonic modes. 
Typical examples for this type of driven bosonic Hamiltonian occur in cavity or circuit QED where the qubit is embedded in the bosonic environment of a photonic cavity or the normal modes of a quantum LC circuit. The contact between the driven system and the reservoir has the form
\begin{equation}
    \hat{H}_{\rm coup} = \sum_k   \left(\hat{a}^{\dagger}_k +   \hat{a}_k \right) \boldsymbol{V}_k \cdot \hat{\boldsymbol{\sigma}},
\end{equation}
and it describes a process of absorption or emission of a bosonic excitation upon a flip between the ground state and the single excited state of the qubit. 
Unlike the problem of the quantum capacitor, this problem cannot be exactly solved for arbitrary coupling between the reservoir and the qubit, because of the nonlinear nature of the coupling. For weak coupling it can be solved by perturbation theory and quantum master equations. As in the case of the quantum capacitor, the power introduced by the driving protocol 
produces dissipation into the reservoir. However, in the present case, there is no conserved particle transport and there is no associated Joule law as in Eq. (\ref{joule}). 

This problem has been investigated in detail for cyclic protocols, in the limit of slow driving and considering weak coupling between the qubit and the bath, see Refs. \cite{Scandi2019Oct,Abiuso2020Mar,Brandner2020Jan,Miller2020Dec,Miller2021May,TerrenAlonso2022Feb} and  a review in Ref. \cite{arrachea2023energy}. Here we highlight the main points. Starting from the
expression for the power in the adiabatic linear-response formalism (see Sec. \ref{sec:power}), the net dissipated energy  by the driving after a cycle of period $\cal T$ reads
\begin{equation}\label{qdiss}
    Q_{\rm diss}= 
    \int_0^{\cal T} dt \; \dot{\boldsymbol{B}}(t) \cdot \underline{\Lambda} [\boldsymbol{B}(t)] \cdot \dot{\boldsymbol{B}}(t),
\end{equation}
where $\underline{\Lambda} [\boldsymbol{B}(t)]$ is the matrix defined by the adiabatic coefficients 
${\Lambda}_{j,j'} [\boldsymbol{B}(t)] $. The most remarkable feature of this expression is the fact that
only the symmetric part of this matrix contributes to the dissipated energy. Furthermore, since this 
is proportional to the entropy production, the second law of thermodynamics implies that this quantity is
positive definite. These are mathematical properties compatible with the definition of a metric in the parameter space and this has motivated  the concept of the {\em thermodynamic length}
\cite{Weinhold1975Sep,Salamon1980Jul,Salamon1983Sep,Diosi1996Dec,Crooks2007Sep},
\begin{equation}\label{length}
    {\cal L}=\int_{t_1}^{t_2} dt \sqrt{ \dot{\boldsymbol{B}}(t) \cdot \underline{\Lambda} [\boldsymbol{B}(t)] \cdot \dot{\boldsymbol{B}}(t)},
\end{equation}
which is the length of the curve parametrized by $t$ in the space of parameters $\boldsymbol{B}$ connecting 
$t_1$ and $t_2$. Using the Cauchy-Schwarz inequality $\int_{t_0}^{t_1} dt f^2 \int_{t_1}^{t_2} dt g^2 
\geq \left[\int_{t_1}^{t_2} f g dt \right]^2$, with $g=1$ and $f$ being the argument of Eq. (\ref{length}), one finds
\begin{eqnarray}
     Q_{\rm diss}\geq \frac{{\cal L}^2}{\cal T},
\end{eqnarray}
which means that the dissipated energy is lower-bounded by a geometric quantity that depends on the path.
Interestingly, the lower bound is obtained when the integrand is constant along the path, which is equivalent
to a protocol where the velocity is adjusted in order to satisfy a constant dissipation rate at each point of the trajectory \cite{Salamon1980Jul,Salamon1983Sep}. These ideas have been the basis for many 
studies devoted to finding optimal protocols to minimize the dissipated energy \cite{Scandi2019Oct,Abiuso2020Mar,Brandner2020Jan,Miller2020Oct,
Miller2020Dec,Miller2021May,TerrenAlonso2022Feb,VanVu2023Feb,Zhen2021Nov}.

%%%%%%%%%%%%%%%%%%%%%%%%%%%%%%%%%%%%%%%%%%%%%%%%%%%%%
\subsection{Pumping}\label{intro-pump}
%%%%%%%%%%%%%%%%%%%%%%%%%%%%%%%%%%%%%%%%%%%%%%%%%%%%%
In the previous section we discussed pure ac transport processes that may take place in a driven system in contact to a \textit{single} reservoir.
We now turn to analyze processes where the driven system is in contact to two or more reservoirs and/or sources and where the time-dependent driving can result in a directed (dc) current. Studying dc currents means that \textit{on average} no charge accumulation occurs on the central system, such that charge conservation always leads to  $q \sum_{\alpha} \dot{N}_{\alpha}=0$, where the sum runs over all the reservoirs. 

A rather trivial example for dc transport induced by time-dependent driving is rectification: an ac bias leads to an ac current of which a conductor ``cuts off" one of the current directions. In other words, (a part of) the current induced by the bias during one half of the driving period is selected by for example a conductor acting as an energy filter thereby producing a dc transport current. 
However, dc transport can also truly be \textit{induced} by the time-dependent driving and take place in the absence or even against external stationary biases. This is the case in so-called quantum pumps: A quantum pump is a quantum system under the action of periodically time-dependent parameters which induce a net flux per cycle. Examples are
\begin{description}
\item[(i)] Particle pumps, where an open quantum system that is capable of temporarily storing particles is in contact with two or more reservoirs and subject to an asymmetric cyclic operation. As a consequence of the time-dependent driving, a
net amount of particles is transported between reservoirs. This results in dc particle currents in the absence of a chemical potential bias or even against the chemical potential bias.
\item[(ii)] Heat pumps, where heat is transported between two or more thermal baths as a response to time-dependent driving, in the absence of a temperature bias or against a temperature bias. Here, the open driven quantum system needs to be able to \textit{temporarily} store energy.
\item[(iii)] Power pumps, where power is exchanged between two or more driving sources. These last two examples provide connections between time-dependent quantum transport and the implementation of small-scale (heat) engines and motors, see Sec.~\ref{sec:energy_conversion}.
\end{description}

Most of these mechanisms have a classical counterpart and the term ``quantum'' applies because of the nature of the driven system and because the outcome usually reflects properties like quantum coherence and the quantum statistics of particles. 
Often, the mechanism of pumping is illustrated with an Archimedes pump, namely a pipe with a rotating screw, which pumps water against gravity, or with a peristaltic pump, similar to biological systems, where subsequent modulations of system parts lead to motion (like, e.g., movement of worms). These classical devices properly capture the idea of an asymmetric cyclic operation leading to transport. In fact, breaking the symmetry between the source and drain reservoirs in the cyclic driving protocol is a necessary condition to have a directed flux. This ingredient is the same pointed out in examples of ``quantum ratchets'' \cite{Reimann1997Jul,Kohler2005Feb,Hanggi2009Mar}, where the paradigmatic example, presented in the famous Feynman lectures \cite{Feynman2011Oct}, is a wheel with a sawtooth border contacted by a pawl and connected to a paddle wheel through a gear. 
The operation of the pawl leading to directed motion of the wheel involves work injected from outside, fundamental for the energy balance. This illustrates the possibility of generating a directed particle current in the quantum realm induced by time-periodic driving with a suitable breaking of both spatial and time-reversal symmetries. Hence, many 
systems identified as ``quantum ratchets'' can be also regarded as quantum pumps. 

A quantum pump is hence described by a time-dependent periodic Hamiltonian like the one introduced in Eq. (\ref{hamil0}) while satisfying $\hat H(t)=\hat H(t+{\cal T})$ with the driving period ${\cal T}$. 

%%%%%%%%%%%%%%%%%%%%%%%%%%%%%%%%%%%%%%%%%%%%%%%%%%%%%
\subsubsection{Adiabatic charge and energy pumping}\label{sec:mechanism_chargepumping}
%%%%%%%%%%%%%%%%%%%%%%%%%%%%%%%%%%%%%%%%%%%%%%%%%%%%%
A relevant scenario to understand the pumping mechanism is the adiabatic-response regime, corresponding to a low driving frequency or in other words a long driving period ${\cal T}$ compared to the typical time scale associated to the equilibrium system, as introduced in Secs.~\ref{sec:adiabatic_kubo},~\ref{adia-scatt},~\ref{sec:greens_adiabatic} and~\ref{sec:ME_adiabatic}. As a consequence, adiabatic pumping reveals equilibrium properties of the driven system, instead of for example photon-assisted tunneling, where transport takes place due to absorption of energy quanta from the driving. We briefly discuss the mechanism here
on the basis of the adiabatic linear-response treatment introduced in Sec.~\ref{sec:adiabatic_kubo}.

The most remarkable property of adiabatic pumping is its relation to geometric properties similar to the Berry phase \cite{Victor1984Mar}. From the expression for the
particle current given in Eq. (\ref{eq:adia-n-e}), the net charge transported in a cycle can be written as follows
\begin{equation}\label{pumped-charge}
    q \Delta N_{\alpha}=q \oint_{{\cal C}_{\boldsymbol{X}}} \boldsymbol{\Lambda}_\alpha({\boldsymbol{X}}) \cdot d{\boldsymbol{X}},
\end{equation}
where we have introduced the notation $\boldsymbol{\Lambda}_\alpha(\boldsymbol{X}) =\left(\Lambda_{\alpha,1}(\boldsymbol{X}), \ldots,\Lambda_{\alpha,N}(\boldsymbol{X})\right)$ and we have used the fact that the response function depends parametrically on time, hence, $\Lambda_{\alpha, j }(\boldsymbol{X}) \equiv \Lambda_{I^N_\alpha,F^j} 
     \left[\boldsymbol{X}(t)\right]$.  The structure of Eq. (\ref{pumped-charge}) reflects the fact that the pumped charge  can be calculated as a line integral over the protocol 
${\cal C}_{\boldsymbol{X}}$
defined by the parameters
under their temporal evolution. The vector defined from the response functions has the structure of a Berry connection. This representation is useful to highlight the mathematically necessary conditions to have a non-vanishing value of the pumped charge. These are: (i) a finite area in the parameter space enclosed by ${\cal C}_{\boldsymbol{X}}$, which requires a minimum of two driving parameters with a phase difference in their time dependence (ii) adiabatic response functions $\Lambda_{\alpha,j}(\boldsymbol{X})$ that define a non-vanishing rotation over the protocol ${\cal C}_{\boldsymbol{X}}$. The second condition is difficult to predict and depends on the microscopic details of the driven device, its coupling to the reservoirs, as well as on the  driving protocol. Spatial asymmetries in these quantities are necessary but not sufficient conditions.

A widely studied example is a mesoscopic quantum dot or two serially coupled quantum dots in a two-terminal configuration \cite{Geerligs1990May,
Kouwenhoven1991Sep,
Pothier1992Jan,Switkes1999Mar,DiCarlo2003Dec,Buitelaar2008Sep,Roche2013Mar}, where the driving parameters are 
two gate potentials that change periodically but asynchronously in time. The key of the pumping mechanism is a net transport of charge between the two reservoirs as a consequence of the local driving, without applying any extra voltage at the two terminals. This problem has been addressed in many configurations and realizations, including models of non-interacting electrons and also considering many-body interactions. Consequently,  many theoretical methods have been used to solve it.
The scattering matrix formalism for non-interacting electrons has been particularly illuminating in providing explicit expressions with the structure of Eq. (\ref{pumped-charge}) in particular devices \cite{Brouwer1998Oct,Altshuler1999Mar,Makhlin2001Dec}. However, the geometic structure of adiabatic pumping has been worked out in detail also relying on other methods, for example using generalized quantum master equations~\cite{Calvo2012Dec,Monsel2022Jul} where also the role of the Landsberg phase was highlighted~\cite{Pluecker2017Apr,Pluecker2017Nov}. This has been a crucial step in unveiling the geometrical nature of pumping in the adiabatic-response regime. Further progress was done after the proposal of the Floquet version of this formalism \cite{Moskalets2002Nov}, of relevance in understanding this mechanism beyond the adiabatic regime. Electron quantum pumps have been studied in a significant
number of devices using scattering matrix theory~\cite{Brouwer1998Oct,Polianski2001Jul,Moskalets2004May,Moskalets2005Jul}, non-equilibrium Green's functions\cite{Aono2004Sep,Splettstoesser2005Dec,Sela2006Apr,Fioretto2008Jun,Lopez2003Jan,Arrachea2005Sep,Arrachea2005Sep-1,Stefanucci2008Feb,FoaTorres2005Dec}, renormalization-group techniques~\cite{Eissing2016Dec,Eissing2016Jan,Kennes2012Feb} and master equations~\cite{Splettstoesser2005Dec,Splettstoesser2006Aug,Calvo2012Dec,Yuge2012Dec,Pluecker2017Apr,Monsel2022Jul}, reaction-coordinate mapping for the treatment of non-Markovian effects~\cite{Restrepo2019Jul} among others.

Of interest is not only pumping of charge. Also the topic of spin-pumping in electronic systems has been addressed extensively~\cite{Governale2003Oct,Splettstoesser2008May,Futterer2010Sep,Riwar2010Nov,Winkler2013Apr,Dittmann2016Aug,Zheng2003Sep,Liu2011Sep,Entin-Wohlman2020Aug,Chowdhury2024Apr,Chowdhury2025Jan}. In this review, we will however for conciseness not focus on the topic of spin pumping and refer instead to the Review presented in Ref.~\cite{Rojek2014Sep}.

Since electrons do not only carry charge but also energy, akin to charge pumping, also energy pumping in driven quantum systems is in general expected. Indeed, in the adiabatic-response regime, the adiabatic linear response procedure 
explained before leads to the following relation between the energy flux into the reservoir $\alpha$ and the driving parameters,
\begin{equation}\label{ieal}
    I^E_{\alpha} (t) =  \sum_j \Lambda^E_{\alpha, j }(\boldsymbol{X}) \dot{X}_j(t).
\end{equation}
The response functions are the ones defined in Eq. (\ref{eq:adia-n-e}), $\Lambda^E_{\alpha, j }(\boldsymbol{X}) = \Lambda_{I^E_\alpha,F^j} 
     \left[\boldsymbol{X}(t)\right]  $.
In driven systems where charge transport takes place, we define the heat flux as $ I^Q_{\alpha} (t) = I^E_{\alpha} (t) - \mu_{\alpha} I^N_{\alpha} (t) $, being $\mu_{\alpha}$ the chemical potential of the reservoir $\alpha$. The geometric nature of the heat pumping, in analogy to charge pumping was pointed out  in Refs. \cite{Avron1992Apr,Avron2003Aug,Niu1990Apr}.
In systems where the reservoirs are represented by bosonic excitations, corresponding to normal modes of harmonic oscillators, like phonons, photons, etc, the energy current coincides with the heat current. In both regimes, the net heat transported per cycle between the two reservoirs 
can be expressed in terms of geometric quantities as in Eq. (\ref{pumped-charge}),
\begin{equation}\label{pumped-heat}
    Q_{\alpha}=\oint_{{\cal C}_{\boldsymbol{X}}} \boldsymbol{\Lambda}^Q_\alpha(\boldsymbol{X}) \cdot d{\bf X},
\end{equation}
with $\boldsymbol{\Lambda}^Q_\alpha(\boldsymbol{X})=\boldsymbol{\Lambda}^E_\alpha(\boldsymbol{X})-\mu_{\alpha}\boldsymbol{\Lambda}_\alpha(\boldsymbol{X})$ for electron reservoirs or
$\boldsymbol{\Lambda}^Q_\alpha(\boldsymbol{X})=\boldsymbol{\Lambda}^E_\alpha(\boldsymbol{X})$ for reservoirs of bosonic modes. 

\subsection{Power pumping in adiabatic response}\label{sec:power-pumping}
Extending the discussion of  Section \ref{sec_qubit_dissipation} to a more general context where 
the system is driven by several time-dependent parameters $\boldsymbol{X}(t)$, 
we introduce  $\underline{\Lambda}(\boldsymbol{X})$ for the matrix with elements $\Lambda_{j,j'}(\boldsymbol{X})$ (see Sec. \ref{sec:power}). These coefficients
satisfy Onsager
reciprocal relations: $ \Lambda_{j, j' }(t) =\pm \Lambda_{j',j}(t)$, where the $\pm$ sign depends on
the parity of the operators 
$\hat{F}^j$ under time reversal and on other symmetries of the device. Examples are 
particle density operators in the case of driven electron systems or the probability density in the Bloch sphere, like in the case of the two level system. These operators are even under time-reversal symmetry. However, in other situations, it is possible to have force operators with different parities,
in which case $\underline{\Lambda}(\boldsymbol{X})=\underline{\Lambda}^{\rm s}(\boldsymbol{X})+
\underline{\Lambda}^{\rm a-s}(\boldsymbol{X})$ has symmetric ($\underline{\Lambda}^{\rm s}(\boldsymbol{X})$) 
and antisymmetric ($\underline{\Lambda}^{\rm a-s}(\boldsymbol{X})$)  components. 

We highlight here again that
the  heat production generated by the driving 
is related to the symmetric component this matrix. Eq. (\ref{qdiss}) is in the more general case
expressed as 
\begin{equation}
    Q_{\rm diss}=\int_0^{\cal T} dt \dot{\boldsymbol{X}}(t) \cdot \underline{\Lambda}^{\rm s}(\boldsymbol{X}) \cdot \dot{\boldsymbol{X}}(t).
\end{equation}
This component defines the metric and the thermodynamic length as discussed in  Section \ref{sec_qubit_dissipation},
and it is related to the net entropy production as $T \Sigma= Q_{\rm diss}$.

Instead, the power pumping mechanism is related to the antisymmetric component as follows
\begin{equation}
    P^{\rm (pump)}=\frac{1}{2 \cal T} \sum_{\ell,\ell'}\int_0^{\cal T} dt \dot{\boldsymbol{X}}(t) \cdot \left[\Lambda^{\rm a-s}_{\ell, \ell'}(\boldsymbol{X}) - \Lambda^{\rm a-s}_{\ell', \ell}(\boldsymbol{X})\right]\cdot \dot{\boldsymbol{X}}(t)
\end{equation}
Antisymmetric components of the $\underline{\Lambda}(\boldsymbol{X})$ tensor have been analyzed in Refs. \cite{Bode2011Jul,Bode2012Feb,Bode2012Mar} in the framework of electron systems and in \cite{Campisi2012Sep} for charged harmonic oscillators coupled to a time-dependent electric field.

This mechanism is intimately related to the problem of topological frequency conversion \cite{Martin2017Oct,Nathan2019Mar,Nathan2020Sep,Esin2025Apr} and 
work-work conversion \cite{Cangemi2020Oct,Cavaliere2022Sep}, which have been recently explored in the context of driven qubits beyond the adiabatic regime.

%%%%%%%%%%%%%%%%%%%%%%%%%%%%%%%%%%%%%%%%%%
\subsection{Adiabatic-response energy conversion and quantum machines}\label{sec:energy_conversion}
%%%%%%%%%%%%%%%%%%%%%%%%%%%%%%%%%%%%%%%%%%
Interestingly, as pointed out in Sec.~\ref{sec:mechanism_chargepumping}, the combination of time-dependent driving and dc bias by chemical potential and/or temperature differences imposed at the
reservoirs, results in processes involving energy conversion, such as realized in engines or motors.

One of such mechanisms is identified as a {\em motor}
or a {\em generator} and it is realized when pumping takes place under the presence 
of a dc voltage bias \cite{Bustos-Marun2013Aug,Calvo2017Oct,Fernandez-Alcazar2017Apr,Napitu2015Jun}.  In a motor, power due to the applied bias is converted into power injected into the driving fields; in a generator, power spent due to the time-dependent driving is converted into electrical power with a charge current flowing against a potential bias. 
Here we discuss the mechanism with focus on a two-terminal device with a small voltage bias $\delta \mu$ so that it can be treated in linear response, which 
is combined with the adiabatic linear response in the driving velocities $\dot{\boldsymbol{X}}$.

We consider the reservoir with the lowest $\mu$ as a reference and omit the reservoir label in what follows.
The particle current entering this reservoir and the mean value of the
induced forces in this framework read 
\begin{eqnarray}\label{linear-el}
I^N(t) &= &   \Lambda_{N,N }(\boldsymbol{X}) \delta \mu+ \sum_{j=1}^{M} \Lambda_{N, j }(\boldsymbol{X}) \dot{X}_j(t)  , \nonumber \\
F_j(t) &= &   \Lambda_{j,N }(\boldsymbol{X}) \delta \mu + \sum_{j'=1}^{M} \Lambda_{j, j' }(\boldsymbol{X}) \dot{X}_{j'}(t),\;\;\;j=1, \ldots, N.
\end{eqnarray}
Here, the different coefficients $\Lambda_{\mu,\nu}$ are again susceptibilities evaluated with the equilibrium
Hamiltonian defined by the parameters frozen at time $t$. The coefficient $\Lambda_{N,N}(\boldsymbol{X})$ is related to the electrical conductance. Using the notation of Section \ref{sub:par-en-flux}, it reads
$\Lambda_{N,N}(\boldsymbol{X})=\int dt' (t-t')\chi_{I^{N},I^N}(t-t')$,
while
$\Lambda_{j,N }(\boldsymbol{X})$ is the response of the $j$-th force to the chemical potential bias. In the notation of 
Section \ref{sub:par-en-flux} it reads $\Lambda_{j,N }(\boldsymbol{X})=\int dt' (t-t')\chi_{F^j,I^N}(t-t')$. These coefficients
satisfy Onsager
reciprocal relations: 
%$ \Lambda_{j, j' }(t) =\pm \Lambda_{j',j}(t)$ and
$\Lambda_{j,N }(t)=\pm \Lambda_{N,j}(t)$, where $\pm$ depends on the parity of the operators 
$\hat{F}^j$ under time reversal and on other symmetries of the device. In what follows, we assume a situation where 
$\Lambda_{j,N }(t)=- \Lambda_{N,j}(t)$. This choice is consistent with the fact that the current is odd under time reversal, while $\hat{F}^j$ are usually densities, which are even \cite{Ludovico2016Feb,TerrenAlonso2019Mar} (see Section \ref{sec:power-pumping}).

The relevant quantities that determine the performance of the motor
are the electrical power produced (or dissipated) by the transported charge and the power invested (or received) by the external driving sources. Averaged over a cycle they read
\begin{eqnarray}
    W_{\rm el} & = &q^2 \int_0^{\cal T} dt I^N(t) \delta \mu =  q^2 \int_0^{\cal T} dt \Lambda_{N,N }(t) \left(\delta \mu\right)^2+
    q\oint_{{\cal C}_{\boldsymbol{X}}} \boldsymbol{\Lambda}(\boldsymbol{X}) \cdot d \boldsymbol{X} \delta \mu ,\label{wel} \\
    W_{X} & = & \sum_j \int_0^{\cal T} dt F_j(t) \dot{X}_j = 
    - q \oint_{{\cal C}_{{\bf X}}} \boldsymbol{\Lambda}(\boldsymbol{X}) \cdot d\boldsymbol{X} \delta \mu
    +\int_0^{\cal T} dt \dot{\boldsymbol{X}} \cdot \underline{\Lambda}(\boldsymbol{X}) \cdot \dot{\boldsymbol{X}},\label{wx}
\end{eqnarray}
where we have defined $\underline{\Lambda}(\boldsymbol{X})$ for the matrix with elements $\Lambda_{j,j'}(\boldsymbol{X})$
and the vector $\boldsymbol{\Lambda}(\boldsymbol{X}) $ with components $\Lambda_{N, j }(\boldsymbol{X})$. The diagonal terms in Eqs. (\ref{wel}) and (\ref{wx}), namely the ones proportional to $\delta\mu^2$ and $\dot{X}^2$, are positive in the present sign convention and are associated to energy dissipation. In fact, the net dissipated energy associated to the net entropy production is given by 
\begin{equation}
Q_{\rm diss}= W_{\rm el}+  W_X= q^2\int_0^{\cal T} dt \Lambda_{N,N }(t) \left(\delta \mu\right)^2+
\int_0^{\cal T} dt \dot{\boldsymbol{X}} \cdot \underline{\Lambda}(\boldsymbol{X}) \cdot \dot{\boldsymbol{X}}.
\end{equation}
We can identify Eq. (\ref{qdiss}) in the second term, which accounts for the dissipation due to the driving.
The off-diagonal terms, namely the ones proportional to both the bias and the driving parameter derivatives, describe the energy-conversion processes and are the fundamental ones  
for the motor operation. Notice that the terms obtained by the closed integral over the driving protocol, $\oint_{{\cal C}_{\boldsymbol{X}}}$, provide the pumped charge between the two reservoirs
induced by the driving. We can identify two different operations: (i) When the charge is pumped downstream with respect to the bias $\delta \mu$, the electrical work has an extra component, while
the same amount of work is \textit{received} by  the driving sources. This operation is identified as a motor and its efficiency is defined as $\eta^{\rm mot}=-W_X/W_{\rm el}$. (ii) When the charge is pumped against the
$\delta \mu$, we have a generator. In this case, the driving sources must \textit{invest} work
in order to transport charge against the dc bias. The efficiency is defined as
$\eta^{\rm gen}=-W_{\rm el}/W_X$.

When, instead of a chemical potential bias, a temperature bias $\delta T$ is imposed, while the coldest reservoir has temperature $T$, the driven system behaves as a {\em heat engine}
or  as a {\em refrigerator}.  As before, for a small  $\delta T/T$ this problem  can be treated in linear response by relying on the
Luttinger Hamiltonian representation discussed in Section \ref{sec:temperature bias},
in
combination with the adiabatic linear response in the velocities $\dot{\boldsymbol{X}}$. This leads to a set of equations, similar to Eqs. (\ref{linear-el}). For simplicity we focus on the case of exchange of energy without particles with the reservoirs, so that the energy flux is directly interpreted as the heat flux. Considering the coldest reservoir as a reference and omitting the reservoir label for simplicity, we can proceed in analogy to the lines resulting in Eqs.~(\ref{linear-el}) by introducing the response functions
$\Lambda_{E,E}(\boldsymbol{X})=\int dt' (t-t')\chi_{I^{E},I^E}(t-t')$, and $\Lambda_{j,E }(\boldsymbol{X})=\int dt' (t-t')\chi_{F_j,I^N}(t-t')$, satisfying 
the Onsager relations. As before, we assume $\Lambda_{j,E }(\boldsymbol{X})=-\Lambda_{E,j }(\boldsymbol{X})$. In terms of these quantities,
the net heat $Q$ entering the coldest reservoir per cycle, as well as the power invested by the driving sources are written as follows
\begin{eqnarray}
   Q & = & \int_0^{\cal T} dt I^E(t) \frac{\delta T}{T} =   \int_0^{\cal T} dt \Lambda_{E,E }(\boldsymbol{X}) \;
   \left(\frac{\delta T}{T}\right)^2+
    \oint_{{\cal C}_{{\boldsymbol X}}} \boldsymbol{\Lambda}^E({\boldsymbol X}) \cdot d{\boldsymbol X} \frac{\delta T}{T} ,\label{wt} \\
    W_{X} & = & \sum_j \int_0^{\cal T} dt \; F_j(t) \dot{X}_j = 
    - \oint_{{\cal C}_{{\boldsymbol X}}} \boldsymbol{\Lambda}^E({\boldsymbol X}) \cdot d{\boldsymbol X} \frac{\delta T}{T}
    +\int_0^{\cal T} dt \dot{{\bf X}} \cdot \underline{\Lambda}({\boldsymbol X}) \cdot \dot{{\boldsymbol X}}.\label{wx-t}
\end{eqnarray}
Here, we have introduced the notation $\boldsymbol{\Lambda}^E(\boldsymbol{X}) $ as a vector with components $\Lambda_{E, j }(\boldsymbol{X})$.
 As in the case of motors and generators, the diagonal terms contribute to the dissipation and the entropy production.
Instead, the off-diagonal components are related to heat-work conversion mechanisms, and depend on the geometrically  pumped heat. The two operational modes are: (iii) heat engine, corresponding to the pumped heat entering the cold reservoir coming from the hot reservoir. In such a case, the first term of Eq. (\ref{wx-t}) describes the work released by the driven system into the driving sources. The efficiency of this machine is defined as $\eta^{\rm he}=-W_X/Q $. (iv) The refrigerator corresponds to
heat extracted from the coldest reservoir, in which case, the second term of Eq. (\ref{wt}) has a larger absolute value than the
first one with an opposite sign. This implies that the first term of Eq. (\ref{wx-t}) contributes in the same direction
as the one provided by the pure driving sources, namely the power provided by the driving is dissipated and used to extract heat from the cold reservoir. The efficiency of the refrigerator, typically called the coefficient of performance in this case, reads ${\rm COP}=-Q/W_X $. 
As usual, the second law imposes bounds for the efficiency or coefficient of performance of the machine: it can be shown that
\begin{equation}
\eta^{\rm he} \leq \eta_\mathrm{Carnot}, \;\;\;\;\; {\rm COP} \leq {\rm COP}_\mathrm{Carnot}, \;\;\;\;\; \eta^{\rm mot}, \; \eta^{\rm gen} \leq 1
\end{equation}
where $\eta_\mathrm{Carnot}=\delta T/T$ and ${\rm COP}_\mathrm{Carnot}=T/\delta T$ are, respectively the  efficiency of the Carnot heat engine and the
coefficient of performance of the Carnot refrigerator\footnote{The expression for $\eta_{\mathrm{Carnot}}$ is valid in linear response.}. For the case of the motor and the generator mechanisms, the bound does not depend on the temperature and the mechanism may exist even at zero temperature \cite{Bustos-Marun2013Aug,Calvo2017Oct,Fernandez-Alcazar2017Apr,Ludovico2018Dec,Bruch2018May}.

There are many examples of heat engines and refrigerators operating without particle fluxes. These are basically continuous finite-time versions of the four-stroke Carnot, Otto and Stirling cycles.
Pioneering works in this direction are \cite{Esposito2010Oct,Schmiedl2007Dec,Geva1992Sep}, more recent contributions are Refs.
\cite{Juergens2013Jun,Cavina2017Aug,Brandner2020Jan,Abiuso2020Mar,TerrenAlonso2022Feb}, and recent reviews are Refs. \cite{arrachea2023energy,Myers2022Apr,Cangemi2024Oct}.

%%%%%%%%%%%%%%%%%%%%%%%%%%%%%%%%%%%%%%%%%%
\subsection{Adiabatic-response quantum transport induced by classical mechanical and magnetic dynamics}
%%%%%%%%%%%%%%%%%%%%%%%%%%%%%%%%%%%%%%%%%%

Up to here, we have considered quantum transport induced by slow external control of system parameters modifying the potential landscape of electrons. The applied driving fields---typically realized by gates---have therefore been treated as classical. There are however other scenarios where the quantum particles that intervene in the transport process are coupled to different  
degrees of freedom obeying a slow classical dynamics, which follows its own equation of motion. Two examples are nanomechanical electronics as well as quantum transport due to spin exchange with a magnetic moment.

In nanomechanical electronics, electron quantum transport  takes place with the electrons coupled
to vibrational degrees of freedom of a structure that is described by one or more classical displacements $X_j(t)$.
These effects are relevant in experimental scenarios taking place in transport in nanoelectromechanical systems \cite{Craighead2000Nov,Naik2006Sep,Steele2009Jul,Tabanera-Bravo2024Mar},
suspended quantum dots in carbon nanotubes \cite{Weig2004Jan,LeRoy2004Nov,Lassagne2009Jul,Urgell2020Jan},
graphene sheets \cite{Bunch2007Jan} and molecular systems \cite{Park2000Sep}.
This problem has been studied in many theoretical works \cite{Brandbyge1995Aug,Verdozzi2006Jul,Lu2012Jun,Lu2016May,Hussein2010Oct,Ohm2012Apr,Arrachea2014Sep,Rudge2023Mar,Rudge2024May,Ribetto2022Oct,Stefanucci2023Sep}.

Here, we present the treatment of Refs. \cite{Pistolesi2008Aug,Bode2011Jul,Bode2012Feb} formulated in terms of the adiabatic linear response. 
 The emerging picture in nanoelectromechanical systems is that of vibrational modes represented by classical coordinates $X_j$, whose dynamics is affected by the non-equilibrium environment generated by the
electron currents. On general grounds, we expect those modes to be described by a Langevin equation of the form
\begin{equation}\label{lange}
    M_j \ddot{X}_j + \frac{\partial U}{\partial X_j}= F^{\rm cl}_j -\sum_{j'} \gamma_{j,j'}\dot{X}_{j'}+\xi_j,
\end{equation}
where the left-hand side is the equation of motion for the free vibrational modes, while the effect of the environment provided by the electrons is taken into account on the right-hand side.
The term $F^{\rm cl}_j$ is the {\em classical} force acting on the mode $j$.  The matrix $\gamma_{j,j'}$  accounts for the effect of friction generated by the environment and $\xi_j$  represents the noise, characterized
by a correlation 
\begin{equation}\label{cl-noise}
\langle \xi_j(t) \xi_{j'}(t')\rangle =D_{j,j'}(t,t'), 
\end{equation}
which is usually local in the mode coordinate and time, $D_{j,j'}(t,t')= D_0 \delta_{j,j'} \delta(t-t')$. The interesting
interplay comes from the joint treatment of the quantum-mechanical description for the electron dynamics, which leads to microscopic expressions for the forces and the matrices $\gamma_{j,j'}$ and
$D_{j,j'}$. 

The starting point is a Hamiltonian for the full  system of the form 
\begin{equation}\label{ham-el-x}
    \hat{H}= \hat{H}_X + \hat{H}_{\rm el}+ \hat{H}_{{\rm el}-X}.
\end{equation}
The first term describes the free vibrational modes,
\begin{eqnarray}
    \hat{H}_X= \sum_j \frac{\hat{P_j}^2}{2M_j}+ U(\hat{\boldsymbol{X}}),
\end{eqnarray}
where $\hat{\boldsymbol{X}}$ is a vector with components $\hat{X}_j$. These are operators describing quantum-mechanical vibrational modes. 
The term $\hat{H}_{\rm el}$ describes the free electrons, including their coupling to the reservoirs, which may have a voltage or temperature bias. The last term describes the coupling between the electrons and the vibrational modes. For sake of concreteness, we assume here the following coupling,
\begin{equation}
    \hat{H}_{{\rm el}-X}= \sum_{n,n'}  \hat{c}^{\dagger}_n \left[V \left(\hat{\boldsymbol{X}}\right)\right]_{n,n'} \hat{c}_{n'} ,
\end{equation}
where $\hat{c}^{\dagger}_n $ and $\hat{c}_n$ are creation and annihilation electron operators in state
$n$. We focus on situations where the time scale for the electronic dynamics is much shorter than the time scale of the vibrational dynamics. Having in mind this assumption, 
we proceed with the derivation of Eq. (\ref{lange}) as follows. The equation of motion for the components $ X_j(t)=\langle\hat{X}_j\rangle (t)$ of the vector $\boldsymbol{X}(t)$
is
\begin{equation}
    M_j \ddot{X}_j(t)+ \left\langle \frac{\partial U}{\partial \hat{X}_j}\right\rangle(t) =-
      \sum_{n,n'}\left\langle \hat{c}^{\dagger}_n \left[\partial_{\hat{X}_j} V \left(\hat{\boldsymbol{X}}\right)\right]_{n,n'} \hat{c}_{n'} \right\rangle(t),
\end{equation}
with $\ddot{X}_j(t)=\langle \ddot{\hat{X}}_j(t) \rangle$, and $\hat{X}_j(t)$ is expressed in the Heisenberg representation with respect to the full Hamiltonian. This equation is approximated by decoupling the dynamics of $\hat{\boldsymbol{X}}$ from that of $\hat{c}^{\dagger}_n$ and $\hat{c}_n$ making use of the assumption of separated time scale. 
Consequently, we express the term in the right-hand side as 
\begin{equation}
\left\langle \hat{c}^{\dagger}_n \left[\partial_{\hat{X}_j} V \left(\hat{\boldsymbol{X}}\right)\right]_{n,n'} \hat{c}_{n'} \right\rangle(t)\simeq\left\langle \left[\partial_{\hat{X}_j} V \left(\hat{\boldsymbol{X}}\right)\right]_{n,n'} \right\rangle_{\!\!X}(t)\, N_{n,n'}(\boldsymbol{X}(t)),
\end{equation}
where the first mean value is taken with respect to $\hat{H}_X$. The mean values of the fermionic operators are instead calculated with respect to the effective Hamiltonian $\hat{H}^{\rm eff}\left(\boldsymbol{X}(t)\right) $, which is obtained from Eq. (\ref{ham-el-x}) by replacing the operator $\hat{\boldsymbol{X}}(t)$ with the mean value $\boldsymbol{X}(t)=
\langle \hat{\boldsymbol{X}}(t)\rangle_X
$.
We introduce the definition of fermionic densities $\hat{N}_{n,n'}=\hat{c}^{\dagger}_n  \hat{c}_{n'}$ and
$N_{n,n'}(\boldsymbol{X}(t))= {\langle \hat{N}_{n,n'}(t)\rangle}_{\rm eff}$ 
for these mean values.
With these approximations, one finds
\begin{eqnarray}\label{eq-mot}
     M_j \ddot{X}_j(t)+ \left\langle \frac{\partial U}{\partial \hat{X}_j}\right\rangle(t)  &\simeq &
      -
      \sum_{n,n'}\left\langle \left[\partial_{\hat{X}_j} V \left(\hat{\boldsymbol{X}}\right)\right]_{n,n'} \right\rangle(t)\, N_{n,n'}(\boldsymbol{X}(t)).
\end{eqnarray}
As a next step, and taking into account the slow dynamics of the vibrational modes, it is possible to implement the adiabatic linear response treatment for the fermionic densities
\begin{eqnarray}
       N_{n,n'}\left(\boldsymbol{X}(t)\right) & \simeq &
       {\langle \hat{c}^{\dagger}_n  \hat{c}_{n'} \rangle}_t (\boldsymbol{X}) + \sum_{j'} \Lambda_{n,n',j'} (\boldsymbol{X}) \;  \dot{X}_{j'}(t),
\end{eqnarray}
where the coefficients are
\begin{eqnarray}
\Lambda_{n,n',j }(t)& =& \int_{-\infty}^t dt' (t-t') \chi_{n,n',j}(t-t') \\\chi_{n,n',j}(t-t')&=&-i \theta(t-t') {\left\langle \left[\hat{N}_{n,n'}(t), \hat{N}_{n,n'}(t')\right]\right\rangle}_{\!t}\, \left[\partial_{X_j} V(\boldsymbol{X})\right]_{n,n'}\,.
\end{eqnarray}
Here, the mean values ${\langle\bullet\rangle}_t$ are calculated with respect to the effective frozen Hamiltonian $\hat{H}_{ t}^{\rm eff}$, which corresponds to the parameters $\boldsymbol{X}(t)$ frozen at time
$t$ in the effective Hamiltonian
$\hat{H}^{\rm eff}\left(\boldsymbol{X}(t)\right) $.
 
Comparing with Eq. (\ref{lange}), we can identify the forces and the friction coefficients, 
\begin{eqnarray}
F^{\rm cl}_j(\boldsymbol{X})&=& - \sum_{n,n'} \left[\partial_{X_j} V(\boldsymbol{X})\right]_{n,n'}{\langle \hat{c}^{\dagger}_n  \hat{c}_{n'} \rangle}_t (\boldsymbol{X})\,, \nonumber \\
\gamma_{j,j'}(\boldsymbol{X}) &=& \sum_{n,n'} \left[\partial_{X_j} V(\boldsymbol{X})\right]_{n,n'}\Lambda_{n,n',j'} (\boldsymbol{X})\,. 
\end{eqnarray}
 Interestingly, this description enables a microscopic calculation for the damping
coefficient, representing  the environment of electrons coupled to the nanomechanical degree of freedom. The corresponding noise $\xi_j$  and noise correlation
defined in Eq. (\ref{cl-noise})
can also be derived in a similar way from microscopic calculations. The environment 
in this type of problems is out-of equilibrium and is generated by coherent transport of electrons through the nanostructure.
Simultaneously, the effect of the nanomechanical motion on the electron transport can also be taken into account by means of the adiabatic response
induced to the parameters $\boldsymbol{X}(t)$, as explained in Sec. \ref{sec:adiabatic_kubo}. In Ref. \cite{Bode2011Jul,Bode2012Feb}, the response
functions are calculated with scattering matrix and non-equilibrium Green's functions formalism, respectively. Further interesting ways in which pumping is induced by mechanical motion have for example been identified in twisted bilayer graphene, where mechanically induced sliding motion induces electrical currents~\cite{Zhang2020Jan} or via the movement of a kink in buckled graphene~\cite{Suszalski2020Oct}.

The other interesting example of combined classical and quantum dynamics corresponds to electron systems coupled via spin exchange interaction with a magnetic moment $\boldsymbol{M}(t)$ that can be described classically. The dynamics of the precessional motion of a magnetic moment with damping can be described by the Landau-Lifshitz-Gilbert equation.
This equation has been the basis for describing the dynamics of spintronics in thin films~\cite{Tserkovnyak2002Feb} and molecular magnetic systems~\cite{Delgado2010Jan}. It is akin to the Langevin equation and reads
\begin{equation}
\label{llg}
    \dot{\boldsymbol{M}}=\boldsymbol{M} \times \left[-\partial_{\boldsymbol{M}}U-J \boldsymbol{s}^0-\gamma\dot{\boldsymbol{M}} +\delta \boldsymbol{B} \right],
\end{equation}
where the first term represents the effect of the conservative forces, and the other terms represent, respectively, the spin torque, the friction and
the noise generated by the electrons. 
As in the case of the nanomechanical system, it is possible to calculate the explicit expressions for $\boldsymbol{s}^0, \; \gamma, \; \delta \boldsymbol{B}$ in terms of a microscopic model under the assumption of a different time scale for the dynamics of the electrons and $\boldsymbol{M}$. The derivation of Eq.~\eqref{llg} from a microscopic Hamiltonian for a driven molecular magnet was presented in Ref.~\cite{Bode2012Mar} following 
the adiabatic expansion of time-dependent Green's functions and scattering matrix. Similar approaches were followed in Refs.~\cite{Brataas2008Jul}. Landau-Lifshitz-Gilbert-like equations have also been derived for generalized master equations for pumping of electronic spins through metallic islands and quantum dots induced by rotating magnetizations~\cite{Winkler2013Apr,Rojek2014Sep}. Other models, including effects like spin-orbit coupling~\cite{Petrovic2018Nov,Xu2023Oct,Mahfouzi2012Oct} and bosonic baths~\cite{Anders2022Mar} were also considered.
The case of the strong driving regime was addressed in Ref.~\cite{Bajpai2020Oct}, while related geometric properties were analyzed in Ref.~\cite{Shnirman2015Apr}.

%%%%%%%%%%%%%%%%%%%%%%%%%%%%%%%%%%%%%%%%%%%%%%%%%%%%%
\section{Focus areas and concrete challenges}\label{sec_focus_areas}
%%%%%%%%%%%%%%%%%%%%%%%%%%%%%%%%%%%%%%%%%%%%%%%%%%%%%

%%%%%%%%%%%%%%%%%%%%%%%%%%%%%%%%%%%%%%%%%%%%%%%%%%%%%
\subsection{Single-particle control and transport spectroscopy}\label{sec_single-particle-control}
%%%%%%%%%%%%%%%%%%%%%%%%%%%%%%%%%%%%%%%%%%%%%%%%%%%%%

Single-electron control in quantum conductors is motivated by questions ranging from fundamental aspects to applications. Having control over single electrons in a conductor is the basis for observing tunable single- and few-particle quantum and interaction effects, but it is also a crucial ingredient for metrology or the realization of flying qubits.
With single-particle control, we mean the ability to investigate and manipulate physical observables, such as electrical currents, down to the single-electron level. For instance, in pumps and turnstiles, the transfer of one electron per cycle allows quantized currents to be obtained at a high level of precision, finding applications in metrology. Another aspect is the ability to generate on-demand single-electron excitations in quantum conductors, realizing the electronic equivalent of single-photon sources. Finally, transport quantities induced by time-dependent drives can also be used as spectroscopic tools to acquire information on some properties of the conductor itself, such as screening properties and decay rates. Often, such properties are not accessible by steady-state techniques, making time-dependent transport a valuable tool.

\begin{figure}[t]
    \centering
    \includegraphics[width=0.8\textwidth]{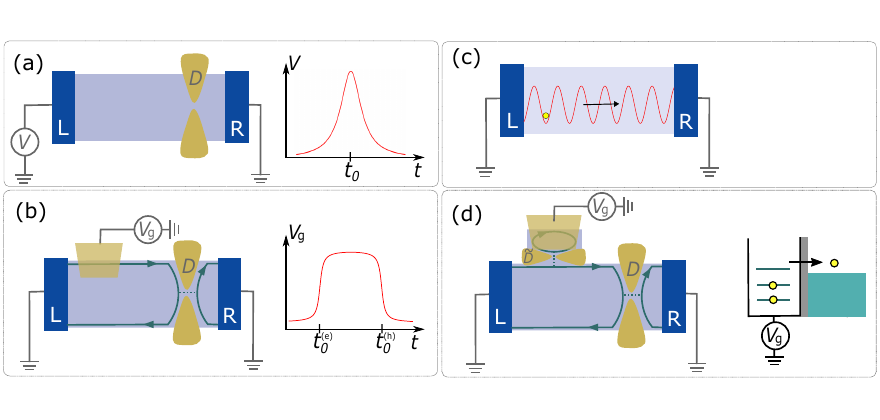}
    \caption{Principles of single-electron injection (a) via a specifically designed bias voltage (b) Time-dependent gate voltage applied to chiral conductors (c) transport of a captured electron in a moving potential minimum (d) injection from a driven discrete spectrum (from a confined conductor region) (e) exploiting energy gaps. In setups (a), (b), and (d), the presence of a quantum point contact (with transmission probability $D$) allows one to perform spectroscopy on the emitted state, e.g., by looking at the noise.}
    \label{fig:injection_principles}
\end{figure}

In this Section, we highlight some of the efforts to realize and investigate this basic step for further implementations of time-dependently controlled devices. In Secs.~\ref{sec:injection_target_drives} and~\ref{sec:injection_confined} we review two types of strategies to achieve controlled, on-demand, single-electron injection, via engineered voltage pulses and exploiting discrete energy levels from confined regions, respectively, summarized in Fig.~\ref{fig:injection_principles}. In Sec.~\ref{sec_spectroscopy}, we discuss how time-dependent transport can be used as a spectroscopic tool.

%%%%%%%%%%%%%%%%%%%%%%%%%%%%%%%%%%%%%%%%%%%%%%%%%%%%%
%%%%%%%%%%%%%%%%%%%%%%%%%%%%%%%%%%%%%%%%%%%%%%%%%%%%%

%%%%%%%%%%%%%%%%
\subsubsection{Single-particle injection via targeted driving signals}
\label{sec:injection_target_drives}
%%%%%%%%%%%%%%%

We review here two strategies that allow one to excite single-electrons via voltage pulses. In the first case, Fig.~\ref{fig:injection_principles}(a), one applies a voltage bias $V(t)$ to one of the contacts connected to the conductor. In general, this induces a phase shift on the propagating electrons, in a similar way to what we have discussed in the context of Floquet theory. The seminal result of Ivanov et al.~\cite{Ivanov1997Sep} is that there is a specific shape of $V(t)$ so that the phase shift corresponds to processes by which electrons only gain energy, in a way that no holes are generated.
In the second case, Fig.~\ref{fig:injection_principles}(b), one considers coupling the conductor to a gate electrode, where a gate potential $V_{\rm g}(t)$ is applied. Following the same logic, in order to excite single electrons the voltage should be engineered to produce a phase shift with the desired property. The additional complication, in this case, is that one must first find the true potential felt by the propagating electrons, which is in general different from $V_{\rm g}(t)$. We start with the first approach, which is conceptually simpler, and discuss in some details how the shape of $V(t)$ can be used to the desired task of generating a single-electron state.\\

%%%%%%%%%%%%%%%%%%%%%%%%%%%%%%%%%%%%%%%%%%%%%%%%%%%%%%%%%%%%%%%%%%%%%%%
\paragraph{Injection by time-dependent voltage biases and Levitons}
%%%%%%%%%%%%%%%%%%%%%%%%%%%%%%%%%%%%%%%%%%%%%%%%%%%%%%%%%%%%
When a time-dependent driving of the \emph{bias} voltage, $V(t)$, is applied to a coherent conductor, the system is brought out of equilibrium and a time-dependent charge current $I^c(t)$ is induced. For the setup in Fig.~\ref{fig:injection_principles}(a), the current flowing towards to quantum point contact simply $I^c(t)=q^2V(t)/h$, assuming that the conductor supports a single channel.
The idea of using voltage pulses to generate single- or few-electron states is rather simple: if the drive is such that $\int dt\,I^c(t)=q$, then a single charge $q$ is transferred. However, for a generic shape of the pulse $V(t)$, this ac driving results in the creation of electron-hole pairs or photo-assisted shot noise~\cite{Reydellet2003Apr,Rychkov2005Oct,Battista2014Aug,Vanevic2016Jan} accompanying the desired single-electron excitation.
This creation of electron-hole pairs therefore counteracts the aim to create controlled and precise single-particle excitations. However, when driving with a Lorentzian-shaped bias voltage 
\begin{equation}
 V_\mathrm{Lor}(t) =  \frac{\hbar}{q}\frac{2\sigma}{\left(t-t_0\right)^2+\sigma^2}  \label{eq:VoltageLeviton}
\end{equation}
the creation of electron-hole pairs is fully suppressed. Here, $t_0$ is the injection time and $\sigma$ the half width at half maximum of the voltage pulse. The state generated by this voltage has later been called a \emph{Leviton}~\cite{Dubois2013Oct}, after the seminal works of Levitov~\cite{Levitov1996Oct,Ivanov1997Sep,Keeling2006Sep}.
The reason for the full suppression of electron-hole pair creation can be understood in the following way. Particles being exposed to a generic time-dependent voltage drive $V(t)$, pick up the phase $e^{-i\phi(t)}$, where
\begin{equation}
\phi(t)=\frac{q}{\hbar}\int_{-\infty}^t dt' V(t')
    \label{eq:Faraday-flux}
\end{equation}
is given by the Faraday flux. This phase determines the probability amplitude
\begin{equation}
    c_\varepsilon=\int_{-\infty}^\infty dt e^{-i\phi(t)}e^{i\varepsilon t/\hbar}\label{eq:picked_up_energy}     
\end{equation}
that electrons are scattered from one energy state to another, with $\varepsilon$ being the energy picked up~\cite{Moskalets2002Nov}. The probability for this process to occur is $p(\varepsilon)=\left|c_\varepsilon\right|^2$. Requiring that no holes are excited amounts to the condition $c_\varepsilon=0$ for all $\varepsilon<0$, i.e., no states below the Fermi energy (here taken at $\varepsilon=0$) are created. This requirement translates into a constraint on the function $e^{-i\phi(t)}$ and, thus, on the voltage $V(t)$. In particular, $e^{-i\phi(t)}$ has to be analytic in the lower complex plane (and have at least one pole in the upper plane in order not to vanish everywhere). Considering that $|e^{-i\phi(t)}|=1$, the simplest possible choice is
\begin{equation}
    e^{-i\phi(t)}=\frac{t-t_0+i\sigma}{t-t_0-i\sigma}\,,
\end{equation}
which leads to the Lorentzian profile $V(t)=V_{\rm Lor}(t)$ of Eq.~\eqref{eq:VoltageLeviton}.

The experimental evidence for Levitons was achieved by exploiting a periodic train of Lorentzian pulses of generic amplitude, namely
\begin{equation}
    V(t)=\mathfrak{a}\sum_{m\in\mathbb{Z}}V_{\rm Lor}(t-m\mathcal{T}),
    \label{eq:periodic_Lorentzian}
\end{equation}
and sending the resulting state to a quantum point contact. The parameter $\mathfrak{a}$ is a tunable, continuous dimensionless quantity, and has a strong impact on the shot noise observed after partitioning the excitations at a quantum point contact.
In fact, the theory predicts that the noise should be minimal whenever $\mathfrak{a}$ is a positive integer. To understand this, let us first take a step back and consider a generic periodic voltage
\begin{equation}
    V(t)=V_{\rm dc}+V_{\rm ac}(t)=V(t+\mathcal{T})\,,\qquad\int_0^{\mathcal{T}}V_{\rm ac}(t)dt=0\,.
    \label{eq:generic_dc+ac_voltage}
\end{equation}
Then, we can use the tools of Floquet theory, slightly adapting the definitions of Sec.~\ref{sec:Floquet_ac_voltages}.
Specifically, the phase factor $e^{-i\phi(t)}$ introduced in Eq.~\eqref{eq:Faraday-flux} can be expanded in a Fourier series of the form
\begin{equation}
    e^{-i\phi(t)}=\exp\left(-i\frac{q}{\hbar}V_{\rm dc}t\right)\sum_{\ell\in\mathbb{Z}}c_\ell(\mathfrak{a}) e^{-i\ell\Omega t}\equiv \exp\left(-i\mathfrak{q}\Omega t\right)\sum_{\ell\in\mathbb{Z}}c_\ell(\mathfrak{a}) e^{-i\ell\Omega t}\,.
\end{equation}
Here, the dimensionless parameter $\mathfrak{q}=qV_{\rm dc}/(\hbar\Omega)$ represents the number of charges (i.e., the charge in units of $q$) per period carried by the drive. Moreover, the coefficients $c_\ell(\mathfrak{a})$ are the same as those defined in Eq.~\eqref{eq:photoassisted_amplitudes}, and depend on the ac component $V_{\rm ac}(t)$ \emph{only}. The functional form depends on the shape of the voltage, while $\mathfrak{a}=qV_{\rm ac}^{(0)}/(\hbar\Omega)$ is a dimensionless parameter encoding the characteristic amplitude $V_{\rm ac}^{(0)}$ of $V_{\rm ac}(t)$, see for instance below Eq.~\eqref{eq:photoassisted_amplitudes}. The coefficients $c_\ell$ replace the amplitudes $c_\varepsilon$ introduced in Eq.~\eqref{eq:picked_up_energy} for non-periodic sources. Similarly, $p_\ell(\mathfrak{a})=|c_\ell(\mathfrak{a})|^2$ replace the probabilities $p(\varepsilon)$.
If $V_{\rm ac}^{(0)}=V_{\rm dc}$, such as in Eq.~\eqref{eq:periodic_Lorentzian}, then $\mathfrak{a}=\mathfrak{q}$ also represents the number of charges (i.e., the charge in units of $q$) per period carried by the drive $V(t)$.

It is possible to show that $c_{\ell}(\mathfrak{a})=0$ for all $\ell<-\mathfrak{a}$, whenever $\mathfrak{a}\in\mathbb{N}$. This property is once again a consequence of the analytic structure of $e^{-i\phi(t)}$ and is the key ingredient determining the anticipated minimal-noise feature of Lorentizan pulses. Indeed, the zero-temperature current fluctuations across the quantum point contact of the setup in Fig.~\ref{fig:injection_principles}(a) can be evaluated as~\cite{Dubois2013Aug}
\begin{equation}
    \Delta\mathcal{S}\equiv\mathcal{S}-\mathcal{S}^{\rm dc}=\frac{2q^2}{h}D(1-D)\hbar\Omega\sum_{\ell<-\mathfrak{q}}p_\ell(\mathfrak{q})|\ell+\mathfrak{q}|\propto N_{\rm e/h}^{\rm ex}\overset{V_{\rm Lor}}{=}0,
\label{eq:minimal_noise_Lor}
\end{equation}
showing that Lorentzian pulses with integer $\mathfrak{q}$ do not produce any excess noise above the dc component $\mathcal{S}^{\rm dc}$, and no extra particle-hole pairs $N_{\rm e/h}^{\rm ex}$ are created. It is also important to notice that the excess noise $\Delta\mathcal{S}$ has local minima at integer $\mathfrak{q}$ for \emph{any} kind of driving signal. This is related to the so-called dynamical orthogonality catastrophe~\cite{Ivanov1997Sep}, which implies that current fluctuations are enhanced whenever the Faraday flux~\eqref{eq:Faraday-flux} is not a multiple of $2\pi$. As a result, the excess noise is minimized for all pulses at integer $\mathfrak{q}$, when the flux is instead a multiple of $2\pi$. The Lorentzian drive is the optimal drive, i.e., the only one leading to $\Delta\mathcal{S}=0$ at integer $\mathfrak{q}$.

Another curious property of the Lorentzian drive, is that even the Faraday flux $\phi=\pi$ is special. In this case, it was shown that strictly zero-energy quasiparticle excitations with fractional charge can be generated~\cite{Moskalets2016Jul}. They are always accompanied by the emission of particle-hole pairs (as the drive is non optimal), but they produce the minimal noise in junctions between a normal metal and a superconductor~\cite{Belzig2016Jan}.

From a practical point of view, Lorentzian pulses are generated in the lab by Fourier synthesis, namely by properly combining monochromatic signals of different frequencies. The minimal-noise property, hints of which are already visible for signals approaching the shape of a Lorentzian, but composed of few harmonics only~\cite{Vanevic2007Aug,Vanevic2012Dec}, was first demonstrated in Ref.~\cite{Dubois2013Aug}.
%%%%%%%%%%%%%%%%%%%%%%%
\begin{figure}[t]
    \centering
    \includegraphics[width=0.5\linewidth]{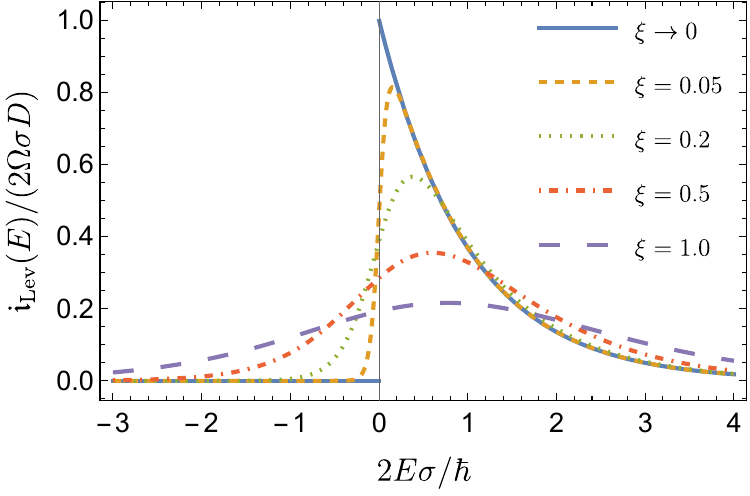}
    \caption{Temperature dependence of the spectral current $\mathfrak{i}_{\rm Lev}(E)$ for a Leviton with unit charge, $\mathfrak{q}=\mathfrak{a}=1$. The temperature dependence is encoded in the parameter $\xi=2k_{\rm B}T\sigma/\hbar$. Figure adapted from Ref.~\cite{Dashti2019Jul}.}
    \label{fig:spectral_current_Lev}
\end{figure}
%%%%%%%%%%%%%%%%%%%%%%%%%%%%%%

Another way to confirm that only excitations above the Fermi energy are created by the drive~\eqref{eq:VoltageLeviton} is by looking at the spectral current $\mathfrak{i}(E)$ of the emitted signal. This quantity has been introduced in Eq.~\eqref{eq:spectral_current_general}, which in the two-terminal configuration of Fig.~\ref{fig:injection_principles}(a) reduces to 
\begin{equation}
    \mathfrak{i}(E)=D\sum_{\ell=-\infty}^\infty \left|c_\ell(\mathfrak{a})\right|^2\left[f_\mathrm{L}(E_\ell)-f_\mathrm{R}(E)\right] = D\sum_{\ell=-\infty}^\infty \left|c_\ell(\mathfrak{a})\right|^2\left[f_0(E_\ell-\mathfrak{q}\hbar\Omega)-f_0(E)\right]
    \label{eq:generic_spectral_current}
\end{equation}
where $f_0(E)=[1+\exp(\beta E)]^{-1}$ is the equilibrium Fermi function. As demonstrated in~\cite{Dubois2013Aug} for Levitons and a harmonic drive, shot noise spectroscopy can be used to access information on the spectral distribution by varying $\mathfrak{a}$ and $\mathfrak{q}$ independently.
For a Leviton with unit charge, Eq.~\eqref{eq:generic_spectral_current} reduces to
\begin{equation}
    \mathfrak{i}_{\rm Lev}(E)= 2\Omega \sigma D\,e^{-2E\sigma/\hbar}\theta(E)\,,
\end{equation}
where this expression holds at zero temperature and when the period $\mathcal{T}$ is the largest time scale in the problem. The finite-temperature expression has been obtained in~\cite{Dashti2019Jul} and the result is illustrated in Fig.~\ref{fig:spectral_current_Lev}, showing a thermal smearing, with a decay eventually dominated by temperature rather than the pulse width $\sigma$.

Before moving to the description of a different injection strategy, we mention some additional features of minimal excitations generated by voltage pulses.  Ivanov et al.\ showed that overlapping pulses appear to behave independently (as solitons), and their overlap does not affect the total charge noise~\cite{Ivanov1997Sep}. However, this feature does not survive when energy transport is considered. In that case, $N$ overlapping Lorentzian pulses produce energy currents and energy fluctuations that scale as $N^2$ and $N^3$ times the single-particle quantities, respectively~\cite{Battista2014Aug}, showing that the excited pulses do not strictly behave independently (unless of course their overlap vanishes). Despite this difficulty, it is possible to define a proper excess energy noise, inspired by the Schottky relation, that vanishes even for overlapping pulses~\cite{Vannucci2017Jun}.
Second, reference~\cite{Glattli2018Mar} has extended the study of multiple-pulse excitations by considering a pseudorandom binary injection protocol, characterized by a binary pattern $\{\pi_k\}$ with entries $0$ or $1$, such that at each time $k\mathcal{T}$ a particle is injected if $\pi_k=1$ and no particle is injected if $\pi_k=0$. It was shown that (i) Levitons maintain the minimal noise property and (ii) the number of extra electron/hole pairs generated by other drives is not only a property of the driving function shape, but also of the injection protocol specified by the probabilities $\pi_k$. \\

%%%%%%%%%%%%%%%%%%%%%%%%%%%%%%%%%%%%%%%%%%%%%%%%%%%%%%%%%%%%%%%%%%%%%%%%%%%%%%%%%%%
\paragraph{Injection by gate driving}
%%%%%%%%%%%%%%%%%%%%%%%%%%%%%%%%%%%%%%%%%%%%%%%%%%%%%%%%%%%%%%%%%%%%
\begin{figure}[t]
    \centering
    \includegraphics[width=0.6\textwidth]{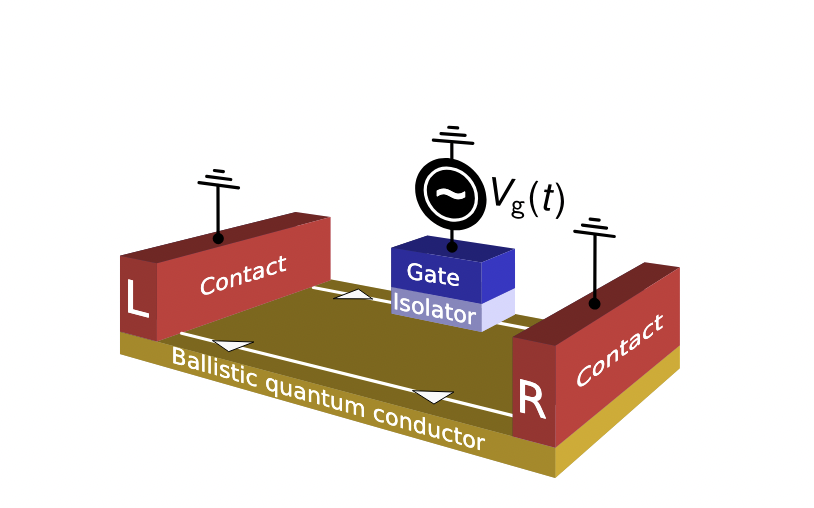}
    \caption{Injection scheme where electronic excitations are locally created in a conductor in the quantum Hall regime by time-dependently driving a capacitively coupled gate. Figure taken from Ref.~\cite{Misiorny2018Feb}.}
    \label{fig:gate drive}
\end{figure}
%%%%%%%%%%%%%%%%%%%%%%%%%%%%%%%%%%%%%%%%%%%%%%%%%%%%%%%%%%%%%%%%%%%
An alternative to applying a time-dependent drive to the voltage \textit{bias}, is by driving a time-dependent \textit{gate} voltage, as shown in Fig.~\ref{fig:injection_principles}(b) and in Fig.~\ref{fig:gate drive} in greater detail. This is particularly relevant for chiral systems, such as conductors in the quantum Hall regime, where the propagation direction of electron excitations is given by the chirality of the edge channels. In this injection scheme, the voltage signal applied to the gate is transmitted to the conductor via capacitive coupling. The externally controlled drive is the gate voltage $V_{\rm g}(t)$, but what matters for the excitations in the conductor is the \emph{local} potential $V_{\rm loc}(t)$. The relation between the two is nontrivial and mainly depends on screening effects and the interplay between the driving frequency and the response timescales of the system. There are three relevant time scales:
\begin{description}
    \item[(i)] The charge relaxation time $\tau_{\rm c}$ set by the capacitive coupling between the gate and the conductor, which is given by $\tau_{\rm c}=hC/q^2$, with $C$ the geometrical capacitance.
    \item[(ii)] The traversal time $\tau_{\rm f}$ that a particle in the conductor takes to go across the gated region.
    \item[(iii)] The dwell time $\tau_{\rm g}$ on the gate plate that forms the external part of the capacitor.
\end{description}
All these quantities combine in the admittance $G(\omega)$ of the conductor which gives the current response in the frequency domain
\begin{equation}
    I^c(\omega)=G(\omega)V_{\mathrm{g}}(\omega)\,.
\label{eq:current_response_gate_frequency}
\end{equation}
Assuming a perfect transmission between the right and left reservoirs, as well as chiral propagation, the result is~\cite{Misiorny2018Feb}
\begin{equation}
    G(\omega)=-i\omega Cg(\omega)\,,\qquad
    \frac{1}{g(\omega)}=1-\frac{i\omega\tau_{\rm c}}{1-\exp(i\omega\tau_{\rm f})}-\frac{1}{M_{\rm g}}\frac{i\omega\tau_{\rm c}}{1-\exp(i\omega\tau_{\rm g})}\,,
\end{equation}
where $M_{\rm g}$ is the number of transport channels in the gate electrode. For a metallic conductor, with $M_{\rm g}\to\infty$, the above result was also found in Ref.~\cite{Litinski2017Aug}.
Performing a low-frequency expansion, one has
\begin{equation}
    G(\omega)=-i\omega C_\mu(1+i\omega R C_\mu)+\mathcal{O}(\omega^3)\,,
\end{equation}
where
\begin{align}
    R&=\frac{h}{2q^2}\left(1+\frac{1}{M_{\mathrm{g}}}\right)\,,\\
    C_\mu&=\left(\frac{1}{C}+\frac{h}{q^2\tau_{\mathrm{f}}}+\frac{1}{M_{\mathrm{g}}}\frac{h}{q^2\tau_{\mathrm{g}}}\right)^{-1}
\end{align}
are the equivalent charge relaxation resistance and the electrochemical capacitance, respectively. Together, they combine in an effective $RC$ time for the system, $\tau_{RC}=RC_\mu$. When $M_{\rm g}\to\infty$, $R$ becomes the B\"uttiker resistance $R_\mathrm{qu}=h/(2q^2)$~\cite{Buttiker1993Sep,Pretre1996Sep,Gabelli2006Jul}, see also Eq.~\eqref{rq}.

Taking the Fourier transform of~\eqref{eq:current_response_gate_frequency}, we find the time-resolved current
\begin{equation}
    I^c(t)=C\frac{d}{dt}\int_{-\infty}^{+\infty}dt' g(t-t')V_{\rm g}(t')\,.
\end{equation}
It is challenging to obtain analytical expressions, due to the complicated form of the function $g(\omega)$. However, it is possible to make progress by using an expansion of the gate drive in harmonics to efficiently obtain numerical results at arbitrary driving frequency $\Omega$. Referring to~\cite{Misiorny2018Feb} for details, we here discuss two limiting cases.

In the adiabatic driving regime, defined by $\Omega\tau_{RC}\ll 1$, the relation between injected current and gate-voltage signal is
\begin{equation}
    I^c(t)=C_\mu\frac{dV_{\rm g}}{dt}\,,
\label{eq:gating_derivative}
\end{equation}
resulting in a purely capacitive response. Therefore, one must engineer $V_{\rm g}$ so that its derivative produces the desired signal. This means that clean, minimal excitations corresponding to those generated by a Lorentzian voltage bias are obtained by a gate voltage of the form
\begin{equation}
    V_{\rm g}^{\rm box}(t)=\frac{q}{2\pi C_\mu}\mathrm{Re}\left(i\ln\left\{\frac{\sin[\Omega(t-t_0+i\sigma)/2]}{\sin[\Omega(t-t_0-\mathcal{T}/2+i\sigma)/2]}\right\}\right)-\frac{q}{2C_\mu}\,,
\label{eq:gate_box}
\end{equation}
where $\sigma$ corresponds to the half width at half maximum of the Lorentzian pulses arising from $V_{\rm g}^{\rm box}$ via~\eqref{eq:gating_derivative}. Note that~\eqref{eq:gate_box} approaches a square-wave drive in the limit $\sigma\to 0$.

In the fast driving regime $\Omega\tau_{RC}\gg 1$, the equivalent circuit is essentially a short circuit, so the current profile directly follows the gate voltage
\begin{equation}
    I^c(t)\approx 2\frac{q^2}{h}\frac{\tau_{\rm f}}{\mathcal{T}}V_{\rm g}^{\rm odd}(t)\,,
\end{equation}
where only the odd harmonics of $V_{\rm}(t)$ enter in the response. As a result, the response features an alternating pattern of pulse/anti-pulse in each consecutive period of the drive.
The gate driving technique discussed in this section has recently been exploited in experiments~\cite{Ruelle2024Sep}.

%%%%%%%%%%%%%%%%%%%%%%%%%%%%%%%%%%%%%%%%%%%%%%%%%%%%%%%%%%%%%%%%%%%%%%%
\subsubsection{Single-particle injection from confined regions} 
\label{sec:injection_confined}
%%%%%%%%%%%%%%%%%%%%%%%%%%%%%%%%%%%%%%%%%%%%%%%%%%%%%%%%%%%%%%%%%%%%%%%
A complementary approach to the one described above is to release electrons into a conductor from a confined region, like a quantum dot. In this case, the strategy is to exploit the discrete spectrum of the confined region to extract electrons one by one, or to engineer clever loading and unloading protocols to achieve the same goal. The development of single-electron pumps was strongly motivated by metrological applications aimed at a redefinition of the current standard. In this section we present an overview of such approaches, discussing the mesoscopic capacitor, electron pumping and turnstiles in more detail.

%%%%%%%%%%%%%%%%%%%%%%%%%%%%%%%%%%%%%%%%%%%%%%%%%%%%%%%%%%%%
\paragraph{Injection from a mesoscopic capacitor}
%%%%%%%%%%%%%%%%%%%%%%%%%%%%%%%%%%%%%%%%%%%%%%%%%%%%%%%%%%%%
In order to describe single-electron injection relying on the driving of a \textit{confined region}, we start from the most simple realization, namely the mesoscopic capacitor~\cite{Buttiker1993Jun,Buttiker1993Sep,Pretre1996Sep}, see Fig.~\ref{fig:injection_principles}(d). The main idea is that charging and discharging of a small capacitor plate can result in the controlled emission of single electrons into a conductor, when the addition energy of electrons is large with respect to temperature. The emission process also relies on a discrete spectrum of the confined region~\cite{Moskalets2008Feb} or Coulomb blockade, possibly in combination with a superconducting energy gap. 

One important application of the mesoscopic capacitor as single-electron source is in quantum Hall systems. In fact, the realization of an on-demand and coherent single-electron source in this regime~\cite{Feve2007May} has been a major experimental achievement initiating the development of electronic quantum optics, discussed in further detail in Sec.~\ref{sec_QOE}.
Therefore, we describe here a model of a capacitor-based single-electron source in the quantum Hall regime, following~\cite{Moskalets2008Feb}. The confined region consists of an edge mode in a closed loop of circumference $L$, coupled via a quantum point contact to an open edge channel constituting the target conductor where excitations will be injected from the dot. The quantum point contact tunes the reflection $(r)$ and transmission $(d)$ probability amplitudes, connecting the confined region to the open edge channel, see Fig.~\ref{fig:injection_principles}{(d)}. Finally, the confined region is coupled to a top gate to which a potential $V_{\rm g}(t)$ can be applied. Starting from the \textit{stationary} case, and assuming without loss of generality $V_{\rm g}=0$, we can write a scattering matrix that has a Fabry-P\'erot form
\begin{equation}
    S_0(E)= r+{d}^{2}\sum_{\ell=1}^{\infty}r^{\ell-1}\exp\left[i\ell\frac{E\tau_d}{\hbar}\right]\,,
\label{eq:scatt_mes_cap_full}
\end{equation}
where $\tau_d=L/v_F$ is the time it takes to go travel a full circumference, and we have assumed a linear energy-momentum relation, which can always be done close enough to the Fermi level. Note that the scattering matrix $S_0$ of this simple system is a scalar. If we next compute the density of states $\nu_0(E)=(2\pi i)^{-1}S_0^*\,dS_0/dE$, we find a structure with peaks separated by $\Delta=h/\tau_d$, with broadening $\gamma=D\Delta/2\pi$. Therefore, at $D\ll 1$, we have discrete, well-defined energy levels in the confined region.

Let us turn now to the driven regime. Also in this case, the local Coulomb interaction in the confined region does often  not play a crucial role, due to screening by the gate(s) and a very small geometrical capacitance of the gate-dot system~\cite{Feve2007May} (see, however, Refs.~\cite{Litinski2017Aug,Dittmann2018Apr,Kashuba2012Jun,Mora2010Sep,Filippone2012Sep} and Sec.~\ref{intro-ac} for the treatment of interactions). Therefore, a description based on a scattering matrix, which still allows to include interaction effects self-consistently, is justified. Using the mixed time-energy representation introduced in Sec.~\ref{t-per-scatt}, we have
\begin{equation}
    S(t,E)= r+{d}^{2}\sum_{\ell=1}^{\infty}r^{\ell-1}\exp\left[i\ell\frac{E\tau_d}{\hbar}-i\Phi_\ell(t)\right]\,.
\label{eq:scatt_mes_cap_full}
\end{equation}
The structure is the same as in the stationary case: each term for a given $\ell$ in the above sum represents the scattering amplitude of a particle entering the dot at time $t-\ell\tau_d$, executing $\ell$ round trips and exiting the dot at time $t$. In this time interval, the particle accumulates the dynamical phase and the one induced by the action of the gate potential, encoded in the function
\begin{equation}
    \Phi_\ell(t)=\frac{q}{\hbar}\int_{t-\ell\tau_d}^t dt' V_g(t')\,.
\end{equation}
Using~\eqref{eq:scatt_mes_cap_full}, the time-resolved charge current emitted from the 
quantum dot, can be evaluated via Eq.~\eqref{eq:current_floquet_time-energy}, considering a two-terminal geometry without any additional voltage bias between the contacts. The result is~\cite{Moskalets2008Feb}
\begin{align}
    I^c(t)&=I^{c}_{\rm lin}(t)+I^{c}_{\rm nonlin}(t)\label{eq:curr_mes_full}\\
    I^c_{\rm lin}(t)&=\frac{q^2}{h}D^2\sum_{\ell=1}^{\infty}R^{\ell-1}[V_{\rm g}(t)-V_{\rm g}(t-\ell\tau_d)]\label{eq:curr_mes_full_lin}\\
    I^{c}_{\rm nonlin}(t)&=\frac{qD^2}{\pi\tau_d}\mathrm{Im}\left\{\sum_{\ell,m}\frac{2\pi^2k_{\rm B}T/\Delta}{\sinh(2\pi^2mk_{\rm B}T/\Delta)}r^mR^{\ell -1}\left[e^{-i\Phi_m(t-\ell\tau_d)}-e^{-i\Phi_m(t)}\right]\right\}\label{eq:curr_mes_full_nonlin}\,,
\end{align}
where the nonlinear term mainly contributes in the low-temperature regime $k_{\rm B}T\ll\Delta$, and the linear term is independent of temperature. This rather complicated expression gives the current response at arbitrary gate-voltage amplitude and frequency.

An insightful analytic result can be found in the limit where the driving frequency is small, $\Omega\tau_d\ll 1$, also referred to as the adiabatic-response regime. By expanding~\eqref{eq:curr_mes_full} at first order in the driving frequency, the time-resolved current reduces to a purely capacitive response
\begin{equation}
I^c(t) = q^2\frac{\partial V_\mathrm{g}(t)}{\partial t}\int dE\left(-\frac{\partial f(E)}{\partial E}\right)\nu(t,E)\,.
\label{eq:mes_curr_adiab}
\end{equation}
This expression represents basically the response of the occupation number on the quantum dot/mesoscopic capacitor to a time-dependent change of the gate voltage. Here, an important quantity is the \emph{instantaneous} density of states of the quantum dot
\begin{equation}
\nu(t,E) = \frac{1}{2\pi i}S_0^*(t,E)\frac{\partial S_0(t,E)}{\partial E}
\end{equation}
obtained from the frozen scattering matrix, see also Sec.~\ref{adia-scatt}. One easily finds $\nu(t,E)=\nu_0(E-qV_\mathrm{g}(t))$. Equation~\eqref{eq:mes_curr_adiab} tells us that, in the adiabatic-response regime, the frozen density of states defines a capacitance just like a stationary density of states gives rise to a quantum capacitance.
The capacitive response~\eqref{eq:mes_curr_adiab} is very similar to the adiabatic response of~\eqref{eq:gating_derivative}.
Indeed, the protocol of injection by local gating can be recovered by considering a fully open mesoscopic capacitor, with $D\to 1$. Consider now a monochromatic drive $V_{\rm g}(t)=V_{\rm g}^0\cos(\Omega t)$, with $|qV_{\rm g}^0|\lesssim\Delta$ so that we can consider a single level of the dot only. In the first half of the period $\mathcal{T}$ Eq.~\eqref{eq:mes_curr_adiab} gives~\cite{Dashti2019Jul}
\begin{equation}
    I^c(t)=qD\int dE\left(-\frac{\partial f}{\partial E}\right)\frac{\sigma/\pi}{t^2+\sigma^2},\qquad\sigma=\frac{\gamma}{qV_{\rm g}^0\Omega}=\frac{D\Delta}{2\pi qV_{\rm g}^0\Omega}\,.
\end{equation}
At low temperature $k_{\rm B}T\ll\Delta$ we thus have the emission of a Lorentzian pulse that reminds us of the Leviton source. Indeed, in this regime, the latter and the mesoscopic capacitor produce the same type of excitation. However, the presence of the density of states in Eq.~\eqref{eq:mes_curr_adiab} is such that the two sources do have different spectral properties and temperature dependence of the transport quantities~\cite{Dashti2019Jul}.
Finally, it can be shown directly from Eq.~\eqref{eq:mes_curr_adiab} that the excited charge (per half-period) is quantized to $Mq$ whenever $\max(V_{\rm g})-\min(V_{\rm g})=M\Delta/q$, independently of temperature~\cite{Moskalets2008Feb}.
\begin{figure}[t]
    \centering
    \includegraphics[width=0.6\linewidth]{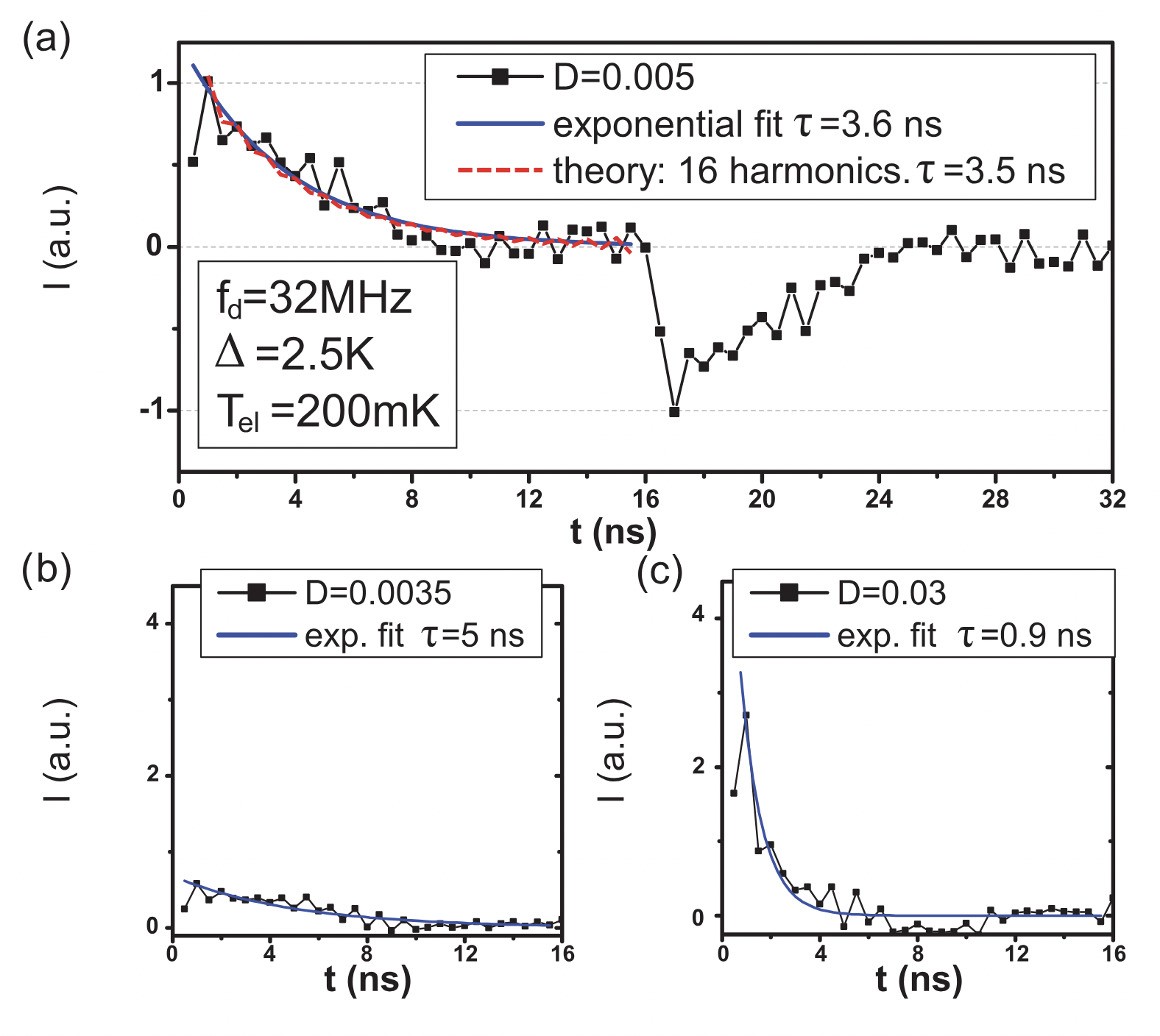}
    \caption{Current emission from a mesoscopic capacitor in the non-adiabatic regime. A square pulse with amplitude equal to one level spacing $\Delta$ is applied to the top gate of a mesoscopic capacitor, resulting in a current exhibiting an exponential decay. Reprinted figure with permission from~\cite{Parmentier2012Apr}. Copyright 2012 by the American Physical Society.}
    \label{fig:mes_cap_square}
\end{figure}

The other relevant situation is the fast emission regime $\Omega\tau_d\gg 1$, and particularly a square-wave driving protocol, first implemented in~\cite{Feve2007May}. Consider then a gate potential $V_{\rm g}(t)=0$ for $t\in(m\mathcal{T},m\mathcal{T}/2)$ and $V_{\rm g}(t)=V_{\rm g}^0$ for $t\in(m\mathcal{T}/2,m\mathcal{T})$, with $m\in\mathbb{Z}$. Let us also suppose that the transitions happen on a time scale $\delta t\ll\tau_d$, making the protocol non-adiabatic.
Focussing on the interval $t\in(0,\mathcal{T}/2)$, yet $t\gg\tau_d$, Eq.~\eqref{eq:curr_mes_full_lin} predicts an exponential current profile
\begin{equation}
    I_{\rm lin}^c(t)\propto e^{-t/\tau},\quad \tau=-\frac{\tau_d}{\ln(1-D)}=\tau_d\left(\frac{1}{D}-\frac{1}{2}\right)+\mathcal{O}(D)\,.
\end{equation}
For $k_\mathrm{B}T\gg\Delta$ this is the only contribution to the current, which is however associated to a non-quantized transferred charge. Quantization is recovered in the regime $k_{\mathrm{B}}T\ll\Delta$, thanks to the nonlinear term~\eqref{eq:curr_mes_full_nonlin}. Once again, the result is that quantization requires an excitation amplitude equal to an integer multiple of the level spacing $\Delta$. Overall, the current retains an exponential profile, as observed in experiments and shown in Fig.~\ref{fig:mes_cap_square}. Notice that the injected charge is quantized as long as the escape time $\tau$ is not comparable to a half-period of the driving signal, i.e., for not too small transmissions $D$.

Until here, charge quantization refers to the \textit{average} injected charge per half-period. Clearly, this ``quantization" does not exclude spurious emission events where more than one charge is emitted in a cycle and no charges are emitted in other cycles, still keeping the average quantized. Therefore, a further characterization of the system is required to convincingly show that the mesoscopic capacitor is a reliable single-electron emitter. Such a study requires considering the current fluctuations in the device, see also Sec.~\ref{sec_noise}. This was carried out experimentally in Ref.~\cite{Mahe2010Nov}. It turns out that having a driving amplitude equal to one level spacing is not enough to guarantee a robust charge quantization. In particular, a resonant emission, corresponding to having the emission energy in resonance with the Fermi level, generates extra fluctuations that do not make the source a reliable single-electron emitter. 
The optimal regime to achieve this goal is instead achieved when the Fermi level is exactly halfway between two dot levels, so the emission energy is $\Delta/2$ above the Fermi level. A detailed discussion, comparing theory and experimental data, can be found in~\cite{Parmentier2012Apr}. A typical emission energy for the driven mesoscopic capacitor is around 0.1 meV, making this source the least energetic among those presented in this section. This also means that the effect of the Fermi sea is relevant, and relaxation processes can be important~\cite{Wahl2014Jan,Ferraro2014Oct}.\\

%%%%%%%%%%%%%%%%%%%%%%%%%%%%%%%%%%%%%%%%%%%%%%%%%%%%%%%%%%
\paragraph{Single-electron pumping}
%%%%%%%%%%%%%%%%%%%%%%%%%%%%%%%%%%%%%%%%%%%%%%%%%%%%%%%%%%%

%%%%%%%%%%%%%%%%%%%%%%%%%%%%%%%%%%%%%%%%%%%%%%%%%%%%%%%%%%%
\begin{figure}
    \centering
 \includegraphics[width=0.65\linewidth]{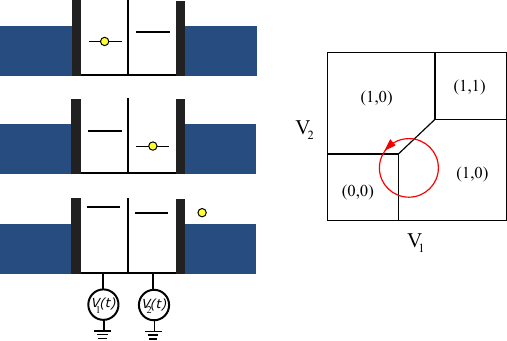}
    \caption{(a) Sketch of the energy landscape of a driven double quantum dot. The three snapshots show different level configurations. As response to the changes between these configurations a quantized charge current is induced. (b) Cycle in parameter space encircling one triple point in the stability diagram. 
    \label{fig:double_dot_pump}}
\end{figure}
%%%%%%%%%%%%%%%%%%%%%%%%%%%%%%%%%%%%%%%%%%%%%%%%%%%%%%%%%%%

The concept of quantum pumping has been introduced in Sec.~\ref{sec:mechanism_chargepumping}. Under specific conditions, it is possible to use pumping to realize controlled single- or multi-particle emission. One strategy, as first demonstrated in Ref.~\cite{Pothier1992Jan} is adiabatic pumping around the triple point in the stability diagram of a double dot or of two metallic islands coupled in series, see Fig.~\ref{fig:double_dot_pump}. Such a double-dot system is described by the Hamiltonian
\begin{eqnarray}
    \hat{H}_\mathrm{ddot}= \sum_{j=\mathrm{1,2}}\epsilon_j(t)\hat{N}_j-\frac{v}{2}\sum_{\sigma={\uparrow,\downarrow}}\left(\hat{d}_{1\sigma}^\dagger \hat{d}_{2\sigma}+\hat{d}_{2\sigma}^\dagger \hat{d}_{1\sigma}\right)+U\hat{N}_1\hat{N}_2\label{eq:Hddot}
\end{eqnarray}
with the interdot coupling $v$ and the single-dot occupations $\hat{N}_j=\sum_{\sigma}\hat{N}_{j\sigma}=\sum_{\sigma}\hat{d}^\dagger_{j\sigma}\hat{d}^{}_{j\sigma}$. Here, we set the onsite interaction to be the largest energy scale such that double occupation on a single dot is excluded. For a double-island system, the Hamiltonian is equivalent but with continuous energy states. The stability diagram of this type of systems has a honeycomb structure, where a section including one triple point between neighboring stable regions is shown in Fig.~\ref{fig:double_dot_pump}{(b)}, as function of the gate voltages applied to the dots. By \textit{slowly} modulating the gates, electrons can be transferred in adiabatic response between dot 1 and the left contact, between dot 2 and the right contact, or between the dots. These dynamics can be described by a classical rate equation in the sequential tunneling regime
\begin{equation}
    \frac{d}{dt}\left(\begin{array}{c}P_0\\P_\mathrm{b}\\P_\mathrm{a}\end{array}\right) = \left(\begin{array}{ccc}-2\Gamma_\mathrm{a}f(\epsilon_\mathrm{a})-2\Gamma_\mathrm{b}f(\epsilon_\mathrm{b}) & \Gamma_\mathrm{b}\left(1-f(\epsilon_\mathrm{b})\right) & \Gamma_\mathrm{a}\left(1-f(\epsilon_\mathrm{a})\right)\\
    2\Gamma_\mathrm{a}f(\epsilon_\mathrm{a}) &- \Gamma_\mathrm{b}\left(1-f(\epsilon_\mathrm{b})\right) & 0\\
    2\Gamma_\mathrm{b}f(\epsilon_\mathrm{b}) &0 & -\Gamma_\mathrm{a}\left(1-f(\epsilon_\mathrm{a})\right)
    \end{array}\right)\left(\begin{array}{c}P_0\\P_\mathrm{b}\\P_\mathrm{a}\end{array}\right)\ .
\end{equation}
 Here the eigenstates $\eta=\mathrm{b,a}$ of the double-dot are the bonding and anti-bonding states, which---sufficiently far away from the triple points---equal the local occupation states of dots 1 and 2. Their energies are given by $\epsilon_\mathrm{b/a}=(\epsilon_1+\epsilon_2)/2\pm\sqrt{v^2+(\epsilon_1-\epsilon_2)^2/4}$ and their effective couplings to the left and right reservoirs, $\alpha=\mathrm{L,R}$, which are time-dependent via the level energies, are given by
\begin{eqnarray}
    \Gamma_{\eta\alpha}=\frac{1}{2}\left(1-\alpha\eta\frac{\epsilon_1-\epsilon_2}{\sqrt{4v^2+(\epsilon_1-\epsilon_2)^2}}\right)\,,
\end{eqnarray}
where $\alpha=\mathrm{L/R}$ and $\eta=\mathrm{b,a}$ take the values $\pm 1$ when used as variables. We also introduce the abbreviations $\Gamma_\mathrm{\alpha}=\sum_{\eta=\mathrm{b,a}}\Gamma_{\eta\alpha}$ and $\Gamma_\mathrm{\eta}=\sum_{\alpha=\mathrm{L,R}}\Gamma_{\eta\alpha}$. The induced charge current is given by
\begin{eqnarray}
    I_\alpha^c = -q\sum_{\eta=\mathrm{b,a}}\frac{\Gamma_{\alpha\eta}}{\Gamma_\eta}\left(\frac{d}{dt}P^{(0)}_\eta(t)\right)\,.
\end{eqnarray}
When driving the gates around a triple point as indicated by the example cycle in Fig.~\ref{fig:double_dot_pump}{(b)}, exactly one charge is transferred \textit{peristaltically} through the system per cycle from left to right. A quantized current is found when the amplitudes and phases of the driving parameters are chosen such that the distance between the parameter cycle and the triple point is much larger than the temperature and if the time between the crossing of two lines is long enough to allow the system to reach the stable occupation state.

Such peristaltic single-electron pumps have for example been experimentally realized in double metallic islands~\cite{Pothier1992Jan}, in single-dopants in silicon~\cite{Roche2013Mar}, in graphene~\cite{Connolly2013Jun}, and in carbon nanotubes~\cite{Buitelaar2008Sep}.
The accuracy of such devices is typically limited by cotunneling effects~\cite{Jensen1992Nov}.\\

%%%%%%%%%%%%%%%%%%%%%%%%%%%%%%%%%%%%%%%%%%%%%%%%%%%%%%%%%%%%
\paragraph{Injection via nonadiabatic single-parameter pumps}
%%%%%%%%%%%%%%%%%%%%%%%%%%%%%%%%%%%%%%%%%%%%%%%%%%%%%%%%%%%%

%%%%%%%%%%%%%%%%%%%%%%%%%%%%%%%%%%%%%%%%%%%%%%%%%%%%%%%%%%%%
\begin{figure}[b]
    \centering
    \includegraphics[width=0.5\linewidth]{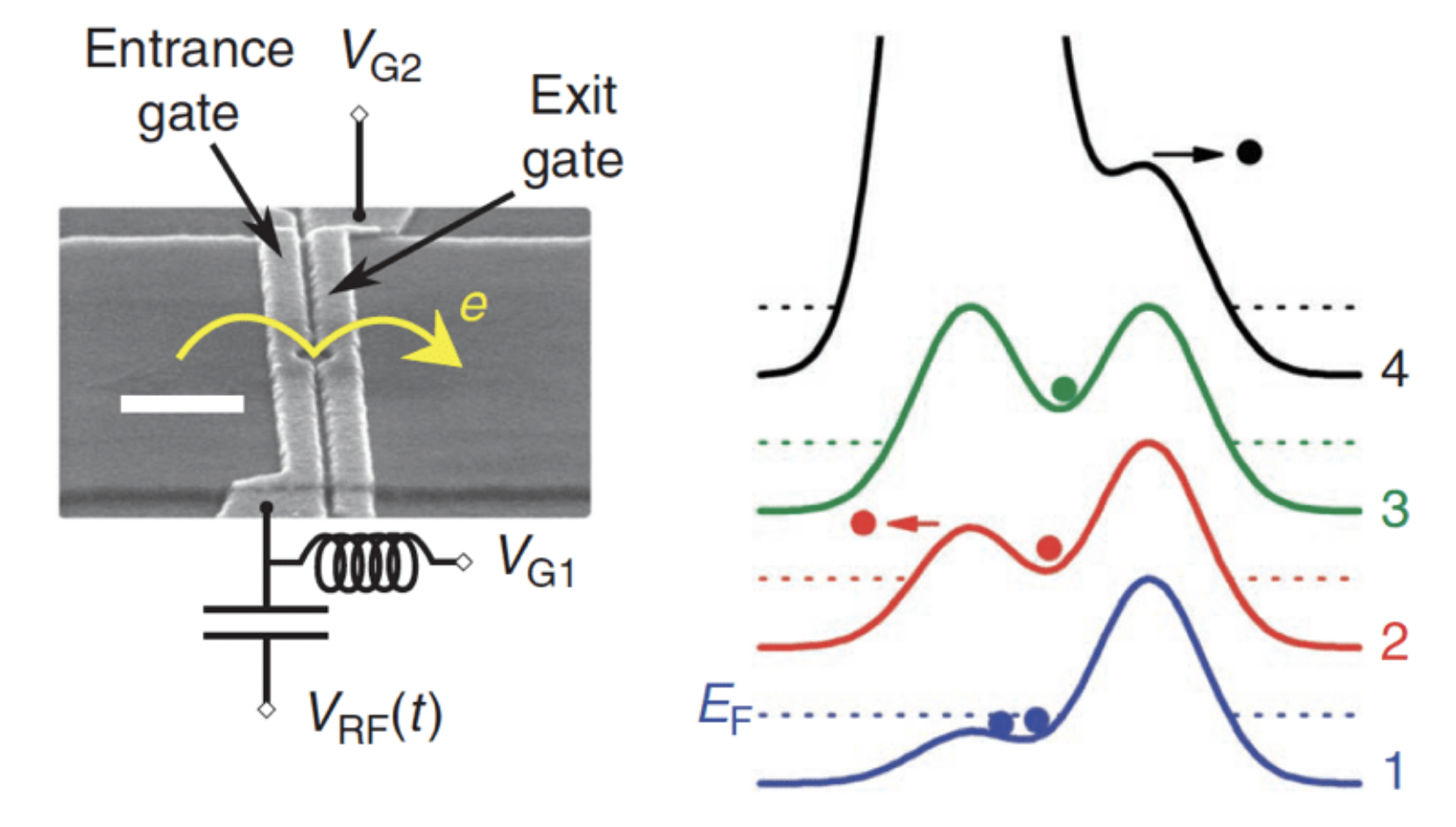}
    \caption{Working principle of a single-electron source based on a dynamical quantum dot. This is also referred to as a single-parameter nonadiabatic pump. Figure reproduced with permission from Springer Nature Ref.~\cite{Giblin2012Jul}.}
    \label{fig:hot_electron_pump}
\end{figure}
%%%%%%%%%%%%%%%%%%%%%%%%%%%%%%%%%%%%%%%%%%%%%%%%%%%%%%%%%%%%

 With the goal to meet the high requirements for using single-electron sources for the implementation of a current standard, a different strategy has therefore recently become more promising, exploiting nonadiabatic pumping going along with so-called hot-electron emission~\cite{Kaestner2008Apr,Blumenthal2007May}. These nonadiabatic single-electron pumps are typically realized in a gated wire, where the modulation of a potential barrier leads to the asymmetric creation and depletion of a quantum dot which is charged and discharged by electron transmission from opposite sides of the conductor. The working principle is shown in Fig.~\ref{fig:hot_electron_pump}.

The left side of Fig.~\ref{fig:hot_electron_pump} illustrates an example of a single-electron pump implemented in a nanostructure based on GaAs, and exploiting a dynamical quantum dot. The latter is defined by two parallel gates with a small space between them, where electrons can be trapped. By periodically modulating the two gate potentials the emission of one electron per repetition cycle can be achieved, as originally demonstrated in~\cite{Blumenthal2007May}. A schematic of the working principle is shown on the right side of Fig.~\ref{fig:hot_electron_pump}: first, electrons are loaded by lowering the barrier on the left gate below the Fermi level (1). Then, the barrier is gradually increased (2), allowing for back tunneling until the dot is isolated and one electron remains trapped (3). Next, the left barrier is elevated until it exceeds the right one (4), resulting in the emission of the loaded electron. Using such devices a quantized current with a precision of one part per million has been demonstrated~\cite{Giblin2012Jul}. Furthermore, by placing an additional barrier downstream with respect to the source (see Fig.~\ref{fig:hot-carrier_barrier-response} in Sec.~\ref{sec_energy_spectroscopy}), it is also possible to measure the energy distribution and the wavepacket width of the emitted electrons~\cite{Fletcher2013Nov,Ubbelohde2015Jan}, perform tomography~\cite{Fletcher2019Nov}, as well as time-resolved interference~\cite{Pavlovska2023Apr,Ubbelohde2023Jul,Fletcher2023Jul}. 
For more details we refer the reader to the review
\cite{Kaestner2015Sep,Edlbauer2022Dec}. Here, it is sufficient to highlight that single-electron pumps of this kind result in large emission energies (order of 100 meV, that is much larger than the Fermi level in the system, of order 10 meV) and for this reason the emitted particles are often called hot electrons.\\

%%%%%%%%%%%%%%%%%%%%%%%%%%%%%%%%%%%%%%%%%%%%%%%%%%%%%%%%%%%
\paragraph{Injection and transport via surface acoustic waves}
%%%%%%%%%%%%%%%%%%%%%%%%%%%%%%%%%%%%%%%%%%%%%%%%%%%%%%%%%%%
%%%%%%%%%%%%%%%%%%%%%%%%%%%%%%%%%%%%%%%%%%%%%%%%%%%%%%%%%%%
\begin{figure}[b]
    \centering
    \includegraphics[width=0.65\linewidth]{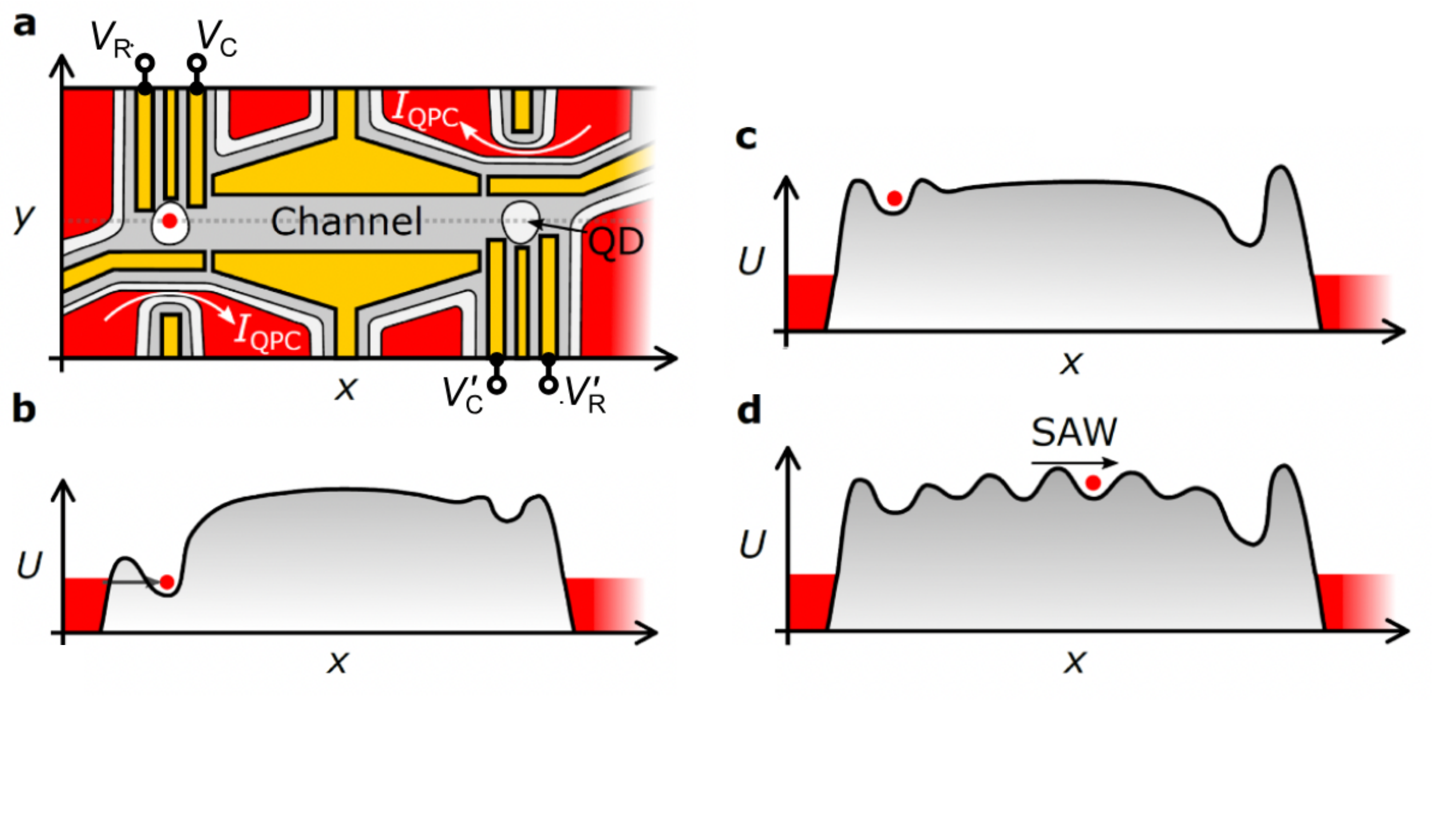}
    \caption{Schematics of the SAW single-electron source. Grey areas in (a) indicate fully depleted regions. (b-c) Loading of an electron in the source quantum dot. (d) Capture of the loaded electron in a SAW minimum and transfer over the depleted channel. Figure adapted with permission from Ref.~\cite{Edlbauer2019Jul}.}
    \label{fig:SAW}
\end{figure}
%%%%%%%%%%%%%%%%%%%%%%%%%%%%%%%%%%%%%%%%%%%%%%%%%%%%%%%%%%%
Another approach to emit and transport single-electrons relies on a combination of quantum dots as discussed in the previous paragraph and the action of a surface acoustic wave (SAW). This principle is similar to the Thouless pump, discussed in Sec.~\ref{sec:Thouless}.
The key idea is to exploit the piezoelectric properties of GaAs to generate an electrically induced perturbation of the  potential landscape that propagates along certain crystal directions at a speed of about 3000 m/s, that is much slower than the Fermi velocity. Details on how the SAW is generated can be found for instance in~\cite{Edlbauer2019Jul}, with recent developments  and optimization reported in~\cite{Wang2022Sep}.
The important point is that the SAW creates a moving quantum dot that is able to transport single electrons over micrometer-scale distances. The protocol is illustrated in Fig.~\ref{fig:SAW}. The setup (a) features two quantum dots, the left one being the source and the right one the receiver. Each of them is equipped with a quantum point contact acting as a charge sensor, allowing one to detect the presence of electrons in the dot, with single-electron sensitivity.
At first, an electron is captured (b) in the left dot and isolated from the Fermi sea (c) by acting on the gate voltages $V_{\rm R}$ and $V_{\rm C}$. Then a SAW is sent through the device and creates a train of moving quantum dots. The electron is thus loaded in a minimum of the SAW and transported over the depleted channel (d).
This approach was pioneered in~\cite{Hermelin2011Sep,McNeil2011Sep}, where electrons were transported over micrometer distances with more than 90\% efficiency. Recent developments~\cite{Takada2019Oct,Wang2024Oct} have greatly improved the accuracy, also allowing for controlled interference experiments~\cite{Wang2023Jul}. We will come back to this point in Sec.~\ref{sec:interferometry}.
The technique is nowadays very advanced, to the point where there is complete control over the number of emitted electrons, as well as which minimum of the SAW electrons are loaded in.
One key advantage of the SAW single-electron source compared to other approaches is the possibility of performing a \emph{single-shot detection} of the emitted electrons, enabling counting statistics experiments~\cite{Wang2023Jul,Shaju2024Aug}. This is achieved by using a receiver quantum dot, equipped with a nearby quantum point contact, as a very sensitive charge detector. Similarly to the nonadiabatic single-electron pumps, and differently from the mesoscopic capacitor and the Leviton source, the SAW source injects electrons that are well separated from the Fermi sea.\\

%%%%%%%%%%%%%%%%%%%%%%%%%%%%%%%%%%%%%%%%%%%%%%%%%%%%%%%%%%
\paragraph{Superconducting turnstiles}
%%%%%%%%%%%%%%%%%%%%%%%%%%%%%%%%%%%%%%%%%%%%%%%%%%%%%%%%%%%

%%%%%%%%%%%%%%%%%%%%%%%%%%%%%%%%%%%%%%%%%%%%%%%%%%%%%%%%%%%
\begin{figure}[b]
    \centering
    \includegraphics[width=0.8\linewidth]{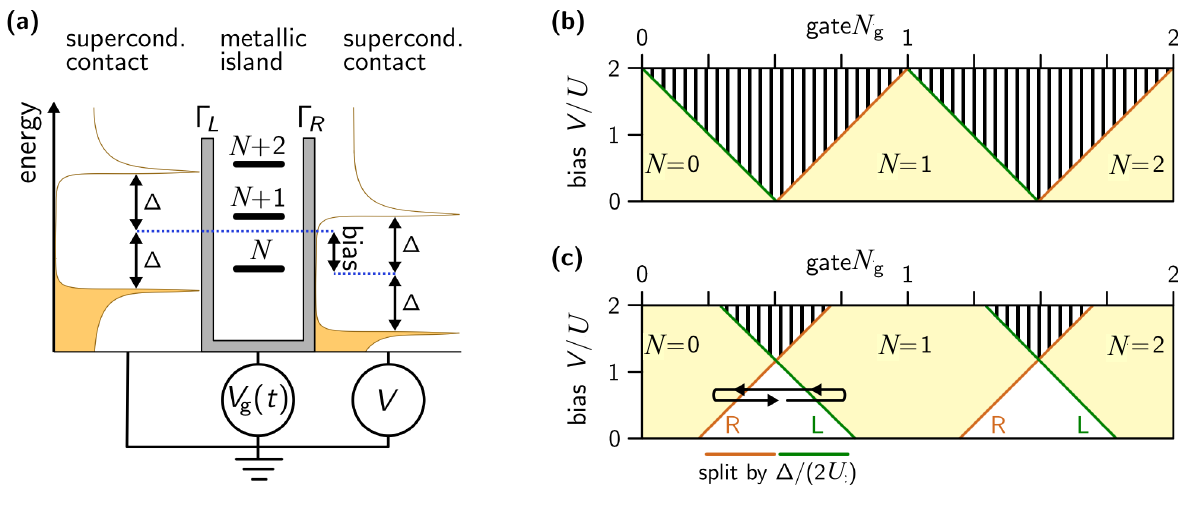}
    \caption{Working principle of a superconducting turnstile, here shown at the example of an SINIS system. (a) Sketch of the normal conducting island contacted to two superconducting reservoirs via insulator barriers (SINIS). (b) Stability diagram of a normal-conducting island for comparison. (c) Stability diagram of the island in contact with superconducting reservoirs. A possible operation cycle of the turnstile is indicated by a black line. Figure adapted with permission from Ref.~\cite{Dittmann2018}.}
    \label{fig:turnstile}
\end{figure}
%%%%%%%%%%%%%%%%%%%%%%%%%%%%%%%%%%%%%%%%%%%%%%%%%%%%%%%%%%%

The single-electron pumps introduced above exploit Coulomb interaction and tunable (effective) confinement for the realization of precise single-electron emission. An alternative strategy is the implementation of so-called turnstiles in hybrid superconducting devices. A major difference with respect to quantum pumps is that these turnstiles are operated at finite bias, which sets the direction of single-electron transport. Furthermore, in these hybrid turnstile systems, in addition to Coulomb interaction, the superconducting gap is used to selectively charge and discharge a central Coulomb-blockaded island. An example for such a device, here a normal-metal island with superconducting contacts, is shown in Fig.~\ref{fig:turnstile}. Also realizations with superconducting islands have been proposed and realized, see for example~\cite{Pekola2013Oct}.

The superconducting turnstile shown in Fig.~\ref{fig:turnstile} is described by the Hamiltonian
\begin{eqnarray}
    \hat{H}_\mathrm{turnstile} & = & \sum_{k,\sigma}\epsilon_k \hat{d}_{k\sigma}^\dagger \hat{d}_{k\sigma}^{}+U\left(\hat{N}-N_\mathrm{g}(t)\right)^2\nonumber\\
    && +\sum_{\alpha,k,\sigma}\epsilon_{\alpha k} \hat{a}_{\alpha k\sigma}^\dagger \hat{a}_{\alpha k\sigma}^{}+\sum_{\alpha k} \left(\Delta \hat{a}_{\alpha k\uparrow}\hat{a}_{\alpha -k\downarrow}+\Delta^*\hat{a}^\dagger_{\alpha k\uparrow}\hat{a}_{\alpha -k\downarrow}^\dagger\right)\nonumber\\
    &&+\sum_{\alpha k \sigma}\left(w_{\alpha k}\hat{a}^\dagger_{\alpha k\sigma}\hat{d}_{\alpha\sigma} +\mathrm{H.c.}\right)\ .\label{eq:Hturnstile}
\end{eqnarray}
The first line describes the island, containing the gate-induced island occupation in the charging energy, which is time-dependent through gate driving, $N_\mathrm{g}(t)=C_\mathrm{g}V_\mathrm{g}(t)$. The contacts are BCS superconductors, which hence have a spectral density $\nu(E)=\nu_0|\mathrm{Re}\left[(E+i\gamma_\mathrm{D})/\sqrt{(E+i\gamma_\mathrm{D})^2-|\Delta|^2}\right]|$. Here, the Dynes parameter $\gamma_\mathrm{D}$ models smoothening of the sharp peaks in the spectral density, $\Delta$ is the superconducting gap, and $\nu_0$ the density of states of the contacts in the normal conducting state. Finally, weak tunneling between contacts and central island is described by the last line of Eq.~\eqref{eq:Hturnstile}. Across the turnstile a symmetric bias $\pm V/2$ is applied.

The working principle of the turnstile is the following. Assume that the island is initialized with a particle number $N$. Then the energy required to add or to remove a particle from the island via tunneling from/to the left or right contact is 
\begin{eqnarray}
    \delta E^{\mathrm{L/R},N}_+ & = & U\left[\left(N+1-N_\mathrm{g}\right)^2-\left(N-N_\mathrm{g}\right)^2\right]\mp\frac{V}{2}\\
    \delta E^{\mathrm{L/R},N}_- & = & U\left[\left(N-1-N_\mathrm{g}\right)^2-\left(N-N_\mathrm{g}\right)^2\right]\pm\frac{V}{2}
\end{eqnarray}
In the normal-conducting case this would lead to a stability diagram (Coulomb diamonds) as shown for positive $V$ in Fig.~\ref{fig:turnstile}{(b)}. However, in the presence of superconducting leads the Coulomb diamonds are enlarged by the white regions where two different charge states are stable simultaneously. In the sequential tunneling regime transfer of single particles is strongly suppressed by the superconducting gap in these regions, where transport would require the breaking of a Cooper pair. This results in highly asymmetric loading and unloading rates when the driving of the gate voltages makes the driving cycle (an example is indicated by the black line in Fig.~\ref{fig:turnstile}{(c)}) cross the lines confining different charge states. Concretely, when crossing the green line the island can go from charge state $N=0$ to $N=1$ via coupling to the left lead (but not the other way around), while it can go from  charge state $N=1$ to $N=0$ via coupling to the right lead when crossing the orange line (but not the other way around).

A similar working principle underlies the functioning of turnstiles with superconducting islands, both with normal-conducting and with superconducting leads.
As described above, for the operation of these turnstiles, in particular the \textit{gap} of the superconductor was exploited for realizing single-particle control. However, also the superconducting phase can play play an important role in pumping, as for example discussed in Sec.~\ref{sec_spectroscopy} below. 

Furthermore, the oscillating phase of a biased superconductor was used to implement time-dependent driving of a mesoscopic conductor, thereby realizing an adiabatic quantum pump~\cite{Giazotto2011Nov} following the theoretical proposal of Ref.~\cite{Russo2007Aug}.

%%%%%%%%%%%%%%%%%%%%%%%%%%%%%%%%%%%%%%%%%%%%%%%%%%%%%
\subsubsection{Time-dependent transport as spectroscopy ``knob''}\label{sec_spectroscopy}
%%%%%%%%%%%%%%%%%%%%%%%%%%%%%%%%%%%%%%%%%%%%%%%%%%%%%

In the previous section, we have mostly focused on the \textit{control} over single particles in quantum conductors that can be achieved with time-dependent driving. In this context we have highlighted also research on the dynamical properties resulting from this time-dependent control, such as the decay dynamics in response to a gate switch. 
However, quantum transport induced by time-dependent driving provides further intriguing opportunities, since it can be used as an additional ``knob" in order to perform transport spectroscopy that goes beyond the possibilities that arise from steady-state transport spectroscopy. 
 In this section, we will show some examples: we start by demonstrating how the inherent properties of the dynamics can be accessed, in particular the decay rates of the central system, as well as the geometric nature of adiabatic pumping. In a second step, we show how time-dependent transport spectroscopy can be used to reveal fundamental many-body properties of the central system, which are not accessible from steady-state measurements.
For this purpose, pumping in the adiabatic-response regime is particularly appropriate, since the slow driving minimally excites the system, and at the same time it leads to a dc signal that is more straightforward to detect than purely time-dependent currents.\\
%%%%%%%%%%%%%%%%%%%%%%%%%%%%%%%%%%%%%%%%%%%%%%%%%%%%%%%%%%%%%%%%%
\begin{figure}[b]
    \centering
    \includegraphics[width=0.65\linewidth]{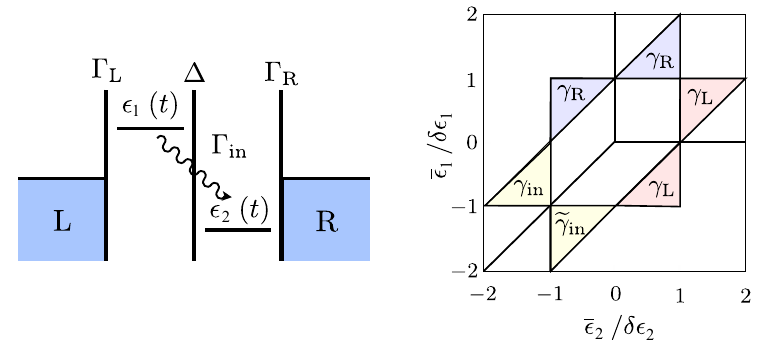}
    \caption{ Left panel: Sketch of a driven double dot. Stability diagram of the double dot. The central white region around the triple point of three coexisting double-dot states is the region where quantized charge can be pumped. In the surrounding triangles charge is transferred in the nonadiabatic pumping regime and allows for readout of the indicated rates. Figure adapted from Ref.~\cite{Riwar2016Jun}.}
    \label{fig:spectroscopy}
\end{figure}
%%%%%%%%%%%%%%%%%%%%%%%%%%%%%%%%%%%%%%%%%%%%%%%%%%%%%%%%%%%%%%%%%

%%%%%%%%%%%%%%%%%%%%%%%%%%%%%%%%%%%%%%%%%%%%%%%%%%%%%%%%%%
\paragraph{Readout of relaxation times}
%%%%%%%%%%%%%%%%%%%%%%%%%%%%%%%%%%%%%%%%%%%%%%%%%%%%%%%%%%%

A main ingredient to understand transport in the presence of time-dependent driving are clearly the decay dynamics of a system in response to an excitation. Here we show, how the different relaxation times of a double quantum dot---of high relevance in the context of single-particle control or in the context of charge qubits to name two examples---can be read out by analyzing the transition between adiabatic and nonadiabatic charge pumping~\cite{Riwar2016Jun}. We consider two quantum dots that are weakly coupled to each other and to one electronic reservoir each. 
The double dot is described by the Hamiltonian given in Eq.~\eqref{eq:Hddot}, where we now choose $\epsilon_j(t)=\bar{\epsilon}_j+\delta\epsilon_j\cos\left(\Omega t+\phi_j\right)$ as the time-dependently driven local dot levels. The coupling $\Delta$ between the dots, as well as the tunnel coupling $\Gamma_\alpha$ between the dots and the neighbouring contacts are weak, $\beta\Delta,\beta\Gamma_\alpha\ll 1$. Double occupations of the double dot is forbidden by strong Coulomb interaction. This setting is ideal for adiabatic pumping of quantized charges, since driving of the separate dots via gates is experimentally straightforwardly realized and in the limit of large driving amplitudes $\delta \epsilon_j\gg\Delta$ the working principle of a peristaltic pump allows to "squeeze" one electron after the other through the device. This happens when the gates drive the system around the triple point indicated in the stability diagram in Fig.~\ref{fig:spectroscopy}, see also Sec.~\ref{sec:injection_confined}. In the adiabatic-response regime, the pumped charge has geometric properties, leading to the fact that the current pumped per cycle $I^c$ is related to the current pumped in the reversed cycle ${I^c}'$ by $I^c=-{I^c}'$. This is different when the driving is fast, such that the response is delayed due to the response times of the system. Relevant time scales are here the charge relaxation rate of the double dot due to coupling to the left and right contact, namely $\gamma_\mathrm{L}\equiv\Gamma_\mathrm{L}(1+f_\mathrm{L}(E))$ and $\gamma_\mathrm{R}=\Gamma_\mathrm{R}(1+f_\mathrm{R}(E))$, as well as the inelastic relaxation rate enabling transitions between the dot levels at different energies. It turns however out that in the nonadiabatic regime the ratio of the pumped currents due to reversed driving protocols always has the shape 
\begin{eqnarray}
    \frac{I^c}{{I^c}'}=e^{\gamma_x\delta t_x}
\end{eqnarray}
with some relaxation rate $\gamma_x$ and a specific time scale $\delta t_x$, which depend on the region of the stabilty diagram, we are considering, see Fig.~\ref{fig:spectroscopy}.
For example, in the upper violet regions, the ratio between currents reads
\begin{eqnarray}
    \frac{I^c}{{I^c}'}=e^{\gamma_\mathrm{R}\left(\Delta t-\Delta t'\right)}
\end{eqnarray}
where $\Delta t$ is the time between the crossing between left and right dot level and between the right dot level and the Fermi energy of the right contact (and vice versa for $\Delta t'$). Interestingly, the relaxation rate $\gamma_\mathrm{R}$ detected in this way, will be found to take two different values in the two violet regions, namely $\gamma_\mathrm{R}\simeq\Gamma_\mathrm{R}$ and $\gamma_\mathrm{R}\simeq2\Gamma_\mathrm{R}$ depending on whether the discharging or the charging is regarded. This factor 2, stemming from the value of the Fermi function in the considered regions is a result of spin degeneracy of the single dot levels. This strategy to read out relaxation rates from nonadiabatic pumping suggests an alternative to time-resolved measurements~\cite{Feve2007May}, counting experiments~\cite{Gustavsson2006Feb}, measurement of finite-frequency noise~\cite{Parmentier2012Apr} or radio-frequency reflectometry~\cite{House2015Nov}. Relaxation rates due to inelastic scattering have also been detected from spontaneous emission spectra~\cite{Fujisawa1998Oct}.\\

%%%%%%%%%%%%%%%%%%%%%%%%%%%%%%%%%%%%%%%%%%%%%%%%%%%%%%%%%%
\paragraph{Geometric phase of pumping}
%%%%%%%%%%%%%%%%%%%%%%%%%%%%%%%%%%%%%%%%%%%%%%%%%%%%%%%%%%%

Furthermore, the geometric nature of pumping is particularly evident in superconducting systems, where transport takes place via non-dissipative Cooper-pair transfer~\cite{Russomanno2011Jun}. This was for example analyzed both theoretically~\cite{Aunola2003Jul,Mottonen2006Jun} and experimentally~\cite{Mottonen2008Apr} in so-called superconducting sluices, which are flux-assisted Cooper-pair pumps. The transferred charge in a two-terminal device is then directly related to the quantum state $\Psi(t)$ describing the system via
\begin{equation}
    2q\,\Delta N_\alpha = \frac{2q}{h}\int_0^\mathcal{T}\langle\Psi(t')|\left(\partial_\varphi \hat H(t)|\Psi(t')\rangle\right),
\end{equation}
with the phase-difference $\varphi$ across the central system. The Hamiltonian $\hat H(t)=\hat H(\boldsymbol{X}(t),\varphi)$ depends on time through the driving parameters $\boldsymbol{X}(t)$ and will have additional dependencies on local charges and their conjugated phases of elements of the specific central circuit realization. This transferred charge is directly related to the Berry curvature. Interestingly, it has been shown that this concept can even be extended to revealing non-abelian holonomies in superconducting sluices~\cite{Brosco2008Jan}. This is possible when the periodic adiabatic driving takes place in a degenerate sub-space $\mathcal{H}_n$ of the Hilbert space. A quantum state $\psi_{n\alpha}\in\mathcal{H}_n$ then follows the time evolution $|\psi_{n\alpha}(\mathcal{T})\rangle=\left[U_n(\mathcal{T})\right]_{\alpha\gamma}|\psi_{n\gamma}(\mathcal{T})\rangle$ (plus higher-order non-adiabatic corrections) with the time-evolution matrix of the dimension of $\mathcal{H}_n$ given by
\begin{equation}
    U_n(\mathcal{T}) = e^{-\frac{i}{\hbar}\int_0^\mathcal{T}E_n(t)dt}\mathscr{T}e^{-\int_0^\mathcal{T}\Gamma_n(t)dt}
\end{equation}
where $E_n(t)$ is the energy eigenvalue of the degenerate subspace and $\Gamma_n(t)=\langle\psi_{n\alpha}(t)|\dot{\psi}_{n\gamma}(t)\rangle$ is the geometric connection. We recall that $\mathscr{T}$ denotes time-ordering. If the system now starts the evolution in a superposition of these degenerate eigenstates $c_{n\gamma}|\psi_{n\gamma}\rangle$, then the transferred charge after one cycle is given by
\begin{eqnarray}
    2q\Delta N & = & \frac{2q}{\hbar}\sum_{\gamma\gamma'}c_{n\gamma}^*c_{n\gamma'}\int_0^\mathcal{T}dt\Big\{\partial_\varphi E_n \delta_{\gamma\gamma'}\\
    && -i\hbar\sum_{\alpha\alpha'}\Big([U_n^\dagger]_{\gamma\alpha}\left(\partial_\varphi[\Gamma_n]_{\alpha\alpha'}\right)[U_n]_{\alpha'\gamma'}
    -\partial_t\left([U_n^\dagger]_{\gamma\alpha}\langle\psi_{n\alpha}|\partial_\varphi|\psi_{n\alpha'}\rangle[U_n]_{\alpha'\gamma'}\right)\Big)\Big\}\nonumber\ .
\end{eqnarray}
It consists of the usual contribution from the supercurrent (first term) as well as geometrical contributions due to pumping which reveal qualitatively new effects for non-Abelian dynamics. They occur as a modified periodicity in the pumped charge as well as a dependence on the starting point of the driving cycle. In order to even show the non-commutativity arising from the non-Abelian dynamics, a three-island system has been proposed in Ref.~\cite{Brosco2008Jan}. 

A number of different opportunities for transport spectroscopy with adiabatic pumping have been theoretically proposed in hybrid systems with normal- and superconducting contacts containing for example quantum dots as central driven region~\cite{Blaauboer2002May,Splettstoesser2007Jun}. Concretely, it was shown that the adiabatically pumped charge reveals distinct features of Andreev interference~\cite{Taddei2004Aug} or of crossed Andreev reflection~\cite{Hiltscher2011Oct}. Also features of a Higgs-like pair amplitude have been elucidated~\cite{Kamp2021Jan}. Hybrid superconducting devices have furthermore recently come to the focus since they can host Majorana fermions.
Even here, adiabatic pumping has been suggested as a tool to reveal features of Majorana states that are distinct from the ones potentially observable in steady-state transport measurement~\cite{Tripathi2019Feb}, and are expected in the noise of a quantum pump~\cite{Herasymenko2018Jun}.\\

%%%%%%%%%%%%%%%%%%%%%%%%%%%%%%%%%%%%%%%%%%%%%%%%%%%%%%%%%%
\paragraph{Revealing level renormalization}
%%%%%%%%%%%%%%%%%%%%%%%%%%%%%%%%%%%%%%%%%%%%%%%%%%%%%%%%%%%

These latter examples show that measurement of pumped charge can serve as a tool to access the otherwise hidden intrinsic properties of the central many-body quantum system. Further examples, in which the advantages of using adiabatic pumping as  ``enhanced transport spectroscopy" become visible, show the possibility to read out Coulomb interaction effects. This is for example the case when pumping charge through a single-level quantum dot with \textit{strong onsite} Coulomb interaction $U$ and weakly coupled to two electronic contacts, $\alpha=\mathrm{L,R}$. The Hamiltonian for this setup reads
\begin{equation}
    \hat H=\sum_\sigma\epsilon(t)\hat d^\dagger_\sigma \hat d_\sigma+U\hat{N}_\uparrow \hat{N}_\downarrow+\sum_{\alpha k\sigma}\epsilon_{\alpha k\sigma}\hat c^\dagger_{\alpha k\sigma}\hat c_{\alpha k\sigma}
+\sum_{\alpha k \sigma}\left(w_{\alpha k}(t)\hat d^\dagger_\sigma \hat c_{\alpha k\sigma}+\mathrm{h.c.} \right)
\end{equation}
with the dot number operators for spin $\sigma$ given by $\hat{N}_\sigma=\hat{d}^\dagger_\sigma \hat{d}_\sigma$ and $\sum_\sigma\hat{N}_\sigma=\hat{N}$. Here, the single-level energy $\epsilon(t)$ and the couplings to the contacts $w_{\alpha k}(t)$ can be time-dependently driven. In the sequential-tunneling regime the current pumped into the left the contacts takes the particularly simple form
\begin{equation}\label{eq:lowest_order_I}
    I_\mathrm{L}^c(t)=q\frac{\Gamma_\mathrm{L}(t)}{\Gamma(t)}\frac{d}{dt}\langle \hat{N}\rangle^{(0)}(t)
\end{equation}
where $\langle \hat{N}\rangle^{(0)}(t)$ is the expectation value of the dot occupation in zeroth order in the tunnel coupling. This expression shows in a clear way the working principle of the single-dot adiabatic pump: a current arises due to the time-dependent variation of the dot occupation. If this results in a dc current after time-averaging and in which direction this current flows, depends on the (relative) time-dependence of the coupling parameters $\Gamma_\mathrm{L}(t)$ and $\Gamma_\mathrm{R}(t)$, where here $\Gamma(t)=\Gamma_\mathrm{L}(t)+\Gamma_\mathrm{R}(t)$. In the weak-coupling regime, this pumping mechanism, in addition to the usual requirements of a minimum of two out-of-phase time-dependently driven parameters, needs a time-dependent variation of the level energy $\epsilon(t)$ (or of the interaction strength) since the time-dependent couplings do not impact the dot occupation, which in lowest order is given by
\begin{equation}
    \langle \hat{N}\rangle^{(0)}(t)=\frac{f(\epsilon(t))(1-f(\epsilon(t)+U))}{1+f(\epsilon(t))-f(\epsilon(t)+U)}.
\end{equation}
Higher-order effects in the tunnel coupling lead to a finite life-time broadening as well as to a renormalization of the level position, also referred to as Lamb-shift. These renormalization effects directly impact the expectation value of the dot-occupation
\begin{eqnarray}
    \langle \hat{N}\rangle^{(1)} & = & \langle \hat{N}\rangle^\mathrm{broad}+\langle \hat{N}\rangle^\mathrm{ren}\\
   \langle \hat{N}\rangle^\mathrm{broad} & = & \left(2-\langle \hat{N}\rangle^{(0)}\right)\phi'(\epsilon)+\langle\hat{N}\rangle^{(0)}\phi'(\epsilon+U)\\
   \langle \hat{N}\rangle^\mathrm{ren} & = & \frac{d\langle \hat{N}\rangle^{(0)}}{d\epsilon}\sigma(\epsilon,U)
\end{eqnarray}
but they are typically hard to access experimentally, since they appear as small corrections to the lowest-order result. Here, $\phi(\epsilon)=\frac{\Gamma}{2\pi}\mathrm{Re}\Psi(\frac{1}{2}+\frac{i\beta\epsilon}{2\pi})$ with the digamma function $\Psi$ and $\sigma(\epsilon,U)=\phi(\epsilon)-\phi(\epsilon+U)$ is the level renormalization. A rigorous calculation in second order in the tunnel coupling, based on the gerneralized master equation approach of Sec.~\ref{sec:ME_perturbative} shows that the pumping current is directly related to these corrections of the dot occupation
\begin{equation}
 I_\mathrm{L}^c(t)=q\frac{\Gamma_\mathrm{L}(t)}{\Gamma(t)}\frac{d}{dt}\left(\langle \hat{N}\rangle^{(0)}(t)+ \langle \hat{N}\rangle^\mathrm{ren}(t)\right)+q\frac{d}{dt} \langle \hat{N}\rangle^\mathrm{broad,L}.
\end{equation}
Interestingly, the effect of line-width broadening (entering here only due to coupling to the left contact) contributes in terms of a total time-derivative. In the time-integrated charge pumped per period it hence cancels out. When choosing the coupling contacts $\Gamma_\mathrm{L}(t)$ and $\Gamma_\mathrm{R}(t)$ as pumping parameters, the lowest order contribution of Eq.~\eqref{eq:lowest_order_I} is zero, as explained above. This means that the pumped charge as a result of driving $\Gamma_\mathrm{L}(t)$ and $\Gamma_\mathrm{R}(t)$ is uniquely due to level renormalization. In fact in the regime of weak driving (small driving amplitudes), the pumped charge is even directly proportional to the level renormalization, $\Delta N_\mathrm{L}=q\mathcal{A}\sigma(\epsilon,U)d\langle \hat{N}\rangle^{(0)}/d\epsilon$, with the area $\mathcal{A}$ enclosed by the pumping parameters. This pumped charge \textit{induced} by level renormalization provides hence a tool for its direct readout.\\

%%%%%%%%%%%%%%%%%%%%%%%%%%%%%%%%%%%%%%%%%%%%%%%%%%%%%%%%%%
\paragraph{Read-out of screening effects}
%%%%%%%%%%%%%%%%%%%%%%%%%%%%%%%%%%%%%%%%%%%%%%%%%%%%%%%%%%

Time-dependent driving of a conductor has also been proposed as a tool to read out screening effects, see Sec.~\ref{sec:screening}, resulting from the pile-up of charges in a biased conductor~\cite{Dashti2021Dec}. In the absence of Coulomb interaction, transport through a conductor can be described within scattering theory, characterizing the conductor by the energy-dependent transmission probability $D(E)$. Pile-up of charges and resulting screening effects modify this transmission probability as function of the applied potential or even temperature biases
\begin{equation}
    D(E,\left\{qV,\Delta T\right\})=D_0(E)-\frac{1}{2}\frac{dD_0(E)}{dE}\left(\xi qV-\chi k_\mathrm{B} \Delta T\right)
\end{equation}
with the screening coefficients $\xi,\chi$, which contribute linearly in the regime of weak screening. These screening effects (such as quantified via the coefficients $\xi$ and $\chi$) are again difficult to read out and typically occur as corrections in nonlinear response~\cite{Christen1996Sep,Pedersen1998Nov,Polianski2006Apr,Polianski2007Oct,Sanchez2013Jan,Meair2013Jan}, while linear-response coefficients are not affected by these screening effects. However, the injection of a pure ac signal onto such a conductor, results in modifications of the (thermoelectric) linear-response coefficients, 
\begin{equation}
    \left(\begin{array}{c} I^c \\ I^Q \end{array}\right)=\left(\begin{array}{c} I^\text{dir}_\text{ac} \\I_\text{ac}^{E,\text{dir}} \end{array}\right)+\left(\begin{array}{cc} G+G_\text{ac} & L+L_\text{ac} \\ M+M_\text{ac} & K+K_\text{ac} \end{array} \right) \left(\begin{array}{c} V  \\ \Delta T \end{array}\right),
\end{equation}
where the corrections $G_\text{ac},L_\text{ac},M_\text{ac},K_\text{ac}$ are \textit{induced} by screening effects. Concretely, the corrections read
\begin{align}
    G_\text{ac}&=\xi \frac{q^2}{2h}\mathcal{J}_0, & \quad L_\text{ac}&=-\chi \frac{k_\text{B}q}{2h} \mathcal{J}_0,\nonumber \\
    M_\text{ac}&=\xi \frac{k_\text{B}q}{2h} T_0 \mathcal{J}_1, & \quad K_\text{ac}&=-\chi \frac{k^2_\text{B}T_0}{2h}  \mathcal{J}_1, 
\label{eq:screening_coeffs}
\end{align}
where ${\mathcal J}_\ell=
\sum_{n}  \int dE   {|S_{F}(E,E_n)|}^2 \frac{dD_0(E)}{dE} \left(E/k_\text{B}T_0\right)^\ell\left[f(E)-f(E+\hbar\Omega n)\right]$ with the Floquet scattering matrix (here a scalar) characterizing the ac source, see Sec.~\ref{t-per-scatt}.
The modifications of this set of thermoelectric response coefficients given in Eq.~\eqref{eq:screening_coeffs} can be switched on and off by the driving and their combined readout hence gives direct access to the screening coefficients~\cite{Dashti2021Dec}.

%%%%%%%%%%%%%%%%%%%%%%%%%%%%%%%%%%%%%%%%%%%%%%%%%%%%%
\subsection{Dissipation and noise}\label{sec:noise and energy}
%%%%%%%%%%%%%%%%%%%%%%%%%%%%%%%%%%%%%%%%%%%%%%%%%%%%%

The charge current, as discussed in the previous sections, is the most frequently analyzed observable in quantum transport. However, also energy currents are of high-interest in the context of transport due to time-dependent driving, not only  because the time-dependent drive acts as a source of energy for the system, but also because energy currents give complementary access to the spectral properties of transported particles. In addition, fluctuations of charge and energy currents can limit the performance, but they also elucidate the properties of quantum transport due to time-dependent driving for example due to particle-particle correlations. These aspects are the focus of the present subsection.

%%%%%%%%%%%%%%%%%%%%%%%%%%%%%%%%%%%%%%%%%%%%%%%%%%%%%
\subsubsection{Energetic and thermodynamic properties of time-dependent transport}\label{sec_energy_spectroscopy}
%%%%%%%%%%%%%%%%%%%%%%%%%%%%%%%%%%%%%%%%%%%%%%%%%%%%%

It has been discussed in Sec.~\ref{sec:mechanism_energy_dynamics} at the most simple example of the mesoscopic capacitor  that the energy response of a time-dependently driven system contains complementary features compared to the charge response.

\begin{figure}
    \centering
    \includegraphics[width=0.85\linewidth]{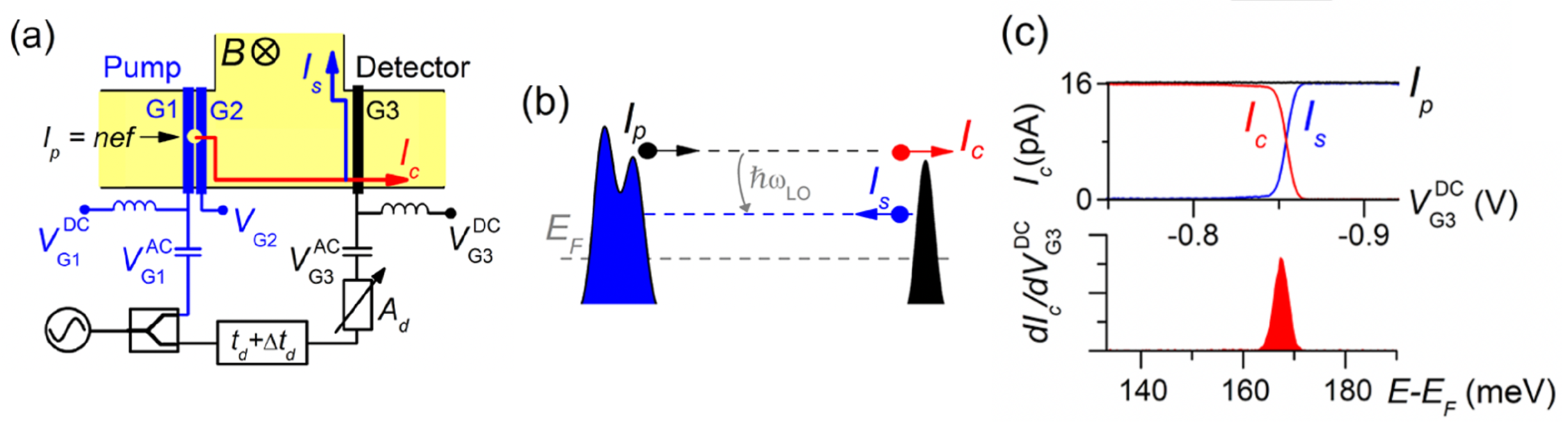}
    \caption{Hot-carrier single-electron pump (blue) injecting particles onto a driven barrier, which filters the injected signal in energy space. The conductance in response to the barrier-gate provides the spectral properties of the injected signal. Reprinted figure with permission from \cite{Fletcher2013Nov}. Copyright 2013 by the American Physical Society.}
    \label{fig:hot-carrier_barrier-response}
\end{figure}

As a result, energy currents and other thermodynamic properties~\cite{Caso2010Jan,Nello2024Oct} are an additional means to study the properties of time-dependently driven single- or few electron sources, as introduced in Secs.~\ref{sec:injection_target_drives} and~\ref{sec:injection_confined}. Indeed, when sigle-electron sources are realized with the goal to provide single-particle excitations as input for quantum experiments, such as flying qubits~\cite{Edlbauer2022Dec} or for fundamental experiments in quantum optics with electrons as introduced in the following Sec.~\ref{sec_QOE}, the \textit{spectral} properties of the signal are often relevant. Information about these spectral properties~\cite{Battista2012Feb,Dashti2019Jul,Schulenborg2024Mar} can for example be accessed via the measurement of energy currents or energy-resolved currents. For the analysis of single-electron sources, such energy-resolved measurements have been carried out using (driven) detector barriers, placed behind the single-electron source~\cite{Fletcher2013Nov,Ubbelohde2015Jan} or via tomographic techniques~\cite{Jullien2014Oct,Thibierge2016Feb,Bisognin2019Jul,Locane2019Sep}. 
A sketch of the driven-detector barrier setup, exploiting the influence of the barrier-gate driving on the charge response is shown in Fig.~\ref{fig:hot-carrier_barrier-response}. It demonstrates how the signal---emitted here from a hot-carrier single-electron pump (blue, see also Sec.~\ref{sec:injection_confined})---is injected onto a driven barrier (red), where the conductance in response to the barrier drive yields the spectral properties of the source current.
%%%%%%%%%%%%%%%%%%%%%%%%%%%%%%%%%%%%%%%%%
%%%%%%%%%%%%%%%%%%%%%%%%%%%%%%%%%%%%%%%%%

The complementary behavior of charge and energy currents of time-dependently driven systems can go that far to lead to counterintuitive effects of charge-energy separation. For example, it has been shown that---depending on delay times between different single-electron signals in electron interferometers, interference effects can be fully suppressed in charge currents, while the energy current carried by these particles displays coherent oscillations and vice versa~\cite{Rossello2015Mar,Hofer2014Dec}. Another related of this phenomenon is the generation of
charge pulses be means of heat pulses in interferometers~\cite{Portugal2024Jun}.

It has furthermore been shown in Sec.~\ref{sec:energy_conversion} that time-dependently driven quantum systems, such as quantum pumps constitute a possibility to implement heat engines in electronic conductors. 
Here, the time-dependent driving is used to realize the engine cycle, connecting and disconnecting the quantum systems to different heat or work baths. 
In addition to this conventional heat-engine implementation, time-dependent driving has also been proposed to improve the efficiency of thermoelectric devices~\cite{Zhou2015Oct}, where the different time-dependent response of charge and energy currents is exploited.
Time-dependent driving can furthermore be used to prepare unconventional resources for thermoelectrics, which due to their non-thermal properties lead to counter-intuitive results when standard efficiencies are investigated~\cite{Ryu2022May,Aguilar2024Sep}.

%%%%%%%%%%%%%%%%%%%%%%%%%%%%%%%%%%%%%%%%%%%%%%%%%%%%%
\subsubsection{Fluctuations in time-dependent transport}\label{sec_noise}
%%%%%%%%%%%%%%%%%%%%%%%%%%%%%%%%%%%%%%%%%%%%%%%%%%%%%

Up to here, the focus has been on \textit{average} transport observables resulting from or influenced by the time-dependent driving of system parameters. However, the fluctuations of these observables, or in other words their noise, can play an important role in small-scale devices~\cite{Blanter2000Sep,Campisi2011Jul,Landi2024Apr}. 
The noise of a transport current is related to the two-time-correlator of the current operator  $\mathcal{S}_{\alpha\gamma}(t,t') = \langle \delta \hat{I}_\alpha(t')\delta\hat{I}_\gamma(t)\rangle$
with $\delta \hat{I}_\alpha(t) := \hat{I}_\alpha(t) - I_\alpha$.
Note that the noise is often defined via the symmetrized correlator, see Eq.~\eqref{eq:noise-correlator-general}. Furthermore, while we here present an expression for particle-current correlations, the same procedure can be employed for other types of transport currents. An important conceptual difference of the current-current correlator in time-dependently driven systems is that it does depend on two times $t,t'$, compared to the current-current correlator  in stationary systems which only depends on a time difference $\tau\equiv t-t'$. Therefore the noise power in general also depends on two frequencies, and a time-average, as shown in Eq.~\eqref{eq:single-frequency noise}, needs to be employed to obtain a standard, frequency-dependent noise power. 
The most often studied zero-frequency noise, $\mathcal{S}_{\alpha\gamma}$, is hence obtained from a double time integration. 
Zero- and finite-frequency noise of time-dependently driven systems have been analyzed in detail employing different methods. A detailed introduction to the calculation of noise in time-dependently driven systems using scattering theory~\cite{Moskalets2002Jul,Polianski2002Jun,Moskalets2008Jul,Parmentier2012Apr} can be found in this book~\cite{Moskalets2011Sep}. In such a setting, weak interactions effects have been included using a renormalization procedure~\cite{Devillard2008Aug}. Zero- and finite-frequency noise of time-dependently driven quantum dots and metallic islands with a possibly strong onsite interaction has furthermore been analyzed based on a generalized master equation approach~\cite{Riwar2013May,Dittmann2018Sep}. Also Bosonization techniques have been employed to calculate the noise of adiabatic pumps in the presence of strong correlations~\cite{Herasymenko2018Jun} and Green's function techniques for photon-assisted tunneling~\cite{Suzuki2015Apr}.

Beyond the finite- and zero-frequency noise of time-dependently driven conductors, also the \textit{full counting statistics} have been analyzed, giving access not only to the variance but to all cumulants of a transport current~\cite{Muzykantskii2003Oct,Vanevic2007Aug,Benito2016Nov}, possibly at very high driving frequencies~\cite{Hubler2023Jul,Cuetara2015May} or---in the opposite regime---for adiabatic pumping~\cite{Honeychurch2020Nov,Riwar2021Jan,Croy2016Apr}.
In the latter case, full counting statistics reveals nicely the geometric nature of quantum pumping~\cite{Pluecker2017Apr}.
Let us here highlight three aspects that make the noise in time-dependently driven systems a crucial observable. 
%\begin{description}

%\item
(1)~Time-dependently driven conductors are of high interest as single-electron sources for the realization of the current standard~\cite{Pekola2013Oct,Kaneko2016Feb}. Here, precision is of utmost importance which means that fine-tuning of the device parameters or even feedback mechanisms~\cite{DeRanieri2014Jul,Fricke2014Jun,Potanina2019Jan} are employed in order to reduce the noise~\cite{Maire2008Feb}, for example in the charging and discharging process of driven quantum dots~\cite{Fricke2013Mar}. 
Excitations with suppressed noise have recently also been created by shaping an ac driving signal~\cite{Gabelli2013Feb}. The ideal case of a so-called \textit{Leviton}, a Lorentzian shaped pulse~\cite{Keeling2006Sep,Keeling2008Nov}, is characterized by the absence of electron-hole pair creation leading to a noise that is suppressed to the one of a stationary system~\cite{Vanevic2012Dec,Dubois2013Aug,Dubois2013Oct,Vanevic2016Jan,Glattli2018Mar,Assouline2023Dec}; see Sec.~\ref{sec:injection_target_drives} for more details. 

%\item
(2)~Noise of time-dependently driven conductors is furthermore employed as an important spectroscopy tool, in order to reveal quantum statistics and multiparticle quantum effects. One particularly prominent example is the field of quantum optics with electrons, as discussed in detail in Sec.~\ref{sec_QOE} of this review. Importantly, time-dependent driving of a conductor creates
multiparticle correlations that appear in noise~\cite{Moskalets2006Mar}; in particular interference in electron-hole pairs created by driving has been discussed~\cite{Reydellet2003Apr,Rychkov2005Oct,Battista2014Aug}, including dephasing and interaction effects~\cite{Polianski2005Oct}, as well as the possibility of entanglement generation by driving~\cite{Sherkunov2012Feb}. All these phenomena can be straightforwardly revealed by noise, while they are typically hidden in average currents. Moreover, photoassisted shot noise has also been used to provide an alternative determination of the fractional charge in the fractional quantum Hall effect~\cite{Crepieux2004May,Kapfer2019Jan,Bisognin2019Apr}

%\item
(3)~In equilibrium, fluctuation dissipation theorems (FDTs) relate the noise of a system to linear-response coefficients, see also Sec.~\ref{sec:kubo}. Extensions of FDTs beyond equilibrium have been the focus of extensive research, see for example Ref.~\cite{Esposito2009Dec} for a review. Nonequilibrium due to time-dependent driving has been one of the fields of interest, which we want to focus on here. Note that nonequilibrium FDTs can also serve as guideline for noise reduction in driven systems. In the following, we focus on some instances of fluctuation-dissipation theorems of time-dependently driven systems.
%\end{description}
\\

%%%%%%%%%%%%%%%%%%%%%%%%%%%%%%%%%%%%%%%%%%%%%%%%%%%%%%%%%
\paragraph{Fluctuation-dissipation theorem for adiabatic quantum-dot pump}
%%%%%%%%%%%%%%%%%%%%%%%%%%%%%%%%%%%%%%%%%%%%%%%%%%%%%%%%%

While in general the fluctuation-dissipation theorem is violated in the presence of a time-dependent driving, which can directly be brought in connection with the heat pumped into the system~\cite{Ren2010Apr}, one can find modified fluctuation-dissipation theorems for adiabatic quantum pumping, here focussing on the regime of \textit{weak tunnel coupling}~\cite{Riwar2021Jan}. A convenient starting point for the derivation of fluctuation dissipation theorems is to exploit fluctuation relations of the cumulant generating function of the transport full counting statistics, here for the driven system~\cite{Hino2020Jul}. The generating function $\mathcal{F}$ can be found from the time-evolution of the system in the presence of counting fields, where in the long-time limit $\mathcal{F}(\left\{\chi_\alpha,\xi_\alpha\right\})$ depends on the charge counting field for exchange with contact $\alpha$, $\chi_\alpha$, and on the energy counting field $\xi_\alpha$, but not on initial or final measurement times. From the generating function, the charge current and its noise as well as the energy current can be found
\begin{eqnarray}
    I^c_\alpha & = & -i\partial_{\chi_\alpha}\mathcal{F}|_{\left\{\chi_\alpha,\xi_\alpha\right\}\rightarrow 0}\,,\\
     I^E_\alpha & = & -i\partial_{\xi_\alpha}\mathcal{F}|_{\left\{\chi_\alpha,\xi_\alpha\right\}\rightarrow 0}\,,\\
    \mathcal{S}_{\alpha\gamma} & = & -\partial_{\chi_\alpha}\partial_{\chi_\gamma}\mathcal{F}|_{\left\{\chi_\alpha,\xi_\alpha\right\}\rightarrow 0}\,.
\end{eqnarray}
For a slowly driven system, the generating function can be expanded in orders of the driving frequency with an instantaneous and an adiabatic-response contribution. Those can be written in terms of the smallest eigenvalue of the time-evolution operator in the presence of the counting fields $\lambda_0(\{\chi_\alpha,\xi_\alpha\})$ and the related eigenvectors, $|0(\{\chi_\alpha,\xi_\alpha\}))$ and $(0(\{\chi_\alpha,\xi_\alpha\})|$, which are operators in Liouville space,
\begin{eqnarray}
    \mathcal{F}^{(0)}(\left\{\chi_\alpha,\xi_\alpha\right\}) & = & \int_0^{\mathcal{T}} \frac{dt}{\mathcal{T}}\lambda_0(\{\chi_\alpha,\xi_\alpha\})\\
    \mathcal{F}^{(1)}(\left\{\chi_\alpha,\xi_\alpha\right\}) & = & -\int_0^{\mathcal{T}} \frac{dt}{\mathcal{T}}(0(\{\chi_\alpha,\xi_\alpha\})|\partial_t|0(\{\chi_\alpha,\xi_\alpha\}))\ .
\end{eqnarray}
The adiabatic response contribution clearly shows the discussed geometric properties. One of the consequences are fluctuation relations which differ by a sign for the instantaneous,  $\mathcal{F}^{(0)}(\left\{\chi_\alpha,\xi_\alpha\right\})= \mathcal{F}^{(0)}(\left\{i\beta_\alpha\mu_\alpha-\chi_\alpha,i\beta_\alpha-\xi_\alpha\right\})$, and adiabatic response, $\mathcal{F}^{(1)}(\left\{\chi_\alpha,\xi_\alpha\right\})= -\mathcal{F}^{(1)}(\left\{i\beta_\alpha\mu_\alpha-\chi_\alpha,i\beta_\alpha-\xi_\alpha\right\})$, since micro-reversibility for the adiabatic response also requires a change in the current direction. Based on these modified fluctuation relations and on gauge invariance and the fact that the energy current is \textit{not} conserved in the presence of driving, one furthermore finds for the generating functions
\begin{eqnarray}
      \mathcal{F}^{(0)}(\{\chi_\alpha\}) & = &  \mathcal{F}^{(0)}(\{i\beta\mu_\alpha-\chi_\alpha\})\\
      \mathcal{F}^{(1)}(\{\chi_\alpha\}) & = &  \mathcal{F}^{(1)}(\{i\beta\mu_\alpha-\chi_\alpha\})+\beta \mathcal{Q}^{(1)}(\{\chi_\alpha\}) 
\end{eqnarray}
where we here focus on charge transport. These relations are valid for all contacts being taken at the same inverse temperature $\beta$. The important function $\mathcal{Q}^{(1)}(\{\chi_\alpha,\xi_\alpha\})=-\int \frac{d\mathcal{T}}{dt}(0(\{\chi_\alpha,\xi_\alpha\})|\mathbf{\dot{e}}|0(\{\chi_\alpha,\xi_\alpha\}))$ is related to energy transport, where $\mathbf{e}$ is the energy superoperator. These relations are the starting point for deriving a fluctuation-dissipation theorem for weakly coupled
 and slowly driven quantum systems, which reads
\begin{equation}
    \mathcal{S}_{\alpha\gamma}^{(1)}|_{\{\mu_\alpha\rightarrow\mu\}}=k_\mathrm{B} T\left(\frac{\partial I^{c(1)}_\alpha}{\partial\mu_\gamma}+\frac{\partial I_\gamma^{c(1)}}{\partial\mu_\alpha}-\frac{\partial I^{E(1)}_\mathrm{pump}}{\partial\mu_\gamma\partial\mu_\alpha}\right)\ .
\end{equation}
 Here, $I^{E(1)}_\mathrm{pump}$ is the total energy current pumped into the system by the driving fields. This is a first instance of the important role that energy currents play for dissipation relations of driven systems.\\

%%%%%%%%%%%%%%%%%%%%%%%%%%%%%%%%%%%%%%%%%%%%%%%%%%%%%%%%%
\paragraph{Fluctuation relations from perturbation theory}
%%%%%%%%%%%%%%%%%%%%%%%%%%%%%%%%%%%%%%%%%%%%%%%%%%%%%%%%%

Also for more general driving schemes, including the time-dependent driving of a bias voltage, fluctuation relations can be found, which relate the noise to the current~\cite{Safi2011Nov,Safi2014Jan}. Starting from a general Hamiltonian
\begin{equation}
    \hat H(t)=\hat H_0+\left(e^{i\frac{qV_{\rm dc}}{\hbar}t}A(t)\hat{\mathcal{V}} +\text{h.c.}\right)
\end{equation}
with a tunneling term composed of a tunneling operator $\hat{\mathcal{V}}$, a time-dependent driving $A(t)$, and a dc voltage $V_{\rm dc}$, a perturbative approach in the tunneling with respect to the unperturbed Hamiltonian $\hat H_0$ has been applied~\cite{Safi2009Apr,Safi2011Nov,Safi2014Jan}. This approach shows that the noise of the charge current in a two-terminal setup\footnote{Here, we simply write $\mathcal{S}$, without specifying the indices $\mathcal{S}_{\alpha\gamma}$ because in a two-terminal setups all correlation functions are the same (up to a sign) due to charge-current conservation $I_{\mathrm{R}}^c=-I_{\mathrm{L}}^c$.} can be expressed as
\begin{equation}
    \mathcal{S} = q\int \frac{d\omega}{2\pi}|A(qV_{\rm dc}-\hbar\omega)|^2\coth\left(\frac{\hbar\omega}{2k_\mathrm{B}T}\right)I^c(\omega)
\end{equation}
where the driving term $A$ explicitly appears in the relation (here written in frequency space). For a non-driven system, this expression reduces to the stationary non-equilibrium fluctuation relations~\cite{Rogovin1974Jul,Levitov2004Sep}. It involves the statement that charge fluctuations are super-poissonian, while specific driving schemes, such as the previously introduced Lorentzian voltage pulses have poissonian noise.\\

%%%%%%%%%%%%%%%%%%%%%%%%%%%%%%%%%%%%%%%%%%%%%%%%%%%%%%%%%
\paragraph{Mesoscopic capacitor: Fluctuation theorem for higher harmonics}
%%%%%%%%%%%%%%%%%%%%%%%%%%%%%%%%%%%%%%%%%%%%%%%%%%%%%%%%%

In general, the noise power of a driven conductor depends on two frequencies. For conductors subject to periodic driving with frequency $\Omega$, this noise power can be expressed as the sum over noise-frequency dependent Fourier components, $\mathcal{S}(\omega,\omega')=\sum_{\ell=0}^\infty\delta(\omega+\omega'-\ell\Omega)\mathcal{S}_\ell(\omega)$.  
For (a set of) mesoscopic capacitors described by scattering theory, the $\ell$th Fourier component is given by~\cite{Moskalets2007Jan,Moskalets2009Aug}
\begin{eqnarray}
        &&\mathcal{S}_\ell(\omega)  =  \frac{q^2}{2h}\sum_{n=-\infty}^\infty\int dE\left[f(E)(1-f(E_n-\hbar\Omega))+f(E_n-\hbar\Omega)(1-f(E))\right]\\
        &&\qquad\qquad\times\int_0^\mathcal{T}\frac{dt}{\mathcal{T}}e^{in\Omega t}\left[S^*(t,E_n-\hbar\omega)S(t,E)-1\right]
        \int_0^\mathcal{T}\frac{dt'}{\mathcal{T}}e^{i(\ell-n)\Omega t'}\left[S^*(t',E)S(t',E_n-\hbar\omega)-1\right]\nonumber
\end{eqnarray}
with the scattering matrix of the capacitor $S(t,E)$. Also here, a relation with the energy current can be established, which for the driven capacitor is written as
\begin{equation}
    I^E = -\frac{i}{2\pi}\int dE f(E)\int_0^\mathcal{T}\frac{dt}{\mathcal{T}}S(t,E)\frac{\partial S^*(t,E)}{\partial t}\,.
\end{equation}
In the adiabatic-response limit and for small amplitudes of the gate potential $U(t)$ driving the mesoscopic capacitor, both the Fourier-components of the finite-frequency noise and of the energy current can be expressed in terms of the density of states
\begin{equation}
\nu(t,E) = \frac{i}{2\pi}S(t,E)\frac{\partial S^*(t,E)}{\partial E}.
\end{equation}
Concretely one finds
\begin{eqnarray}
    \mathcal{S}_\ell(\omega) & = & \pi q^2\hbar^2\omega(\omega-\ell\Omega)\coth\left(\frac{\hbar\omega}{2k_\mathrm{B}T}\right)\int dE\left(-\frac{\partial f(E)}{\partial E}\right)\left\{\nu^2\right\}_\ell\\
    I^E & = & \frac{hq^2U^2\Omega^2}{8}\int dE \left(-\frac{\partial f(E)}{\partial E}\right)\left[\left\{\nu^2\right\}_0-\left\{\nu^2\right\}_{-2}-\left\{\nu^2\right\}_2\right]
\end{eqnarray}
with $\left\{\nu^2\right\}_\ell$ being the $\ell$-th Fourier component of $\nu^2(t,E)$. Starting from these expressions, a fluctuation-dissipation theorem ccan be found for the driven mesoscopic apacitor, relating Fourier components of the finite-frequency noise to the energy current due to the driving
\begin{equation}
    \mathcal{S}_0(\omega)-\frac{\mathcal{S}_2(\omega)+\mathcal{S}_{-2}(\omega)}{2}=\frac{2I^E}{U^2}\frac{\hbar\omega^3}{\Omega^2}\coth\left(\frac{\hbar\omega}{2k_\mathrm{B}T}\right).
\end{equation}
The study of Fourier-components of the finite-frequency noise has also been promoted as a tool to study transfer processes in weakly coupled capacitors and to identify features of Coulomb interaction impacting  the capacitor's finite-frequency noise~\cite{Dittmann2018Sep}.\\

%%%%%%%%%%%%%%%%%%%%%%%%%%%%%%%%%%%%%%%%%%%%%%%%%%%%%%%%%
\paragraph{Trade-off relations}
%%%%%%%%%%%%%%%%%%%%%%%%%%%%%%%%%%%%%%%%%%%%%%%%%%%%%%%%%

These examples show the relation between noise in driven systems and the energy (or heat) currents provided by the driving fields. However, also these energy currents fluctuate and it is hence of interest to study the \textit{fluctuations} of energy or heat currents~\cite{Battista2013Mar}, which are directly related to \textit{power fluctuations}~\cite{Moskalets2014May,Battista2014Aug,Ronetti2019May,Dashti2018Aug} or even the correlations between charge and energy currents~\cite{Battista2014Dec}.

One of the scopes of studying power fluctuations comes from the interest in realizing heat engines at the nanoscale, producing power due to coupling to different heat baths and to external driving. In these nanoscale engines, not only the output power, but also the \textit{precision} of the output power is relevant. The attainable precision of an engine is limited by a trade-off, known as thermodynamic uncertainty relation~\cite{Barato2015Apr,Gingrich2016Mar,Pietzonka2018May}, with the efficiency and the average output power. Recently, these thermodynamic uncertainty relations valid for arbitrary transport currents have been extended to time-dependent driving~\cite{Koyuk2019Jun} also in the context of quantum transport~\cite{Potanina2021Apr,Lu2022Mar}. In particular, for adiabatic pumping, the role of the geometric contribution to currents or produced work has been highlighted~\cite{Lu2022Mar}
\begin{equation}
    \frac{\langle\langle J^2\rangle\rangle}{\langle J\rangle}\langle\Sigma\rangle\geq 2\left[\frac{1}{1+\frac{\langle J\rangle_\text{geo}}{\langle J\rangle_\text{dyn}}}\right]^2
\end{equation}
In the steady state limit, the bracket on the right-hand side of the inequality equals 1 and the standard thermodynamic uncertainty relation is found, bounding the average of a current $\langle J\rangle$ and its variance $\langle\langle J^2\rangle\rangle$ together with the entropy production $\langle\Sigma\rangle$ by 2. In the case of an additional time-dependent driving, the current does not only contain the dynamical contribution, given by the time-average over instantaneous currents, but also the geometric contribution, which lowers the imposed bound. Also beyond adiabatic-response driving and beyond linear response, bounds on the precision of a driven conductor can be found, which if subject to a large temperature bias is directly related to the power provided by (or transferred to) the external driving~\cite{Tesser2025May}.

%%%%%%%%%%%%%%%%%%%%%%%%%%%%%%%%%%%%%%%%%%%%%%%%%%%%%
\subsection{Electronic quantum optics
}\label{sec_QOE}
%%%%%%%%%%%%%%%%%%%%%%%%%%%%%%%%%%%%%%%%%%%%%%%%%%%%%
\subsubsection{General context}
The experimental progress in controlling single-electron excitations, from their emission to coherent propagation in quantum conductors, and possibly single-shot detection, has triggered the development of the subfield known as electronic quantum optics (EQO)~\cite{Grenier2011May}. In this section, we describe the main achievements within this field over the past 20 years and we highlight some of the challenges that are still being faced.

Electron quantum optics exploits the coherent manipulation of few-electron states in electronic quantum conductors to achieve a sort of signal processing with electric currents at the quantum level~\cite{Roussel2017Mar,Roussel2021May}. To this aim, a theoretical description of the coherence of quantum electric signals, as well as the circuit elements that allow one to manipulate them in various ways is needed. The key elements in electron quantum optics are (i) reliable single-electron sources, (ii) waveguides for electron propagation, and (iii) beamsplitters with which interferometers can be built. By combining these elements, it is possible to perform controlled (and in principle time-resolved) electronic interferometry and achieve signal processing with quantum electrical currents~\cite{Roussel2021May}.

(i) We have described in Sec.~\ref{sec_single-particle-control} some of the different strategies that have been proposed to implement reliable, on-demand single-electron sources. The success in implementing these sources has been a major step forward towards the development of electron quantum optics. The available single-electron sources are complementary to each other, insofar as they work in distinct energy ranges and operate in different platforms that play the role of waveguides for electronic propagation. 

(ii) 
Most of EQO experiments have been performed in 2-dimensional electron gases (2DEGs) in the integer quantum Hall regime (see Ref.~\cite{Bocquillon2014Jan} for an early review). Here, currents are carried by chiral edge states, along which propagation occurs according to a direction fixed by the applied magnetic field. Edge states in the integer quantum Hall effect are protected against backscattering, behaving as ideal waveguides for electron propagation. This is why integer quantum Hall edge states have been chosen as a platform to be combined with low-energy single-electron sources, such as the Leviton voltage source~\cite{Dubois2013Oct} and the mesoscopic capacitor~\cite{Feve2007May}. Another strategy is to use 2DEGs in combination with surface acoustic waves. Here, the surface acoustic wave acts both as a source and as a carrier, by loading a single electron from a properly initialized quantum dot in one of its minima and transporting it along a channel that is defined in the 2DEG by electrostatic gating~\cite{Bauerle2018Apr,Edlbauer2022Dec}.\\
In addition to these well-established platforms, theoretical proposals have been put forward, suggesting other systems where additional effects can play a crucial role and enrich the physics of EQO. Very early, the strongly-correlated fractional quantum Hall edge modes have been proposed~\cite{Rech2017Feb}, suggesting that the generation of clean, on-demand \emph{electronic} excitations can also be achieved there, and spurring several studies where the impact of strong correlation on the charge and energy dynamics, and interference of few-electron states were addressed~\cite{Vannucci2017Jun,Ronetti2018Aug,Ferraro2018Dec,Ronetti2019May,Ronetti2019Jul,Safi2019Jan,Safi2020Jul,Safi2022Nov,Bertin-Johannet2024Jan,Taktak2025Feb}. One of the exciting features of this platform is the presence of exotic anyonic excitations with both fractional charge and statistics. Recent developments in this regime have been reported, and we will come back to this point at the end of this section. In addition, the interplay between single-electron sources (Levitons in particular) and superconductivity has been explored in several studies~\cite{Ferraro2015Feb,Belzig2016Jan, Acciai2019Aug,Bertin-Johannet2022Mar,Bertin-Johannet2023May,Burset2023Dec,Bertin-Johannet2024May,Ronetti2024Dec,Burset2025Mar}, highlighting the opportunities for the on-demand generation of entangled states.
Finally, the generation and propagation of few-electron states has also been studied in the helical edge states of two-dimensional topological insulators~\cite{Inhofer2013Dec,Hofer2013Dec,Calzona2016Jul,Acciai2017Aug,Acciai2019Oct,Ronetti2020Feb}, where the presence of spin-momentum locking offers additional opportunities for interferometry~\cite{Ferraro2014Feb}. The fate of Levitons in the presence of interacting systems of different kind has been a major focus, leading to the conclusion that their minimal noise property (recall Sec.~\ref{sec:injection_target_drives}) is very robust~\cite{Vanevic2016Jan,Rech2017Feb,Acciai2019Aug,Acciai2019Oct,Bertin-Johannet2022Mar,Fukuzawa2023Sep}.

(iii) Beam splitters in EQO are typically implemented by relying on quantum point contacts. They are usually defined by electrostatic gates deposited on the heterostructure hosting the 2DEG where the propagation of the few-electron states of interest occurs. In a typical configuration, quantum point contacts allow electrons to tunnel from an edge state to another, such that an incoming excitation is partitioned in two outgoing channels. In this way, interferometers can be realized. Notable examples in EQO are the Mach-Zehnder and Hong-Ou-Mandel interferometers. They are described in more detail in Sec.~\ref{sec:interferometry}.

%%%%%%%%%%%%%%%%%%%%%%%%%%%%%%%%%%%%%%%%%%%%%%%%%%%%%%%
\subsubsection{Description of single-electron sources using the theory of electronic coherence}
%%%%%%%%%%%%%%%%%%%%%%%%%%%%%%%%%%%%%%%%%%%%%%%%%%%
A powerful method to describe generic states in EQO is the theory of electronic coherence, which is introduced in Sec.~\ref{sec_electron-coherence}.
The definitions given therein are very general and can be employed to characterize an arbitrary non-equilibrium state, which is in general a highly non-trivial task.
Let us therefore begin by discussing the properties of the electron coherence function for some idealized cases that help our intuition.

\paragraph{Ideal single- and multi-electron states}
We consider a situation where the non-equilibrium state is generated by an \emph{ideal single-electron source} that emits a single-electron state on top of a (effectively non-interacting) Fermi sea. We therefore assume the many-body state $\rho=\ket{\Phi}\bra{\Phi}$, where
\begin{equation}
    \ket{\Phi}=\int dt\,\varphi(t)\hat{\psi}^\dagger(t)\ket{F},
\label{eq:single-electron-state}
\end{equation}
and $\varphi(t)$ is a normalized wave function\footnote{Wave functions are expressed in the time domain; the assumption of chiral evolution allows us to always trade the position $x$ with $v_Ft$.}, with Fourier transform
\begin{equation}
    \tilde\varphi(\omega)=\int_{-\infty}^{+\infty}dt\,e^{i\omega t}\varphi(t)\,.
\end{equation}
In Eq.~\eqref{eq:single-electron-state}, $\ket{F}$ is the unperturbed Fermi sea, whose correlation function is given by
\begin{equation}
    \mathcal{G}^<_F(t'-t)\equiv\Braket{\hat{\psi}^\dagger(t')\hat{\psi}(t)}_F= \frac{i\pi k_{\rm B}T}{2\pi v_F\hbar\sinh\left(\frac{t'-t}{\hbar}\pi k_{\rm B}T+i0^+\right)}\,,
\end{equation}
where $T$ is the electronic temperature, and we have considered the electrochemical potential $\mu$ as a reference energy, namely we have set $\mu=0$. Given this choice, the wavefunction $\varphi(t)$ represents an electron above the Fermi sea if all its frequency components $\tilde\varphi(\omega)$ lie at $\omega>0$, when $T=0$, or well above the thermal excitation scale in the finite-temperature case. In the following, we will assume that this is the case, unless otherwise specified.

Then, one finds
\begin{equation}
    \mathcal{G}^<(t,t')=\mathcal{G}^<_F(t'-t)+\frac{1}{v_F}\varphi^*(t')\varphi(t)\,.
\label{eq:electron_coherence_time_wavefunction}
\end{equation}
Equation~\eqref{eq:electron_coherence_time_wavefunction} shows that the injection of a single-electron state on top of the Fermi sea by an ideal source leads to an excess coherence function that is factorized in a part that depends on $t$ and one that depends on $t'$. This crucial property extends to the case of an $M$-electron state formed with orthonormal wavefunctions
\begin{equation}
\Braket{\varphi_k|\varphi_j}=\int_{-\infty}^{+\infty}dt\,\varphi^*_k(t)\varphi_j(t)=\delta_{j,k}\qquad (j,k=1,\dots,M)
\label{eq:wavefunction_orthogonality}
\end{equation}
in a Slater determinant, for which one finds
\begin{equation}
    \mathcal{G}^<(t,t')=\mathcal{G}^<_F(t'-t)+\frac{1}{v_F}\sum_{k=1}^{M}\varphi_k^*(t')\varphi_k(t)\,,
\label{eq:electron_coherence_multielectron}
\end{equation}
showing the same factorization with respect to $t$ and $t'$ as in the single-electron case. Similarly, the energy representation reads
\begin{equation}
    \tilde{\mathcal{G}}^<(\omega,\omega')=\tilde{\mathcal{G}}_F^<(\omega)\delta(\omega-\omega')+\frac{1}{v_F}\sum_{k=1}^{M}\tilde{\varphi}^*_k(\omega')\tilde{\varphi}_k(\omega)\,,
\label{eq:electron_coherence_energy_wavefunction}
\end{equation}
where $\tilde\varphi_k(\omega)$ are the Fourier-transformed wavefunctions.
The presence of the Fermi sea contribution $\mathcal{G}^<_F(t'-t)$ constitutes one of the major differences compared to standard photonic quantum optics, as the electronic excitations of interest are not in the vacuum, but in a condensed-matter system with many electrons. For this reason, it is customary to define \textit{excess} coherence functions
\begin{equation}
    \Delta \mathcal{G}^\gtrless(t,t')=\mathcal{G}^\gtrless(t,t')-\mathcal{G}^\gtrless_F(t'-t)
\end{equation}
that encode the variations in the system due to the presence of the injected, time-dependent few-electron state. From this excess coherence function, all relevant quantities characterizing the state of interest can be obtained. For example, the excess charge current is easily obtained as the diagonal part of the excess electron coherence. Taking the pure $M$-electron state introduced above, we have the excess current
\begin{equation}
    \Delta I^c(t)=q\sum_{k=1}^{M}|\varphi_k(t)|^2\,,
    \label{eq:excess_current_eqo}
\end{equation}
and the excess electron distribution function
\begin{equation}
    \Delta f_e(\omega)=\frac{1}{2\pi}\sum_{k=1}^{M}|\tilde{\varphi}_k(\omega)|^2\,.
\end{equation}
Finally, the case of a pure $M$-hole state is described by
\begin{equation}
    \Delta\mathcal{G}^<(t,t')=-\frac{1}{v_F}\sum_{k=1}^{M}\varphi_k^*(t')\varphi_k(t)\,,
\end{equation}
and an identical discussion follows along the lines of the pure-electron case illustrated above.

In the following, we apply the formalism of this section to three relevant injection schemes, that are closely connected to the single-electron sources described in Secs.~\ref{sec:injection_target_drives} and~\ref{sec:injection_confined}.\\

\begin{figure}[t]
    \centering
    \includegraphics[width=0.65\linewidth]{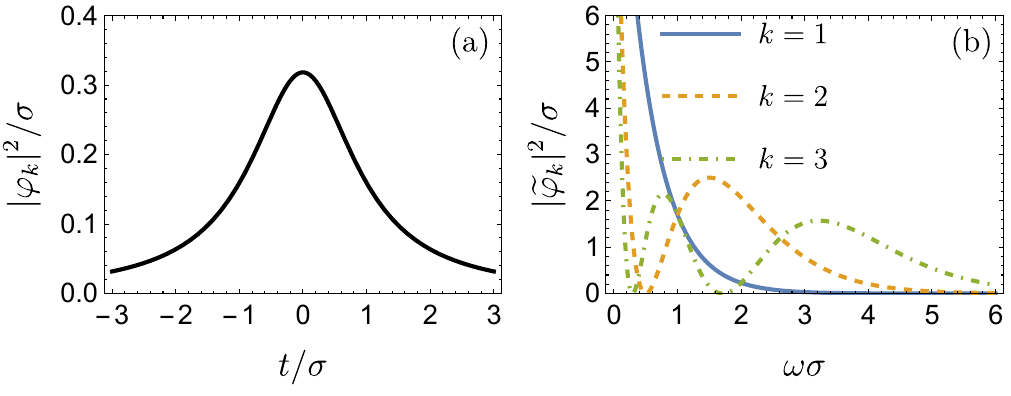}
    \caption{(a) Time representation and (b) energy representation of the Leviton wavefunctions, see Eqs.~\eqref{eq:wavefunctions_levitons_time} and~\eqref{eq:wavefunctions_levitons_energy}, with $t_0=0$.}
    \label{fig:Levitons_wf}
\end{figure}

\paragraph{Levitons}
Let us start with $M$-Leviton states, that can be obtained by applying voltage pulses of the form $V(t)=M V_{\rm Lor}(t)$, with $V_{\rm Lor}(t)$ in Eq.~\eqref{eq:VoltageLeviton}. For these states, we have
\begin{equation}
    \varphi_k(t)=\sqrt{\frac{\sigma}{\pi}}\frac{(t-t_0+i\sigma)^{k-1}}{(t-t_0-i\sigma)^{k}},
\label{eq:wavefunctions_levitons_time}
\end{equation}
where $t_0$ is the emission time and $\sigma$ is the characteristic width of the pulse. This gives rise to the characteristic Lorentzian current profile
\begin{equation}
    \Delta I^c_{\mathrm{Lor}}(t)=Mq\frac{\sigma/\pi}{(t-t_0)^2+\sigma^2}\,.
\end{equation}
Notice that injecting $M$ Levitons per pulse affects the current by a multiplicative factor only\footnote{This is because these excitations are generated by voltage pulses, and the current can then simply be obtained as ${I}^c(t)=q^2 V(t)/h$, instead of using Eq.~\eqref{eq:excess_current_eqo}.}, and we gain no information about how the different states are filled. This can be seen by considering the energy representation of the wavefunctions, given by
\begin{equation}
    \tilde{\varphi}_k(\omega)=2ie^{i\omega t_0}\sqrt{\pi \sigma}\,\theta(\omega)L_{k-1}(2\omega \sigma)e^{-\omega \sigma}\,,
\label{eq:wavefunctions_levitons_energy}
\end{equation}
where $L_k$ are the Laguerre polynomials. This shows that the energy distribution has an overall exponential decay governed by the time width $\sigma$. The vanishing of $\tilde{\varphi}_k(\omega)$ at negative energies is a consequence of Lorentzian pulses being pure-electron states that do not excite any hole contribution ($\omega=0$ is the reference energy determined by the Fermi level). Moreover, the structure of the wavefunctions~\eqref{eq:wavefunctions_levitons_energy} shows that energy states are occupied starting at very low energy above the Fermi level, with wavefunctions with larger $k$ ($k$ numbering the electrons contained in the pulse) contributing at higher energies, see Fig.~\ref{fig:Levitons_wf}.

Finally, the Wigner function of an $M$-Leviton state can be obtained explicitly~\cite{Ferraro2013Nov} (see Appendix~\ref{app:wigner} for the derivation)
\begin{equation}
    \Delta W^<(t,\omega)=2\sqrt{\pi}\theta(\omega)e^{-2\omega \sigma}\sum_{k=0}^{M-1}\sum_{p=0}^{k}\frac{1}{p!}\left(\frac{2\omega \sigma}{\sqrt{\omega t}}\right)^{2p+1}L_{k-p}^{(2p)}(4\omega \sigma)J_{p+1/2}(2\omega t)\,.
\label{eq:wigner_levitons}
\end{equation}
\begin{figure}[t]
    \centering
    \includegraphics[width=0.8\linewidth]{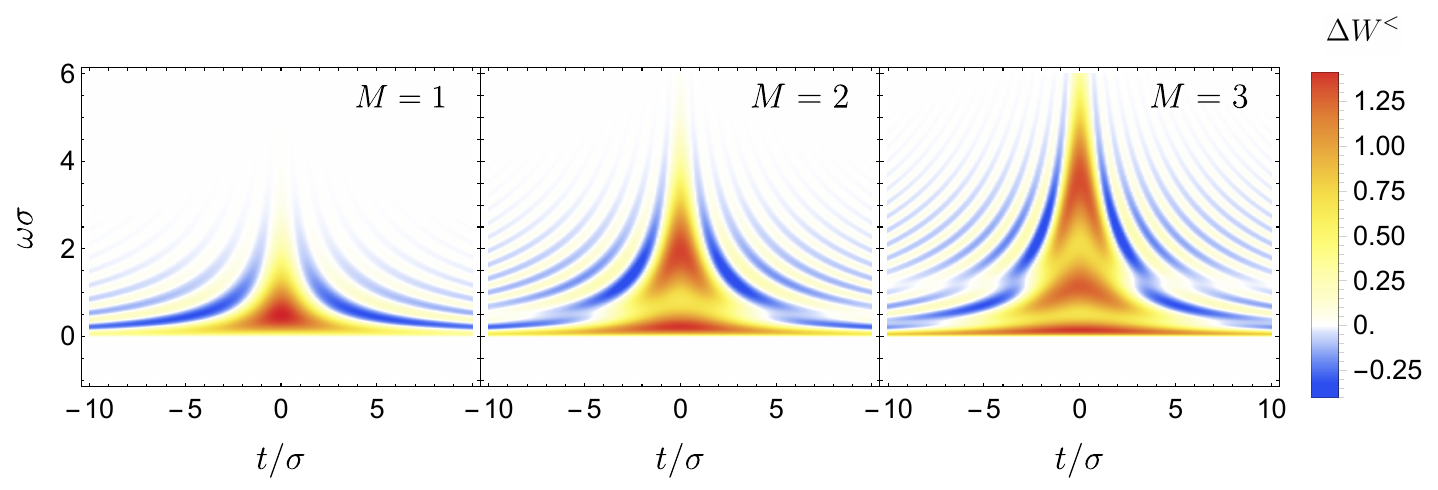}
    \caption{Wigner function representation~\eqref{eq:wigner_levitons} of $M$-Leviton states for the first few values of $M$.}
    \label{fig:Levitons_Wigner}
\end{figure}
Once again, the presence of $\theta(\omega)$ stems from the purely electronic character of Leviton states. Furthermore, the Wigner function combines the time and energy information previously observed separately in the excess current $\Delta I^c(t)$ and in the distribution function $\Delta f_e(\omega)$, as depicted in Fig.~\ref{fig:Levitons_Wigner}.

\paragraph{Single-electron excitations with a well-defined energy}
As discussed above, Leviton states have a well-defined time width, while their energy is rather spread, with the largest contributions arising from just above the Fermi level.
Another relevant example of single-electronic excitation is the one generated by the driven mesoscopic capacitor~\cite{Feve2007May}, consisting of a quantum dot capacitively coupled to a top gate and tunnel-coupled to a chiral conductor, cf. Fig.~\ref{fig:injection_principles}(d). In this case, the emitted state can be rather complicated and requires a full description of the scattering matrix of the driven dot, cf Sec.~\ref{sec:injection_confined}. However, in the so-called optimal operation regime~\cite{Parmentier2012Apr,Mahe2010Nov}, the emitted state is a single-electron excitation (thus $M=1$) well described by the following wavefunction
\begin{equation}
    \tilde{\varphi}(\omega)=\sqrt{\frac{\gamma}{\mathcal{N}}}\frac{\theta(\omega)}{\omega-\omega_0+i\gamma/2}\,,\qquad\mathcal{N}=\frac{1}{2}+\frac{1}{\pi}\arctan\left(\frac{2\omega_0}{\gamma}\right)\,.
\label{eq:wavefunction_Landau}
\end{equation}
\begin{figure}[t]
    \centering
    \includegraphics[width=0.5\linewidth]{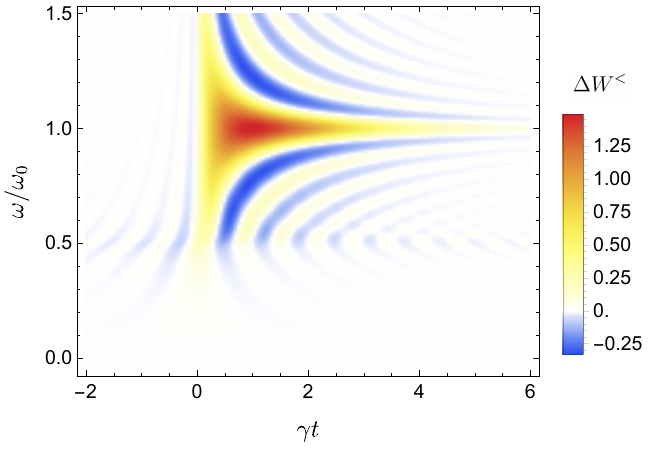}
    \caption{Wigner function representation of a single-electron excitation with a well-defined energy, whose wavefunction is given in Eq.~\eqref{eq:wavefunction_Landau}. We have chosen $\omega_0=10\gamma$.}
    \label{fig:Landau_Wigner}
\end{figure}
Here, $\gamma$ is the level broadening as introduced in Sec.~\ref{sec:injection_confined}. This wavepacket represents an excitation with a well-defined energy $\varepsilon_0=\hbar\omega_0$ above the Fermi level. In a sense, it is a ``conjugate'' of the Leviton with $M=1$, as it has a Lorentzian shape in energy rather than in time. Therefore, the current profile $\propto|\varphi(t)|^2$ has an exponential decay with a characteristic time $\sigma=1/\gamma$. This is confirmed by the explicit calculation of the wavefunction in the time domain, which reads~\cite{Ferraro2013Nov}
\begin{equation}
    \varphi(t)=-i\sqrt{\frac{\gamma}{\mathcal{N}}}e^{-\omega_0|t|-i\omega_0t}\left\{\theta(t)-\frac{i}{2\pi}\mathrm{Ei}\left[t\left(\frac{\gamma}{2}+i\omega_0\right)\right]\right\}\,,
\label{eq:wavefunction_ses_well-defined-energy}
\end{equation}
with $\mathrm{Ei}[\bullet]$ the exponential integral function. The wavefunction~\eqref{eq:wavefunction_Landau} is valid in the regime $\gamma\ll\omega_0$, as $\gamma$ represents the escape rate from the driven dot. So, the mentioned condition is required to have a well-defined energy in the dot. The Wigner function computed by using Eqs.~\eqref{eq:Wigner_definition} and~\eqref{eq:electron_coherence_energy_wavefunction} is shown in Fig.~\ref{fig:Landau_Wigner}. The cutoff at $\omega\lesssim\omega_0/2$ is a consequence of the $\theta$ function in the wavefunction~\eqref{eq:wavefunction_Landau}. This feature shows that the Wigner function of the single-electron excitation of Eq.~\eqref{eq:wavefunction_ses_well-defined-energy} is not exactly obtained by rotating the corresponding one of the 1-Leviton state.\\

\paragraph{Single level driven at constant speed across the Fermi sea}
\begin{figure}[t]
    \centering
    \includegraphics[width=0.65\linewidth]{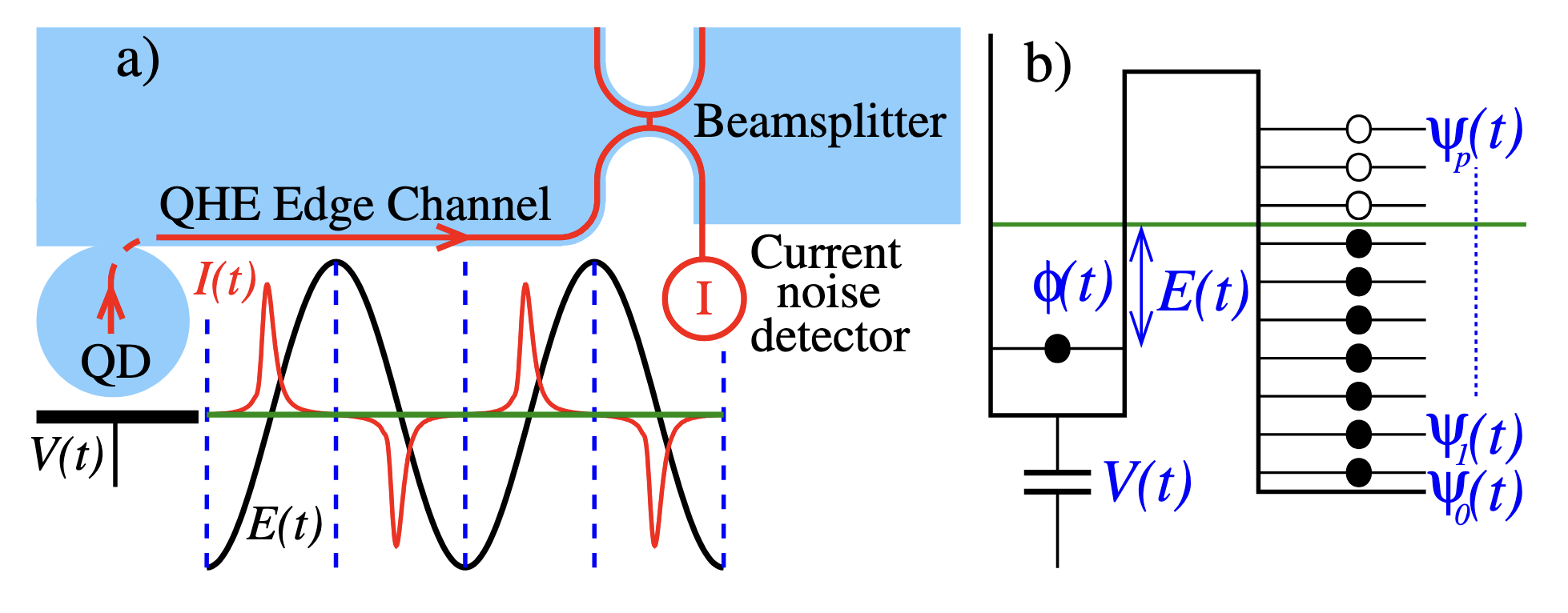}
    \caption{Single-electron emission from a localized level driven at constant speed across the Fermi level of a continuum. (a) Sketch of the setup. (b) Illustration of the working principle: the dot level follows a linear evolution $E(t)=c\,t$ due to the application of a gate voltage $V(t)$ to the dot. For appropriate voltages (see main text), the state obtained via the linear behavior of $E(t)$ is a good approximation in each half-period of the drive. Reprinted figure with permission from Ref.~\cite{Keeling2008Nov}. Copyright 2008 by the American Physical Society.}
    \label{fig:Keeling2008}
\end{figure}
Another interesting case is that of a single energy level coupled to a continuum of states, and subject to a driving such that the crossing of the level with the Fermi energy off the continuum happens with constant speed. This situation has been studied in Ref.~\cite{Keeling2008Nov,Moskalets2017Jul}, and is linked, for instance, to the response of a harmonically driven mesoscopic capacitor, where a single level of a quantum dot is periodically driven above and below the Fermi level of a conductor coupled to the dot. In this way, when the dot level crosses the Fermi level, it evolves linearly in time as $E(t)=ct$ (assuming without loss of generality that the Fermi level is set at zero energy and the crossing occurs at $t=0$). This linear behavior describes the evolution of the dot level in the first half-period of the drive and is a good approximation when the period of the drive is long compared to $\max(\hbar\gamma^{-1},\gamma c^{-1})$ and when the extremal value of $E(t)$ is larger than $\gamma$, the linewidth of the localized level, as depicted in Fig.~\ref{fig:Keeling2008}.
With this particular protocol, Ref.~\cite{Keeling2008Nov} has shown that a single-electron state is emitted, whose wavefunction is given by
\begin{equation}
    \varphi(t)=\sqrt{\frac{\gamma}{2\pi c}}\int_{0}^{+\infty}d\omega e^{-i\omega t-\frac{\gamma\omega}{2c}+i\frac{\hbar\omega^2}{2c}}\iff\tilde{\varphi}(\omega)=\sqrt{\frac{2\pi\gamma}{c}}\theta(\omega)e^{-\frac{\gamma\omega}{2c}+i\frac{\hbar\omega^2}{2c}}\,.
\label{eq:wavefunction_level_constant_rapidity}
\end{equation}
From the energy representation, it is easy to check that this state is properly normalized. Reference~\cite{Moskalets2017Jul} has obtained the same result with a different approach, starting from the scattering matrix of a driven mesoscopic capacitor.

\begin{figure}[t]
    \centering
    \includegraphics[width=0.65\linewidth]{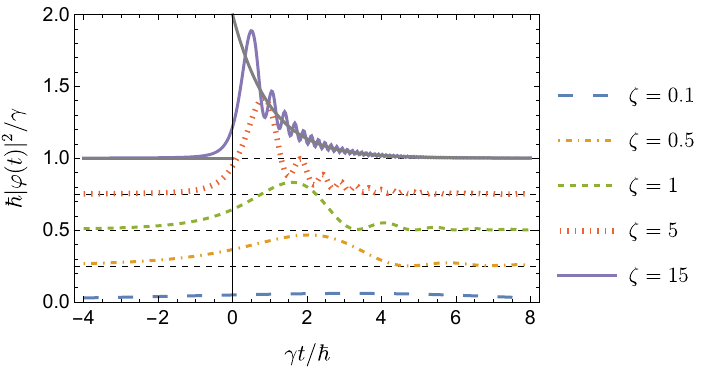}
    \caption{Current profile obtained from the wavefunctions~\eqref{eq:wavefunction_level_constant_rapidity} for different values of the adiabaticity parameter $\zeta=\hbar c/\gamma$. For illustration purposes, different curves are vertically shifted by an amount $0.25/\gamma$.}
    \label{fig:phi_level_c}
\end{figure}

Let us now analyze the main features of the single-electron state~\eqref{eq:wavefunction_level_constant_rapidity}. It is useful to introduce the time scale $\tau_\gamma=2\gamma/c$ and the adiabaticity parameter $\zeta=\hbar c/\gamma^2$, with which we can rewrite
\begin{equation}
    \varphi(t)=\frac{1}{\sqrt{\pi\tau_\gamma}}\int_0^{+\infty}ds\, e^{-is\frac{t}{\tau_\gamma}-s+2is^2\zeta}\,.
\end{equation}
This form makes the analysis of the adiabatic regime rather simple. Indeed, when $\zeta\to 0$, the above integral can be easily solved and one finds (up to a phase) the same wavefunction as the one of a Leviton, namely Eq.~\eqref{eq:wavefunctions_levitons_time}, with $k=1$, and $\sigma=\tau_\gamma$. As a result, the current profile $\Delta I^c(t)\propto|\varphi(t)|^2$ is a Lorentzian of width $\tau_\gamma$, and the energy distribution decays exponentially as expected. This shows that the driven mesoscopic capacitor in the adiabatic-response regime generates the same state as the Leviton voltage source.
Note that this exponential decay follows from the frequency representation $\tilde{\varphi}(\omega)$ in~\eqref{eq:wavefunction_level_constant_rapidity} for any value of $\zeta$. 
The current profile, however, strongly depends on the emission regime, as shown in Fig.~\ref{fig:phi_level_c}, illustrating the evolution of the current with increasing values of $\zeta$. Starting from the Lorentzian profile of the adiabatic-response regime, there is a crossover where the current develops interference fringes with oscillations that are more and more rapid as $\zeta$ increases, while the asymptotic shape for $\zeta\to\infty$ is exponential with a decay rate $\gamma$. 
The origin of the oscillations can be attributed to interference among the (energy-resolved) scattering amplitudes describing the escape process from the dot to the continuum~\cite{Moskalets2017Jul}. Indeed, when the time it takes for the dot level to cross the Fermi level is much larger than the dwell time, all energy components of the dot's initial wavefunction escape the dot independently, so that the current profile simply reflects the Breit-Wigner resonance (with the energy width $\gamma$ rescaled by $c$ to obtain a time width $\tau_\gamma$). When the speed of the driving increases, the evolution $E(t)$ is fast enough so that several energy-resolved components escape the dot in time windows that overlap, causing interference. 

\begin{figure}[t]
    \centering
    \includegraphics[width=0.8\linewidth]{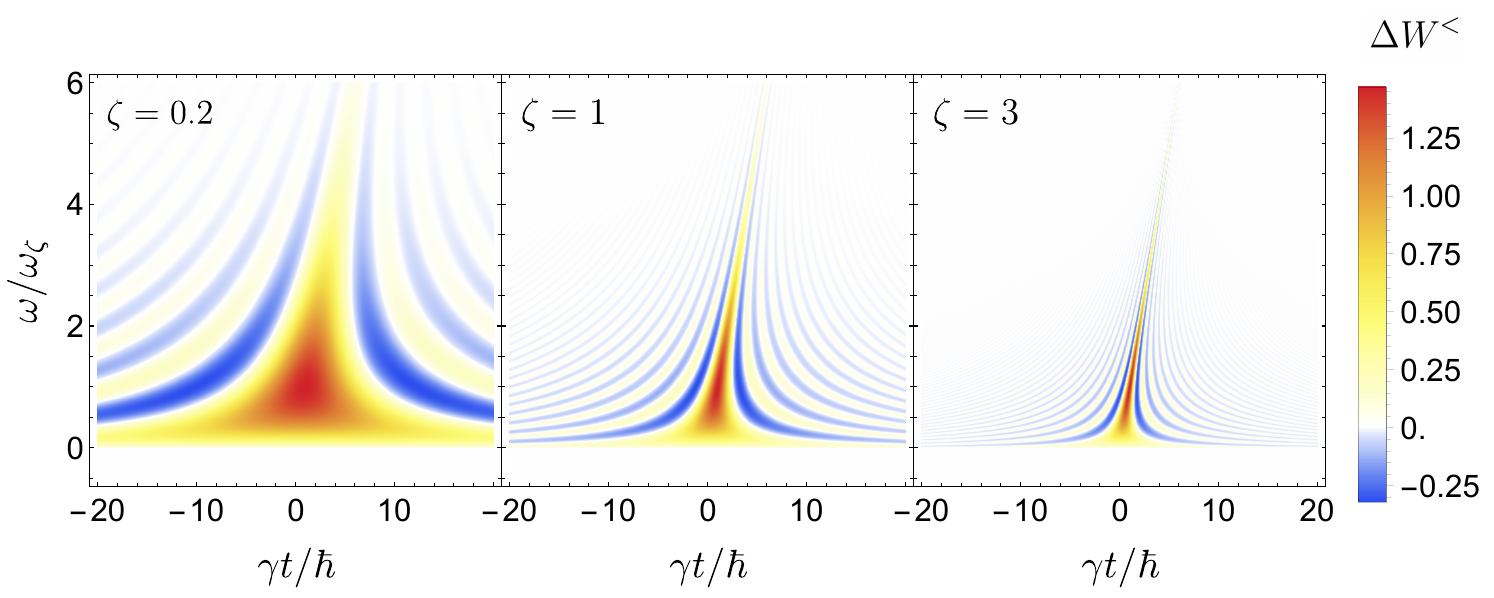}
    \caption{Wigner function of the single-electron excitation described by the wavefunction~\eqref{eq:wavefunction_level_constant_rapidity}. Here, we show three different values of the adiabaticity parameter $\zeta$, and the frequency axis has been rescaled to $\omega_\zeta=\zeta\gamma/\hbar=\omega\gamma/c$. So, for larger $\zeta$, larger values of the frequency $\omega$ are involved. Moreover, increasing $\zeta$ leads to a Wigner function that is more and more concentrated along the line $\omega=ct/\hbar$, reflecting the evolution $E(t)=ct$ of the dot level.}
    \label{fig:wigner_level_c}
\end{figure}

While the non-adiabatic emission regime has an exponentially decaying current profile that reminds of the state emitted by the mesoscopic capacitor with square voltage drive~\cite{Feve2007May}, there is a crucial difference. Indeed, the state~\eqref{eq:wavefunction_level_constant_rapidity} has always an exponential profile in energy, meaning that it is not an energy-resolved excitation. In order to show this more clearly, we compute the Wigner function, which is shown in Fig.~\ref{fig:wigner_level_c}. Comparing this with the Wigner function in Fig.~\ref{fig:Landau_Wigner} we observe a clear difference.
Indeed, increasing the speed $\zeta$ does not lead to the appearance of a well-defined energy like in Fig.~\ref{fig:Landau_Wigner}. Instead, we observe a structure that is more similar to the Wigner function of a Leviton, except for a tilt along the line $\omega=ct/\hbar$. The effect of this tilt is negligible in the adiabatic regime $c\to 0$, and thus one recovers exactly the Leviton Wigner function in this limit, consistently with the above analysis at the wavefunction level. We also observe that increasing $\zeta$ makes the Wigner function more and more concentrated on the line $\omega=ct/\hbar$, reflecting the evolution $E(t)=ct$ of the dot level. Moreover, higher energy components become relevant in this regime (notice that the frequency axis in Fig.~\ref{fig:wigner_level_c} is rescaled with respect to $\omega_\zeta=\zeta\gamma/\hbar$). Finally, the Wigner function being less spread is also connected to the sharper rise of the current profile after $t=0$ that we observed in Fig.~\ref{fig:phi_level_c} at large $\zeta$.

%%%%%%%%%%%%%%%%%%%%%%%%%%%%%%%%%%%%%%%%%%%%%%%%%%
\paragraph{Finite-temperature effects}
%%%%%%%%%%%%%%%%%%%%%%%%%%%%%%%%%%%%%%%%%%%%%%%%%%
Until now, we have discussed the case of thermal excitations with energies below the characteristic frequency components of the electronic wavefunctions $\varphi(t)$. When this condition is not fulfilled, the simple representation~\eqref{eq:electron_coherence_time_wavefunction} of the coherence function no longer holds. However, for ideal single-electron emitters, it is possible to show that introducing a finite temperature does not completely destroy the structure of the electron coherence and simply turns pure states into mixed states~\cite{Moskalets2017Jul}:
\begin{equation}
    \Delta\mathcal{G}^<(t,t')=\int_{-\infty}^{+\infty}d\varepsilon \left(-\frac{\partial f}{\partial\varepsilon}\right)\varphi^*_\varepsilon(t')\varphi_\varepsilon(t)\,,
\label{eq:el_coherence_finiteT}
\end{equation}
where $f(\varepsilon)$ is the Fermi function at temperature $T$. Technically, this expression is valid if the scattering matrix of the source has the symmetry property $S(t,\varepsilon+\delta\varepsilon)=S(t-\delta\varepsilon/c,\varepsilon)$, for some constant $c$. Examples in which this condition is met are the mesoscopic capacitor source, the level moving at constant speed (see the earlier discussion in this section), and the Leviton source. In this last case, $S(t,\varepsilon)$ is energy independent, so the symmetry property is formally recovered as $c\to\infty$. The components $\varphi_\varepsilon$ characterizing the mixed state in~\eqref{eq:el_coherence_finiteT} are given by $\varphi_\varepsilon(t)=e^{-i\varepsilon t/\hbar}\varphi(t-\varepsilon/c)$.\\

%%%%%%%%%%%%%%%%%%%%%%%%%%%%%%%%%%%%%%%%%%%%%%%%%%
\paragraph{Ideal periodic sources and general decomposition}
%%%%%%%%%%%%%%%%%%%%%%%%%%%%%%%%%%%%%%%%%%%%%%%%%%
Because of the lack of single-shot detection capabilities in many situations, the most common experimental conditions involve periodic single-electron sources. For the description of this scenario, the periodicity properties discussed in Eqs.~\eqref{eq:electron_coherence_periodicity} and the following are very useful.

In the case of an ideal source that emits purely electronic states above the Fermi sea, we expect to recover a representation of the form~\eqref{eq:electron_coherence_multielectron} even in the periodic case. This can be done by introducing an infinite family of wavefunctions $\{\varphi_{k,\ell}\}$, called atoms of signal~\cite{Roussel2017Mar} or electronic wavelets, such that
\begin{equation}
    \Delta\mathcal{G}^<(t,t')=\frac{1}{v_F}\sum_{k=1}^{M}\sum_{\ell\in\mathbb{Z}}\varphi_{k,\ell}^*(t')\varphi_{k,l}(t)\equiv \frac{1}{v_F}\sum_{k=1}^{M}\sum_{\ell\in\mathbb{Z}}\varphi_{k}^*(t'-\ell\mathcal{T})\varphi_{k}(t-\ell\mathcal{T})\,,
\label{eq:atoms_signal}
\end{equation}
with $\varphi_k(t)\equiv\varphi_{k,0}(t)$\,. This expression can also be thought of as a definition of an ideal periodic source. The orthogonality relation~\eqref{eq:wavefunction_orthogonality} is generalized to $\langle\varphi_{k,\ell}|\varphi_{j,\ell'}\rangle=\delta_{j,k}\delta_{\ell,\ell'}$. The choice of wavefunctions enabling the above representation is not unique, but it is possible to choose them in such a way that $\varphi_{k,\ell}$ represents a localized state on the $\ell$-th period, for each $k$.

Even though this representation only applies to ideal sources, the concept of electronic wavelets (or atoms of signal) is extremely useful to characterize the properties of an arbitrary state generated by a periodic source, making a direct connection with experimental data possible. 
The question is whether it is possible to think of a generic periodic coherence function as being built  by simple blocks of the form~\eqref{eq:atoms_signal}. The remarkable result, proven in Ref.~\cite{Roussel2021May} is that this is indeed the case. Of course, a non-ideal source that injects $M$ electrons per cycle also generates a cloud of particle-hole excitations. As a result, the decomposition of Ref.~\cite{Roussel2021May} relies on a set $\{\varphi_{k,\ell}^{(e)}\}$ of electron-like single-particle wavefunctions (corresponding to states with energy above the Fermi level) and a set $\{\varphi_{j,\ell'}^{(h)}\}$ of hole-like wavefunctions (corresponding to states with energy below the Fermi level). Here, the index $\ell$ labels the period, and $k=1,\dots M'_e$, $j=1,\dots M'_h$, with the constraint $M=M_e'-M_h'$. For an ideal electron source, $M_h'=0$ and one recovers Eq.~\eqref{eq:atoms_signal}.
The functions $\{\varphi_{k,\ell}^{(e/h)}\}$ are maximally localized states on the period $\ell$ and provide complete information on the electron/hole content in a generic state. They are analogous of the Wannier functions in solid state theory. 
For a stationary source, they reduce to the Martin-Landauer wavepackets~\cite{Martin1992Jan}
\begin{equation}
    \varphi_k(t)=\frac{1}{\sqrt{\mathcal{T}}}\frac{\sin(\Omega t/2)}{\Omega t/2}e^{-i\Omega\left(k+\frac{1}{2}\right)t}
\label{eq:ML}
\end{equation}
with energy bandwidth $\hbar\Omega$, originally introduced to study the stationary transport of a continuous stream of electrons due to a constant voltage bias. Roughly speaking, the wavefunctions~\eqref{eq:ML} are suitable to decompose states with large inter-period overlap.

The general framework presented in Ref.~\cite{Roussel2021May} is also of practical use because it allows one to numerically extract the wavefunctions $\{\varphi_{k,\ell}^{(e/h)}\}$ from experimental signals, thereby determining how close a given source in experimental conditions is to an ideal one~\cite{Bisognin2019Jul}.\\

%%%%%%%%%%%%%%%%%%%%%%%%%%%%%%%%%%%%%%%%%%%%%%%%%%
\paragraph{Periodic injection of Levitons}
%%%%%%%%%%%%%%%%%%%%%%%%%%%%%%%%%%%%%%%%%%%%%%%%%%
For trains of Levitons the  wave functions in the decomposition~\eqref{eq:atoms_signal} are explicitly known.
They can be obtained analytically for a train of Levitons with a unit charge $q$, namely $M=1$. In this case, it was first shown in Ref.~\cite{Moskalets2015May} that the representation~\eqref{eq:atoms_signal} is satisfied with
\begin{equation}
    \varphi_{1,\ell}(t)=\sqrt{\frac{\sigma}{\pi}}\frac{1}{t-\ell\mathcal{T}+i\sigma}\frac{\Gamma[(t-\ell\mathcal{T}-i\sigma)/\mathcal{T}]}{\Gamma[(t-\ell\mathcal{T}+i\sigma)/\mathcal{T}]}\,,
\end{equation}
where $\Gamma$ denotes Euler's gamma function. For each $\ell$, these wave functions have width $\sigma$ and are localized in the $\ell$-th period. So, when the source injects well-separated Levitons (i.e., $\sigma\Omega\ll 1$), each $\varphi_{1,\ell}$ tends to the single-Leviton wavefunction $\sim(t+i\sigma)^{-1}$, properly shifted in time.

For $M>1$, we know from the general theory of Ref.~\cite{Roussel2021May} that the representation~\eqref{eq:atoms_signal} holds, but the wave functions are not known analytically. Instead, it is possible to show that the following expression is valid~\cite{Ronetti2018Aug}
\begin{equation}
    \Delta\mathcal{G}^<(t,t')=-2i\mathcal{G}_F^<(t'-t)\sin\left(\pi\frac{t'-t}{\mathcal{T}}\right)\sum_{k=1}^{M}\phi_k^*(t')\phi_k(t)\,,
\label{eq:coherence_function_periodic_Levitons}
\end{equation}
where ${\{\phi_k\}}_k$ is a set of $2\mathcal{T}$-periodic functions satisfying the orthogonality relation $\int_0^{\mathcal{T}}\phi_k(t)\phi^*_j(t)dt/\mathcal{T}=\delta_{j,k}$. Notice the difference between this condition and Eq.~\eqref{eq:wavefunction_orthogonality}. They read~\cite{Glattli2016Feb,Ronetti2018Aug}
\begin{equation}
    \phi_k(u)=\sqrt{\frac{\sinh(2\pi\eta)}{2}}\frac{\sin^{k-1}[\pi(u+i\eta)]}{\sin^k[\pi(u-i\eta)]},
\label{eq:wavefunctions_glattli}
\end{equation}
with $u=t/\mathcal{T}$ the dimensionless time and $\eta=\sigma/\mathcal{T}$ the width-to-period ratio. Equation~\eqref{eq:coherence_function_periodic_Levitons} can be proved by direct calculation in the case of a Leviton with unit charge $M=1$ and extended to every $M>1$ by induction.
Notice that this expression does not have the form of Eq.~\eqref{eq:electron_coherence_multielectron} for a multi-electron state. Nonetheless, the functions in Eq.~\eqref{eq:wavefunctions_glattli} are useful to express the noise in a Hong-Ou-Mandel geometry, see ~Sec.~\ref{sec:interferometry}.
To the best of our knowledge, a closed formula for the Wigner function stemming from~\eqref{eq:coherence_function_periodic_Levitons} has not been reported.

\subsubsection{Interferometry of few-electron states}
\label{sec:interferometry}

One of the distinctive features of mesoscopic physics is the possibility of observing phenomena that are intrinsically linked to the phase coherence of electrons, thereby demonstrating quantum interference effects in transport observables for systems as large as tens of micrometers. Prominent examples are the observation of the Aharonov-Bohm effect (see, e.g., the textbooks~\cite{Ihn2009Nov,Imry2008Oct,Nazarov2009May,Datta2005Jun,Ferry2009Aug}), and the implementation of electronic Mach-Zehnder interferometers, pioneered in Ref.~\cite{Ji2003Mar}. Early experiments in this context focused on time-independent transport configurations. Here, we are interested in discussing electron interferometers as spectroscopy tools to probe the dynamical properties of few-electron states. We mainly address two kinds of interferometers, namely the Mach-Zehnder and the Hong-Ou-Mandel ones, and we discuss how they can be used to access important information on the coherence properties of few-electrons states.\\

%%%%%%%%%%%%%%%%%%%%%%%%%%%%%%%%%%%%
\paragraph{Mach-Zehnder interferometer}
%%%%%%%%%%%%%%%%%%%%%%%%%%%%%%%%%
\begin{figure}[b]
\centering
\includegraphics[width=0.5\textwidth]{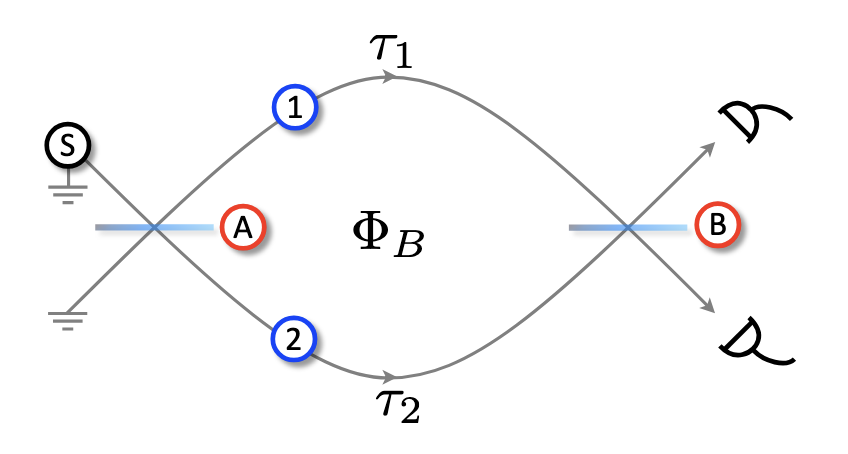}
\caption{Sketch of a Mach-Zehnder interferometer. A single-electron source S emits a state that gets split in a superposition at the beam splitter A, after which propagation can occur on channels 1 and 2, with different propagation times, before reaching a second beam splitter B. The interferometer encloses a magnetic flux $\Phi_B$. Reprinted figure with permission from Ref.~\cite{Ferraro2013Nov}. Copyright 2013 by the American Physical Society.}
\label{fig:MZI_skecth}
\end{figure}
A Mach-Zehnder (MZ) interferometer is depicted in Fig.~\ref{fig:MZI_skecth}, and consists of two channels that are initially mixed at a beam splitter (A), then propagate for different lengths, enclosing a magnetic flux $\Phi_B$, and finally recombine at a second beam splitter (B) before the current they carry is detected. In the simplest configuration, which we will focus on, there is a single source S, located before the first beam splitter at one of the input channels. In addition to the magnetic flux $\Phi_B$, the interferometer is characterized by different propagation times in the upper and lower arms. They can stem either from an asymmetric interferometer, or from the presence of additional mechanisms inducing a relative phase shift between the two arms.
The first study that considered the problem of a MZ interferometer driven by a single-electron source was Ref.~\cite{Haack2011Aug}, that mainly focused on the adiabatic emission regime. Further studied provided extensions to the non-adiabatic regime~\cite{Haack2013May}, multiple sources~\cite{Rossello2015Mar,Juergens2011Oct}, and the effect of noise~\cite{Rossello2015Mar,Hofer2014Dec}. Finally, a study with numerical techniques aimed at describing more realistic devices was presented in Ref.~\cite{Gaury2014May}. Recently, a Mach-Zehnder interferometer fed by ultra-short pulses has been realized experimentally~\cite{Ouacel2025May}.

Evaluating the current at the output of the interferometer, one finds~\cite{Haack2011Aug,Haack2013May}
\begin{equation}
\begin{split}
    I_{\rm out}(t)&=R_{\rm A}R_{\rm B}I_{\rm S}(t-\tau_1)+D_{\rm A}D_{\rm B}I_{\rm S}(t-\tau_2)\\
    &\quad -2qv_F\sqrt{R_{\rm A}R_{\rm B}D_{\rm A}D_{\rm B}}\,\mathrm{Re}\left[e^{-i2\pi\Phi_B/\Phi_0+k_Fv_F\Delta\tau}\mathcal{G}^<(t-\tau_1,t-\tau_2)\right]
\end{split}
\end{equation}
with the flux quantum $\Phi_0=h/q$, and $\Delta\tau=\tau_1-\tau_2$. In this expression, $I_{\rm S}(t)$ is the time-dependent current generated by the single-electron source, so the first line corresponds to classical terms associated with a simple current partitioning due to the beamsplitter. Conversely, the second line is associated with quantum interference, and indeed it contains the single-electron coherence function. Since this term is sensitive to the magnetic flux, a measurement of the time-resolved current $I_\mathrm{out}(t)$ for different magnetic fluxes and/or time of flight differences would allow to reconstruct the single-electron coherence and thus completely characterize the emitting source. This clearly illustrates that interference can be used as a spectroscopic tool. Moreover, as noted in the original proposal~\cite{Haack2011Aug}, it is possible to extract the single-particle coherence length of the injected electrons by looking at the (less demanding) time-averaged current $\overline{I_{\rm out}(t)}$ and the visibility of the associated Aharonov-Bohm oscillations. In essence, the result is that the visibility decays as a function of the interferometer imbalance $\Delta\tau$, on a scale that is precisely the coherence length $\Lambda$ set by the source. This is because for $\Delta\tau>\Lambda$ the current pulses propagating in the different arms do not overlap at the second beamsplitter, determining a decay of $\mathcal{G}$ and therefore suppressing the interference.

As emphasized in Refs.~\cite{Ferraro2013Nov,Roussel2017Mar}, an ideal MZ interferometer produces a linear filter on the incoming single-electron coherence function (or the Wigner function) of the source. However, this linearity holds as long as interactions can be neglected, which in practice is often not justified, due to decoherence and dephasing effects, that could be attributed to neighboring channels~\cite{Tewari2016Jan,Slobodeniuk2016Jan,Goremykina2018Mar,Rodriguez2020May}. These difficulties make the use of MZ interferometers to reconstruct the electronic coherence function rather challenging. For this task, it has turned out that two-particle interference is more suitable, as we discuss in the following. Before moving on to this topic, we would like to close this part by mentioning that a recent proposal has suggested to invert the logic and use a known single-electron state propagating in a MZ interferometer to probe electromagnetic fields, enabling an on-chip detection of nonclassical radiation states~\cite{Souquet-Basiege2024May}. A first step in this direction has been recently achieved~\cite{Bartolomei2025Mar}, relying on a Fabry-P\'erot interferometer.

Another promising avenue employing the MZ interferometer as a key building block is the use of ultrafast electronic excitations to implement flying qubits. In this concept, information is stored in the states of propagating electrons, rather than in localized objects. Explicitly, the state $|0\rangle$ or $|1\rangle$ of a flying qubit can be defined by the presence or absence of the electron in a given propagation channel. The key observation is that a MZ interferometer, combining the effect of the magnetic field and that of a relative phase shift can implement an arbitrary rotation of the qubit state on the Bloch sphere, rendering it a general 1-qubit gate. The vision for more challenging two-qubit gate is to exploit the Coulomb interaction to couple two qubits, in order to achieve a combined operation on both.
Of course, the idea of a flying qubit architecture is viable only if there is enough time to perform a sufficient number of gates before decoherence intervenes. This results in the necessity of implementing picosecond-scale single-electron pulses, which is at the forefront of current technological capabilities.
For details on the flying quibit vision, we refer to the review~\cite{Edlbauer2022Dec}.\\

\begin{figure}[tbp]
\centering
\includegraphics[width=0.45\textwidth]{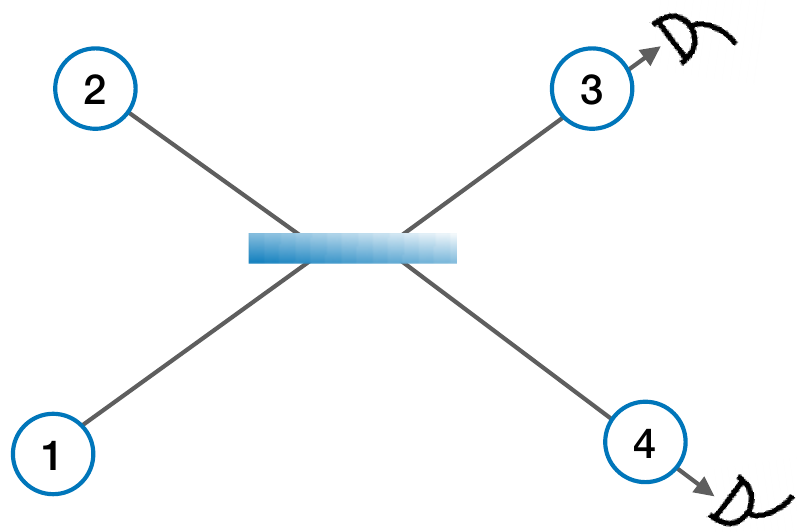}
\caption{Sketch of a HOM interferometer. Two sources, 1 and 2, emit time-dependent excitations that are combined at a beam splitter with a tunable delay. The two output ports, 3 and 4, collect the signal, that could be photon counts in the case of photon sources, or current fluctuations in the case of electron sources.}
\label{fig:HOM_skecth}
\end{figure}

%%%%%%%%%%%%%%%%%%%%%%%%%%%%%%%
\paragraph{Hong-Ou-Mandel interferometer}
%%%%%%%%%%%%%%%%%%%%%%%%%%%%%
We now discuss two-particle interference schemes and how they can be used to reconstruct the state of a single-electron source. The processes we are going to present are related to intensity interferometry in optics. The main character of this overview is the Hong-Ou-Mandel (HOM) experiment, owing its name to the authors of the seminal paper~\cite{Hong1987Nov}. 
It was originally performed with photons, using the setup sketched in Fig.~\ref{fig:HOM_skecth}. It consists of a beam splitter to which two different single-particle states are sent with a tunable time delay, and the number of particles at the output contacts 3 and 4 are monitored, in particular the coincidence counts. When the time delay is much larger than the wave packet extension, the incoming particles are partitioned independently and so coincidence counts occur with probability 1/2. 
At time delays that are comparable to the wave packet extension the behavior is different based on the statistics of the particles. Classical particles are still partitioned independently. Bosons exhibit a \emph{bunching} effect, such that it is much more probable that the two particles exit the interferometer in the same channel than in separate ones. 
As a result, the coincidence counts drop, and vanish completely when the time delay between the incoming particles is zero. This observation was used in the original experiment to determine the extension of single-photon wave packets~\cite{Hong1987Nov}. Fermions, instead, behave in the opposite way and exhibit an \emph{antibunching} effect, which is due to the Pauli principle, leading to an increase in the coincidence counts.
%%%%%%%%%%%%%%%%%%%%%%%%%%
%%%%%%%%%%%%%%%%%%%%%%%%%%%%%

From this description, it is clear that the HOM experiment requires single-particle detectors to monitor the coincidence counts. Unfortunately, in EQO experiments, a single-shot detection with single-electron sensitivity has not yet been achieved (with the exception of SAW-based schemes). However, it is possible to extract equivalent information by looking at the current fluctuations at output 3 and 4, which are experimentally accessible via low-frequency noise measurements, exploiting periodic sources and averaging over a long measurement time. Now, the expectations are reversed: an increase of fluctuations should happen for bosons and a suppression for fermions, indicating that the output state of the latter is always the same (one fermion in channel 3 and one in channel 4). Both autocorrelations and crosscorrelations can be considered. Here, we focus on the latter. They are obtained from Eq.~\eqref{eq:noise-correlator-general}, setting $I=I'=I^c$. Following the notation of Ref.~\cite{Ferraro2013Nov}, one has
\begin{equation}
    \mathcal{S}_{34}(t,t')=q^2v_F^2RD[\mathcal{S}_{11}(t,t')+\mathcal{S}_{22}(t,t')-\mathcal{Q}(t,t')]\,,
\end{equation}
where, as usual, $R$ and $D$ are the reflection and transmission probabilities of the beam splitter. The terms $\mathcal{S}_{11}$ and $\mathcal{S}_{22}$ are the fluctuations of the incoming states that get transmitted to the output fluctuations by the beam splitter. It is clear that $\mathcal{S}_{\alpha\alpha}$ only depends on source $\alpha=1,2$. In some cases, the sources are noiseless and such terms thus vanish. The interesting information on the interference of the incoming states is contained in the correlator $\mathcal{Q}$, which can be expressed as
\begin{equation}
    \mathcal{Q}(t,t')=q^2v_F^2\left[\mathcal{G}_1^<(t',t)\mathcal{G}^>_2(t',t)+\mathcal{G}_2^<(t',t)\mathcal{G}^>_1(t',t)\right]\,.
\end{equation}
The task is to isolate this quantity and also make it experimentally accessible. This can be done in two steps. First, we specify which of the two sources is switched on: for example, $\mathcal{S}_{34}^{\rm{on/off}}$ means that the first source is on and the second is off. We then define $\Delta\mathcal{S}_{34}^{\bullet/\star}=\mathcal{S}_{34}^{\bullet/\star}-\mathcal{S}_{34}^{\rm{off/off}}$, for any given configuration. Finally, we define the HOM noise as
\begin{equation}
    \mathcal{S}_{\rm HOM}\equiv\Delta\mathcal{S}_{34}^{\rm on/on}-\Delta\mathcal{S}_{34}^{\rm on/off}-\Delta\mathcal{S}_{34}^{\rm off/on}\,.
\end{equation}
By construction, this quantity eliminates unwanted contributions and only contains the interesting two-particle interference terms.
Now, we consider the zero-frequency, time-averaged correlators, meaning that we implement the prescription of Eq.~\eqref{eq:time-averaging-def}. One finds
\begin{equation}
    \mathcal{S}_{\rm HOM}=2q^2v_F^2RD\left[\sum_{\alpha=1}^{2}\int\frac{d\omega}{2\pi}\overline{\Delta W_\alpha^<(t,\omega)}[1-2f(\omega)]-\int\frac{d\omega}{\pi}\overline{\Delta W_1^<(t,\omega)\Delta W_2^<(t,\omega)}\right]\,,
\end{equation}
with the excess Wigner functions we described earlier in this section.
The first two terms are called the Hanbury-Brown Twiss (HBT) contribution as they represent the independent partition noise of the two sources, taken separately. In the description given at the beginning of this paragraph, the HBT term corresponds to the situation of large time delay. The electronic version of the HBT experiment was implemented in Ref.~\cite{Bocquillon2012May}. The second term encodes the two-particle interference contribution of the two sources. It clearly shows that the HOM experiment probes the overlap between the Wigner functions of the incoming states. This overlap is typically suppressed when the time delay between the injections is much larger than the typical time width of the Wigner functions themselves. It is a common procedure to express the final result in terms of the HOM ratio
\begin{equation}
    \mathcal{R}=\frac{S_{\rm HOM}}{S_{\rm HBT}}=1-\frac{\int\frac{d\omega}{\pi}\overline{\Delta W_1^<(t,\omega)\Delta W_2^<(t,\omega)}}{\sum_{\alpha=1}^{2}\int\frac{d\omega}{2\pi}\overline{\Delta W_\alpha^<(t,\omega)}[1-2f(\omega)]}\,.
\label{eq:HOM_general}
\end{equation}
Note that the minus sign is related to the fermionic statistics.

An even more transparent interpretation is possible when the sources are ideal single-electron emitters, so that the incoming states are characterized by electron wavefunctions $\varphi^{(1,2)}$. In this case, in the zero-temperature limit, one can show that
\begin{equation}
    \mathcal{R}=\left[1-\left|\Braket{\varphi^{(1)}|\varphi^{(2)}}\right|^2\right]\,,
\label{eq:HOM_SES}
\end{equation}
which demonstrates that the HOM noise is directly related to the (temporal) overlap between the two incoming wave packets $\varphi^{(1)}(t)$ and $\varphi^{(2)}(t)$. For identical wave packets emitted with some time delay $\delta t$, namely $\varphi^{(1)}(t)=\varphi(t)$ and $\varphi^{(2)}(t)=\varphi(t+\delta t)$, the HOM noise vanishes exactly at $\delta t=0$, due to maximal wave-packet overlap. This is consistent with the expected perfect antibunching for the simultaneous arrival of identical particles. The proposal to perform HOM interferometry with single electrons was put forward in Ref.~\cite{Ol'khovskaya2008Oct,Feve2008Jan}, where the HOM noise for two ideal single-electron states based on the adiabatically driven mesoscopic capacitor (cf.~Sec.~\ref{sec:injection_confined}) was calculated. A following work~\cite{Jonckheere2012Sep} extended the result to generic but ideal single-electron emitters and investigating the HOM noise due to electron-hole interferences, showing that this scenario leads to an increase of the fluctuations, rather than a decrease.

If the sources emit an $M$-electron state, a generalization is known in the case of periodic Levitons, and reads
\begin{equation}
    \mathcal{R}=\left[1-\frac{1}{M}\sum_{j,k=1}^{M}\left|\int_0^{\mathcal{T}}\frac{dt}{\mathcal{T}}\phi_j^{(1)}(t)\phi_{k}^{(2)*}(t)\right|^2\right]\,,
\end{equation}
with the functions  $\{\phi_k\}$ defined in Eq.~\eqref{eq:wavefunctions_glattli}. This relation was first conjectured in Ref.~\cite{Glattli2016Feb} and then rigorously proven in~\cite{Ronetti2018Aug}, where a generalization to the fractional quantum Hall regime was also given.

For generic, non-ideal sources, there is no simpler expression than Eq.~\eqref{eq:HOM_general} and the interpretation of the HOM noise as single-particle wavefunction overlap is not justified. However, the result~\eqref{eq:HOM_general} shows that a HOM experiment can be used to perform a tomography protocol and reconstruct the Wigner function of an unknown source, provided the properties of the other source are known. Indeed, as elaborated in~\cite{Grenier2011Sep}, it is possible to devise a protocol in which a series of measurements, using as reference signal properly chosen voltage drives, achieves the desired tomography. Experimental implementations have been demonstrated in~\cite{Jullien2014Oct,Bisognin2019Jul,Chakraborti2025May}.\\

\paragraph{Role of interactions}
Until now, we have completely disregarded the effect of Coulomb interactions during the propagation of the injected states along the conductor. However, typically the propagation channels (integer quantum Hall edge modes mainly) do interact with other neighboring channels, having an impact on transport properties.
In the context we are discussing, an indication came from the first implementation of the HOM experiment with single-electron sources based on the mesoscopic capacitor~\cite{Bocquillon2013Jan}. 
In the experiment, the source is operated in the optimal regime, so that the emission of a pure electron state is expected, thereby producing a HOM noise of the form~\eqref{eq:HOM_SES}, which predicts a full dip at zero time delay. However, the experiment showed a nonzero noise even for identical emission at $\delta t=0$. 
It was soon after shown via a Luttinger liquid description~\cite{Wahl2014Jan}, that Coulomb interactions between co-propagating edge modes can partially reduce the indistinguishability of the emitted states due to decoherence. This model is able to account for a nonzero HOM noise, and further extensive comparison between theory and experiments have been reported in~\cite{Marguerite2016Sep}.
One of the most striking features of interacting one-dimensional systems is charge fractionalization, by which an incoming electron is split into collective excitations that propagate at different velocities. This phenomenon produces side dips in the HOM noise as a function of the time delay, where the additional dips correspond to the partial overlap between excitations propagating at different velocities. This feature has been indeed observed, and Ref.~\cite{Freulon2015Apr} has exploited it to estimate the interaction parameters of the theory. In the context of EQO in integer quantum Hall edge channels, interactions have also been addressed in Refs.~\cite{Degiovanni2009Dec,Degiovanni2010Mar,Grenier2013Aug,Bocquillon2013May,Ferraro2014Oct,Hashisaka2017Jun,Cabart2018Oct,Slobodeniuk2016Jan,Goremykina2018Mar,Rebora2020Jun,Rebora2021Jan}.

The analysis of Coulomb interaction reveals that
Levitons should be more robust to decoherence, compared to the states generated by the mesoscopic capacitor. This is because of their many-body structure: indeed, a feature shared by all the states generated by a voltage pulse is that they are bosonic coherent states when expressed in terms of the particle-hole bosonic operators that are used in the bosonized description of the edge mode Hamiltonian. Then, interactions leave this structure unchanged, because their effect is equivalent to an energy-dependent beam splitter for the bosonic modes~\cite{Grenier2013Aug}. As a result, for this class of states, the HOM dip should always be maximal even in the presence of interactions~\cite{Cabart2018Oct,Rebora2020Jun}. 
In a recent experiment~\cite{Taktak2022Oct}, however, a non-vanishing noise at zero time delay has been observed and was attributed to channel mixing. Unlike Coulomb interactions, mixing induces tunneling processes connecting neighboring channels. The full HOM noise then splits in a sum of several ideal HOM noises (i.e., with no mixing), each governed by a different time delay (in addition to the bare $\delta t$), related to the typical mixing length and the velocity mismatch between the propagating channels. The competition among the different delays is such that the full noise never vanishes as a function of $\delta t$. A simple model of channel mixing with these ingredients~\cite{Acciai2022Mar,Acciai2023May} was able to explain the experimental data~\cite{Taktak2022Oct}.

Finally, we would like to briefly mention recent activity about extending HOM interferometry to anyons in the fractional quantum Hall (FQH) regime. 
The interest in time-resolved interferometry in this context stems from the success in using this tool as a probe of quantum statistics in non-interacting systems. 
As discussed above, HOM interference provides clear signatures that distinguish bosons from fermions and it is therefore natural to ask whether some unique features of exotic anyonic excitations hosted in the FQH can be discovered by noise. 
The analysis of current fluctuations in the FQH regime has a long history, but renewed interest in the field originated from the observation of anyonic statistics in Fabry-P\'erot~\cite{Nakamura2020Sep}
and HOM-like interferometers~\cite{Bartolomei2020Apr}, with the latter generating a lot of theoretical follow-up works. Most of these consider a stationary regime, with no time-dependent transport observables, which is not the focus of this review. However, a few recent works have addressed time-resolved transport of anyons~\cite{Jonckheere2023May}, proposing some observables from which the fractional statistics can be extracted~\cite{Ronetti2025Mar}, and inspiring experimental implementations~\cite{Ruelle2024Sep}. Nevertheless, HOM-like setups where the incoming states have anyonic properties are dominated by a different mechanism compared to fermionic or bosonic HOM interference. Indeed, the most relevant processes are not ``collisions'' between the incoming states, but rather time-domain braiding with anyons that are locally excited at the beam splitter, as originally introduced in~\cite{Lee2019Jul}. So, a naive generalization of fermionic HOM exclusion probability (linked to the magnitude of the dip) is not possible for these setups. Even more, the standard HOM ratio contains no information on the anyonic statistics~\cite{Varada2025May}.
The recent theoretical and experimental activity demonstrates interest in extending concepts from EQO to the FQH regime (and other strongly correlated systems). However, in these systems, the ability to identify a decomposition of generic time-dependent currents into elementary constituents with a well-defined meaning, as discussed in~\cite{Roussel2021May}, is currently lacking. Answering this challenging question would represent a major progress.

%%%%%%%%%%%%%%%%%%%%%%%%%%%%%%%%%%%%%%%%%%%%%%%%%%%%%%%
\subsection{Time-dependent quantum transport in atomic systems, Bose Einstein condensates and phononic systems}
%%%%%%%%%%%%%%%%%%%%%%%%%%%%%%%%%%%%%%%%%%%%%%%%%%%
In this section, we aim to briefly address time-dependent transport in systems that are not based on electrons.
The 
most important difference between quantum electron transport and quantum transport in cold atoms is the fact that, while electronic systems are connected
to the external environment, ultracold gases are well isolated from it due to the
confinement and extremely low temperature and density.  Small  perturbations on ultracold atomic systems 
can easily drive them out of equilibrium and a steady state is not 
always achieved. 

The confinement of the atoms is engineered artificially using optical or
magnetic means. Usually, these systems are not in contact to macroscopic reservoirs and therefore
theoretical microcanonical descriptions of transport have been proposed \cite{Chien2014Aug}. 
Experimentally,
 it has been recently possible to simulate “reservoirs” and ensure that the dynamics in the small region of interest  is in a quasi-steady state for a finite period
of time \cite{Brantut2012Aug,Krinner2015Jan,Krinner2013Mar,Chien2015Dec}. The interaction between atoms can also be tuned by
using magnetic fields. This enabled the realization of lattice models 
with many-body interactions in many geometries~\cite{Greiner2002Jan,Struck2011Aug,Soltan-Panahi2011May,Tarruell2012Mar} and motivated the study of the non-equilibrium dynamics of 
 strongly-interacting closed systems \cite{Polkovnikov2011Aug,Hofferberth2007Sep}.  
 It is also possible to induce complex tunneling
coefficients either by using artificial gauge fields or by modulating the lattice, which realizes the
Peierls substitution for lattice systems in the presence of magnetic flux \cite{Jimenez-Garcia2012May,Struck2012May}. This is important
in order to realize topological phases and ring-shaped systems.

Other non-electronic systems are nanomechanical or phononic devices. They are characterized by absence of particle conservation and transport
consists in the exchange of excitations and energy between the driven system and the baths. In systems with stationary driving by temperature differences there exists a large body of literature, which has been reviewed in Refs. \cite{Dhar2008Sep,Li2012Jul}. The experimental study of cooling
in nanomechanical systems \cite{Chan2011Oct} motivated the study of this mechanism in the framework of driven oscillators and we briefly discuss this effect below.\\

\paragraph{Non-linear transport in bosonic condensates}

An important achievement has been the formation and manipulation of a bosonic condensate using a microscopic magnetic trap on a chip \cite{Folman2000May,Hansel2001Oct}. In these platforms, waveguide geometries are implemented,
which enables the investigation of interference and transport phenomena of the condensate. This motivated to explore analogous phenomena
to those observed in mesoscopic electron systems \cite{Paul2005Jan,Ernst2010Mar}. The appropriate framework to analyze these systems is 
the time-dependent Gross-Pitaevskii equation.
Since this equation has a non-linear  term, it prevents the description of the transport process by means of scattering states as in the theory
of particles described by a linear Schr\"odinger equation. The non-linearity usually generates instabilities to the steady-state solutions and the problem must be solved numerically.

Another very active and interesting direction is the investigation of quantum fluids realized in optical systems. A review article presenting the basics is Ref. \cite{Carusotto2013Feb}, while recent advances are covered in Ref. \cite{Bloch2022Jul}.
These quantum fluids are Bose-Einstein condensates realized in lasers, non-linear optical devices, excitons and polaritons. They are characterized by the macroscopic quantum coherence of usual Bose-Einstein condensates. However they exist in non-equilibrium conditions under the effect of driving and dissipation.  The full quantum mechanical models for these systems are based on the coupling between the quantum field describing the photons in a cavity and quantum well excitons. Many problems in this context are studied by means of a mean-field approximation where instead of describing in parallel the dynamics of photons and excitons, a classical polariton field 
$\Psi({\boldsymbol r},t)$  is introduced, which obeys 
the following modified
Gross-Pitaevskii equation,
\begin{equation}
    i \hbar\partial_t \Psi = \left[\omega^0-\frac{\hbar^2}{2m} \nabla^2 \right]\Psi +
    V_{\rm ext}({\boldsymbol r}) \Psi + g|\Psi|^2 \Psi - i \gamma \Psi+
    i \eta E^{\rm inc}({\boldsymbol r},t).
\end{equation}
Here, $\omega^0$  is the frequency of the lower polariton band, $m$ is the effective mass describing the kinetic term,  $V_{\rm ext}(\boldsymbol r) $ is the external potential felt by the polaritons and $g$ is the effective polariton-polariton interaction. These
terms are the usual ones in the Gross-Pitaevskii equation for bosonic condensation of particles. The additional terms take into account the driven dissipative nature of the polariton gas and represent
the loss rate ($\gamma$) and the driving with the incident field $E^{\rm inc}(\boldsymbol r,t)$. Many problems in this context are studied under coherent 
driving, where $E^{\rm inc}({\boldsymbol r},t)=E_0 e^{i {\boldsymbol k}\cdot {\boldsymbol r}} e^{-i \omega t}$. The stability and the nature of the steady state solutions depend on the parameters. Many scenarios are possible, which we are not able to review here, but they have been discussed in detail in Ref. \cite{Carusotto2013Feb}. 

In cold atoms as well as in photonic systems there is  a large body of work on lattice models implemented by optical confinement of the atoms or 
in arrays of optical cavities. The Hubbard model is a paradigmatic example. Interestingly, synthetic gauge fields can be also realized in these systems. This enables the investigation of topological states of matter like the quantum Hall and quantum spin Hall state. Examples are \cite{DeBernardis2023Jul,Amico2022Nov,Citro2023Feb} along with several other contributions that will be mentioned below.\\

\paragraph{Rings threaded by time-dependent fluxes}\label{focus-ring}
One of the fundamental concepts  in the field of electron quantum transport in mesoscopic structures in contact with reservoirs 
is the nature of the resistance. The concept of ``contact resistance'' \cite{Buttiker1988Nov} has been coined to stress the fact that in these structures inelastic scattering processes take place in the reservoirs while the propagation is ballistic along the quantum system.
This is at the heart of scattering matrix theory of quantum transport proposed by Landauer and B\"uttiker.
An important step in this theoretical construction has been the analysis of the ballistic quantum system bent and closed to form a ring.
 In linear structures the charge transport is induced by recourse to a bias voltage applied at the reservoirs. Instead, here
 the transport mechanism is implemented through a time-dependent magnetic flux $\Phi(t)$, inducing an
 electric field $E \propto dA/dt$, with $\int_0^L A(t) dl = \Phi(t)$. The first observation is the fact that without any inelastic scattering, for a constant electric field this system undergoes periodic 
 oscillations in time named Bloch oscillations \cite{Buttiker1983Jul}. This can be easily understood in 1D cases by noticing that 
 the ring defines a periodic potential for the particles in space and the states are characterized by wave vectors $k= 2 n \pi/L$, being $L$ the length of the ring. The electric field introduces a time-dependent shift $-q E t/\hbar$ in the instantaneous $k$-values, with the subsequent generation of a time-dependent current $J(t)$. 
 The occupied states under a given value of the Fermi energy is repeated with a period ${\cal T}_B= (2 \pi \hbar)/(|q|LE) $. This is precisely the period of the Bloch oscillations and the periodicity of the current. This mechanism was already known in the context of superconductors and is the basis of the ac Josephson effect. The main differences are, on one hand, the macroscopic coherence of the condensate in the superconductors vs the requirements of mesoscopic ballistic transport of normal conductors. On the other hand, the carriers of the superconductor are Cooper pairs with effective charge $2q$, such that
 the corresponding  period is half of the one in Bloch oscillations. 
 
 In absence of any inelastic scattering process Bloch oscillations take place with a pure ac current $J(t)$. The simplest device to introduce inelastic scattering, resistive behavior  and the resulting energy dissipation  was proposed by B\"uttiker in Ref. \cite{Buttiker1985Aug}. It consists in coupling the ring to  a single lead, a fermionic reservoir, where the electrons can exit from the ring or be injected, loosing the phase coherence and exchanging energy. This generates a dc component of the current in addition to Bloch oscillations. The fact that purely elastic processes introduced by disorder are not enough to introduce such a resistive behavior was discussed in 
 Refs. \cite{Lenstra1986Sep,Landauer1986May,Landauer1987May,Gefen1987Oct,Lubin1990Mar,Blatter1988Mar,Ao1990Mar,Gefen1987Aug}. 
The behavior of the dc current in the dissipative ring was later investigated with different techniques in clean and disordered systems in Refs. \cite{Liu2000Mar,Arrachea2002Jul,Yudson2003Apr,Silva2007Oct,Kravtsov2000Apr,Matos-Abiague2005Apr,Arrachea2004Oct,Foieri2007Dec}. 

Generalizations to rings with harmonic time-dependent fluxes coupled to two or more leads have been studied in Refs. \cite{Shin2004Aug,Ludovico2013Mar}. In those cases, the focus was the generation of currents between the leads generated by the time-dependent fluxes at the rings. This problem is basically a pumping  mechanism. Similar studies  focus on pumping generated by locally driving a quantum dot embedded in an annular system connected to two leads \cite{FoaTorres2005Dec,Arrachea2005Sep,Romeo2008Dec}.

The study of persistent currents 
in cold atoms systems confined to ring-shaped potentials and pierced by a synthetic static magnetic field was motivated in Ref.~\cite{Amico2005Aug}. These and  related advances have been reviewed in Ref.~\cite{Amico2022Nov}. Recently, results on quantum transport  induced by time-dependent fluxes in cold atoms were reported in Ref.  \cite{Lau2023May}.
The considered Hamiltonian is $\hat{H}=\hat{H}_{\rm ring}(t)+\sum_\alpha \left(\hat{H}_{\alpha} + \hat{H}_{\mathrm{coup},\alpha}\right) $, with
\begin{eqnarray}
    \hat{H}_{\rm ring}(t)&=& \sum_{j=1}^M\left[ \frac{U}{2} \hat{N}_j \left(\hat{N}_j-1 \right) - w \left(e^{i\phi(t)}\hat{d}^\dagger_{j+1} \hat{d}_{j} + {\rm h.c.}\right)\right], \nonumber \\
    \hat{H}_{\alpha}&=& \sum_{j=1}^{M_\alpha}\left[ \frac{U_{\alpha}}{2} \hat{N}^{\alpha}_j \left(\hat{N}^{\alpha}_j-1 \right) - w_\alpha \left(\hat{a}_{\alpha,j+1}^\dagger \hat{a}_{\alpha,j} + {\rm h.c.}\right)\right], \nonumber \\
    \hat{H}_{\rm coup, \alpha}&=& - w^\mathrm{coup}_{\alpha} \left(\hat{d}^\dagger_{m_\alpha} \hat{a}_{\alpha,1} + {\rm h.c.}\right),
\end{eqnarray}
where $\hat{d}_{j}, \hat{d}^{\dagger}_{j}$ are, respectively, bosonic annihilation and creation operators acting at the site $j$ of the ring, satisfying
$\hat{d}_{M+1}=\hat{d}_{1}$, while
$\hat{N}_j=\hat{d}^{\dagger}_{j}\hat{d}_{j}$ is the number operator. $U$ is the local interaction strength and $w$ is the hopping amplitude between nearest-neighbour sites. The leads are represented by linear chains of the same type of Hamiltonian $\hat{H}_\alpha$, while
 $\hat{H}_{\mathrm{coup}, \alpha}$ is the Hamiltonian describing the coupling between the first site of the lead chain and one site of the ring, here named $j \equiv m_\alpha$.
 
This problem bears relation  to the mechanism of electron pumping discussed in previous sections. In fact, it focuses on the currents induced at the leads because they are coupled to a driven ring. 

Also in close relation to pumping a cooling cycle was proposed to be realized in a 
Bose-Einstein condensate, including details on how to realize the different stages in one-dimensional cold atomic systems \cite{Gluza2021Jul}.
The protocol is very similar to the one proposed in Ref. \cite{Chamon2011Apr} for a nanomechanical system (see below).\\  

 \paragraph{Quantum oscillators and phononic systems}
 Cooling nanomechanical modes down to the ground state has attracted the attention of both experimental \cite{Chan2011Oct} and theoretical \cite{Marquardt2007Aug,Wilson-Rae2007Aug} communities. The basics of this mechanisms can be analyzed in the context of 
 driven coupled quantum-mechanical oscillators. 

%%%%%%%%%%%%%%%%%%%%%%%%%%%%%%%%%%%%%%%%%%%%%%%%%%%
\begin{figure}[tbp]
\centering
\includegraphics[width=0.8\textwidth]{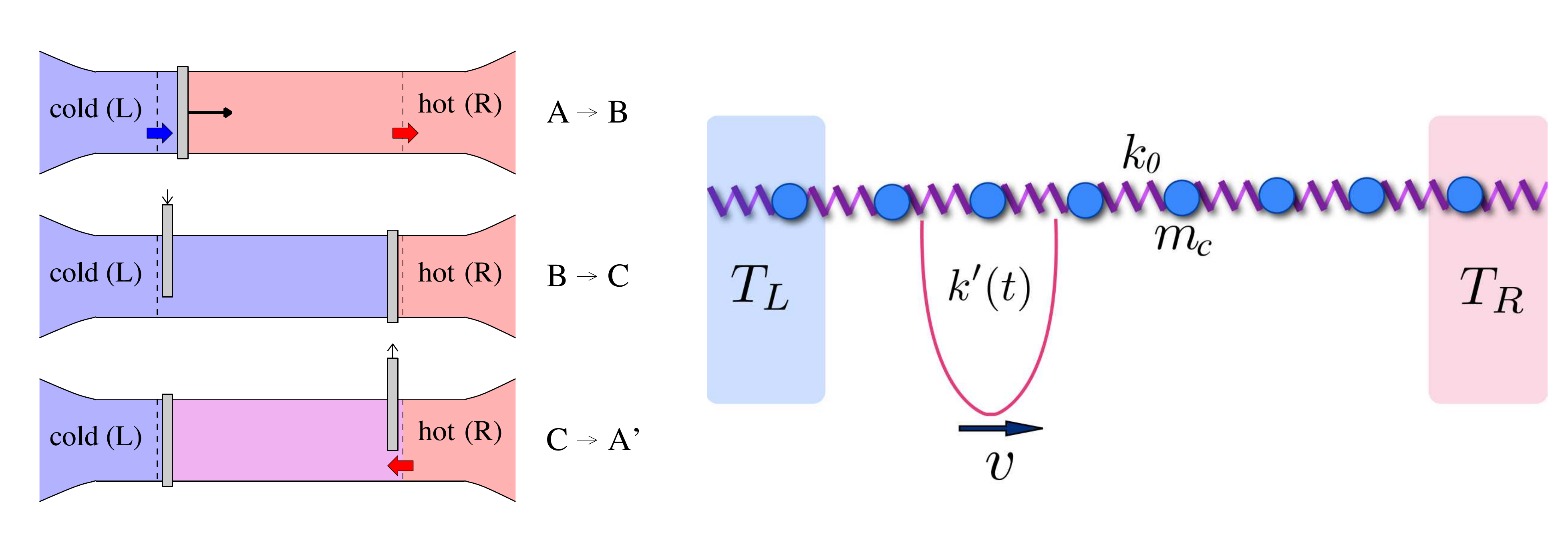}
\caption{Sketch of the cooling cycle proposed in~\cite{Chamon2011Apr}  (left) and microscopic representation in a one-dimensional phononic system (right). A nanomechanical system contains   hot and cold parts. A perturbation is introduced at the boundary of the cold part (A) and it travels slowly until it reaches the hot one (B), a second perturbation is inserted at the hot boundary (C), the first barrier is removed and the cycle is repeated. Figure adapted from Refs. \cite{Chamon2011Apr} and \cite{arrachea2023energy}.}
\label{fig:refri}
\end{figure}
%%%%%%%%%%%%%%%%%%%%%%%%%%%%%%%%%%%%%%%%%%%%%%%%%%%
Figure~\ref{fig:refri} presents the sketch of a cooling cycle proposed for a nanomechanical system in Ref. \cite{Chamon2011Apr}. 
The nanomechanical system targeted for cooling is depicted on the left (cold) side. In this setup, a localized modulation of the phonon velocity or of the pinning potential acts similarly to a semi-reflective, moving barrier for phonons. During stage $A \rightarrow B$, this barrier travels from the cold side to the hot side across a cavity-like area. 
Upon reaching the far end, a second barrier-like perturbation is initiated at the opposite side (stage $B \rightarrow C$). In the step, $C \rightarrow A'$, the initial barrier is turned off, allowing phonons from the hot reservoir to freely expand into the cavity. This sequence is then cyclically repeated. 

A microscopic model for this protocol in the framework of coupled harmonic oscillators has been proposed in Ref. \cite{Arrachea2012Sep}. The Hamiltonian for the full system has the structure of the Hamiltonian Eq. (\ref{hamil0}), with the hot and cold reservoirs modeled by 1D coupled harmonic oscillators. The driven region is described by the same 1D model of harmonic oscillators with a time-dependent perturbation
representing the traveling periodic barrier (see Fig. \ref{fig:refri}). This term of the  Hamiltonian reads
\begin{equation}
    \hat H_{\rm sys}(t)=\sum_{l=1}^{M} \frac{\hat p_l^2}{2m}+\sum_{l=1}^{M-1} \frac{k_0}{2}(\hat x_l-\hat x_{l+1})^2+ 
    \sum_{l=1}^{M}\frac{k'_l(t)}{2}\hat x_l^2.
\end{equation}
This problem was solved by means of non-equilibrium Green's functions. Further discussions on the cooling properties and limits of time-dependent driving in
systems of harmonic oscillators were presented in Ref. \cite{Freitas2018Mar}. More recently, these type of models was regarded as a paradigm to 
engineering dynamical couplings for quantum thermodynamics tasks beyond weak coupling \cite{Carrega2022Feb}.  A related idea has been
explored recently in Ref. \cite{Portugal2022Nov}, where a target oscillator is coupled to a collection of driven oscillators that simulates  an environment with a time-dependent temperature. The problem is analyzed at weak coupling with quantum master equations and driving protocols that cool the target oscillator are identified.

%%%%%%%%%%%%%%%%%%%%%%%%%%%%%%%%%%%%%%%%%%%%%%%%%%%%%
\subsection{Topological effects}\label{focus-topo-pump}
%%%%%%%%%%%%%%%%%%%%%%%%%%%%%%%%%%%%%%%%%%%%%%%%%%%%%
For some years now, the relevance of topology emerged in
many properties of quantum systems. We already highlighted the crucial properties of  the quantum Hall effect - a paradigmatic topological system - to realize quantum optics and other phenomena of quantum transport. In this final section, we would like to briefly survey other activity connecting time-dependent quantum transport and topological effects.

\subsubsection{Topological pumping}
As introduced in Sec.~\ref{intro-pump}, the concept of pumping involves the transport of some entity, such as charge, spin, quasiparticles and/or energy as a consequence of  time-periodic changes of the underlying Hamiltonian. This mechanism takes place in both the classical and quantum realms. In the case of slow driving, this mechanism is typically described by geometric quantities. In certain quantum scenarios, it can also possess {\em topological} properties, meaning that the pumping is
characterized in terms of a topological invariant. This invariant  is directly related to a quantization of the pumped quantity.

The most paradigmatic examples of topological pumping are the mechanisms proposed by Laughlin \cite{Laughlin1981May} and Thouless \cite{Thouless1983May}. In both cases, a single electron is pumped through a fully gapped---insulating---system by a slow time-periodic change in the Hamiltonian. This is an important feature of topological pumping, which makes it fundamentally different from other quantum transport mechanism: here the  particle is not transported through a window defined by two Fermi functions.
Instead, a filled band experiences a peculiar slow change according to which
a particle is spatially moved between two positions.\\ 

\paragraph{Laughlin pump}
In Laughlin's ``gedanken experiment'' two boundaries of a two-dimensional electron gas (2DEG) in the quantum Hall state are connected to form a cylinder \cite{Laughlin1981May}. The interior of the cylinder is threaded by a  magnetic flux that changes linearly in time. The corresponding electromotive force
induces a current and every time the flux is increased by a flux quantum, a single electron is effectively transported between the free edges of the
2DEG through the gapped bulk. The relation between this current and the effective electric field is the Hall conductance, which is proportional to the Chern number. Hence the pumped charge per cycle is related to the Chern number. 
The Laughlin pump has been observed in quantum Hall systems in the Corbino geometry \cite{Dolgopolov1992Nov,Jeanneret1995Apr}, more recently in an anomalous quantum Hall insulator \cite{Kawamura2023Mar} and it has been also realized with cold atoms \cite{Fabre2022Apr}.\\

\paragraph{Thouless pump}\label{sec:Thouless}
The Thouless pump is a sort of  quantum version of Archimedes screw. In the original proposal it is formulated in terms of a Hamiltonian of non-interacting electrons with a slowly varying potential 
which is periodic in space and time \cite{Thouless1983May}. Focusing, for simplicity on 1D, the Hamiltonian has a potential of the form 
$V(x,t)=V_0(x)+V_1(x-vt)$, where $V_0(x+L)=V_0(x)$ and 
$V_1\left(x-v(t+{\cal T})\right)=V_1(x-vt)$, with
$v$ small enough so that the system remains in the ground state. It is also requested that both components of the potential have the same 
spatial periodicity, hence, $L/v = n\mathcal{T}$, with  $n$ being an integer. The particle current integrated over a period gives the number of particles transported in a cycle. Implementing an adiabatic expansion in the wave function we have
\begin{equation}
    |\psi(t)\rangle \simeq \exp \left[ -\frac{i}{\hbar} \int^t dt' \varepsilon_0(t')  \right]\left[|\psi_0(t)\rangle +i \hbar
    \sum_{j\neq 0} |\psi_j(t)\rangle \frac{1}{\varepsilon_j -\varepsilon_0} \langle \psi_j(t)|\dot{\psi}_0(t)\rangle\right],
\end{equation}
where $\hat{H}(t)|\psi_j(t)\rangle =\varepsilon_j |\psi_j(t)\rangle$ defines  the instantaneous eigenenergies and eigenstates for the Hamiltonian
with the time frozen at $t$ and the phase is chosen so that 
$\langle \psi_0(t)|\dot{\psi}_0(t)\rangle =0$. In terms of this, introducing periodic boundary conditions, and calculating 
the instantaneous current within the adiabatic description we get
\begin{equation}
    J(t)=\frac{\hbar^2}{m}\sum_{\mu,\nu} \int \frac{dk}{2\pi}
    \frac{f_\nu (1-f_\mu)}{\varepsilon_\mu(k)-\varepsilon_\nu(k)}
    \left[\langle \partial_x \psi_{\nu,k}|\psi_{\mu,k} \rangle
    \langle \psi_{\mu,k}| \dot{\psi}_{\nu,k} \rangle+
    \langle  \dot{\psi}_{\nu,k}|\psi_{\mu,k} \rangle
    \langle \psi_{\mu,k}| \partial_x{\psi}_{\nu,k} \rangle
    \right],
\end{equation}
where $\mu,\nu$ label the bands of the periodic lattice while
$f_\mu=0, 1$ for empty and filled bands, respectively. Using properties of the periodic boundary conditions, it is possible to express the
 particle transferred per period as follows
 \begin{equation}
 \mathcal{C}=\int_0^{\mathcal{T}} dt \; J(t)=i\sum_\nu f_\nu \int_0^{\mathcal{T}} dt \int \frac{dk}{2\pi}
  \left[\langle \partial_t \psi_{\nu,k}|\partial_k\psi_{\nu,k} \rangle
   -\langle \partial_k \psi_{\nu,k}|\partial_t\psi_{\nu,k} \rangle  
    \right], \label{chern}
 \end{equation}
 which is the 1D version of the definition of the Chern number while the pumped charge per cycle is $\Delta N=q \mathcal{C}$.

There are many studies of models realizing Thouless pumps, which have been reviewed in Refs. \cite{Citro2023Feb} and \cite{Ozawa2019Mar}, including  experimental realizations in cold atoms and optical systems.
Another interesting direction is the study of the role of many-body interactions and a prominent example in this context is the Rice-Mele-Hubbard model in both the fermionic \cite{Viebahn2024Jun} and bosonic versions 
\cite{Lu2016May,Hayward2018Dec}.
Some of these studies focus on the  proper calculation of the invariant $\mathcal{C}$ in terms of many-body functions \cite{Jurgensen2023Mar}.
This can be accomplished in the framework of the theory of polarization \cite{Resta1994Jul,Resta1998Mar,Aligia2023Feb}. In 1D, the polarization in a lattice model with lattice constant $a$ with periodic boundary conditions is defined as 
 \begin{equation}
     {\cal P}(t)=\frac{q a}{2\pi} \mbox{Im} \ln{\langle \Psi(t)|e^{i \frac{2\pi}{Ma}\hat{X}}|\Psi(t)\rangle},
 \end{equation}
 where $\hat{X}=\sum_j^M x_j \hat{N}_j$ is the position operator, being $\hat{N}_j$ the occupation of the site $j$ and $x_j=ja$, for a system
 with a single site per unit cell, while $|\Psi(t)\rangle$ is the many-body wave function. The transported charge per adiabatic cycle is expressed as
 \begin{equation}
   \Delta N=\frac{1}{a}\int_0^{\mathcal{T}} dt \partial_t {\cal P}(t).
 \end{equation}
 It can be verified that this definition recovers the one for non-interacting fermions given by Eq. (\ref{chern}) by  expressing the many-body wave 
 function $|\Psi(t)\rangle$ as a direct product of single-particle states $|\psi_{\nu,k} \rangle $.
 The combined effect of driving and disorder in topological pumping was analyzed in \cite{Wauters2019Dec}.\\

\paragraph{Pumping in the quantum spin Hall effect}
Another topological charge pump has been proposed to take place in a helical Kramers pair of edge states of a 2D topological insulator in the quantum spin Hall state contacted by a magnetic island \cite{Qi2008Apr,Qi2009Jun}. This type of topological pumping motivated several studies \cite{Meng2014Nov,Arrachea2015Nov,Xiao2021Aug,Vinkler-Aviv2017Nov,Tang2022Aug,Tang2024Mar,Silvestrov2016May,Bozkurt2018Jun,Madsen2021Mar}. 
Here, we explain the basic ideas following Ref. \cite{Qi2008Apr}, where 
the topological pumping can be understood in simple terms as follows. 
The free edge states are modeled  by a 
Dirac Hamiltonian 
\begin{equation}
    H_0= i{\rm v}\int dx \Psi^{\dagger}(x) \partial_x \sigma^z \Psi(x),
\end{equation}
where the spinor $\Psi(x)=\left(\psi_{\uparrow}(x), \psi_{\downarrow}(x)\right)^T$ represents the 
pair of edge states with opposite polarization along a given direction $z$ and moving in opposite directions along the edge. The coupled nanomagnet has a magnetization with a perpendicular component $\boldsymbol{M}$, with $\boldsymbol{M}\times \boldsymbol{z}\neq 0$. Such a coupling effectively introduces a mass term and opens a gap in the Dirac spectrum which, in the static problem, prevents the charge transport along the Kramers pairs. The topological pump is introduced by the precession of the magnet,
which is represented by a time-periodic mass in the Hamiltonian
\begin{equation}
    \hat{H}_M= \int dx \Psi^{\dagger}(x) \left[\boldsymbol{M}(x,t) \cdot \boldsymbol{\sigma} \right] \Psi(x).
\end{equation}
This problem has been previously studied in Refs. \cite{Jackiw1976Jun,Goldstone1981Oct}. Introducing the parametrization of the transverse magnetic moment as $M_x=M\cos\theta$ and $M_y=M\sin\theta$, the ground-state density and current induced by the time-dependent mass read
\begin{equation}
    \rho(x,t)=\frac{1}{2\pi}\partial_x \theta(x,t), \;\;\;\;\;\;\;\;j(x,t)=-\frac{1}{2\pi}\partial_t \theta(x,t).
\end{equation}
Hence, the total number of carriers enclosed in a segment confined in $x_1 \leq x \leq x_2$ is given by $\Delta N=[\theta(x_2,t)-\theta(x_1,t)]/(2\pi)$. In particular, a half-charge $\pm q/2$ is carried by a domain wall where $\theta$ changes between $\pi/2$ and $-\pi/2$. Similarly,
the pumped charge crossing a point $x$ as the angle changes in time in an interval $t_1 \leq t \leq t_2$ is
$\Delta N^{\rm pump}(t_1\rightarrow t_2)= [\theta(x,t_2)-\theta(x,t_1)]/(2\pi)$. Hence, under a change of $0\leq \theta \leq 2\pi$ in a cycle,
a charge of $\pm q$ is pumped. 
As in previous cases, the quantized transport takes place through an insulating system that exhibits a gap in the static limit.

\subsubsection{Floquet engineering of topological phases}\label{focus-floquet-eng}
The previous sections focused on time-dependent quantum transport phenomena that arise due to the intrinsic topological nature of certain systems. In this section, we briefly address a complementary and actively developing research direction: the engineering of driving-induced topological phases in systems that are topologically trivial in equilibrium. This area, commonly referred to as Floquet engineering, explores how periodic driving can endow otherwise conventional systems with topologically nontrivial properties.

A wide range of experimental platforms are being investigated in this context, including not only solid-state electronic systems, but also optical lattices and ultracold atomic gases \cite{Goldman2014Aug,Eckardt2017Mar,Oka2019Mar}.
 Reviews covering recent advances are Refs. \cite{Rudner2020May,delaTorre2021Oct}.
 In condensed matter settings, results have been reported in
superconductors \cite{Kundu2013Sep,PeraltaGavensky2018Jun}
as well as in Moir\'e systems \cite{Topp2019Sep,Rodriguez-Vega2021Dec,Calvo2025Jan}.  As in other topological systems, features emerging in the time-dependent transport properties, in particular 
the generated photocurrents,
play a central role  in characterizing and probing  the topological nature of these phases \cite{Cayssol2013Feb,Ma2023Mar}. 

A paradigmatic example is graphene irradiated by a circularly polarized laser of frequency $\Omega$ \cite{Oka2009Feb,Kitagawa2010Dec,Kitagawa2011Dec,Lindner2011Jun,Usaj2014Sep,Sato2019Jun,McIver2020Jan,Nuske2020Dec}. This is described by the following time-dependent tight-binding Hamiltonian
\begin{equation}
    \hat{H}(t)=-w \sum_{\langle l j \rangle }\exp\left[-i\int_{R_l}^{R_j} {\boldsymbol A}_{\rm ac}(t) \cdot d{\boldsymbol r}\right] \hat{c}^{\dagger}_l \hat{c}_j,
\end{equation}
where $\langle i j \rangle$ are nearest-neighbor sites of the graphene lattice, while $\hat{c}^{\dagger}_l$  creates  an electron with
any spin orientation in the site $l$ and $\hat{c}_l$ is the corresponding annihilation operator acting on a lattice site $l$. The effect of the laser field $\boldsymbol{E}_{\rm ac}(t)= \partial_t \boldsymbol{A}_{\rm ac}(t)$ is accounted for by the Peierl's substitution in the time-dependent hopping,
with ${\boldsymbol A}_{\rm ac}(t)=A\left(\cos\Omega t, \sin \Omega t \right)$. 
As a result of the time-dependent term in the Hamiltonian, an effective next-nearest-neighbor dynamical hopping is generated and the model effectively becomes a topological Chern insulator, similar to that predicted by Haldane \cite{Haldane1988Oct}.
The invariant characterizing this phase is the Chern number, which is directly related to the Hall conductivity. In the present problem, such a transport coefficient is defined as the response to an extra transverse dc electric field ${\boldsymbol E}$ represented by an extra vector potential
${\boldsymbol A}_{\rm ex}(t)={\boldsymbol E} t$. In Refs. \cite{Oka2009Feb,Kitagawa2011Dec,Dehghani2015Apr,FoaTorres2014Dec} a Kubo formula has been derived by treating ${\boldsymbol A}_{\rm ex}(t)$ in linear response
for the system in the background of the intense ac field. An overview of later developments in this direction, including the generation of topological quantum pumping by Floquet engineering has been presented 
in Ref. \cite{Oka2019Mar}. Signatures of Floquet states have also been recently observed in graphene \cite{Merboldt2025Jul,Choi2025Jul}.

In addition to models related to real materials, there are a wide variety of proposals in the framework of pure theoretical models. An interesting example is presented in Ref.~\cite{Kolodrubetz2018Apr}, where  a topological phase is generated by Floquet engineering and it is shown to 
enable a Thouless-pump effect as a response to an extra adiabatic driving. 
Another related interesting idea is the generation of a two-dimensional topological phase by driving a single two-level system with two different frequencies~\cite{Martin2017Oct,Nathan2019Mar}. In this case, the topological phase is associated with the mechanism of power pumping and 
the possibility of controlling dissipation by topological driving protocols~\cite{Esin2025Apr}.

%%%%%%%%%%%%%%%%%%%%%%%%%%%%%%%%%%%%%%%%%%%%%%%%%%%%%
\section{Conclusions}\label{sec_conclusion}
%%%%%%%%%%%%%%%%%%%%%%%%%%%%%%%%%%%%%%%%%%%%%%%%%%%%%
In this review, we have provided a comprehensive overview of various aspects related to time-dependent transport in quantum systems. Our primary emphasis has been on time-dependent steady-state regimes, as opposed to transient dynamics, which, although important, fall outside the main scope of this review. We have surveyed the range of theoretical frameworks available to study non-equilibrium quantum systems, deliberately excluding numerics-focused approaches in order to offer analytical insight and physical intuition.

Throughout the discussion, we have identified and analyzed several fundamental mechanisms that are recurrent in quantum-driven systems and are essential to the understanding of quantum transport phenomena. These include the periodic charging and discharging of a mesoscopic capacitor,  dissipation, quantum pumping, noise, and energy conversion, all of which play key roles in the characterization of transport properties and the design of quantum devices.

A significant portion of our review has been devoted to electronic systems, which continue to be a central area of investigation due to their rich phenomenology and relevance to both fundamental physics and practical applications. In particular, we have highlighted recent advances in electron quantum optics, quantum transport spectroscopy, and quantum electrical metrology. Additionally, we have addressed the role of quantum fluctuations, a crucial ingredient in understanding transport and thermodynamic behavior at the quantum scale.

Beyond electronic systems, we have also considered developments in atomic, molecular, and optical systems, as well as in nanomechanical platforms, which are witnessing rapid progress due to advances in experimental control and measurement techniques. These systems also provide versatile testbeds for exploring quantum transport in regimes.

 Furthermore, we have briefly reviewed emerging research that connects time-dependent quantum transport with the topological properties of matter. The interplay between topology and dynamical driving has opened up novel avenues for robust transport control, quantized responses, and the engineering of synthetic dimensions that are at the forefront of contemporary condensed matter and quantum information science.

Taken together, the topics covered in this overview reflect a vibrant and evolving research landscape, where theoretical insights and experimental innovations are continually reshaping our understanding of quantum transport under time-dependent conditions.

%%%%%%%%%%%%%%%%%%%%%%%%%%%%%%%%%%%%%%%%%%%%%%%%%%%%%
\appendix
%%%%%%%%%%%%%%%%%%%%%%%%%%%%%%%%%%%%%%%%%%%%%%%%%%%%%
\section{Technical details}
\subsection{Equivalence between Eqs. (\ref{olr-1}) and (\ref{kuboa}). }
\label{app:kubo_equivalence}
We show here that the response functions obtained in the two different derivations of the linear-response mean value
of the operator $\hat{O}$ at the time $t$ are the same. To this end, we introduce the Lehmann representation in the response functions
of Eqs.~(\ref{olr-1}) and~(\ref{kuboa}). 

In Eq.~(\ref{olr-1}) we get
\begin{align}
    {\left\langle \left[ \hat{{O}}_{H_0}(t), \hat{{F}}_{H_0}^j(t')\right]\right\rangle}_0 &= \frac{1}{Z_0}\sum_{n,m} e^{-\beta E_n } \left\{\langle n| \hat{{O}}_{H_0}(t) |m \rangle
    \langle m| \hat{{F}}^j_{H_0}(t') |n\rangle\right.\nonumber\\
    &\quad\qquad\qquad\qquad\quad\left. -\langle n| \hat{{F}}^j_{H_0}(t') |m \rangle
    \langle m| \hat{{O}}_{H_0}(t)| n\rangle    
    \right\}
    \nonumber \\
    &=\frac{1}{Z_0}\sum_{n,m} e^{-\beta E_n } \left\{ e^{\frac{i}{\hbar}(E_n-E_m)(t-t')} \langle n| \hat{O} |m \rangle
    \langle m| \hat{F}^j |n\rangle\right.\nonumber\\
    &\quad\qquad\qquad\qquad\quad\left. -e^{\frac{i}{\hbar}(E_n-E_m)(t'-t)}\langle n| \hat{F}^j |m \rangle
    \langle m| \hat{O} |n\rangle    
    \right\}\nonumber \\
    &= \frac{1}{Z_0}\sum_{n,m} e^{\frac{i}{\hbar}(E_n-E_m)(t-t')} \langle n| \hat{O} |m \rangle
    \langle m| \hat{F}^j |n\rangle \left(e^{-\beta E_n } - e^{-\beta E_m } \right)
\end{align}

In Eq.~(\ref{kuboa}) we get
\begin{eqnarray}
 & & \int_0^\beta du {\langle \hat{O}_{H_0}(-iu{\hbar}) \dot{\hat{F}}^j_{H_0}(t'-t) \rangle}_0 =-{\frac{i}{\hbar}}\int_0^\beta du {\left\langle  \hat{O}_{H_0}(-iu\hbar)\left[\hat{F}^j_{H_0}(t'-t),\hat{H}_0 \right] \right\rangle}_0
 \nonumber \\
& &  
 = -\frac{i}{{\hbar}Z_0}\int_0^\beta du \sum_{n,m} e^{u(E_n-E_m)} \langle n |\hat{O}|m\rangle \Braket{ m |\left[\hat{F}^j,\hat{H}_0\right]|n} e^{\frac{i}{\hbar}(E_m-E_n)(t'-t)}\nonumber \\
 & & =-\frac{i}{{\hbar}Z_0} \sum_{n,m} e^{-\beta E_n} \frac{\left[e^{\beta(E_n-E_m)}-1\right]}{\left(E_n-E_m\right)} \langle n |\hat{O}|m\rangle \langle m |\hat{F}^j |n\rangle \left(E_n-E_m\right) e^{i(E_m-E_n)(t'-t)}\nonumber \\
 & & = -\frac{i}{{\hbar}Z_0} \sum_{n,m}  \left(e^{-\beta E_m}-e^{-\beta E_n}\right)\langle n |\hat{O}|m\rangle \langle m |\hat{F}^j |n\rangle e^{\frac{i}{\hbar}(E_m-E_n)(t'-t)},
\end{eqnarray}
which proves that the response functions in the two formulations coincide. 

\subsection{Equivalence between Eqs. (\ref{adia-ludo}) and (\ref{adia-cam}). }
We introduce in these two equations the Lehmann representation.
The second term of Eq.~(\ref{adia-ludo}) can be written as
\begin{eqnarray}
\label{adia-ludo-leh}
    & &\frac{i}{{\hbar}} \int_{t_0}^t dt' (t'-t) {\left\langle \left[ {\hat{O}}_{H_0}(t), {\hat{F}}_{H_0}^j(t')\right]\right\rangle}_t \nonumber\\
    & &= \frac{i}{{\hbar}} \int_{t_0}^t dt' (t'-t) \frac{1}{Z_t}\sum_{n,m} e^{\frac{i}{\hbar}(E_n-E_m)(t-t')} \langle n| \hat{O} |m \rangle
    \langle m| \hat{F}^j |n\rangle \left(e^{-\beta E_n } - e^{-\beta E_m } \right)\nonumber \\
 & &  = \int_{t_0}^t dt'  \frac{1}{Z_t}\sum_{n,m} e^{\frac{i}{\hbar}(E_n-E_m)(t-t')} \langle n| \hat{O} |m \rangle
    \langle m| \hat{F}^j |n\rangle \frac{\left(e^{-\beta E_n } - e^{-\beta E_m } \right) }{E_n-E_m},
\end{eqnarray}
where we performed a partial integration and assumed that the exponential vanishes for $t_0$. 
Similarly, Eq. (\ref{adia-cam}) can be written as follows
\begin{eqnarray}
 & &- \int_{t_0}^t dt' \int_0^\beta du \langle  \hat{O}(-iu{\hbar}) \hat{F}^j(t'-t) \rangle_t\nonumber \\
& &  = -\frac{1}{Z_t} \sum_{n,m} \int_{t_0}^t dt' e^{-\beta E_n} \frac{\left[e^{\beta(E_n-E_m)}-1\right]}{\left(E_n-E_m\right)} \langle n |\hat{O}|m\rangle \langle m |\hat{F}^j |n\rangle e^{\frac{i}{\hbar}(E_m-E_n)(t'-t)},
\end{eqnarray}
which is the same as Eq.~(\ref{adia-ludo-leh}).

\subsection{Wigner function of $M$-Leviton states}
\label{app:wigner}
Combining the definition~\eqref{eq:Wigner_definition_energy}, the property~\eqref{eq:electron_coherence_energy_wavefunction}, and the explicit expression~\eqref{eq:wavefunctions_levitons_energy} we have
\begin{align*}
    \Delta W^<(t,\omega)&=\sum_{k=1}^{M}\int_{-\infty}^{+\infty}\frac{d\xi}{2\pi}\tilde{\varphi}_k^*\left(\omega-\frac{\xi}{2}\right)\tilde{\varphi}_k\left(\omega+\frac{\xi}{2}\right)e^{-it\xi}\\
    &=2\sigma e^{-2\omega \sigma}\theta(\omega)\sum_{k=0}^{M-1}\underbrace{\int_{-\omega}^{\omega}du\,L_{k}[2\sigma(\omega-u)]L_{k}[2\sigma(\omega+u)]e^{-2iut}}_{\mathcal{I}_k}\,.
\end{align*}
Next, one exploits the identity
\begin{equation*}
    L_k(x)L_k(y)=\sum_{p=0}^{k}\frac{(xy)^p}{(p!)^2}L_{k-p}^{(2p)}(x+y)\,,
\end{equation*}
where $L_k^{(p)}$ are the generalized Laguerre polynomials, to simplify $\mathcal{I}_k$ to
\begin{align*}
    \mathcal{I}_k&=\sum_{p=0}^{k}\frac{(2\sigma)^{2p}}{(p!)^2}L_{k-p}^{(2p)}(4\omega \sigma)\int_{-\omega}^{\omega}du{(\omega^2-u^2)}^pe^{-2iut}\\
    &=2\omega\sum_{p=0}^{k}\frac{(2\omega \sigma)^{2p}}{(p!)^2}L_{k-p}^{(2p)}(4\omega \sigma)\int_{0}^{1}dy{(1-y^2)}^p\cos(2\omega yt)\\
    &=\frac{\omega\sqrt{\pi}}{p!}\sum_{p=0}^{k}\frac{(2\omega \sigma)^{2p}}{(\omega t)^{p+1/2}}L_{k-p}^{(2p)}(4\omega \sigma)J_{p+1/2}(2\omega \sigma)\,,
\end{align*}
where in the last step we recognized an integral representation of the Bessel functions of the first kind $J_k$. Combining the above results, we obtain the final expression~\eqref{eq:wigner_levitons} reported in the main text.

%%%%%%%%%%%%%%%%%%%%%%%%%%%%%%%%%%%%%%%%%%%%%%%%%%%%%
\section*{Acknowledgements}

L.A.~and J.S.~would like to thank the Institut Henri Poincar\'e (UAR 839 CNRS-Sorbonne Universit\'e) and the LabEx CARMIN (ANR-10-LABX-59-01) for their support, and acknowledge interesting discussions with Alessandro Silva and Iacopo Carusotto.
Furthermore, we acknowledge funding from the European Union’s Horizon 2020 research and innovation programme under grant agreement No.~862683 (FET-OPEN UltraFastNano) (J.S. and M.A.), from the PNRR MUR Project No.~PE0000023-NQSTI (M.A.), from the Knut and Alice Wallenberg foundation through the fellowship program (J.S.), from CONICET and FONCyT through 
PICT 2020-A-03661 from Argentina (L.A.). 

%%%%%%%%%%%%%%%%%%%%%%%%%%%%%%%%%%%%%%%%%%%%%%%%%%%%%
\bibliography{cite.bib}

\end{document}